\documentclass[12pt]{article}
\usepackage{mathrsfs}
\makeatletter \@addtoreset{equation}{section} \makeatother
\renewcommand{\theequation}{\thesection.\arabic{equation}}
\newcommand{\ba}{\begin{array}}
\newcommand{\ea}{\end{array}}
\newcommand{\beq}{\begin{equation}}
\newcommand{\eeq}{\end{equation}}
\newcommand{\bea}{\begin{eqnarray}}
\newcommand{\eea}{\end{eqnarray}}

\def\bce{\begin{center}}
\def\ece{\end{center}}

\def\nonu{\nonumber}

\def\pa{\partial}
\def\al{\alpha}
\def\be{\beta}

\def\de{\delta}

\def\la{\lambda}

\def\si{\sigma}

\def\eps6{{\displaystyle \mathop{\epsilon}^{6}}{}}
\def\g6{{\displaystyle \mathop{g}^{6}}{}}

\def\nab6{{\displaystyle \mathop{\nabla}^{6}}{}}

\def\0{{\sst{(0)}}}
\def\1{{\sst{(1)}}}
\def\2{{\sst{(2)}}}
\def\3{{\sst{(3)}}}
\def\4{{\sst{(4)}}}
\def\5{{\sst{(5)}}}
\def\6{{\sst{(6)}}}
\def\7{{\sst{(7)}}}
\def\8{{\sst{(8)}}}

\def\ba{\begin{array}}
\def\ea{\end{array}}
\def\beq{\begin{equation}}
\def\eeq{\end{equation}}
\def\be{\begin{equation}}
\def\ee{\end{equation}}

\def\la{\lambda}
\def\eps{\epsilon}

\def\ba{\begin{array}}
\def\ea{\end{array}}
\def\beq{\begin{equation}}
\def\eeq{\end{equation}}
\def\be{\begin{equation}}
\def\ee{\end{equation}}

\def\la{\lambda}
\def\eps{\epsilon}

\def\eps6{{\displaystyle \mathop{\epsilon}^{6}}{}}

\def\nab6{{\displaystyle \mathop{\nabla}^{6}}{}}

\newcommand{\bean}{\begin{eqnarray*}}
\newcommand{\eean}{\end{eqnarray*}}

\begin{document}
\thispagestyle{empty} \addtocounter{page}{-1}
   \begin{flushright}
\end{flushright}

\vspace*{1.3cm}
  
\centerline{ \Large \bf
The ${\cal N}=4$  Coset Model
and   the Higher Spin Algebra
}
\vspace*{1.5cm}
\centerline{{\bf  Changhyun Ahn$\dagger$},
  {\bf Dong-gyu Kim$\dagger$} and {\bf Man Hea Kim$\dagger \star$}
} 
\vspace*{1.0cm} 
\centerline{\it 
$\dagger$ Department of Physics, Kyungpook National University, Taegu
41566, Korea} 
\vspace*{0.5cm}
\centerline{\it 
  $\star$ Institut f$\ddot{u}$r Theoretische Physik,
  ETH Zurich, 8093 Z$\ddot{u}$rich, Switzerland}
\vspace*{0.8cm} 
\centerline{\tt ahn, ehdrb430, manhea@knu.ac.kr
} 
\vskip2cm

\centerline{\bf Abstract}
\vspace*{0.5cm}

By computing the operator product expansions
between the first two ${\cal N}=4$ higher spin multiplets
in the unitary coset model,
the (anti)commutators of higher spin currents
are obtained
under the large $(N,k)$ 't Hooft-like limit.
The free field realization with complex bosons and fermions
is presented. The (anti)commutators for generic spins
$s_1$ and $s_2$ with manifest $SO(4)$ symmetry
at vanishing 't Hooft-like coupling constant 
are completely determined. The structure
constants can be written in terms of the ones in the
${\cal N}=2$  ${\cal W}_{\infty}$ algebra found by
Bergshoeff, Pope, Romans, Sezgin and Shen
previously, in addition to the spin-dependent fractional
coefficients and two $SO(4)$ invariant tensors.
We also describe the ${\cal N}=4$ higher spin generators,
by using the above coset construction results,
for general super spin $s$ in terms of oscillators in the matrix
generalization of $AdS_3$ Vasiliev higher spin theory  at nonzero
't Hooft-like coupling constant.
We obtain the ${\cal N}=4$ higher spin algebra for low spins
and present how to determine the structure constants,
which depend on the higher spin algebra parameter, in general,
for fixed
spins $s_1$ and $s_2$.

\baselineskip=18pt
\newpage
\renewcommand{\theequation}
{\arabic{section}\mbox{.}\arabic{equation}}

\tableofcontents


\section{ Introduction}

The large ${\cal N}=4$ holography
in \cite{GG1305} has been proposed by constructing
the matrix generalization of the Vasiliev higher spin theory
on $AdS_3$ \cite{PV9806,PV9812}
and comparing it with the two dimensional minimal model
conformal field theories with large ${\cal N}=4$ superconformal
symmetry. The motivation for this proposal is based on
the fact that the non-abelian version of Vasiliev higher spin theory
might give some hints or evidences
for the better description of type IIB string
theory
on the $AdS_3$ space where the internal seven space is a product of
two three spheres and one sphere.
In this way, the oscillator deformation parameter
in the $AdS_3$ bulk theory plays the role of the free parameter of
the above large ${\cal N}=4$ superconformal algebra.
The complete
dual conformal field
theory is not known as far as we know.
The emergence of higher spin symmetry has not 
been fully clarified from the view point of type IIB string theory,
although there are some recent works in
\cite{GGH,FGJ,EGL,GHKPR,GG1803}, where one observes
that the infinite tower of
modes becomes massless (in the tensionless limit).
One of the findings in \cite{GG1305} is that
the spin contents with their multiplicity in the higher
spin algebra of the Vasiliev higher spin theory,
which contains an exceptional superalgebra
$D(2,1| \frac{\la}{1-\la})$ as
a subalgebra, are found \footnote{See also (\ref{nula})
  for the precise relation between the 't Hooft coupling constant
  $\la$
  and the higher spin algebra parameter $\mu$ (or $\nu$).}.
That is, there exist
seven spin one fields and
eight fields of spin $s =\frac{3}{2}, 2, \frac{5}{2}, 3, \frac{7}{2},
\cdots$.
However, the complete structure of the ${\cal N}=4$
higher spin algebra (i.e.,the (anti)commutators of ${\cal N}=4$
higher spin generators)
is not obtained so far. This fact allows us to further study
this large ${\cal N}=4$ holography more closely and is
one of motivations of the current paper. 

One way to observe the
presence of the higher spin algebra (or its symmetry)
is to study its two dimensional dual conformal field theory,
where the affine Kac-Moody algebra can be obtained
by using the adjoint spin-$1, \frac{1}{2}$ fields. 
In \cite{AK1509}, by writing down the
(higher spin) currents in terms of these fields
living in the ${\cal N}=4$ superconformal coset model, 
the operator product expansion (OPE) between the
lowest (or first) ${\cal N}=4$ higher spin multiplet and itself
of ${\cal W}_{\infty}^{{\cal N}=4 }[\la]$ algebra (in the notation of
\cite{EGR}) is determined.
See also the relevant works in \cite{Ahn1311,GP,BCG,Ahn1408,Ahn1504}.
Note that the currents of the ${\cal N}=4$ (linear) superconformal
algebra become the ones of exceptional superalgebra
$D(2,1|\frac{\la}{1-\la})$ (described by nine bosonic and eight
fermionic generators),
when the ``wedge'' condition
is imposed.
The next (or second) ${\cal N}=4$ higher spin
multiplet occurs in the right hand side of the
above OPE.

\begin{itemize}
\item[]
  We need to further compute the operator product expansions
between these two ${\cal N}=4$ higher spin multiplets
in order to obtain the corresponding higher spin algebra,
the (anti)commutators of ${\cal N}=4$ higher spin generators,
which can be determined by taking large $(N,k)$ 't Hooft-like
limit together with the wedge condition, precisely.
\end{itemize}

We observe that the additional  two (or third and fourth)
${\cal N}=4$
higher spin multiplets arise in this construction.

In \cite{EGR}, the free field construction at $\la=0$
for the large ${\cal N}=4$ holography with
 ${\cal W}_{\infty}^{{\cal N}=4 }[\la]$ algebra
\cite{GG1305} is described
by using $2N$-free complex bosons which transform as
bifundamental $({\bf N}, {\bf \bar{2}} )
\oplus ({\bf \bar{N}}, {\bf 2})$ of $U(N) \times U(K)$ with $K=2$
and
$2N$-free complex
fermions which transform similarly under the $U(N)
\times U(L)$ with $L=2$.
See also similar construction described in \cite{CHR1306}.
Note that in the ${\cal N}=4$ superconformal coset model,
the numerator of the coset
contains $SU(N+2)$, while the denominator of the coset
contains $SU(N)$.
By taking the $U(N)$ invariant combinations \cite{CHR1306,EGR} of
these fields, the higher spin currents can be determined as follows.
$1)$ The four-(higher spin)currents of integer spins
$s=2,3,4, \cdots$ transforming as an adjoint
representation of the $U(K=2)$ Chan-Paton factor
are obtained from the above bosons.
$2)$ 
Other four-(higher spin)currents of integer spins $s=1,2,3, \cdots$
transforming as an adjoint
representation of the $U(L=2)$ Chan-Paton factor
can be determined
from the fermions.
Note that the bilinears of the bosons
start at spin $s=2$, contrary to the ones of the fermions,
which can start at spin $s=1$ \footnote{By realizing
the recent work of \cite{AP1812},
  it is an open problem to check whether the spin-$1$
current can be added or not in the free field construction.}.
By construction, there are only four spin-$1$ currents, compared to
the field contents in the first paragraph.
Note that the higher spin-$1$ current belongs to
one of these spin-$1$ currents.
$3)$ The four-(higher spin)currents of half-integer spins
$s=\frac{3}{2}, \frac{5}{2}, \cdots $,
which transform as bifundamental
representation $({\bf 2},{\bf \bar{2}})$
of the $U(K=2) \times U(L=2)$,
are obtained from the above bosons
and fermions.
$4)$ Finally,
other four-(higher spin)currents of half-integer spins
$s=\frac{3}{2}, \frac{5}{2}, \cdots $
transform as bifundamental
representation $({\bf \bar{2}},{\bf 2})$.
The (anti)commutators between the ${\cal N}=4$
(higher spin) currents
are obtained for low spins \cite{EGR}.

It is natural to ask whether one can determine these
(anti)commutators for two higher spin currents
of {\it general} spin $s_1$ and $s_2$ in  the
${\cal W}_{\infty}^{{\cal N}=4 }[\la=0]$ algebra.
Some time ago, Odake in \cite{Odake}
found the extended super algebra, where
its generators with the particular level condition
are written in terms of bilinears of the
free complex bosonic and fermionic fields.
The bosonic subsector
is given by
the sum of both
$W_{\infty}^K$ algebra \cite{BK}
and $W_{1+\infty}^L$ algebra \cite{OS} (in the notation of
\cite{Odake}). See also
the relevant works in \cite{PRS,PRS1990,PRS1990-1,BPRSS}.
As described in previous paragraph,
the $K$-free complex bosons ($U(N)$ singlet) transform as
the (anti)fundamental ${\bf \bar{K}} 
\oplus  {\bf K}$ of $U(K)$ and
the $L$-free complex
fermions ($U(N)$ singlet)
transform as the (anti)fundamental ${\bf \bar{L}} 
\oplus  {\bf L}$ under the $U(L)$, in the notations of \cite{EGR}. 
In the context of \cite{EGR}, we can interpret that 
the complex free bosons give rise to $W_{\infty}[1]$ algebra,
whose wedge subalgebra is a bosonic
higher spin algebra $hs[1]$ at $\la=1$,
while  complex free fermions produce $W_{\infty}[0]$ algebra,
whose wedge subalgebra is a bosonic
higher spin algebra $hs[0]$ at $\la=0$,
after decoupling the spin-$1$ current \cite{GJL}.
See also \cite{GH}.
The $K^2$-(higher spin)currents of integer spins
$s=2,3,4, \cdots$
transform as an adjoint
representation of the $U(K)$ and 
are obtained from the free bosons.
Similarly, 
the $L^2$-(higher spin)currents from the free fermions
of integer spins
$s=1, 2,3, \cdots$
transform as an adjoint
representation of the $U(L)$.
Moreover, the
$2 K L$-(higher spin)currents of half-integer spins
$s=\frac{3}{2}, \frac{5}{2}, \cdots $
transform as bifundamental
representation $({\bf K},{\bf \bar{L}}) \oplus
({\bf \bar{K}},{\bf {L}})
$
of the $U(K) \times U(L)$,
which can be obtained from the free bosons
and fermions.
The seven nontrivial (anti)commutators
of ${\cal N}=4$ (higher spin)multiplets are given
in terms of a generalized hypergeometric function and
some polynomials  in the modes explicitly \footnote{In this paper,
  we mainly focus on the $K=L=2$ case associated with the large
  ${\cal N}=4$ holography.}. 

\begin{itemize}
\item[] In this paper,
  we observe that the free field construction in \cite{EGR}
(and the higher spin currents
with $K=L=2$) provide
the manifest $SO(4)$ symmetric (anti)commutators
in the ${\cal W}_{\infty}^{{\cal N}=4 }[\la=0]$ algebra
(with {\it arbitrary}
spins $s_1$ and $s_2$)
together with the structure constants
found in \cite{BPRSS,Odake}. In obtaining them, the coset results
are very crucial.
\end{itemize}

What happens for the case of nonzero $\la$?
For the nonzero 't Hooft-like coupling constant $(\la \neq 0)$,
we expect
that the ${\cal N}=4$ higher spin algebra can be 
constructed by oscillators studied in \cite{PV9806,PV9812}. 
In order to obtain this (unknown) higher spin algebra, 
we should resort to the (anti)commutators from the higher spin currents
of ${\cal W}^{{\cal N}=4}_{\infty}[\la]$ algebra in two dimensional
conformal field theory by imposing the ``wedge'' condition
together with the infinity limit of central charge (or infinity
limit of $N$). Then all the nonlinear (and some linear) terms in
${\cal W}^{{\cal N}=4}_{\infty}[\la]$ algebra vanish and we are left
with linear terms in the (anti)commutators.
For the case of ${\cal W}_{\infty}^{{\cal N}=2}[\la]$ algebra (in the
notation of \cite{EGR}),
the ${\cal N}=2$
higher spin algebra for any spins $s_1, s_2$ is found in
\cite{FL}. See also the relevant works in \cite{Korybut,BBB}.
According to \cite{CG}, the generators of wedge subalgebra of
${\cal W}_{\infty}^{{\cal N}=2}[\la]$ algebra match with
the ones of the ${\cal N}=2$ higher spin algebra \cite{FL}.
See also relevant works in
\cite{CHR1111,CG1203,HLPR,HP,Ahn1206,Ahn1208}.
It is straightforward to construct the generators satisfying
the ${\cal N}=2$ higher spin algebra $shs[\la]$
for generic $\la$ from  the
viewpoint of oscillators
by focusing on the (anti)commutators of
the wedge subalgebra of
${\cal W}_{\infty}^{{\cal N}=2}[\la]$ algebra \footnote{The bosonic
  subalgebra of ${\cal N}=2$ higher spin algebra
  $shs[\la]$ contains the two bosonic
  higher spin algebras, $hs[\la]$ and $hs[1-\la]$.
  Then the wedge subalgebra of ${\cal W}_{\infty}^{{\cal N}=2}[0]$
  algebra at $\la=0$
  contains the bosonic higher spin algebras $hs[0]$ and $hs[1]$.}.
Eventually this will lead to the findings in \cite{FL}.
%
However, for the large ${\cal N}=4$ holography associated with
${\cal W}_{\infty}^{{\cal N}=4}[\la]$ algebra with nonzero $\la$, 
there is no known higher spin algebra (as far as we know) from
the beginning.
We will observe that the subalgebra of this
${\cal N}=4$
higher spin algebra $shs_2[\la]$ (in the notation of \cite{EGR}
with $K=L=2$)
should contain the ${\cal N}=2$ higher spin algebra $shs[\la]$
in \cite{FL}
because the wedge subalgebra of
${\cal W}_{\infty}^{{\cal N}=4}[\la]$ algebra
contains the one of 
${\cal W}_{\infty}^{{\cal N}=2}[\la]$ algebra
\footnote{According to the result of \cite{GG1305},
  the higher spin algebra parameter $\mu$, which is a mass
  parameter of scalar field, is equal to $\la$.
  We use  $shs_2[\la]$ rather than  $shs_2[\mu]$. See also
  (\ref{nula}).}.
Although there are oscillator construction in \cite{EGR}
for the ${\cal N}=4$ higher spin algebra, the explicit expression for
this algebra is not known \footnote{In the coset construction of
  \cite{EGR}, the free field construction can be obtained by taking
  large level limit. Recently, by studying the nontrivial adjoint
  currents of $U(K)$ together with the trivial higher spin
  currents, the structure of ``rectangular'' $W$ algebra is found
in \cite{CH1812,CH1906,CHU}.}. We try to
obtain the ${\cal N}=4$ higher spin generators with the help of
oscillator formalism  in the context of the matrix generalization
of $AdS_3$ higher spin theory
by using the corresponding (anti)commutators of
${\cal W}_{\infty}^{{\cal N}=4}[\la]$ algebra in the two dimensional
conformal field theory, via the large ${\cal N}=4$ holography. 

\begin{itemize}
\item[]
  In this paper,
  we explicitly  calculate the $s$-th ${\cal N}=4$
  (higher spin)generators 
  in terms of oscillators. 
  We provide how to determine the (anti)commutators
  of the ${\cal N}=4$ higher spin algebra
$shs_2[\la]$ for fixed spins  $s_1$ and $s_2$.
\end{itemize}

We can write down the possible terms of the
right hand side of the (anti)commutator. Then by using the
general formula of the higher spin generators we explained above,
we can express the (anti)commutator with unknown coefficients
which depend on the $\la$ together with mode dependent factors
(that can be obtained from the $\la=0$ case).
If the spins $s_1$ and $s_2$ are small, then we can do
this computation by hand using the defining relations of the
oscillators with the $2 \times 2$ matrix manipulations.
However, the spins $s_1$ and $s_2$ become large, then this
computation by hand is rather tedious  
and is not possible to express the (anti)commutator in closed form,
although we can obtain the full expressions.
We will provide how to generate $\la$-dependent structure constants
appearing in the right hand side of (anti)commutator systematically,
although
the closed form for these structure constants is not known in this
paper. 

In section $2$, we review on the OPE between the first
${\cal N}=4$ multiplet described in \cite{AK1509}. 
In section $3$, by using the explicit form for the
first and second ${\cal N}=4$ multiplets found in \cite{AK1509,AKK1703},
we calculate the OPE between them and obtain 
the third ${\cal N}=4$ multiplet.
In section $4$, we compute
the OPE between the second ${\cal N}=4$ multiplet.
In section $5$, we take the large $(N,k)$ 't Hooft-like limit
on the two OPEs obtained in previous sections and present
the (anti)commutators. We also describe the subalgebra which has
the ${\cal N}=2$ supersymmetry.
In section $6$, we describe the first and second ${\cal N}=4$
higher spin multiplets by using the free field construction
studied in \cite{EGR}. After writing down the (anti)commutators
satisfied by these fields, we present the (anti)commutators
for general spins $s_1$ and $s_2$.
In section $7$, the oscillator realization for the
first two ${\cal N}=4$ higher spin generators is given
and furthermore, we present the $s$-th ${\cal N}=4$ higher spin
generators in terms of oscillators.  
In section $8$, we summarize what we have obtained in this paper
and the open problems are given.
In Appendices $A,B$, and $C$,
some detailed expressions appeared in the
sections  $2,3$, and $4$ are given \footnote{
In Appendices $D,E$, and $F$, some expressions appeared in the
sections  $2$, $3$, and $4$ for the OPEs are presented.
In Appendix $G$, the (anti)commutators corresponding to
Appendices $D$ and $E$ are described.
In Appendix $H$, we provide
the second ${\cal N}=4$ higher spin multiplet
in section $6$.
In Appendix $I$, the first ${\cal N}=4$ higher spin multiplet
in section $6$ at $\la=1$ is described.
In Appendix $J$,
the remaining (anti)commutators
for the general spins $s_1$ and $s_2$ in section $6$ are given.
In Appendix $K$, the second ${\cal N}=4$ higher spin generators
in section $7$ is presented.
Finally in Appendix $L$,
we explain how the ${\cal N}=2$
higher spin algebra can be obtained from the results of section $7$.}.

We are using the Thielemans package \cite{Thielemans}
with mathematica \cite{mathematica}.
An ancillary mathematica file, {\tt ancillary.nb},
where the missing parts of Appendices are given explicitly,
is included.

\section{Review of ${\cal N}=4$ unitary coset model}

The Wolf space coset in the ``supersymmetric'' version
with groups $G=SU(N+2)$
and $H=SU(N) \times SU(2) \times U(1)$ is given by
\bea
\mbox{Wolf} = \frac{SU(N+2)}{SU(N) \times SU(2) \times U(1)},
\label{Wolf}
\eea
where the group indices
are described by
\bea
G   \,\,  \mbox{indices} \,\, & : & \,\, a, b, \cdots  =
1,2, \cdots, (N+2)^2-1,
\nonu \\
\frac{G}{H}  \,\,  \mbox{indices} \,\, & : &
\,\, \bar{a}, \bar{b},
\cdots =
1,2, \cdots, 4N.
\label{indices}
\eea
In the bosonic version of Wolf space,
there are $4N$-free fermions appearing in an extra
$SO(4N)$ group in the numerator of the coset (\ref{Wolf}) at level $1$
\footnote{In other words,
  we have $\frac{SU(N+2)_{k} \times SO(4N)_1}{SU(N)_{k+2}
    \times SU(2)_{k+N}
  \times U(1)_{2N(N+2)(N+2+k)}}$.
  In \cite{EGR}, these are described by $2N$-complex
  fermions transforming as bifundamental $({\bf N}, {\bf \bar{2}})
  \oplus ({\bf \bar{N}},{\bf 2})$ under the $U(N) \times U(2)$.}.
Note that the ${\cal N}=4$ superconformal coset theory
is given by $\mbox{Wolf} \times SU(2) \times U(1)$.
The ${\cal N}=1$ affine Kac-Moody algebra
can be constructed from the adjoint spin-$1$ current
and the spin-$\frac{1}{2}$
current of group $G=SU(N+2)$.
The operator product expansion
between the (modified) spin-$1$ current $V^a(z)$ and
the spin-$\frac{1}{2}$ current $Q^a(z)$
can be described as follows \cite{KT}:
\bea
V^a(z) \, V^b(w) & = & \frac{1}{(z-w)^2} \, k \, g^{ab}
-\frac{1}{(z-w)} \, f^{ab}_{\,\,\,\,\,\,c} \, V^c(w) 
+\cdots,
\nonu \\
Q^a(z) \, Q^b(w) & = & -\frac{1}{(z-w)} \, (k+N+2) \, g^{ab} + \cdots.
\label{opevq}
\eea
The metric can be obtained from $g_{ab} =
\frac{1}{2c_G} \, f_{a c}^{\,\,\, d}
\, f_{b d}^{\,\,\,c}$, where $c_G$ is the dual Coxeter number of
the group $G$.
The metric $g_{ab}$ which can be represented by
$(N+2)^2-1 \times (N+2)^2-1$ matrix  is given by the generators
of $G$ in the complex basis \cite{AK1506} as follows:
$
g_{ab} = \mbox{Tr} (T_a T_b)=
\left(
\begin{array}{cc}
0 & 1 \\
1 & 0 \\
\end{array}
\right)$, where $
a,b=1,2, \cdots, (N+2)^2-1$ from (\ref{indices}).
 The commutation relation for the $SU(N+2)$ generators
 in the fundamental representation
 is given by $[T_a, T_b] = f_{ab}^{\,\,\,c} \, T_c$.

 \subsection{The $11$ currents of the large ${\cal N}=4$ nonlinear
 superconformal algebra}
 
 The four supersymmetry currents of spin-$\frac{3}{2}$,
 $\hat{G}^0(z)$ and
 $\hat{G}^i(z)$, the six spin-$1$ currents of $SU(2)_k \times SU(2)_N$,
 $A^{\pm i}(z)$ and
 the spin-$2$ stress energy tensor $\hat{T}(z)$ can be
 described in terms of spin-$1,\frac{1}{2}$ currents
 \cite{VanProeyen,GK,AK1411} as follows: 
\bea
\hat{G}^{0}(z) &  = &   \frac{i}{(k+N+2)}  \, g_{\bar{a} \bar{b}} \,
Q^{\bar{a}} \, V^{\bar{b}}(z),
\qquad
\hat{G}^{i}(z)  =  \frac{i}{(k+N+2)} 
\, h^{i}_{\bar{a} \bar{b}} \, Q^{\bar{a}} \, V^{\bar{b}}(z),
\nonu \\
A^{+i}(z) &  = & 
-\frac{1}{4N} \, f^{\bar{a} \bar{b}}_{\,\,\,\,\,\, c} \, h^i_{\bar{a} \bar{b}} \, V^c(z), 
\qquad
A^{-i}(z)  =  
-\frac{1}{4(k+N+2)} \, h^i_{\bar{a} \bar{b}} \, Q^{\bar{a}} \, Q^{\bar{b}}(z),
\nonu \\
\hat{T}(z)  & = & 
\frac{1}{2(k+N+2)^2} \Bigg[ (k+N+2) \, g_{\bar{a} \bar{b}}
\,  V^{\bar{a}} \, V^{\bar{b}} 
+k \, g_{\bar{a} \bar{b}} \, Q^{\bar{a}} \, \pa \, Q^{\bar{b}} 
+f^{\bar{a} \bar{b}}_{ \,\,\,\,\,\, c} \, g_{\bar{a} \bar{c}} \, g_{\bar{b} \bar{d}} \,
Q^{\bar{c}} \, Q^{\bar{d}} \, V^c  \Bigg] (z)
\nonu \\
&&- \frac{1}{(k+N+2)} ( A^{+i}+A^{-i} )^2 (z), \qquad i=1,2,3.
\label{11currents}
\eea
Note that
the Wolf coset indices, $\bar{a}, \bar{b}, \cdots$,
run over  $\bar{a}, \bar{b}, \cdots = 1, 2, \cdots, 4N$ from
(\ref{indices}). 
In the spin-$2$ stress energy tensor, the terms $A^{-i} \, A^{-i}$,
which contain  $ (Q^{\bar{a}} \, Q^{\bar{b}})
( Q^{\bar{c}} \, Q^{\bar{d}})(z)$,
can be further simplified by using the defining relations
(\ref{opevq}). 
The three almost complex structures
$h^{i}_{\bar{a} \bar{b}}$
are given by 
$4N \times 4N$ matrices \cite{AK1506} and 
they are antisymmetric and satisfy the quaternionic algebra
\cite{Saulina}. Note that the Wolf space coset metric
$g_{\bar{a} \bar{b}} \equiv h^0_{\bar{a} \bar{b}}$.

With the $SO(4)$ singlet
$
\hat{T}(z) \rightarrow \hat{L}(z)$,
the spin-$\frac{3}{2}$ currents, transforming as the $SO(4)$ vector
representation,
are given by 
$
\hat{G}^0(z) \rightarrow \hat{G}^2(z)$, 
$\hat{G}^1(z) \rightarrow \hat{G}^3(z)$,
$\hat{G}^2(z) \rightarrow -\hat{G}^4(z)$,
and $\hat{G}^3(z) \rightarrow \hat{G}^1(z)$.
Furthermore, the six spin-$1$ currents, $\hat{T}^{\mu \nu}(z)$
transforming as the $SO(4)$ adjoint representation,
can be obtained from
the corresponding two spin-$1$ currents $A^{\pm i}(z)$ as follows
\cite{AK1509,AKK1703}:
$
A^{\pm 1}(z)  \rightarrow 
    \frac{i}{2} \, \alpha_{\mu\nu}^{\pm 1} \, \hat{T}^{\mu\nu}(z)$,
    $ A^{\pm 2}(z) \rightarrow
   \frac{i}{2} \, \alpha_{\mu\nu}^{\pm 2} \, \hat{T}^{\mu\nu}(z)$,
  $
   A^{\pm 3}(z)  \rightarrow  \pm
   \frac{i}{2} \, \alpha_{\mu\nu}^{\pm 3} \, \hat{T}^{\mu\nu}(z)$,
where the six $4 \times 4$ matrices $\alpha_{\mu \nu}^{\pm i}$ \cite{STVS}
relate the spin-$1$ currents
$A^{\pm i}(z)$ to the spin-$1$ currents $\hat{T}^{\mu \nu}(z)$.

Then the large ${\cal N}=4$ nonlinear superconformal algebra
\cite{GS,VanProeyen,GPTV,GK}
can be obtained explicitly using the above $11$ currents as
follows:$\hat{L}$,
$\hat{G}^{\mu}$, and $\hat{T}^{\mu\nu}$.
The nonlinear structure appears in the OPE
between the spin-$\frac{3}{2}$ currents. Note that
the two levels of $SU(2)$'s are
given by $k$ and $N$ respectively.

 \subsection{The $16$ currents of the large ${\cal N}=4$ linear
 superconformal algebra}

The explicit relation between the
$16$ currents of the large ${\cal N}=4$ linear superconformal
algebra and the $11$ currents  of the
large ${\cal N}=4$ nonlinear superconformal algebra
is described by \cite{GS}
\bea
{T}^{\mu \nu}(z) & = &
 \hat{T}^{\mu \nu}(z)+\frac{2i}{(2+k+N)}\,{ \Gamma^{\mu} \Gamma^{\nu}}(z),
\nonu\\
{G}^{\mu}(z) & = & \hat{ G}^{\mu}(z)+\frac{2i}{(2+k+N)}\,{ U\Gamma^{\mu}}(z)
\nonu \\
& + & \varepsilon^{\mu\nu\rho\si} \Bigg[  
\frac{4i}{3(2+k+N)^2}{ \Gamma^{\nu}\Gamma^{\rho}\Gamma^{\si}}
-\frac{1}{(2+k+N)}{ T^{\nu\rho} \Gamma^{\si}}
\Bigg](z),
\nonu\\
{L}(z) & = & 
\hat{ L}(z)-\frac{1}{(2+k+N)}\,\Bigg[\,{ UU}-{
  \partial \Gamma^{\mu} \Gamma^{\mu}}\,\Bigg](z).
\label{gsformula}
\eea
In the last term of spin-$\frac{3}{2}$ currents in
(\ref{gsformula}), the first
relation of (\ref{gsformula}) should be inserted.
Moreover, 
the four fermionic spin-$\frac{1}{2}$ currents are
given by \cite{Saulina,AK1506}
\bea
{ \Gamma}^0 (z) =-\frac{i }{4(N+1)}
h^j_{\tilde{a} \tilde{b} }
f^{\tilde{a} \tilde{b}}_{\,\,\,\,\,\, \tilde{c}} h^{j \tilde{c}}_{ \,\,\,\, \tilde{d} } Q^{\tilde{d}}(z),
\qquad
{ \Gamma}^j (z) =-\frac{i }{4(N+1)} h^j_{\tilde{a} \tilde{b} }
f^{\tilde{a} \tilde{b}}_{\,\,\,\,\,\, \tilde{c}} Q^{\tilde{c}} (z),
j=1,2,3,
\label{Gamma}
\eea
where there is no sum over $j$ in the first equation of
(\ref{Gamma}). We should change the index structures
  as follows: ${ \Gamma^0} \rightarrow -i { \Gamma^2}$,
  ${ \Gamma^1} \rightarrow
  -i { \Gamma^3}$,
  ${ \Gamma^2} \rightarrow i { \Gamma^4}$ and
  ${ \Gamma^3} \rightarrow -i { \Gamma^1}$ in order to use
  (\ref{gsformula}) from (\ref{Gamma}) with $SO(4)$ vector index.
  We introduce the coset $\mbox{Wolf} \times SU(2) \times U(1)=
  \frac{SU(N+2)}{SU(N)}$ by multiplying
  $SU(2) \times U(1)$ factor in (\ref{Wolf})
  (See also \cite{GG1305} for the bosonic version of the coset)
  and the corresponding notation
  is as follows:
  $\tilde{a}=(\bar{a}, \hat{a})$, 
where the $\bar{a}$ index runs over $4N$ values
as before
and the index $\hat{a}$ associates with the  
$2 \times 2$ matrix corresponding to $SU(2)\times U(1)$
and runs over $4$ values. 
The bosonic spin-$1$ current is given by
\bea
{ U} (z) =-\frac{1}{4(N+1)} h^j_{\tilde{a} \tilde{b} }
f^{\tilde{a} \tilde{b}}_{\,\,\,\,\,\, \tilde{c}} h^{j \tilde{c}}_{ \,\,\,\, \tilde{d} } \left[
 V^{\tilde{d}}
-\frac{1}{2(k+N+2)}  
f^{\tilde{d} }_{\,\,\,\, \tilde{e} \tilde{f}} Q^{\tilde{e}} Q^{\tilde{f}}
\right](z),
\label{ulinear}
\eea
where there is no sum over the index $j$.
Of course, the OPEs between the $11$ currents in (\ref{11currents})
and the spin-$\frac{1}{2}$ currents
${ \Gamma}^{\mu}(z)$ are regular
and similarly,  the OPEs between the $11$ currents and
the spin-$1$ current ${ U}(z)$ do not have any singular terms.

Therefore, the large ${\cal N}=4$ linear superconformal algebra
can be generated by (\ref{gsformula}), (\ref{Gamma}) and
(\ref{ulinear}), together with the $11$ currents obtained
in previous subsection, 
in the ${\cal N}=4$ coset  $\frac{SU(N+2)}{SU(N)}$ model.
Note that
the two levels of $SU(2)$'s are
given by $(k+1)$ and $(N+1)$ respectively contrary to the ones of
nonlinear case.
The central charge is given by
\bea
c = \frac{6(k+1)(N+1)}{(k+N+2)}.
\label{central}
\eea
Under the large $(N,k)$ 't Hooft limit, the
central charge (\ref{central}) becomes $c=6(1-\la)N$,
where the 't Hooft coupling constant is defined as
$\la \equiv \frac{(N+1)}{(k+N+2)}$.
At $\la =0$, the central charge is $c=6N$.
For the infinity limit of $k$, the central charge is given by
$c=6(N+1)$ from (\ref{central}) \cite{GG1406}.

\subsection{Operator product expansion between the first
${\cal N}=4$ higher spin multiplet}

The single ${\cal N}=4$ super OPE studied in
\cite{Schoutens} between the
first ${\cal N}=4$ higher spin multiplet and itself
can be summarized by \cite{AK1509}
\footnote{The notation of $\theta^{4-0}$ stands for
  $\theta^1 \theta^2 \theta^3 \theta^4$. One can read off
  $\theta^{4-i}$ from the relation $\theta^1 \theta^2 \theta^3
  \theta^4 =\theta^{4-i} \theta^i$ (no sum over the index $i$).
  For example, $\theta^{4-i}|_{i=2}=\theta^1 \theta^3 \theta^4$.
Similarly, we have the relation $\theta^1 \theta^2 \theta^3
\theta^4 =\theta^{4-ij} \theta^i \theta^j$.
For example, $\theta^{4-ij}|_{i=1, j=2}=
\theta^3 \theta^4$.}
\bea
{\bf \Phi}^{(1)}(Z_{1})\,{\bf \Phi}^{(1)}(Z_{2})  & = & 
\frac{\theta_{12}^{4-0}}{z_{12}^{4}}\,
\frac{4 \,k \,N \,(k-N)}{(2+k+N)^2}
+\frac{\theta_{12}^{4-i}}{z_{12}^{3}}\,
{\bf Q}^{(\frac{1}{2}),i}_{\frac{3}{2}}
(Z_{2})
+\frac{\theta_{12}^{4-0}}{z_{12}^{3}}\,
{\bf Q}^{(1)}_{2}
(Z_{2})
\nonu\\
&
+& 
\frac{1}{z_{12}^{2}}\,
\frac{2\,k\,N}{(2+k+N)}
+
\frac{\theta_{12}^{4-ij}}{z_{12}^{2}}\,
{\bf Q}^{(1),ij}_{1}(Z_{2})
\nonu \\
& + &
\frac{\theta_{12}^{4-i}}{z_{12}^{2}}\,\Bigg[\,
2\,\partial {\bf Q}^{(\frac{1}{2}),i}_{\frac{3}{2}}
+{\bf Q}^{(\frac{3}{2}),i}_{\frac{3}{2}}
-\frac{(k-N)}{3(2+k+N)}\,{\bf Q}^{(\frac{3}{2}),i}_{\frac{1}{2}}
\,\Bigg](Z_{2})
\nonu\\
&
+&\frac{\theta_{12}^{4-0}}{z_{12}^{2}}\,\Bigg[\,
\frac{3}{2}\, \partial {\bf Q}^{(1)}_{2}
+{\bf Q}^{(2)}_{2}
\nonu \\
&- &
\frac{8(k-N)}{(5+4k+4N+3kN)}
\, ({\bf \Phi^{(1)}} {\bf \Phi^{(1)}}
+
\frac{k\,N}{(2+k+N)}
     {\bf J}^{4-0} \nonu \\
     & - & \frac{k N(k-N) }{(2+k+N)^2} \,\pa^2 \,{\bf J})
\,\Bigg](Z_{2})
+  
\frac{\theta_{12}^{i}}{z_{12}}\,
{\bf Q}^{(\frac{3}{2}),i}_{\frac{1}{2}}(Z_{2})
\nonu \\
& + &
\frac{\theta_{12}^{4-ij}}{z_{12}}\,\Bigg[\,
\partial {\bf Q}^{(1),ij}_{1} 
+  {\bf Q}^{(2),ij}_{1}
\,\Bigg](Z_{2})
\nonu \\
&
+ & \frac{\theta_{12}^{4-i}}{z_{12}}\,\Bigg[\,
\frac{3}{2}\,\partial^{2} {\bf Q}^{(\frac{1}{2}),i}_{\frac{3}{2}}
+\partial {\bf Q}^{(\frac{3}{2}),i}_{\frac{3}{2}}
+ {\bf Q}^{(\frac{5}{2}),i}_{\frac{3}{2}}
\,\Bigg](Z_{2})
\nonu \\ 
&
+& \frac{\theta_{12}^{4-0}}{z_{12}}\,\Bigg[\,
\partial^{2} {\bf Q}^{(1)}_{2}
+\partial {\bf Q}^{(2)}_{2}
+{\bf Q}^{(3)}_{2}
\nonu \\
& - &
\frac{8(k-N)}{(5+4k+4N+3kN)}
\, \partial({\bf \Phi^{(1)}} {\bf \Phi^{(1)}}+
\frac{k\,N}{(2+k+N)}
     {\bf J}^{4-0}
     \nonu \\
&- & \frac{k N(k-N) }{(2+k+N)^2} \,\pa^2 \,{\bf J}
     )
\,
\Bigg](Z_{2})
+  \cdots.
\label{super1super1}
\eea
The nonlinear terms ${\bf \Phi^{(1)}} {\bf \Phi^{(1)}}(Z_2)$ and its
descendant term occur in this OPE (\ref{super1super1}).
In particular,
the dependence of
the first and second
${\cal N}=4$ higher spin multiplets can also arise
in the following quasi (super)
primary fields \footnote{As in \cite{AKP1904},
  the spin of the quasi primary field is given by the number of
  inside of the bracket. The subscript comes from
  the one in the first operator of the OPE in the component approach. See, for example, (\ref{fiveOPEs}).
  We use the simplified notation ${\bf J}^{4-0} \equiv D^1 D^2 D^3 D^4
  \, {\bf J}$.}
\bea
{\bf Q}^{(\frac{5}{2}),i}_{\frac{3}{2}} & = &
\frac{1}{2}\, D^i {\bf \Phi}^{(2)}
-\frac{4(k-N)}{(5+4k+4N+3kN)}
\, {\bf \Phi}^{(1)} D^i {\bf \Phi}^{(1)}
+ {\bf J}\,\mbox{dependent terms},
\nonu\\
{\bf Q}^{(2)}_{2} & = &
2 \,  {\bf \Phi}^{(2)}
+ {\bf J}\, \mbox{dependent terms}.
\label{twoquasiother}
\eea
The stress energy tensor ${\bf J}$ dependent terms in
(\ref{twoquasiother})
are given in Appendix $A$.
All the expressions for the quasi primary (super)fields
are presented in (\ref{quasisuper}) (with {\tt ancillary.nb}).
In obtaining (\ref{super1super1}), the fundamental
OPEs in (\ref{fiveOPEs})
are crucial.
In Appendix $D$, the OPEs between
the first ${\cal N}=4$ multiplet are explicitly given
after the large $(N,k)$ limit is taken. Furthermore, in
(\ref{PhionePhione}), its (anti)commutators are presented explicitly.
See also \cite{AK1509} for more details.
We need to redefine the second ${\cal N}=2$ higher spin multiplet
(\ref{newspintwo})
in next section.

\section{Operator product expansion between the first and second
${\cal N}=4$ higher spin multiplets}

We construct the OPEs
between the first and second
${\cal N}=4$ higher spin
multiplets in component approach and in ${\cal N}=4$
superspace.
The $16=(1+4+6+4+1)$
higher spin currents of superspin $s$ can be 
combined into one single ${\cal N}=4$ super field
as follows \cite{AK1509}:
\bea
 {\bf \Phi^{(s)}} \equiv \Bigg(
\Phi_{0}^{(s)},
\Phi_{\frac{1}{2}}^{(s),i},
\Phi_{1}^{(s),ij},
\Phi_{\frac{3}{2}}^{(s),i},
\Phi_{2}^{(s)} \Bigg), \qquad i,j = 1,2,3,4.
\label{Phiexp}
\eea
The spins of each element are
given by $s, (s+\frac{1}{2}), (s+1), (s+\frac{3}{2})$,
and $(s+2)$ respectively.
The last two component fields in (\ref{Phiexp}) are not quasi primary
fields while the first three component fields are primary fields
under the stress energy tensor $L(z)$ of (\ref{gsformula})
\footnote{We will use the simplified notations for the
  (higher spin)currents as follows:
  \bea
\tilde{ T}^{ij} & \equiv &
\frac{1}{2!}\:\varepsilon_{ijkl}\:{ T}^{kl}, \qquad
  \tilde{G}^{i}  \equiv  G^{i}-
  \frac{(k-N)}{(k+N+2)} \,
  i \, \partial \, \Gamma^{i}, \qquad
  \tilde{L} \equiv L+
 \frac{(k-N)}{2(k+N+2)}\, 
 \partial \,U,
\nonu \\
\tilde{ \Phi}_1^{(s),ij}  & \equiv & 
\frac{1}{2!}\:\varepsilon_{ijkl}\:{ \Phi}_1^{(s),kl}.
\nonu
\eea
Also we have the ${\cal N}=4$ stress energy tensor
${\bf J}=(-\Delta, i \Gamma^i,
-i \, T^{ij}, - \tilde{G}^i, 2 \tilde{L})$ with
$\pa \Delta \equiv-U$.
\label{BigJ}}.
After analyzing the component approach first and then
we will end up with the single OPE in ${\cal N}=4$ superspace. 

\subsection{The new higher spin current $\Phi_0^{(3)}$ of spin $3$ }

Let us consider the OPE between the last component $\Phi_2^{(1)}(z)$
with $s=1$ and the first component $\Phi_0^{(2)}(w)$
with $s=2$ of (\ref{Phiexp}).
By adding the derivative term and other composite field
to the $\Phi_{2}^{(1)}$, we can make it to be a primary field
as follows \cite{AK1509} \footnote{
  We have the following relation
  \bea
\Phi^{(s),i}_{\frac{3}{2}}=\tilde{\Phi}^{(s),i}_{\frac{3}{2}}+\frac{1}{(2s+1)}\frac{(k-N)}{(2+k+N)}\,\partial\Phi^{(s),i}_{\frac{1}{2}}.
\nonu
\eea}:
\bea
\Phi_{2}^{(s=1)} \equiv  \widetilde\Phi_{2}^{(s=1)}+
  p_{1}\,\partial^{2} \, \Phi_{0}^{(s=1)}+p_{2}\, L\,\Phi_{0}^{(s=1)},
  \label{newspin2}
  \eea
  where
  $ \tilde{\Phi}_{2}^{(s=1)} $
  is a primary field under the
  stress energy tensor $L(z)$.
  The coefficients $p_1$ and $p_2$ appearing in (\ref{newspin2})
  were given in \cite{AK1509}
  and they are
  \bea
  p_1 & \equiv & -\frac{(k-N)(29+16k+16N+3k N)}{3(2+k+N)(5+4k+4N+3kN)} , \qquad
  p_2 \equiv
\frac{8(k-N)}{(5+4k+4N+3kN)}.
  \label{p1p2}
  \eea

The higher spin-$2$ current in the second ${\cal N}=4$ multiplet
is redefined from the one (denoted by $\hat{\Phi}_{0}^{(2)}$
in this paper) in \cite{AK1509} (or \cite{AKK1703}) as
follows \footnote{The right hand side of (\ref{newspintwo})
is proportional to the equation $(6.37)$ of \cite{AKK1703}.}:
\bea
\Phi_{0}^{(2)}
& = & 
\hat{\Phi}_{0}^{(2)}
+r_0\,\Phi^{(1)}_0\Phi^{(1)}_0
+r_1\, L
+r_2\, T^{ij}T^{ij}
+r_3\, T^{ij} \tilde{T}^{ij}
+r_4\, T^{ij} \Gamma^i \Gamma^j
+r_5\, \tilde{T}^{ij} \Gamma^i \Gamma^j
\nonu\\
&+&
r_6\, U U
+r_7\, \partial \Gamma^i \Gamma^i
+r_8\, \varepsilon^{ijkl}\,\Gamma^i \Gamma^j \Gamma^k \Gamma^l
\,,
\label{newspintwo}
\eea
where the coefficients are given by
\bea
r_0 & \equiv &
-\frac{3}{2
  (2+N)(2+k+N)^{2}(5+4k+4N+3kN)}(100+187k+118k^{2}+24k^{3}+303N
\nonu
\\
&+ &
505kN+277k^{2}N+46k^{3}N+325N^{2}+455kN^{2}+195k^{2}N^{2}+20k^{3}N^{2}\nonu
\\
& + & 148N^{3}+159kN^{3}
+ 
42k^{2}N^{3}+24N^{4}+16kN^{4}),
\label{rcoeff}
\\
r_1 & \equiv & r_0\,\frac{-4(2+k)(2+N)}{3(4+3k+3N+2kN)}, \qquad
r_2 \equiv r_0\,\frac{(4+k+N)}{6(4+3k+3N+2kN)},
\nonu\\
r_3 & \equiv & r_0\,\frac{(k-N)}{6(4+3k+3N+2kN)},\qquad
r_4 \equiv r_0\,\frac{-2i(4+k+N)}{3(2+k+N)(4+3k+3N+2kN)},
\nonu\\
r_5 & \equiv & r_0\,\frac{-2i(k-N)}{3(2+k+N)(4+3k+3N+2kN)},
r_6 \equiv r_0\,\frac{-4(2+k)(2+N)}{3(2+k+N)(4+3k+3N+2kN)},
\nonu\\
r_7 &  \equiv &
r_0\,\frac{2(20+7k+7N+2kN)}{3(2+k+N)(4+3k+3N+2kN)},
r_8 \equiv r_0\,\frac{-(k-N)}{3(2+k+N)^{2}(4+3k+3N+2kN)}.
\nonu
\eea  
Note that the coefficients $r_1, \cdots, r_8$ in (\ref{rcoeff})
can be written in terms of $r_0$ by using the ${\cal N}=4$
primary condition on the right hand side of (\ref{newspintwo}).
The second order pole of the OPE between $\Phi_2^{(1)}(z)$
and $\Phi_0^{(1)}(w)$ in \cite{AK1509}
has the nonlinear term $\Phi_0^{(1)} \,
\Phi_0^{(1)}(w)$ as well as $\hat{\Phi}_0^{(2)}(w)$.
The coefficient $r_0$ can be fixed further
by absorbing this nonlinear term.
Note that the value $r_0$ appears in the arXiv version of
\cite{AK1509}(and is equal to the value $c_3$ in page 244).
We also analyze  the $\Phi_0^{(3)}$ case similarly. 

Then we can calculate, from the result of the higher spin-$3$
current $\Phi_{2}^{(1)}(z)$ in
\cite{AK1509} and the higher spin-$2$ current (\ref{newspintwo}),
the following OPE explicitly 
\bea
\Phi_{2}^{(1)}(z)\,\Phi_{0}^{(2)}(w)
& = & 
\frac{1}{(z-w)^{4}}\,
\Bigg[\,
 Q^{(1)}_{2}
 -(6\,p_1+4\,p_2)\, Q^{(1)}_{0}
\,\Bigg](w)
\nonu \\ 
& + & 
\frac{1}{(z-w)^{3}}\,
\Bigg[\,
 \partial Q^{(1)}_{2}
 + Q^{(2)}_{2}
 -p_2\, \partial Q^{(1)}_{0}
\,\Bigg](w)
\nonu \\ 
& + & 
\frac{1}{(z-w)^{2}}\,\Bigg[\,
\frac{1}{2}\,\partial^2 Q^{(1)}_{2} 
+\frac{3}{4}\,\partial Q^{(2)}_{2} 
+Q^{(3)}_{2}
-p_2(2 \Phi^{(1)}_0 \Phi^{(2)}_0 + d_{1}^{0,2}\, L \Phi^{(1)}_0)
\,
\Bigg](w)
\nonu \\ 
& + & 
\frac{1}{(z-w)}\,\Bigg[\,
\frac{1}{6}\,\partial^3 Q^{(1)}_{2} 
+\frac{3}{10}\,\partial^2 Q^{(2)}_{2} 
+\frac{2}{3}\,\partial Q^{(3)}_{2} 
+Q^{(4)}_{2}
-p_2(2\, \partial\Phi^{(1)}_0 \Phi^{(2)}_0
\nonu \\ 
& + &
\Phi^{(1)}_0 \partial \Phi^{(2)}_0 + d_{1}^{0,2}\, \partial L \Phi^{(1)}_0)
\,
\Bigg](w)
+\cdots,
\label{2102}
\eea
where the coefficients $p_1$ and $p_2$ are given in (\ref{p1p2}).
The coefficient appearing in the second order pole of (\ref{2102})
is a function of $N$ and $k$ 
as follows:
\bea
d_{1}^{0,2}=
-\frac{16 (k - N) (11 + 16 k + 6 k^2 + 16 N + 17 k N + 4 k^2 N + 6 N^2 + 
   4 k N^2)}{(2 + k + N) (4 + 3 k + 3 N + 2 k N) (5 + 4 k + 4 N + 
   3 k N)}\,.
\label{dcoeff}
\eea
We can calculate this OPE (\ref{2102})
for fixed $N=3$ and after obtaining ${\cal N}=4
$ superspace description, where the fundamental $16$ OPEs
can be generalized to the full $256$ OPEs, we will obtain all
the structure constants which depend on $(N,k)$ explicitly
from the Jacobi identity. 
The various quasi primary fields appearing in (\ref{2102})
are given in (\ref{morequasi1})
\footnote{Note that the quasi primary
  fields $Q_{s_c}^{(s)}$ with $s_c=0,\frac{1}{2}, 1, \frac{3}{2}$
  and $2$, which do not have adjoint indices of
  group $G$,
  are nothing to do with the spin-$\frac{1}{2}$
currents appearing in (\ref{opevq}) directly.}.

In particular, the second order pole of (\ref{2102}) has
the following new primary higher spin-$3$ current,
which cannot be written in terms of the known (higher spin) currents,
\bea
Q_2^{(3)}(w) & = & w_{1,3} \, \Phi_0^{(3)}(w) + \cdots. 
\label{spin3ofthird}
\eea
Note that the subscript $2$ comes from
the index $2$ of $\Phi_2^{(1)}(z)$ in the left hand side of the OPE
\footnote{In the fundamental $16$ OPEs we are describing,
  the higher spin-$2$ current $\Phi_0^{(2)}(w)$ is common. Then by
  specifying only the subscript of the component of
  first ${\cal N}=4$ higher spin
  multiplet (in addition to the spin of quasi primary fields
  which tells us the pole of the singular term in the OPE),
  we can classify all the quasi primary fields
  uniquely (and independently).}.
The remaining $25$ terms in (\ref{spin3ofthird})
are given in (\ref{morequasi1}).
We use the following quasi primary fields
with their spins, $SO(4)$ indices $i,j$ and the subscript
indicating  the number of fermionic coordinates
in front of the quasi primary fields
when we go to the ${\cal N}=4$ superspace
\bea
&& Q_0^{(1)} \,, \cdots ; \,
Q_{\frac{1}{2}}^{(\frac{3}{2}),i}, \,
Q_{\frac{1}{2}}^{(\frac{5}{2}),i}\, , \cdots;
\,
Q_{1}^{(1),ij}, \,
Q_{1}^{(2),ij}, \, Q_{1}^{(3),ij}\, , \cdots;
 \nonu \\
 &&
 Q_{\frac{3}{2}}^{(\frac{1}{2}),i}, \,
 Q_{\frac{3}{2}}^{(\frac{3}{2}),i},
 \, Q_{\frac{3}{2}}^{(\frac{5}{2}),i}, \,  Q_{\frac{3}{2}}^{(\frac{7}{2}),i}\,,
 \cdots;
 \,
 Q_2^{(1)},  \, Q_2^{(2)}, \, Q_2^{(3)},
 \, Q_2^{(4)}\, ,
  Q_2^{(5)} \,, \cdots.
 \label{quasicomponent}
 \eea
Note that the spin of the field in (\ref{quasicomponent})
is given by the number inside the bracket of upper index.
The corresponding ${\cal N}=4$ super fields
can be denoted by using the boldface symbols
as in previous section
(\ref{super1super1}) or (\ref{twoquasiother}).

Therefore, we observe that the lowest component of the third
${\cal N}=4$ higher spin multiplet occurs
in addition to
the components
of the first and second ${\cal N}=4$ higher spin multiplets
in (\ref{2102}).

\subsection{The new higher spin current
$\Phi^{(3),i}_{\frac{1}{2}}$
  of spin $\frac{7}{2}$}

Let us consider the OPE
between the higher spin-$\frac{5}{2}$ current
$\Phi_{\frac{3}{2}}^{(1),i}(z)$
of the first ${\cal N}=4$ multiplet
with $SO(4)$ vector index and the
higher spin-$2$ current $\Phi_{0}^{(2)}(w)$
of the second
${\cal N}=4$ multiplet with $SO(4)$ singlet, which is common to
the fundamental $16$ OPEs.
It turns out that 
the OPE between them satisfies
the following result
\bea
\Phi_{\frac{3}{2}}^{(1),i}(z)\,\Phi_{0}^{(2)}(w)
& = & 
\frac{1}{(z-w)^3}\,Q^{(\frac{3}{2}),i}_{\frac{3}{2}}(w)
\nonu\\
&+ &
\frac{1}{(z-w)^2}\,
\Bigg[\,
\frac{2}{3}\,\partial Q^{(\frac{3}{2}),i}_{\frac{3}{2}}
+Q^{(\frac{5}{2}),i}_{\frac{3}{2}}
-\frac{(k-N)}{3(2+k+N)}\, Q^{(\frac{5}{2}),i}_{\frac{1}{2}}
\,\Bigg](w)
\nonu\\
&+ &
\frac{1}{(z-w)}\,
\Bigg[\,
\frac{1}{4}\,\partial^2 Q^{(\frac{3}{2}),i}_{\frac{3}{2}}
+\frac{3}{5}\,\partial Q^{(\frac{5}{2}),i}_{\frac{3}{2}}
+Q^{(\frac{7}{2}),i}_{\frac{3}{2}}
\,\Bigg](w)
+\cdots\,.
\label{3half102}
\eea
The third and second order poles can be written in terms of
the known (higher spin) currents.
This implies that there is no new primary field in these poles. 
On the other hand,
the first order pole of (\ref{3half102}) has the following 
quasi primary field
\bea
Q^{(\frac{7}{2}),i}_{\frac{3}{2}}
& = &
w_{1,\frac{7}{2}}\, \Phi^{(3),i}_{\frac{1}{2}} + \cdots.
\label{spin7halfquasi}
\eea
A new higher spin-$\frac{7}{2}$ current of the third
${\cal N}=4$ higher spin multiplet  arises in
(\ref{spin7halfquasi}) and the remaining $86$ terms are given in
(\ref{morequasi1}) as before.
We can easily check that the OPE between
the spin-$\frac{3}{2}$ currents $G^i(z)$ and
the higher spin-$3$ current $\Phi_0^{(3)}(w)$
of the third ${\cal N}=4$ multiplet appearing in
(\ref{spin3ofthird}) provides the above higher spin-$\frac{7}{2}$
current
$\Phi^{(3),i}_{\frac{1}{2}}(w)$
at the first order pole (with minus sign).
That is, the relevant OPE for this computation
is given in Appendix $C$ of
\cite{AK1509}. 
See also Appendix $G$ where the corresponding commutator is
presented. Alternatively,
we can read off the corresponding OPEs. More precisely, we have
the third equation of (\ref{N4primary}).
This is one of the consistency checks for the validity of
the OPE in
(\ref{3half102}).

In this case, the OPE described in (\ref{3half102})
contains the
components of the first three ${\cal N}=4$ higher spin multiplets
as before. We expect that due to the $SO(4)$ vector index
$i$ in (\ref{spin7halfquasi}), the ${\cal N}=4$
supersymmetric version
of (\ref{spin7halfquasi}) can combine with the triple product of  
the fermionic coordinates in order to preserve the $SO(4)$ singlet
condition of the OPE.
See (\ref{singleOPE}).

\subsection{Other remaining fundamental eleven OPEs }

Let us calculate further OPEs.
The OPE
between the higher spin-$2$ current
$\Phi_{1}^{(1),ij}(z)$
of the first ${\cal N}=4$ multiplet
with $SO(4)$ adjoint index and the
higher spin-$2$ current $\Phi_{0}^{(2)}(w)$
of the second
${\cal N}=4$ multiplet with $SO(4)$ singlet we described before
can be described as
\bea
\Phi_{1}^{(1),ij}(z)\,\Phi_{0}^{(2)}(w)
& = & 
\frac{1}{(z-w)^{2}}\,
 Q^{(2),ij}_{1}(w)
 + 
\frac{1}{(z-w)}\,\Bigg[
\frac{1}{2}\partial Q^{(2),ij}_{1} 
+Q^{(3),ij}_{1}
\Bigg](w)
+\cdots,
\label{1102}
\eea
where
we have the following quasi primary field with the footnote
\ref{BigJ}
\bea
Q^{(2),ij}_{1}(w)
& = &
w_{1,2}\,\Phi^{(1),ij}_1(w)
 + 
w_{6,2}\, \tilde{\Phi}^{(1),ij}_{1}(w) + \cdots
\,.
\label{quasispintwo}
\eea
The abbreviated six terms of (\ref{quasispintwo})
are given again in (\ref{morequasi1}).
There is no new primary field in this OPE (\ref{1102}). 
Of course, it is obvious that the other components
$\Phi_1^{(3),ij}(w)$
of the third ${\cal N}=4$
higher spin
multiplet cannot appear in this OPE because the spins of the
left hand side are not enough to allow us to have
this higher spin-$4$
currents. 
We will see that the new higher spin-$4$ currents
$\Phi_1^{(3),ij}$ can appear in the OPEs between
the different higher spin
currents.
It is obvious to see
that the composite field of spin-$3$ can arise in the
first order pole of (\ref{1102}).

Let us consider the next OPE.
The OPE
between the higher spin-$
\frac{3}{2}$ current $\Phi_{\frac{1}{2}}^{(1),i}(z)$
of the first ${\cal N}=4$ multiplet
with $SO(4)$ vector index and the
higher spin-$2$ current $\Phi_{0}^{(2)}(w)$ of the second
${\cal N}=4$ multiplet with $SO(4)$ singlet
can be summarized as
follows:
\bea
\Phi_{\frac{1}{2}}^{(1),i}(z)\,\Phi_{0}^{(2)}(w)
& = & \frac{1}{(z-w)}\,Q^{(\frac{5}{2}),i}_{\frac{1}{2}}(w)+\cdots\,.
\label{half102}
\eea
In this case, the first order pole
can be expressed as the known
(higher spin) currents (no new primary field occurs) and
the quasi primary field in (\ref{half102})
contains the higher spin-$\frac{5}{2}$ current
of the first ${\cal N}=4$ higher spin multiplet
as follows:
\bea
Q^{(\frac{5}{2}),i}_{\frac{1}{2}}(w)
& = &
w_{1,\frac{5}{2}}\,\Phi^{(1),i}_{\frac{3}{2}}(w) + \cdots.
\label{spin5halfquasi}
\eea
We can find other $13$ terms of (\ref{spin5halfquasi})
in (\ref{morequasi1}) as before.

Finally, the OPE between the lowest higher spin currents
of the first and second ${\cal N}=4$ higher spin multiplets
can be written in terms of
\bea
\Phi_{0}^{(1)}(z)\,\Phi_{0}^{(2)}(w)
& = & \frac{1}{(z-w)^{2}}\,Q^{(1)}_{0}(w)+\cdots\,,
\label{0102}
\eea
where the quasi primary field  appears in (\ref{0102})
and is nothing but
the higher spin-$1$ current
as follows:
\bea
Q^{(1)}_{0}(w)= d_{1}^{0,2}\,\Phi_{0}^{(1)}(w)\,,
\label{Q0}
\eea
where  the structure constant appearing in (\ref{Q0})
is given by (\ref{dcoeff}).

Then we observe
that the OPEs in (\ref{1102}), (\ref{half102}) and (\ref{0102})
contain the components of the first ${\cal N}=4$ higher spin
multiplet. No new primary fields occur.

Therefore, the $16$ fundamental OPEs, which are
the OPEs between the first ${\cal N}=4$ higher spin multiplet
and the lowest component of the second ${\cal N}=4$
higher spin multiplet,
can be
obtained from (\ref{2102}),
(\ref{3half102}), (\ref{1102}), (\ref{half102}) and (\ref{0102}).
The total number of poles in these OPEs is given by $11$. 
The ${\cal N}=4$ supersymmetry allows us to read off the
remaining $240=(16^2-16)$ OPEs
by going to ${\cal N}=4$ superspace in next
subsection \footnote{Compared to the total
  $136$ OPEs in section $2.3$,
  there are $256=16^2$ OPEs because we are considering {\it two}
different ${\cal N}=4$ higher spin multiplets.}.

\subsection{The ${\cal N}=4$ OPE between the 
first and the second ${\cal N}=4$ higher spin multiplets}

We would like to express the OPEs in component approach
by using the ${\cal N}=4$ superspace approach.
Again the fundamental $16$ OPEs (five kinds of OPEs
between the first ${\cal N}=4$ higher spin multiplet
and the lowest component of the second ${\cal N}=4$
higher spin multiplet)
are given by 
(\ref{2102}),
(\ref{3half102}), (\ref{1102}), (\ref{half102}) and (\ref{0102})
and will determine the remaining $240$ OPEs by using the
${\cal N}=4$ supersymmetry described before. 
That is, we can generalize them in ${\cal N}=4$ superspace
by taking the following replacements \cite{AK1509}
for the components of both
${\cal N}=4$ stress energy tensor ${\bf J}$ in the footnote \ref{BigJ}
and
the higher spin multiplet ${\bf \Phi}^{(s)}$ in
(\ref{Phiexp})
\bea
U(w) & \rightarrow &  \pa \, {\bf J}(Z_2),
\qquad
\Gamma^i(w)  \rightarrow  -i \, D^i {\bf J}(Z_2)
\equiv -i \, {\bf J}^i(Z_2),  
\nonu \\
T^{ij}(w) & \rightarrow & -\frac{i}{2!} \, \varepsilon^{ijkl} \,
D^k D^l {\bf J}(Z_2) \equiv -\frac{i}{2!} \,  \varepsilon^{ijkl} \,
{\bf J}^{kl}(Z_2),
\nonu \\
\tilde{G}^i(w) & \rightarrow & \frac{1}{3!} \,
\varepsilon^{ijkl} \, D^j D^k D^l {\bf J}(Z_2)
\equiv \frac{1}{3!} \,\varepsilon^{ijkl}  \, {\bf J}^{jkl}(Z_2),
\nonu \\
\tilde{L}(w)  & \rightarrow & \frac{1}{2 \cdot 4!} \,
\varepsilon^{ijkl} \, D^{i} D^j D^k D^l {\bf J}(Z_2)
\equiv \frac{1}{2 \cdot 4!} \,\varepsilon^{ijkl}  \, {\bf J}^{ijkl}(Z_2),
\nonu \\
\Phi_{0}^{(s)}(w) & \rightarrow &  
    {\bf \Phi}^{(s)}(Z_2),
    \qquad
\Phi_{\frac{1}{2}}^{(s),i}(w)  \rightarrow   D^i
    {\bf \Phi}^{(s)}(Z_2),  
\nonu \\
\Phi_{1}^{(s),ij}(w) & \rightarrow & -\frac{1}{2!} \,
 \varepsilon^{ijkl} \, D^k D^l {\bf \Phi}^{(s)}(Z_2),
\qquad
\Phi_{\frac{3}{2}}^{(s),i}(w)  \rightarrow  -\frac{1}{3!} \,
\varepsilon^{ijkl} \, D^j D^k D^l {\bf \Phi}^{(s)}(Z_2),
\nonu \\
\Phi_{2}^{(s)}(w)  & \rightarrow & \frac{1}{ 4!} \,
\varepsilon^{ijkl} \, D^{i} D^j D^k D^l {\bf \Phi}^{(s)}(Z_2),
\label{comptosuper}
\qquad s=1,2,3, \cdots,
\eea
and putting the relevant fermionic coordinates together with
singular terms.
Due to the different number of fermionic coordinates,  
the total $11$ singular terms in the above fundamental OPEs
arise in ${\cal N}=4$ superspace independently
\footnote{We obtain the components of the
  ${\cal N}=4$ (higher spin)multiplet as follows \cite{AK1509}.
  For the lowest component, we simply put the fermionic coordinates
  to zero. For the second component with index $i$,
  we multiply the super derivative
  $D^i$ and then take the fermionic coordinates as zero.
  For the third component with $i=1,j=2$, we act
  the super derivatives $D^3 \,D^4$ and
  take $\theta_1^i=0=\theta_2^i$ with minus sign.
  For the fourth component with $i=1$, we can multiply
  $ D^2 \,D^3 \,D^4$ and take $\theta_1^i=0=\theta_2^i$
  with minus sign.
For the last component,  we can multiply
$ D^1\, D^2 \,D^3 \,D^4$ and take $\theta_1^i=0=\theta_2^i$.
For the other indices of
the third and fourth compnents, similar analysis can be done. 
\label{componentreduce}}.

Then the single ${\cal N}=4$ super OPE
between the first and the second  ${\cal N}=4$
higher spin multiplets, like as in (\ref{super1super1}),
can be summarized by 
\bea
{\bf \Phi}^{(1)}(Z_{1})\,{\bf \Phi}^{(2)}(Z_{2})  &=& 
\frac{\theta_{12}^{4-0}}{z_{12}^4}\,\Bigg[\,
 {\bf Q}^{(1)}_{2}
 -(6\,p_1+4\,p_2)\,{\bf Q}^{(1)}_{0} 
\,\Bigg](Z_2)
+\frac{\theta_{12}^{4-i}}{z_{12}^3}\, {\bf Q}^{(\frac{3}{2}),i}_{\frac{3}{2}}(Z_2)
\nonu\\
&
+& \frac{\theta_{12}^{4-0}}{z_{12}^3}\,
\Bigg[\, 
\partial {\bf Q}^{(1)}_{2} + {\bf Q}^{(2)}_{2}
 -p_2\, \partial {\bf Q}^{(1)}_{0}
\,\Bigg](Z_2) 
+\frac{1}{z_{12}^2}\, {\bf Q}^{(1)}_{0}(Z_2)
+\frac{\theta_{12}^{4-ij}}{z_{12}^2}\, {\bf Q}^{(2),ij}_{1}(Z_2)
\nonu\\
&
+ & \frac{\theta_{12}^{4-i}}{z_{12}^2}\,\Bigg[\, 
\frac{2}{3}\,\partial   {\bf Q}^{(\frac{3}{2}),i}_{\frac{3}{2}}
+ {\bf Q}^{(\frac{5}{2}),i}_{\frac{3}{2}}
-\frac{(k-N)}{3(2+k+N)}\,{\bf Q}^{(\frac{5}{2}),i}_{\frac{1}{2}}
\,\Bigg](Z_2)
\nonu\\
&
+& \frac{\theta_{12}^{4-0}}{z_{12}^2}\,
\Bigg[\, 
\frac{1}{2}\,\partial^2 {\bf Q}^{(1)}_{2} + \frac{3}{4}\,\partial{\bf Q}^{(2)}_{2} +{\bf Q}^{(3)}_{2}
-p_2 (
2 {\bf \Phi}^{(1)} {\bf \Phi}^{(2)} + d_{1}^{0,2}\,
{\bf J}^{4-0} {\bf \Phi}^{(1)}
)
\,\Bigg](Z_2)
\nonu\\
&
+&\frac{\theta_{12}^{i}}{z_{12}}\, {\bf Q}^{(\frac{5}{2}),i}_{\frac{1}{2}}(Z_2)
+\frac{\theta_{12}^{4-ij}}{z_{12}}\,\Bigg[\, 
\frac{1}{2}\,\partial   {\bf Q}^{(2),ij}_{1}
+   {\bf Q}^{(3),ij}_{1}
\,\Bigg](Z_2)
\nonu \\
& + & \frac{\theta_{12}^{4-i}}{z_{12}}\,\Bigg[\, 
\frac{1}{4}\,\partial^2   {\bf Q}^{(\frac{3}{2}),i}_{\frac{3}{2}}
+ \frac{3}{5}\,\partial {\bf Q}^{(\frac{5}{2}),i}_{\frac{3}{2}}
+{\bf Q}^{(\frac{7}{2}),i}_{\frac{3}{2}}
\,\Bigg](Z_2)
\nonu\\
&
+&\frac{\theta_{12}^{4-0}}{z_{12}}\,
\Bigg[\, 
  \frac{1}{6}\,\partial^3 {\bf Q}^{(1)}_{2} + \frac{3}{10}\,\partial^2 {\bf Q}^{(2)}_{2} +\frac{2}{3}\, \pa \,
       {\bf Q}^{(3)}_{2} +{\bf Q}^{(4)}_{2}
\nonu\\
&
-& p_2(2\, \partial {\bf\Phi}^{(1)} {\bf\Phi}^{(2)}
+{\bf\Phi}^{(1)} \partial {\bf\Phi}^{(2)} + d_{1}^{0,2}\, \partial {\bf J}^{4-0} {\bf\Phi}^{(1)}
)
\,\Bigg](Z_2)
+\cdots.
\label{singleOPE}
\eea
As before, the coefficients $p_1$ and $p_2$ appearing in several
places are given in (\ref{p1p2})
and the structure constant $d_{1}^{0,2}$ is given in (\ref{dcoeff}). 
We present the quasi primary super fields in ${\cal N}=4$ superspace
in (\ref{morequasi}).
The new primary  third ${
  \cal N}=4$ higher spin multiplet in (\ref{singleOPE})
arise in the following quasi primary field
\bea
    {\bf Q}_2^{(3)}(Z_2) & = & d_{1}^{2,2} \, {\bf \Phi}^{(3)}(Z_2)
    + \cdots,
\label{Spin3quasi}
\eea
where
the abbreviated $25$ terms are given in (\ref{morequasi}). 
This is ${\cal N}=4$ supersymmetric
version of previous component result in
(\ref{spin3ofthird}).
Moreover, we have
the descendant field of ${\bf \Phi}^{(3)}(Z_2)$
\bea
{\bf Q}^{(\frac{7}{2}),i}_{\frac{3}{2}}(Z_2) & = &
d_{1}^{\frac{3}{2},1}\,  D^i {\bf \Phi}^{(3)}(Z_2) + \cdots,
\label{Spin7halfquasi}
\eea
which is the ${\cal N}=4$ supersymmetric
version of the component result in (\ref{spin7halfquasi}).

Therefore, the OPE in (\ref{singleOPE})
contains the first three ${\cal N}=4$ higher spin multiplets
in addition to the stress energy tensor ${\bf J}$:
${\bf \Phi}^{(1)}$, ${\bf \Phi}^{(2)}$ and ${\bf \Phi}^{(3)}$.
Note that the structure constants appearing in (\ref{singleOPE})
are fixed by using the various Jacobi identities in the component
approach.
As explained before, all the $256$ OPEs can be read off
from the OPE (\ref{singleOPE}) with the help of the footnote
\ref{componentreduce}. We will describe some of them very briefly
in next
subsection.

\subsection{The new higher spin currents $\Phi_1^{(3),lm}$,
  $ \Phi_{\frac{3}{2}}^{(3),l}$ and $\Phi_2^{(3)}$, 
  of spin $4, \frac{9}{2},5$ }

So far, we have discussed about the $16$ fundamental OPEs.
How do we observe (or obtain) the remaining components of
the third ${\cal N}=4$ higher spin multiplet?
Because the third ${\cal N}=4$ higher spin
multiplet appears both in (\ref{Spin3quasi}) and in
(\ref{Spin7halfquasi}), we can focus on these poles associated with
them in the OPE
(\ref{singleOPE}).

For example,
we can observe that
the higher spin-$4$ currents $\Phi_1^{(3),mn}(w)$
arise the OPEs between the higher spin-$2$ currents $\Phi_1^{(1),ij}(z)$
and
the higher spin-$3$ currents  $\Phi_1^{(2),kl}(w)$
\footnote{In this subsection,
  we do not care about the exact numerical factors and signs. We will
  demonstrate how we can observe
  the existence of the remaining components of
  the third ${\cal N}=4$ higher spin multiplet we do not
  see  so far. We refer to \cite{AK1509} for further detailed
descriptions.}.
Let us fix the indices: $m=1,n=2$ and $i=k=3, j=1,l=2$.
The former higher spin currents can be obtained by multiplying
the super derivatives $D_1^{2} \, D_1^{4}$
(up to the overall numerical
factor) into
${\bf \Phi}^{(1)}(Z_1)$ and putting the fermionic coordinates to
zero according to (\ref{comptosuper}).
On the other hand,
the latter higher spin currents can be obtained by multiplying
the super derivatives $D_2^{1} \, D_2^{4}$
(up to the overall numerical
factor) into
${\bf \Phi}^{(2)}(Z_2)$ and putting the fermionic coordinates to
zero.
Then it is easy to see, from the right hand side of the OPE in
(\ref{singleOPE}),
that by splitting the two
super derivatives $D_2^{1} \, D_2^{4}$
into the piece of $\frac{\theta_{12}^{4-p}}{z_{12}}$
and the quasi primary field (\ref{Spin7halfquasi}) respectively,
we have  $\frac{\theta_{12}^{4-13}}{z_{12}}$ and
$D_2^{4} \, D_2^{p=3} \, {\bf \Phi}^{(3)}(Z_2)$.
Then the pole term can be reduced to $\frac{1}{(z-w)}$
by acting $D_1^{2} \, D_1^{4}$ and 
we arrive at 
$\Phi_1^{(3),12}(w)$ after putting
the vanishing fermionic coordinates,
where we ignore all the numerical factors
as well as signs.
We can see the corresponding OPE in (\ref{ope2}), where
all the nonlinear terms (and some linear terms) disappear.

Similarly,
the higher spin-$\frac{9}{2}$ currents
$\Phi_{\frac{3}{2}}^{(3),l=1}(w)$
 can be determined from
 the OPEs between the higher spin-$2$ currents $\Phi_1^{(1),ij}(z)$
and
the higher spin-$3$ currents  $\Phi_{\frac{3}{2}}^{(2),k}(w)$,
where we fix the indices $i=k=4$ and $j=1$.
As before, 
the former can be obtained by multiplying
the super derivatives $D_1^{2} \, D_1^{3}$
into
${\bf \Phi}^{(1)}(Z_1)$ and putting the fermionic coordinates to
zero.
The latter can be obtained by multiplying
the super derivatives $D_2^{1} \, D_2^{2}\, D_2^{3}$
 into
${\bf \Phi}^{(2)}(Z_2)$ and putting the fermionic coordinates to
zero.
After we act the super derivatives
$D_1^{2} \, D_1^{3} \, D_2^{1}$ on 
the pole $\frac{\theta_{12}^{4-n}}{z_{12}}$ and
the super derivatives $ D_2^{2}\, D_2^{3}$ on
the quasi primary field (\ref{Spin7halfquasi}), we
are left with 
$\Phi_{\frac{3}{2}}^{(3),1}(w)$ (the index $n$ becomes $4$)
at the first order pole.
The corresponding OPE can be found in (\ref{ope2}) as before
\footnote{
Finally, 
 the higher spin-$5$ current  $\Phi_2^{(3)}(w)$
 can be obtained from
 the OPEs between the higher spin-$\frac{5}{2}$ currents
 $\Phi_{\frac{3}{2}}^{(1),i}(z)$
and
the higher spin-$\frac{7}{2}$ currents  $\Phi_{\frac{3}{2}}^{(2),j}(w)$.
Let us fix the index as $i=j=1$.
The former can be obtained by multiplying
the super derivatives $D_1^{2} \, D_1^{3}\, D_1^{4}$
into
${\bf \Phi}^{(1)}(Z_1)$ and putting the fermionic coordinates to
zero.
The latter can be obtained by multiplying
the super derivatives $D_2^{2} \, D_2^{3}\, D_2^{4}$
 into
${\bf \Phi}^{(2)}(Z_2)$ and putting the fermionic coordinates to
 zero.
After we act the super derivatives
$D_1^{2} \, D_1^{3} \, D_1^{4}$ on 
$\frac{\theta_{12}^{4-n}}{z_{12}}$ and
the super derivatives $ D_2^{2}\, D_2^{3}\, D_2^{4}$ on
the quasi primary field (\ref{Spin7halfquasi}), where the index $n$
becomes $1$, we
are left with 
$\Phi_{2}^{(3)}(w)$ at the first order pole. 
We can find the corresponding OPE in (\ref{ope2}).}.

Of course, these new higher spin-$4, \frac{9}{2}, 5$
currents appear in other OPEs in (\ref{ope2}).
All the component results can be obtained from (\ref{singleOPE})
by applying the super derivatives to both sides
and putting the fermionic coordinates to zero.
We present the $256$ OPEs under the large $(N,k)$ limit
in (\ref{ope2}).
The $(N,k)$ dependent
structure constants will be attached in the {\tt ancillary.nb}
file.
Its (anti)commutators appear in (\ref{PhionePhitwo}).
When we compare the results \cite{AKP1904} in an
orthogonal coset model with the results of this paper,
so far we do not observe any $SO(4)$ non-singlet
${\cal N}=4$ higher spin multiplet.



\section{Operator product expansion between
  the second ${\cal N}=4$ multiplet and itself with $N=5$}

Since we do not complete this OPE for general $N$,
we will present some parts of
the OPE with fixed $N=5$. The reason why we consider
this particular $N=5$ case is that this is the smallest value
for the existence of the new higher spin-$4$ current. 

\subsection{The new higher spin current of spin $4$ }

Compared to $N=3$ case where there is no new higher spin-$4$
primary current, 
the $N=5$ case leads to the following 
new higher spin-$4$ current with the corresponding OPE
\bea
\Phi_{2}^{(2)}(z)\,\Phi_{0}^{(2)}(w)
& = & \frac{1}{(z-w)^{6}}\,
8 \, \alpha \, e_{0}^{0,4}
+\frac{1}{(z-w)^{5}}\,
Q^{(1)}_{2}(w)
+\frac{1}{(z-w)^{4}}\,\Bigg[\,
\frac{3}{2}\,\partial Q^{(1)}_{2}
+Q^{(2)}_{2}
\,\Bigg](w)
\nonu\\
& + &
\frac{1}{(z-w)^{3}}\,\Bigg[\,
\partial^{2} Q^{(1)}_{2}
+\partial Q^{(2)}_{2}
+Q^{(3)}_{2}
\,\Bigg](w)
\nonu\\
& + &
\frac{1}{(z-w)^{2}}\,\Bigg[\,
\frac{5}{12}\,\partial^{3} Q^{(1)}_{2}
+\frac{1}{2}\, \partial^{2} Q^{(2)}_{2}
+\frac{5}{6}\,\partial Q^{(3)}_{2}
+Q^{(4)}_{2}
\,\Bigg](w)
\nonu\\
& + &
\frac{1}{(z-w)}\,\Bigg[\,
\frac{1}{8}\,\partial^{4} Q^{(1)}_{2}
+\frac{1}{6}\,\partial^{3} Q^{(2)}_{2}
+\frac{5}{14}\,\partial^{2} Q^{(3)}_{2}
+\frac{3}{4}\,\partial Q^{(4)}_{2}
+ Q^{(5)}_{2}
\,\Bigg](w)
\nonu 
\\
& + &
p_{1}\,\sum\limits _{n=3}^{4}
\frac{1}{(z-w)^{n}} \, \Big\{
\partial^{2} \Phi_{0}^{(2)} \,\Phi_{0}^{(2)}
\Big\}_{n}(w)+
p_{2}\,\sum\limits _{m=1}^{4}
\frac{1}{(z-w)^{m}}\, \Big\{
(L \Phi_{0}^{(2)})\,\Phi_{0}^{(2)}
\Big\}_{m}(w)
\nonu \\
& + & \cdots,
\label{section4ope}
\eea
where $p_1$ and $p_2$
are in \cite{AK1509} and by substituting the value of $s$
they are given by
\bea
p_1 & = & -\frac{(k - N) (55 + 29 k + 29 N + 3 k N)}
{(2 + k + N) (59 + 37 k + 
    37 N + 15 k N)}\,,
p_2  =  \frac{36 (k - N)}{(59 + 37 k + 37 N + 15 k N)}\,.
\label{p1p2sect4}
\eea
In particular, the new higher spin-$4$ current which is the
lowest component of the fourth ${\cal N}=4$ higher spin multiplet
arises
in the
quasi primary field appearing in (\ref{section4ope})
\bea
Q^{(4)}_{2}  & = &  
w_{0,4}\,\Phi_{0}^{(4)} + \cdots,
\label{chapter4new}
\eea
where the abbreviated parts of (\ref{chapter4new})
are given in (\ref{quasiappendixc}) with (\ref{coeffexp}).
The equivalent expression of (\ref{section4ope})
is found in (\ref{ope4}).
Other quasi primary fields can be found in Appendix $C$.
Compared to the one in \cite{AKP1904} where the orthogonal
coset model is described, the OPE in (\ref{section4ope})
looks similar (in the context of
the quasi primary fields and their coefficients in the OPE)
but the first order pole in $p_2$ term is new
in this paper.

\subsection{Other OPEs }

The remaining four kinds of fundamental OPEs, by following
the method in previous section,
are presented in Appendix $C$.
In order to apply the Jacobi identities used in previous sections,
we need to calculate the OPE between ${\bf \Phi}^{(1)}(Z_1)$ and 
${\bf \Phi}^{(3)}(Z_2)$ which is beyond the scope of this paper.
This is because the Jacobi identity between
the three higher spin currents, ${\bf \Phi}^{(1)}$,
${\bf \Phi}^{(2)}$ and ${\bf \Phi}^{(1)}$ leads to the above OPE
because the OPE between ${\bf \Phi}^{(1)}(Z_1)$ and 
${\bf \Phi}^{(2)}(Z_2)$ produces ${\bf \Phi}^{(3)}(Z_2)$.
As long as the ${\cal N}=4$ higher spin algebra is concerned,
we will obtain the higher spin algebra associated
with  the OPE ${\bf \Phi}^{(2)}(Z_1)$ and 
${\bf \Phi}^{(2)}(Z_2)$ and the OPE
 ${\bf \Phi}^{(1)}(Z_1)$ and 
${\bf \Phi}^{(3)}(Z_2)$ in section $7$.





\section{Summary of coset construction }

According to the observation of \cite{GG1305},
the appropriate limit on the parameters $N$ and $k$
should be taken from the  OPEs we have obtained so far
in order to relate them to the classical asymptotic symmetry
algebra in the $AdS_3$ higher spin theory with matrix
generalization.
If we further restrict the modes of the higher spin operators
in the (nonlinear)
classical asymptotic symmetry algebra to the wedge modes,
then we obtain the ${\cal N}=4$ higher spin algebra
which is called ``wedge'' subalgebra of ${\cal W}_{\infty}^{{\cal N}=4}
[\la]$ algebra.

\subsection{The large $(N,k)$ 't Hooft-like limit and
the nonlinear ${\cal W}_{\infty}^{{\cal N}=4}[\la]$ algebra}

From the explicit OPEs found in previous sections $2,3$ (and $4$),
we can take the large $(N,k)$ 't Hooft limit into the
various structure constants by keeping the
't Hooft-like coupling constant $\la \equiv \frac{(N+1)}{(k+N+2)}$
fixed.
Before we are taking the infinity limit of $N$
after writing down the $k$ in terms of $\la$ and $N$ first,
we are left with the nonlinear (and linear)
terms together with the power of
$\frac{1}{N}$
factor.
Now we take the infinity limit of $N$, then all the
nonlinear and some of the linear terms vanish. 
We present them in Appendices $D$ and $E$
and its (anti)commutators version will appear in Appendix $G$.
From these results of Appendices, we restrict to use the
wedge condition for the modes described before.
Then all the expressions with typewriter font in Appendix
$G$ disappear. 
Later, we will use these (anti)commutators to obtain
the ${\cal N}=4$ higher spin generators
in terms of oscillators which will satisfy the above
${\cal N}=4$ higher spin algebra.

There exists other limiting procedure which is the infinity limit
of level $k$. In this case, the central charge is given by
$c =6(N+1)$ around the discussion of (\ref{central}).
We obtain the free field construction \cite{EGR}
from the coset construction
we have described so far.
We will obtain the free field construction in section $6$,
where the central charge $c =6N$ which is different from the above.
Then we can expect that the extra piece in the coset construction
should reflect the extra central charge $c=6$ which is the
diffence between above two central charges.

\subsection{The ${\cal N}=2$
  wedge subalgebra of ${\cal W}_{\infty}^{{\cal N}=4}[\la]$ algebra
 }

Before we take the $U(2)$ matrix generalization of the
$AdS_3$ Vasiliev higher spin theory,
the original $AdS_3$
higher spin theory has ${\cal N}=2$ supersymmetry.
This implies that
we should observe the corresponding ${\cal N}=2$
higher spin algebra from the ${\cal N}=4$ higher spin algebra,
which obtained from the original higher spin theory
by adding Chan-Paton factors (or $U(2)$ matrix generalization).
In terms of oscillator formalism with explicit four $U(2)$
matrices, it will be obvious that we can choose 
the right candidate for the
${\cal N}=2$ higher spin generators,
among the ${\cal N}=4$ higher spin generators, which will
satisfy the above ${\cal N}=2$ higher spin algebra.

We expect that by taking the particular combinations
among the currents and the higher spin currents,
\bea
\Phi^{(1)}_0 & \rightarrow &    J,
\qquad
\frac{1}{\sqrt{2}} \left(G^2+\Phi^{(1),2}_{\frac{1}{2}} \right)
\rightarrow  G^+,
\nonu\\
\frac{1}{\sqrt{2}} \left( G^2-\Phi^{(1),2}_{\frac{1}{2}}\right) & \rightarrow
& G^-,
\qquad
L  \rightarrow    L,
\label{n2sca}
\eea
then the
${\cal N}=2$ wedge subalgebra of ${\cal W}_{\infty}^{{\cal N}=4}[\la]$
algebra can be realized.
The $SO(4)$ index $2$ in the above (\ref{n2sca})
is one of the choices
among four values $i=1,2,3,4$.
Once we take the spin-$1$ current $J$ of
${\cal N}=2$ superconformal algebra as the higher spin-$1$
current $\Phi_0^{(1)}$, then the two spin-$\frac{3}{2}$ currents
$G^{\pm}$ of ${\cal N}=2$ superconformal algebra should be the linear
combination like as (\ref{n2sca}) because the commutator
between the higher spin-$1$ current $\Phi_0^{(1)}$ and
the spin-$\frac{3}{2}$
currents $G^{i}$ gives the higher spin-$\frac{3}{2}$ currents
$\Phi_{\frac{1}{2}}^{(1),i}$ and vice versa.
See also (\ref{N4primary}).
We can easily check that other (anti)commutators are satisfied
by using the left hand side of (\ref{n2sca}). See also
(\ref{n2superconformal}).
In \cite{GG1305}, the choice for two supersymmetry generators
$G^{\pm}$ is taken from the ones having the off diagonal $U(2)$
matrices. 

Moreover, the first ${\cal N}=2$ higher spin multiplet
in the context of ${\cal N}=2$ wedge subalgebra
of ${\cal W}_{\infty}^{{\cal N}=4}[\la]$ algebra
can be obtained from the following combinations
\bea
-\frac{\sqrt{3}}{8}
\, \Phi^{(2)}_0 &  \rightarrow  & W^{2\, 0},
\qquad
\frac{\sqrt{3}}{4 \sqrt{2}}
\, \left( \tilde{\Phi}^{(1),2}_{\frac{3}{2}}  -
\frac{1}{2}\,\Phi^{(2),2}_{\frac{1}{2}} \right)   \rightarrow   W^{2\, +},
\nonu\\
\frac{\sqrt{3}}{4 \sqrt{2}}
\, \left( \tilde{\Phi}^{(1),2}_{\frac{3}{2}}  +
\frac{1}{2}\,\Phi^{(2),2}_{\frac{1}{2}} \right) &  \rightarrow & W^{2\, -},
\qquad
-\frac{\sqrt{3}}{8}\, \tilde{\Phi}^{(1)}_{2}  \rightarrow  W^{2\, 1}.
\label{Spin2identification}
\eea
Here we are using the notation of \cite{Romans}.
Note that the fermionic components of first and the second of
${\cal N}=4$ higher spin multiplets are mixed.
Similarly,
 the second ${\cal N}=2$ higher spin multiplet
can be obtained from the following combinations
\bea
-\frac{1}{64} \, \Phi^{(3)}_0&  \rightarrow &  c^{3}_{22} \, W^{3\, 0},
\qquad
\frac{1}{64\sqrt{2}}\, \left(3\, \tilde{\Phi}^{(2),2}_{\frac{3}{2}}  -
\Phi^{(3),2}_{\frac{1}{2}}
 \right)   \rightarrow   c^{3}_{22}\, W^{3\, +},
\nonu\\
\frac{1}{64\sqrt{2}}\, \left(3\, \tilde{\Phi}^{(2),2}_{\frac{3}{2}}  +
\frac{1}{2}\,\Phi^{(3),2}_{\frac{1}{2}} \right) &  \rightarrow &  c^{3}_{22}\, W^{3\, -},
\qquad
-\frac{3}{64}
\, \tilde{\Phi}^{(2)}_{2}  \rightarrow    c^{3}_{22} \, W^{3\, 1},
\label{Spin3identification}
\eea
where $c_{22}^3$ is the structure constant appearing in
the higher spin-$3$ current in the OPE between the higher spin-$2$
current and itself.
We will come back to this issue together with
Appendix $L$ in section $7$.

Then from these observations,
we can further generalize to the next ${\cal N}=2$ higher spin
multiplet from the ${\cal N}=4$ higher spin multiplets as follows: 
\bea
&&(s=0,\,\,\, \frac{1}{2},\frac{1}{2},\frac{1}{2},\frac{1}{2},\,\,\,\,
1,1,1,1,1,1,
\,\,\,\, \frac{3}{2},\frac{3}{2}:G^2,\frac{3}{2},\frac{3}{2},\,\,\,\,
2:L),
\nonu \\
&&(s=1:\Phi_{0}^{(1)},\,\,\,\,
\frac{3}{2},\frac{3}{2}:\Phi_{\frac{1}{2}}^{(1),2},
\frac{3}{2},\frac{3}{2}, \,\,\,\, 2,2,2,2,2,2, \,\,\,\,
\frac{5}{2},\frac{5}{2}:\tilde{\Phi}_{\frac{3}{2}}^{(1),2},\frac{5}{2},
\frac{5}{2},\,\,\,\,3:\tilde{\Phi}_{2}^{(1)}),
\nonu \\
&&(s=2:\Phi_{0}^{(2)},\,\,\,\,
\frac{5}{2},\frac{5}{2}:\Phi_{\frac{1}{2}}^{(2),2},
\frac{5}{2},\frac{5}{2}, \,\,\,\, 3,3,3,3,3,3, \,\,\,\,
\frac{7}{2},\frac{7}{2}:\tilde{\Phi}_{\frac{3}{2}}^{(2),2},\frac{7}{2},
\frac{7}{2}, \,\,\,\,4:\tilde{\Phi}_{2}^{(2)}), \nonu \\
&&(s=3:\Phi_{0}^{(3)}, \,\,\,\,
\frac{7}{2},\frac{7}{2}:\Phi_{\frac{1}{2}}^{(3),2},
\frac{7}{2},\frac{7}{2}, \,\,\,\, 4,4,4,4,4,4, \,\,\,\,
\frac{9}{2},\frac{9}{2}:\tilde{\Phi}_{\frac{3}{2}}^{(3),2},\frac{9}{2},
\frac{9}{2},\,\,\,\,5:\tilde{\Phi}_{2}^{(3)}),\nonu \\
&&(s=4:\Phi_{0}^{(4)}, \,\,\,\,
\frac{9}{2},\frac{9}{2}:\Phi_{\frac{1}{2}}^{(4),2},
\frac{9}{2},\frac{9}{2}, \,\,\,\, 5,5,5,5,5,5, \,\,\,\,
\frac{11}{2},\frac{11}{2}:\tilde{\Phi}_{\frac{3}{2}}^{(4),2},
\frac{11}{2},\frac{11}{2}, \,\,\,\,6:\tilde{\Phi}_{2}^{(4)}), \,\,\,\,
\nonu \\
&& \cdots \,\,\,\, .
\label{list}
\eea
We list the $16$ (higher spin)currents with its spins
and we denote the corresponding ${\cal N}=2$ quantities
by the fields.
Each ${\cal N}=2$ higher spin multiplet
of superspin $s$ can be obtained from
two different ${\cal N}=4$ higher spin multiplets
of superspin $s$ and $(s-1)$ in (\ref{list}).
The first two, $\Phi_0^{(s)}$ and $\Phi_{\frac{1}{2}}^{(s),2}$
can be taken from the former,   
while the last two $\tilde{\Phi}_{\frac{3}{2}}^{(s-1),2}$
and $\tilde{\Phi}_{2}^{(s-1)}$
 can be obtained from the latter.
 As observed before, the choice of $SO(4)$ vector index
in the fermionic field 
 is arbitrary.
In our convention, we take the index as $2$.  
In section $7$, we will see that the corresponding
fermionic quantity has a $2 \times 2$ identity matrix.

\section{Free field realization with vanishing 't Hooft-like
  coupling constant }

In this section, we describe the free field realization
associated with the ${\cal W}_{\infty}^{{\cal N}=4}[\la]$ algebra
at the vanishing 't Hooft coupling constant.
After identifying some of the $16$ currents of
the large ${\cal N}=4$ linear superconformal algebra  in terms of
free fields, the first and second ${\cal N}=4$ multiplets
are written in terms of free fields.
Then the (anti)commutators for general spins $s_1, s_2$
with manifest $SO(4)$ symmetry are obtained  explicitly \footnote{In this section we are using
  the notations for the spins $h_1$, $h_2$ and $h$,
  instead of using $s_1$, $s_2$, and $s$.}. 

\subsection{ Some of the $16$ currents in terms of free fields}

As in the section $1$,
there are
$2N$-free complex bosons transforming as
bifundamental $({\bf \bar{N}}, {\bf 2} )
\oplus ({\bf N}, {\bf \bar{2}})$ of $U(N) \times U(2)$
and
$2N$-free complex
fermions transforming similarly under the $U(N)
\times U(2)$.
We follow the notations used in \cite{EGR} and \cite{Odake}. 
The complex bosons are denoted by
$\pa \, \phi^{\bar{i},a}$ and $\pa \, \bar{\phi}^{i,\bar{a}}$
where $i, \bar{i} = 1, 2, \cdots, N$ and $a, \bar{a}=1,2$.
The complex fermions are denoted by
$\psi^{\bar{i},\alpha}$ and $\bar{\psi}^{i,\bar{\al}}$
where $i, \bar{i} = 1, 2, \cdots, N$ and $\al, \bar{\al}=1,2$.
By constructing the bilinear terms between these
free fields, we can obtain the following quasi (eight bosonic
and eight fermionic) primary fields
of spin $h$ as follows:
\bea
W^{\bar{\alpha} \beta}_{\mathrm{F},h}(z) 
&
=& n_{W_{\mathrm{F},h}}
\sum_{k=0}^{h-1} \sum_{i,\, \bar{\imath}=1}^N
\delta_{i,\bar{\imath}}(-1)^k
\left(\begin{array}{c}
h-1 \\  k \\
\end{array}\right)^2
\,(\,\partial^{h-k-1}\bar{\psi}^{i,\bar{\alpha}} \,
\partial^k\psi^{\bar{\imath},\beta}\,)(z) \, ,
\nonu \\
W^{\bar{a} b}_{\mathrm{B},h}(z) 
&
=& n_{W_{\mathrm{B},h}}
\sum_{k=0}^{h-2}\sum_{i,\, \bar{\imath}=1}^N 
\delta_{i,\bar{\imath}}\frac{(-1)^k}{(h-1)}
\left(\begin{array}{c}
h-1 \\  k \\
\end{array}\right)
\left(\begin{array}{c}
h-1 \\  k+1 \\
\end{array}\right)
\,(\,\partial^{h-k-1}\bar{\phi}^{i,\bar{a}}  \, \partial^{k+1}\phi^{\bar{\imath},b} \,)(z)\, , \nonu \\
Q^{\bar{ a} \alpha}_h(z) 
&
=& n_{Q_h} \sum_{k=0}^{h-\frac{3}{2}}\sum_{i,\, \bar{\imath}=1}^N 
\delta_{i,\bar{\imath}}(-1)^k
\left(\begin{array}{c}
h-\frac{3}{2} \\  k \\
\end{array}\right)
\left(\begin{array}{c}
h-\frac{1}{2} \\  k \\
\end{array}\right)
\,(\,\partial^{h-k-\frac{1}{2}}\bar{\phi}^{i,\bar{a}}\partial^k \psi^{\bar{\imath},\alpha}\,)(z)\, , \nonu \\
\bar{Q}^{a \bar{\alpha}}_h(z) 
&=& n_{\bar{Q}_h} \sum_{k=0}^{h-\frac{3}{2}} \sum_{i,\, \bar{\imath}=1}^N
\delta_{i,\bar{\imath}}(-1)^{h-\frac{3}{2}+k}
\left(\begin{array}{c}
h-\frac{3}{2} \\  k \\
\end{array}\right)
\left(\begin{array}{c}
h-\frac{1}{2} \\  k \\
\end{array}\right)
\,(\,\partial^{h-k-\frac{1}{2}}\phi^{\bar{\imath},a} \partial^k  \bar{\psi}^{i,\bar{\alpha}}\,)(z)\, .
\label{WWQQ}
\eea
Each field has four components, $11,12,21$, $22$.
As described in the section $1$, there are only four
spin-$1$ currents and  there are eight currents
of spins $s=\frac{3}{2}, 2, \frac{5}{2}, 3, \cdots$.
The overall normalizations are
taken from the ones in \cite{BPRSS} \footnote{The $q$ can be
  arbitrary value.}.
\bea
n_{W_{\mathrm{F},h}} =  \frac{2^{h-3}(h-1)!}{(2h-3)!!}\,q^{h-2},
n_{W_{\mathrm{B},h}}=\frac{2^{h-3}\,h!}{(2h-3)!!}\,q^{h-2},
n_{Q_h}=n_{\bar{Q}_h}=\frac{2^{h-1}(h-\frac{1}{2})!}{(2h-2)!!}\,q^{h-\frac{3}{2}}.
\label{nor}
\eea
The binomial coefficients used in (\ref{WWQQ}) are denoted by
$\left(\begin{array}{c}
n \\  k \\
\end{array}\right) \equiv \frac{n!}{k! \, (n-k)!}$.
The OPEs between the free fields are
given by
\bea
\pa \, \bar{\phi}^{i,\bar{a}}(z) \, \pa \,
\phi^{\bar{j}\,b}(w) = \frac{1}{(z-w)^2}\, \de^{i\bar{j}}
\, \de^{\bar{a} b} + \cdots,
\qquad
\bar{\psi}^{i,\bar{\al}}(z) \, \psi^{\bar{j},\beta}(w) =\frac{1}{(z-w)}\, \de^{i,\bar{j}} \, \de^{\bar{\al},\beta}+ \cdots.
\label{OPE}
\eea
The indices of (anti)fundamental representations of
$U(N)$ in (\ref{WWQQ}) are summed.

The stress energy tensor of spin $2$
is given by the usual Sugawara
construction from (\ref{WWQQ}) and (\ref{nor}) as follows:
\bea
L & = &
W^{11}_{\mathrm{B},2}+W^{22}_{\mathrm{B},2}+W^{11}_{\mathrm{F},2}+
W^{22}_{\mathrm{F},2}.
\label{Lterm}
\eea
The central charge is
\bea
c = N (2K+L)\Bigg|_{K=L=2} = 6N.
\label{centralfree}
\eea
This can be seen from the discussion of (\ref{central}).

We can think of the four spin-$\frac{3}{2}$ currents
of the large ${\cal N}=4$ superconformal algebra as the linear
combinations of $Q^{\bar{ a} \alpha}_{\frac{3}{2}}$ and
$\bar{Q}^{ a \bar{\alpha}}_{\frac{3}{2}}$.
By using the fact that they are primary fields under the stress energy
tensor (\ref{Lterm}) (corresponding to the second relation of
(\ref{16commanticomm})) and the defining equations for the
OPEs between the spin-$\frac{3}{2}$ currents
(corresponding to the fourth relation of (\ref{16commanticomm}))
which contain the
stress energy tensor in the first order pole,
we obtain the following four spin-$\frac{3}{2}$ currents
of the large ${\cal N}=4$ linear superconformal algebra
\bea
G^1
&
=&
-\frac{1}{2}\,(
Q^{11}_{\frac{3}{2}}
+i\sqrt{2}\,Q^{12}_{\frac{3}{2}}
+2i \sqrt{2}\,Q^{21}_{\frac{3}{2}}
-2\,Q^{22}_{\frac{3}{2}}
-2\,\bar{Q}^{11}_{\frac{3}{2}}
-2i \sqrt{2}\,Q^{12}_{\frac{3}{2}}
-i\sqrt{2}\,\bar{Q}^{21}_{\frac{3}{2}}
+\bar{Q}^{22}_{\frac{3}{2}}
)\,,
\nonu\\
G^2
&
=&
\frac{i}{2}\,(
Q^{11}_{\frac{3}{2}}
+2i\sqrt{2} \,Q^{21}_{\frac{3}{2}}
-2 \,Q^{22}_{\frac{3}{2}}
-2\, \bar{Q}^{11}_{\frac{3}{2}}
-2i\sqrt{2} \, \bar{Q}^{12}_{\frac{3}{2}}
+\bar{Q}^{22}_{\frac{3}{2}}
)\,,
\nonu\\
G^3
&
=&
\frac{i}{2}\,(
Q^{11}_{\frac{3}{2}}
+i\sqrt{2} \,Q^{12}_{\frac{3}{2}}
-2\,Q^{22}_{\frac{3}{2}}
-2\, \bar{Q}^{11}_{\frac{3}{2}}
-i \sqrt{2} \, \bar{Q}^{21}_{\frac{3}{2}}
+\bar{Q}^{22}_{\frac{3}{2}}
)\,,
\nonu\\
G^4
&
=&
\frac{1}{2}\,Q^{11}_{\frac{3}{2}}
+Q^{22}_{\frac{3}{2}}
+\bar{Q}^{11}_{\frac{3}{2}}
+\frac{1}{2} \, \bar{Q}^{22}_{\frac{3}{2}}
\,.
\label{fourG}
\eea
Of course, there are $4!$
different choices for $SO(4)$ vector indices.  
One of them is given by (\ref{fourG}).

The six spin-$1$ currents
of the large ${\cal N}=4$ superconformal algebra
appear
as the linear
combinations of $W^{\bar{\alpha} \beta}_{\mathrm{F},1}
$.
We can obtain the following spin-$1$ currents from the
previous OPEs between the spin-$\frac{3}{2}$ currents 
by focusing on the second order pole
\bea
\frac{1}{4q}\,(T^{14}+T^{23})
&
=&
i\,W^{11}_{\mathrm{F},1}
-\sqrt{2}\,W^{12}_{\mathrm{F},1}
-\sqrt{2}\,W^{21}_{\mathrm{F},1}
-i\,W^{22}_{\mathrm{F},1}
\,,
\nonu\\
\frac{1}{4q}\,(T^{13}+T^{42})
&
=&
-W^{11}_{\mathrm{F},1}
-i\,\sqrt{2}\,W^{21}_{\mathrm{F},1}
+W^{22}_{\mathrm{F},1}
\,,
\nonu\\
\frac{1}{4q}\,(T^{12}+T^{34})
&
=&
W^{11}_{\mathrm{F},1}
+i\,\sqrt{2}\,W^{12}_{\mathrm{F},1}
-W^{22}_{\mathrm{F},1}
\,.
\label{twoTT}
\eea
We can check that the remaining relations (the third, fifth and sixth in (\ref{16commanticomm}))
are satisfied.
Note that there are only four independent spin-$1$ currents
in the free field construction as mentioned before.
Three of them are given in
(\ref{twoTT}).  
We will observe the remaining one in next subsection soon.

\subsection{The first ${\cal N}=4$ higher spin multiplet
  in terms of free fields}

We consider the first ${\cal N}=4$ higher spin multiplet  
in terms of the free fields.
We can construct the following higher spin-$1$ current
by summing over the indices
\footnote{\label{lowestspincase}
  This can be generalized to $
  \Phi^{(s)}_{0}= 
q^{-s+2}\,
(\,
(s-1)\,W^{a a}_{\mathrm{B,s}}
-s\,W^{\alpha \alpha}_{\mathrm{F,s}}
\,)\,$ for general spin $s$ where we added the $q$ dependent rescaled
overall factor and the first term which vanishes for $s=1$.}
\bea
\Phi^{(1)}_{0}
=
4q\,(W^{11}_{\mathrm{F},1}+W^{22}_{\mathrm{F},1})\,.
\label{spinone}
\eea
The overall factor can be determined by the commutator
between this higher spin-$1$ current and the higher spin-$\frac{3}{2}$
current, which leads to the spin-$\frac{3}{2}$ current. 
See the second relation of (\ref{PhionePhione}).
Then the four components of $W_{F,1}^{\bar{\alpha}\beta}$
have the relations in (\ref{twoTT}) and (\ref{spinone}).

Because we have seen the four spin-$\frac{3}{2}$ currents
from (\ref{fourG}),
the remaining four spin-$\frac{3}{2}$ higher spin currents 
can be obtained by using the OPE between
the spin-$\frac{3}{2}$ currents (\ref{fourG}) and the
higher spin-$1$ current in (\ref{spinone}) (or the third relation of
(\ref{N4primary}))
as follows \footnote{In this case, we have
  the general $s$ dependent expression by multiplying the factor
  $-q^{-s+1}\,\frac{(2s-1)}{4}$ in (\ref{higherspin3half})
  and changing the upper index $(1)$ to $(s)$
  and the subscript $\frac{3}{2}$ to $(s+ \frac{1}{2})$.
  For example, we have
$
\Phi^{(s),1}_{\frac{1}{2}}
=-q^{-s+1}\,\frac{(2s-1)}{8}
\,(
Q^{11}_{s+\frac{1}{2}} + \cdots)$ where the abbreviated part
is the same as the seven terms with the subscript $s+\frac{1}{2}$ in
the
first relation of (\ref{higherspin3half})
and so on.}:
\bea
\Phi^{(1),1}_{\frac{1}{2}}
&
=&
\frac{1}{2}\,(
Q^{11}_{\frac{3}{2}}
+i\sqrt{2}\,Q^{12}_{\frac{3}{2}}
+2i \sqrt{2}\,Q^{21}_{\frac{3}{2}}
-2\,Q^{22}_{\frac{3}{2}}
+2\,\bar{Q}^{11}_{\frac{3}{2}}
+2i \sqrt{2}\,Q^{12}_{\frac{3}{2}}
+i\sqrt{2}\,\bar{Q}^{21}_{\frac{3}{2}}
-\bar{Q}^{22}_{\frac{3}{2}}
)\,,
\nonu\\
\Phi^{(1),2}_{\frac{1}{2}}
&
=&
-\frac{i}{2}\,(
Q^{11}_{\frac{3}{2}}
+2i\sqrt{2} \,Q^{21}_{\frac{3}{2}}
-2 \,Q^{22}_{\frac{3}{2}}
+2\, \bar{Q}^{11}_{\frac{3}{2}}
+2i\sqrt{2} \, \bar{Q}^{12}_{\frac{3}{2}}
-\bar{Q}^{22}_{\frac{3}{2}}
)\,,
\nonu\\
\Phi^{(1),3}_{\frac{1}{2}}
&
=&
-\frac{i}{2}\,(
Q^{11}_{\frac{3}{2}}
+i\sqrt{2} \,Q^{12}_{\frac{3}{2}}
-2\,Q^{22}_{\frac{3}{2}}
+2\, \bar{Q}^{11}_{\frac{3}{2}}
+i \sqrt{2} \, \bar{Q}^{21}_{\frac{3}{2}}
-\bar{Q}^{22}_{\frac{3}{2}}
)\,,
\nonu\\
\Phi^{(1),4}_{\frac{1}{2}}
&
=&
-\frac{1}{2}\,Q^{11}_{\frac{3}{2}}
-Q^{22}_{\frac{3}{2}}
+\bar{Q}^{11}_{\frac{3}{2}}
+\frac{1}{2}\,\bar{Q}^{22}_{\frac{3}{2}}
\,.
\label{higherspin3half}
\eea
The defining relations (\ref{OPE}) are used in this computation.
Compared to the spin-$\frac{3}{2}$ currents in (\ref{fourG}),
they are almost the same except some different signs in front of
fields.
In other words, 
the four components of $Q_{\frac{3}{2}}^{\bar{a} \alpha}$
and the four components of $\bar{Q}_{\frac{3}{2}}^{a \bar{\alpha}}$
have relations with eight (higher spin)currents 
via (\ref{fourG}) and (\ref{higherspin3half}).

We can proceed to obtain the next higher spin currents similarly.
The six spin-$2$ higher spin currents 
are determined by
 using the OPE between
the spin-$\frac{3}{2}$ currents (\ref{fourG}) and the
higher spin-$\frac{3}{2}$ current in (\ref{higherspin3half})
(or the fourth relation of
(\ref{N4primary}))
and it turns out that \footnote{
By multiplying the overall factor
  $-q^{-s+1}\,\frac{(2s-1)}{4}$ in (\ref{spintwo})
  and changing the upper index $(1)$ to $(s)$
  and the subscript $2$ to $s+1$, the general expression
  for the spin $s$ with subscript $1$
can be obtained.}
\bea
\Phi^{(1),12}_{1}
&
=&
2i\,W^{11}_{\mathrm{B},2}
-\sqrt{2}\,W^{12}_{\mathrm{B},2}
-2i\,\,W^{22}_{\mathrm{B},2}
+2i\,W^{11}_{\mathrm{F},2}
-2\sqrt{2}\,W^{12}_{\mathrm{F},2}
-2i\,W^{22}_{\mathrm{F},2}\,,
\nonu\\
\Phi^{(1),13}_{1}
&
=&
-2i\,W^{11}_{\mathrm{B},2}
+4\sqrt{2}\,W^{21}_{\mathrm{B},2}
+2i\,\,W^{22}_{\mathrm{B},2}
-2i\,W^{11}_{\mathrm{F},2}
+2\sqrt{2}\,W^{21}_{\mathrm{F},2}
+2i\,W^{22}_{\mathrm{F},2}\,,
\nonu\\
\Phi^{(1),14}_{1}
&
=&
2\,W^{11}_{\mathrm{B},2}
+i\sqrt{2}\,W^{12}_{\mathrm{B},2}
+4i\sqrt{2}\,\,W^{21}_{\mathrm{B},2}
-2\,W^{22}_{\mathrm{B},2}
-2\,W^{11}_{\mathrm{F},2}
-2i\sqrt{2}\,W^{12}_{\mathrm{F},2}
\nonu \\
& - & 2i\sqrt{2}\,W^{21}_{\mathrm{F},2}
+2\,W^{22}_{\mathrm{F},2}\,,
\nonu\\
\Phi^{(1),23}_{1}
&
=&
-2\,W^{11}_{\mathrm{B},2}
-i\sqrt{2}\,W^{12}_{\mathrm{B},2}
-4i\sqrt{2}\,\,W^{21}_{\mathrm{B},2}
+2\,W^{22}_{\mathrm{B},2}
-2\,W^{11}_{\mathrm{F},2}
-2i\sqrt{2}\,W^{12}_{\mathrm{F},2}
\nonu \\
& - & 2i\sqrt{2}\,W^{21}_{\mathrm{F},2}
+2\,W^{22}_{\mathrm{F},2}\,,
\nonu\\
\Phi^{(1),24}_{1}
&
=&
-2i\,W^{11}_{\mathrm{B},2}
+4\sqrt{2}\,W^{21}_{\mathrm{B},2}
+2i\,\,W^{22}_{\mathrm{B},2}
+2i\,W^{11}_{\mathrm{F},2}
-2\sqrt{2}\,W^{21}_{\mathrm{F},2}
-2i\,W^{22}_{\mathrm{F},2}\,,
\nonu\\
\Phi^{(1),34}_{1}
&
=&
-2i\,W^{11}_{\mathrm{B},2}
+\sqrt{2}\,W^{12}_{\mathrm{B},2}
+2i\,\,W^{22}_{\mathrm{B},2}
+2i\,W^{11}_{\mathrm{F},2}
-2\sqrt{2}\,W^{12}_{\mathrm{F},2}
-2i\,W^{22}_{\mathrm{F},2}\,.
\label{spintwo}
\eea
We have seen the spin-$2$ current from (\ref{Lterm})
and we will see the remaining higher spin-$2$ current  
in next subsection.
Therefore, we have explicit relations of four components of
$W_{F,2}^{\bar{\alpha}\beta}$ and four components of
 $W_{B,2}^{\bar{a}b}$ with eight (higher spin)currents.  

We can use
the OPE between
the spin-$\frac{3}{2}$ currents (\ref{fourG}) and the
higher spin-$2$ currents in (\ref{spintwo})
(or the fifth relation of
(\ref{N4primary})) and 
the four spin-$\frac{5}{2}$ higher spin currents 
are found \footnote{We can generalize them by
multiplying  $q^{-s+2}\,\frac{(2s-1)}{4}$ in (\ref{spin5half})
  and changing the upper index $(1)$ to $(s)$
  and the subscript $\frac{5}{2}$ to $s+\frac{3}{2}$.}
\bea
\tilde{\Phi}^{(1),1}_{\frac{3}{2}}
&
=&
-\frac{1}{2q}\,(
Q^{11}_{\frac{5}{2}}
+i\sqrt{2}\,Q^{12}_{\frac{5}{2}}
+2i\sqrt{2}\,Q^{21}_{\frac{5}{2}}
-2\,Q^{22}_{\frac{5}{2}}
-2\,\bar{Q}^{11}_{\frac{5}{2}}
-2i\sqrt{2}\,\bar{Q}^{12}_{\frac{5}{2}}
-i\sqrt{2}\,\bar{Q}^{21}_{\frac{5}{2}}
+\bar{Q}^{22}_{\frac{5}{2}}
)\,,
\nonu\\
\tilde{\Phi}^{(1),2}_{\frac{3}{2}}
&
=&
\frac{i}{2q}\,(\,
Q^{11}_{\frac{5}{2}}
+2i\sqrt{2}\,Q^{21}_{\frac{5}{2}}
-2\,Q^{22}_{\frac{5}{2}}
-2\,\bar{Q}^{11}_{\frac{5}{2}}
-2i\sqrt{2}\,\bar{Q}^{12}_{\frac{5}{2}}
+\bar{Q}^{22}_{\frac{5}{2}}
)\,,
\nonu\\
\tilde{\Phi}^{(1),3}_{\frac{3}{2}}
&
=&
\frac{i}{2q}\,(
Q^{11}_{\frac{5}{2}}
+i\sqrt{2}\,Q^{12}_{\frac{5}{2}}
-2\,Q^{22}_{\frac{5}{2}}
-2\,\bar{Q}^{11}_{\frac{5}{2}}
-i\sqrt{2}\,\bar{Q}^{21}_{\frac{5}{2}}
+\bar{Q}^{22}_{\frac{5}{2}}
)\,,
\nonu\\
\tilde{\Phi}^{(1),4}_{\frac{3}{2}}
&
=&
\frac{1}{2q}\,
(
Q^{11}_{\frac{5}{2}}
+2\,Q^{22}_{\frac{5}{2}}
+2\,\bar{Q}^{11}_{\frac{5}{2}}
+\bar{Q}^{22}_{\frac{5}{2}}
)\,.
\label{spin5half}
\eea
Note that the relative numerical coefficients
are the same as the ones in (\ref{higherspin3half}).
The remaining half of the higher spin-$\frac{5}{2}$ currents can be
obtained later.

Finally, the single higher spin-$3$ current 
of the first ${\cal N}=4$ higher spin multiplet
is obtained from
the OPE between
the spin-$\frac{3}{2}$ currents (\ref{fourG}) and the
higher spin-$\frac{5}{2}$ currents in (\ref{spin5half})
(or the sixth relation of
(\ref{N4primary})) \footnote{
\label{highestspincase}
  By
multiplying  $-q^{-s+1}\,\frac{(2s-1)}{4}$ in (\ref{freespin3})
  and changing the upper index $(1)$ to $(s)$
  and the subscript $3$ to $s+2$,
  we obtain the general expression
$\tilde{\Phi}^{(s)}_{2}= 
q^{-s}\,\frac{(2s-1)}{2}\,
(\,
W^{a a}_{\mathrm{B,s+2}}
+W^{\alpha \alpha}_{\mathrm{F,s+2}}
\,)\,$.}
\bea
\tilde{\Phi}^{(1)}_{2}
&
=&
-\frac{2}{q}\,(
W^{11}_{\mathrm{B},3}
+W^{22}_{\mathrm{B},3}
+W^{11}_{\mathrm{F},3}
+W^{22}_{\mathrm{F},3}
).
\label{freespin3}
\eea
This higher spin-$3$ current (\ref{freespin3})
looks like as the one in
(\ref{Lterm}).
The remaining seven higher spin-$3$ currents can be found in next
subsection.

The $16$
higher spin currents at $\la=1$ are given in Appendix $I$
\footnote{By considering the currents,
  $L$, 
$G^i$ ($i=1,3,4$),   
$-G^2$,  
$A^{-1}=\frac{1}{4q}(T^{14}-T^{23})$,   
$A^{-2}=\frac{1}{4q}(T^{13}-T^{42})$,   
$A^{-3}=-\frac{1}{4q}(T^{12}-T^{34})$,   
and higher spin currents,
  $-(-1)^s\,\Phi^{(s)}_0$, 
$-(-1)^s\,\Phi^{(s),i}_{\frac{1}{2}}$ ($i=1,3,4$), 
$(-1)^s\,\Phi^{(s),2}_{\frac{1}{2}}$, 
$(-1)^s\,\Phi^{(s),ij}_{1}$ ($i,j=1,3,4$),
$-(-1)^s\,\Phi^{(s),2i}_{1}$ ($i=1,3,4$), 
$(-1)^s\,\tilde{\Phi}^{(s),i}_{\frac{3}{2}}$ ($i=1,3,4$), 
$-(-1)^s\,\tilde{\Phi}^{(s),2}_{\frac{3}{2}}$,
and $(-1)^s\,\tilde{\Phi}^{(s)}_2$, we obtain the same higher
spin algebra
for $\la=0$.}.

\subsection{The second ${\cal N}=4$ higher spin multiplet
  in terms of free fields}

We can also obtain the second ${\cal N}=4$
higher spin multiplet which can be realized by the free
fields.

For the higher spin-$2$ current $\Phi_0^{(2)}$, we can take 
the four terms appearing in (\ref{Lterm}), where the
first two terms have the same coefficient and the last two
terms have the same coefficient.
We should use the primary condition and the defining relations
in (\ref{PhionePhitwo}) which contain
this higher spin-$2$ current with the known expressions in the
right hand side. For example, there are
the first, sixth and eleventh terms in (\ref{PhionePhitwo}).
Then we obtain the higher spin-$2$ current explicitly
and we present it in Appendix $H$.
Or we can use the fourth relation of (\ref{PhionePhione})
to obtain the higher spin-$\frac{5}{2}$ current first
and then use the fourth relation of (\ref{N4primary})
to obtain the higher spin-$2$ current.

For the other higher spin-$\frac{5}{2},3,\frac{7}{2},4$ currents,
$\Phi_{\frac{1}{2}}^{(2),i}$, $\Phi_{1}^{(2),ij}$,
$\tilde{\Phi}_{\frac{3}{2}}^{(2),i}$
and $\tilde{\Phi}_2^{(2)}$,
we apply the procedure done in previous subsection to those
higher spin currents. We can determine the higher
spin-$\frac{5}{2}$ currents from the third relation of
(\ref{N4primary}). Once we find these currents, we can substitute
them into the fourth relation of (\ref{N4primary}) and obtain
six higher spin-$3$ currents.
Now we can use the fourth relation to obtain the higher
spin-$\frac{7}{2}$ currents. Finally,
the sixth relation of (\ref{N4primary}) leads to the higher spin-$4$
current. The complete expressions for these $16$ higher spin currents
are given in Appendix $H$.

In this way, we can determine any ${\cal N}=4$ higher spin
multiplet in terms of free fields.

We can check that the subalgebra
from the complex bosons gives rise to $W_{\infty}[1]$ algebra
\cite{GG1205}.
We obtain the following spin-$2,3,4$ currents (where we fix $N=3$
and it is straightforward to consider the general $N$ case)
\bea
T
&=&
W^{11}_{\mathrm{B},2}+W^{22}_{\mathrm{B},2}\,,
\qquad
W
= \frac{1}{2 \sqrt{6} q} \,
\Bigg( W^{11}_{\mathrm{B},3}+W^{22}_{\mathrm{B},3}\Bigg)\,,
\nonu\\
U
&=& \frac{\sqrt{\frac{41}{42}}}{32 q^2} \,\Bigg(
W^{11}_{\mathrm{B},4}+W^{22}_{\mathrm{B},4}
-\frac{96\, q^2}{41}
\,\Bigg(TT-\frac{3}{10}\partial^2 T \Bigg) \Bigg)\,.
\label{gen}
\eea
Then it is easy to observe that the square of structure constant
$C_{33}^4$, which is defined as the coefficient in front of
spin-$4$ current in the OPE between the spin-$3$ current and itself,
is given by $(C_{33}^4)^2=\frac{3584}{123}$.
From the general expression appearing in
\cite{GG1205,HornfeckPLB,Hornfeck,Blumenhagenandother},
\bea
(C_{33}^4)^2=
\frac{64(c+2)(\lambda-3)(c(\lambda+3)+2(4\lambda+3)(\lambda-1))}
{(5c+22)(\lambda-2)(c(\lambda+2)+(3\lambda+2)(\lambda-1))}\,,
\label{334structureconstant}
\eea
we obtain $\la =1$ at $c=2\,N\,K =12$. Therefore, the nonlinear
$W_{\infty}[1]$
algebra (\ref{gen})
is realized by complex free bosons. See also the relevant
paper \cite{BKnpb}  
\footnote{Similarly,
  the subalgebra from the complex fermions is equivalent to
  $W_{\infty}[0]$ algebra.
  With the following spin-$1,2,3,4$ currents at the central charge $c=
  N \, L-1=5$ (by subtracting the contribution from the
  spin-$1$ current),
  \bea
J
&=&
W^{11}_{\mathrm{F},1}+W^{22}_{\mathrm{F},1}\,,
\qquad
T
=
W^{11}_{\mathrm{F},2}+W^{22}_{\mathrm{F},2}
-\frac{4\,q^2}{3}\,J J\,,
\nonu\\
W
&=& \frac{\sqrt{3}}{8 q}
\Bigg(W^{11}_{\mathrm{F},3}+W^{22}_{\mathrm{F},3}
-\frac{16\,q^2}{3}\,J T
-\frac{64\,q^2}{27}\,J J J \Bigg)\,,
\nonu\\
U
&=& \frac{\sqrt{\frac{47}{21}}}{32 q^2}
\Bigg( W^{11}_{\mathrm{F},4}+W^{22}_{\mathrm{F},4}
-\frac{64\,q^2}{15}\,J \partial^2 J
+\frac{32\,q^4}{5}\, \partial J \partial J
-8\,q^2\, J W
-\frac{64\,q^2}{3}\, J J T
\nonu\\
&-&
\frac{128\,q^6}{27}\, JJJJ
-\frac{296\,q^2}{47}
( T T
-\frac{3}{10}\,\partial^2 T)
\Bigg)
\,,
\label{gen1}
\eea
we can obtain 
$(C_{33}^4)^2=\frac{756}{47}$ by using (\ref{334structureconstant})
which leads to $\la=0$ at $c=5$.
Note that the OPEs between the spin-$1$ current and
the spin-$2,3,4$ currents do not have any singular terms.
 Therefore, the nonlinear
$W_{\infty}[0]$
 algebra (\ref{gen1}) generated by
 spins $2,3,4, \cdots$
 is realized by complex free fermions.
 There is a free boson realization for $W_{1+\infty}$ algebra
 \cite{Prochazka}.}.

\subsection{The (anti)commutators in
  ${\cal W}_{\infty}^{{\cal N}=4}[\la=0]$ algebra}

Note that the quasi primary fields in (\ref{WWQQ})
have $N$ independent terms for fixed $k$.
Each term has its own nontrivial OPE according to (\ref{OPE}).
This implies that there are
$N$ independent (anti)commutators between the higher spin
currents.
We can apply the result of Odake in \cite{Odake} to
obtain $N$ copies of the sum of the field dependent terms and the
central term. Then again we can use the expressions in (\ref{WWQQ})
to reexpress the above field dependent terms by collecting them
in terms of the original
quasi primary fields. The central terms can be added and it will
contribute to the overall factor $N$.
Then we obtain the following (anti)commutators for general
spin $h_1, h_2$ 
\bea
\big[(W^{\bar{\alpha} {\beta}}_{\mathrm{F},h_1})_m,(W^{\bar{\gamma}
    {\delta}}_{\mathrm{F},h_2})_n\big] 
\!&=& \!
\sum_{h\geq-1}p_{\mathrm{F}}^{h_1h_2 h}(m,n)\frac{q^{h}}{2} 
 \Big( 
 \delta^{\bar{\gamma}
     {\beta}}  W^{\bar{\alpha} {\delta}}_{\mathrm{F},h_1+h_2-2-h}
+ (-1)^{h} \delta^{\bar{\alpha} {\delta}}
W^{\bar{\gamma} {\beta}}_{\mathrm{F},h_1+h_2-2-h}
\Big)_{m+n}\
\nonu\\
&
+& c_{W_{\mathrm{F},h_1}} \,
\delta^{\bar{\alpha} {\delta}}\delta^{{\beta} \bar{\gamma}}
\delta^{h_1 h_2}\,q^{2(h_1-2)}\,\delta_{m+n}\ ,
\nonu\\
\big[(W^{\bar{a} {b}}_{\mathrm{B},h_1})_m,(W^{\bar{c} {d}}_{\mathrm{B},h_2})_n\big] 
\!& = &\!
\sum_{h\geq-1}  p_{\mathrm{B}}^{h_1h_2 h}(m,n) \frac{q^{h}}{2}  
\Big( 
\delta^{\bar{c} {b}}  W^{\bar{a} {d}}_{\mathrm{B},h_1+h_2-2-h}
+ (-1)^{h} 
\delta^{\bar{a} {d}}  W^{\bar{c} {b}}_{\mathrm{B},h_1+h_2-2-h} 
\Big)_{m+n}
\nonu\\
&
+& c_{W_{\mathrm{B},h_1}} \,
\delta^{\bar{a} {d}}\delta^{{b} \bar{c}}
\delta^{h_1 h_2}\,q^{2(h_1-2)}\,\delta_{m+n}\ ,
\nonu\\
\big[(W^{\bar{\alpha} {\beta}}_{\mathrm{F},h_1})_m,(Q^{\bar{a} {\gamma}}_{h_2})_r\big] 
\!&= &\!
\sum_{h\geq-1}\, q_{\mathrm{F}}^{h_1h_2 h}(m,r) \, 
q^{h} 
\,\delta^{\bar{\alpha} {\gamma}} 
\, (Q^{\bar{a} {\beta}}_{h_1+h_2-2-h})_{m+r}\ ,
\nonu\\
\big[(W^{\bar{\alpha} {\beta}}_{\mathrm{F},h_1})_m,(\bar{Q}^{{a} \bar{\gamma}}_{h_2})_r\big] 
\!&= &\!
\sum_{h\geq-1}\, q_{\mathrm{F}}^{h_1h_2 h}(m,r) \, 
q^{h} (-1)^h
\,\delta^{ \beta \bar{\gamma}} 
\, (\bar{Q}^{a \bar{\alpha}}_{h_1+h_2-2-h})_{m+r}\ ,
\nonu\\
\big[(W^{\bar{a} {b}}_{\mathrm{B},h_1})_m,(Q^{\bar{c} {\alpha}}_{h_2})_r \big] 
\!&=&\!
 \sum_{h\geq-1 }\, q_{\mathrm{B}}^{h_1h_2h}(m,r)\,
q^{h}
\, \delta^{\bar{c} {b}} 
\, (Q^{\bar{a} {\alpha}}_{h_1+h_2-2-h})_{m+r}\ ,
\nonu\\
\big[(W^{\bar{a} {b}}_{\mathrm{B},h_1})_m,(\bar{Q}^{c \bar{\alpha}}_{h_2})_r\big] 
\!&=&\!
 \sum_{h\geq-1 }\, q_{\mathrm{B}}^{h_1h_2h}(m,r)\,
q^{h} (-1)^{h}
\, \delta^{\bar{a} c} 
\, (\bar{Q}^{{b}\bar{\alpha}}_{h_1+h_2-2-h})_{m+r}\ ,
\nonu\\
\{(Q^{\bar{a} {\alpha}}_{h_1})_r,(\bar{Q}^{{b} \bar{\beta}}_{h_2})_s\} 
\!&=&\!
\sum_{h\geq0} \,
q^{h}
\Big(o_{\mathrm{F}}^{h_1h_2h}(r,s) \, \delta^{\bar{a} {b}} \,
W^{\bar{\beta} \alpha}_{\mathrm{F},h_1+h_2-1-h}
\nonu \\
& + & 
o_{\mathrm{B}}^{h_1h_2h}(r,s) \delta^{\alpha \bar{\beta}}
W^{\bar{a} {b}}_{\mathrm{B},h_1+h_2-1-h}\Big)_{r+s}
\nonu \\
& + & c_{Q\bar{Q}_{h_1}}  
\delta^{\bar{a} {b}}\delta^{{\alpha} \bar{\beta}} \delta^{h_1 h_2} q^{2(h_1-1)}
\delta_{m+n}.
\label{seven}
\eea
Here the central terms in (\ref{seven}) are given by
\bea
c_{W_{\mathrm{F},h}}(m)&=&
N \,\frac{2^{2(h-3)}((h-1)!)^2}{(2h-3)!!\,(2h-1)!!}\prod_{j=1-h}^{h-1}(m+j),
\nonu\\
c_{W_{\mathrm{B},h}}(m)&=&
N \,\frac{2^{2(h-3)}(h-2)!\,h!}{(2h-3)!!\,(2h-1)!!}\prod_{j=1-h}^{h-1}(m+j),
\nonu\\
c_{Q\bar{Q}_{h_1}}(r)&=&
N \,\frac{2^{2(h-\frac{3}{2})}(h-\frac{3}{2})!\,(h-\frac{1}{2})!}{((2h-2)!!)^2}\prod_{j=\frac{1}{2}-h}^{h-\frac{3}{2}}(r+j+\frac{1}{2})\, .
\label{centralterms}
\eea
In (\ref{centralterms}), there is an overall factor $N$.

The mode dependent expressions appearing in (\ref{seven})
are described as follows:
\bea
p_{\mathrm{F}}^{h_1h_2 h}(m,n)
&
=&\frac{1}{2(h+1)!}\,\phi^{h_1,h_2}_{h}(0,\textstyle{-\frac{1}{2}})
\,N^{h_1,h_2}_{h}(m,n),
\nonu\\
p_{\mathrm{B}}^{h_1h_2 h}(m,n)
&
=&\frac{1}{2(h+1)!}\,\phi^{h_1,h_2}_{h}(0,0)
\,N^{h_1,h_2}_{h}(m,n),
\nonu\\
q_{\mathrm{F}}^{h_1h_2 h}(m,r) 
&
=&\frac{(-1)^h}{4(h+2)!}\Big(
(h_1-1)\,\phi^{h_1,h_2+\frac{1}{2}}_{h+1}(0,0)
\nonu \\
& - & (h_1-h-3)\,\phi^{h_1,h_2+\frac{1}{2}}_{h+1}(0,\textstyle{-\frac{1}{2}})
\Big)
\,N^{h_1,h_2}_{h}(m,n),
\nonu\\
q_{\mathrm{B}}^{h_1h_2 h}(m,r) 
&
=&\frac{-1}{4(h+2)!}\Big(
(h_1-h-2)\,\phi^{h_1,h_2+\frac{1}{2}}_{h+1}(0,0)-(h_1)\,\phi^{h_1,h_2+\frac{1}{2}}_{h+1}(0,\textstyle{-\frac{1}{2}})
\Big)
\,N^{h_1,h_2}_{h}(m,n),
\nonu\\
o_{\mathrm{F}}^{h_1h_2h}(r,s) 
&
=&\frac{4(-1)^h}{h!}\Big(
(h_1+h_2-1-h)\,\phi^{h_1+\frac{1}{2},h_2+\frac{1}{2}}_{h}(
\textstyle{\frac{1}{2}},\textstyle{-\frac{1}{4}})
\nonu \\
& - & (h_1+h_2-\frac{3}{2}-h)\,\phi^{h_1+\frac{1}{2},h_2+\frac{1}{2}}_{h+1}(
\textstyle{\frac{1}{2}},\textstyle{-\frac{1}{4}})
\Big)\,  N^{h_1,h_2}_{h-1}(m,n),
\nonu\\
o_{\mathrm{B}}^{h_1h_2h}(r,s) 
&
=&-\frac{4}{h!}\Big(
(h_1+h_2-2-h)\,\phi^{h_1+\frac{1}{2},h_2+\frac{1}{2}}_{h}(
\textstyle{\frac{1}{2}},\textstyle{-\frac{1}{4}})
\nonu \\
& - &
(h_1+h_2-\frac{3}{2}-h)\,\phi^{h_1+\frac{1}{2},h_2+\frac{1}{2}}_{h+1}(\textstyle{\frac{1}{2}},\textstyle{-\frac{1}{4}})
\Big)
 \,  N^{h_1,h_2}_{h-1}(m,n).
\label{modedependence}
\eea
Moreover, we introduce
following quantities
\bea
N^{h_1,h_2}_{h}(m,n)
&
=&
\sum_{l=0 }^{h+1}(-1)^l
\left(\begin{array}{c}
h+1 \\  l \\
\end{array}\right)
[h_1-1+m]_{h+1-l}[h_1-1-m]_l
\nonu \\
      & \times & [h_2-1+n]_l [h_2-1-n]_{h+1-l},
\nonu\\
\phi^{h_1,h_2}_{h}(x,y)
&
= &
{}_4 F_3
 \Bigg[
\begin{array}{c}
  -\frac{1}{2}-x-2y, \frac{3}{2}-x+2y, -\frac{h+1}{2}+x,
  -\frac{h}{2} +x \\
-h_1+\frac{3}{2},-h_2+\frac{3}{2},h_1+h_2-h-\frac{3}{2}
\end{array} ; 1
  \Bigg].
\label{Nphi}
\eea
The falling Pochhammer symbol
$[a]_n \equiv a(a-1) \cdots (a-n+1)$
is used in (\ref{Nphi}).
The previous notation for the binomial coefficients 
is used.
The generalized hypergeometric function, 
with $4$ upper arguments $a_i$,  $3$ lower arguments $b_i$
and variable $z$, is 
defined as the series
\bea
 {}_4 F_3
 \Bigg[
\begin{array}{c}
a_1, a_2, a_3, a_4 \\
b_1,b_2,b_3
\end{array} ; z
  \Bigg] 
=
\sum_{n=0 }^{\infty}
\frac{(a_1)_n (a_2)_n (a_3)_n (a_4)_n}
{(b_1)_n (b_2)_n (b_3)_n}
\frac{z^n}{n!}\,,
\label{defF43}
\eea
where the rising Pochhammer symbol
$(a)_n \equiv a(a+1) \cdots (a+n-1)$ is used in (\ref{defF43}).
Due to the fact that there is a relation between
the arguments and the variable, $b_1+b_2+b_3 = a_1+a_2+a_3+a_4
+z$ for the $\phi^{h_1,h_2}_{h}(x,y)$ in (\ref{Nphi}),
the infinite series (\ref{defF43}) for this
particular case terminates \cite{PRS}. 
The quantity $N^{h_1,h_2}_{h}(m,n)$
in (\ref{Nphi})
can be written in terms of Clebsch Gordan coefficients
\cite{PRS}.

In summarizing
\footnote{ As observed in \cite{Odake}
  where $N$ is fixed by $1$ in the context of coset model,
  we can check that by contracting the indices
  in the second relation of (\ref{seven}), the bosonic realization
  provides the $W_{\infty}$ algebra \cite{PRS1990},
  where the central charge
  is given by $c=2$,
  with increased central charge by a
  factor $N \, K$ according to (\ref{centralfree}).
  The fermionic realization with the first relation of (\ref{seven})
  contains
  the subalgebra $W_{1+\infty}$ algebra \cite{PRS1990-1},
  where the central charge
  is given by $c=1$,
  with increased central charge by a
  factor $N \, L$. Note that the variable $y$ of
  $p_{\mathrm{F}}^{h_1h_2 h}(m,n)$ is equal to
  $-\frac{1}{2}$ while
  the one of $p_{\mathrm{B}}^{h_1h_2 h}(m,n)$ is equal to $0$
  in (\ref{modedependence}) and (\ref{Nphi}). The variable $y$
  corresponds to the variable $s$ in \cite{PRS,PRS1990-1}.},
the (anti)commutators  
between the ${\cal N}=4$ higher spin multiplets
are basically described by (\ref{seven}). More precisely,
they with manifest $SO(4)$ symmetry 
are written in terms of the linear combinations of
(\ref{seven}).
Because we have the $s$-th ${\cal N}=4$ higher spin multiplet
in terms of free fields 
from the footnotes \ref{lowestspincase}-\ref{highestspincase}
(see also Appendix $J$ where the inverse relations between them
are given),
we can calculate the (anti)commutators between them
by using the relations in (\ref{seven}).
Then we arrive at the following results
\footnote{In practice, we calculate
  the (anti)commutators for $h_1, h_2 \leq 5$
  and read off the $h_1, h_2$ and $h$-dependences
  in the right hand side of the (anti)commutators.},
with simplified notations
where the mode indices are ignored,
\bea
\Big[\Phi^{(h_1)}_{0},\Phi^{(h_2)}_{0}\Big]
&=&
\sum_{h = 0}^{\frac{1}{2}(h_1+h_2-3)} \frac{1}{ 2 (h_1+h_2-2h)-5 }
\nonu\\
&
\times&
\Bigg(
\Big(\!
-h_1 \, h_2\,p_{\mathrm{F},2h}^{h_1 h_2 }
+(h_1-1) \, (h_2-1)\,p_{\mathrm{B},2h}^{h_1 h_2 }
\Big)\,\Phi^{(h_1+h_2-2h-2)}_{0}
\nonu\\
&
+&
\frac{2}{2(h_1+h_2-2h)-9}
\Big(
h_1 \, h_2\,(h_1+h_2-2h-3)\,p_{\mathrm{F},2h}^{h_1 h_2 }
\nonu\\
&+
&
(h_1-1) \, (h_2-1)\,(h_1+h_2-2h-2)\,p_{\mathrm{B},2h}^{h_1 h_2 }
\Big)\,
\tilde{\Phi}^{(h_1+h_2-2h-4)}_{2}
\Bigg)
\nonu\\
&
+&
\delta^{h_1 h_2}\,\frac{N\,2^{2h_1-5}h_1!(2h_1!-(h_1-1)!)}{(2h_1-3)!!(2h_1-1)!!}\prod_{j=-h_1+1}^{h_1-1}(m+j)
\,,
\nonu\\
\Big[\Phi^{(h_1)}_{0},\Phi^{(h_2),i}_{\frac{1}{2}}\Big]
&=&
\sum_{h =0}^{\frac{1}{2}(h_1+h_2-3)}
\!\!\!\frac{-(2h_2-1)}{2 (h_1+h_2-2h)-5}
\Big(
h_1\,q_{\mathrm{F}, 2h}^{h_1(h_2+\frac{1}{2})}
-(h_1-1)\, q_{\mathrm{B}, 2h}^{h_1(h_2+\frac{1}{2})}
\Big)
\nonu \\
&\times & \Phi^{(h_1+h_2-2h-2),i}_{\frac{1}{2}}
+
\sum_{h =-1}^{\frac{1}{2}(h_1+h_2-4)}
\!\!\!\frac{(2h_2-1)}{2 (h_1+h_2-2h)-9 }
\nonu \\
& \times & \Big(
h_1\,q_{\mathrm{F},2h+1}^{h_1(h_2+\frac{1}{2}) }
-(h_1-1)\, q_{\mathrm{B},2h+1}^{h_1(h_2+\frac{1}{2})}
\Big)
\tilde{\Phi}^{(h_1+h_2-2h-4),i}_{\frac{3}{2}}
\,,
\nonu\\
\Big[\Phi^{(h_1)}_{0},\Phi^{(h_2),ij}_{1}\Big]
&=&
\sum_{h = 0}^{\frac{1}{2}(h_1+h_2-2)}
\!\!\!\!\!\!\frac{-(2h_2-1)}{ 4(h_1+h_2-2h)\!-\!10}
\nonu \\
& \times & \Bigg(\!
\Big(
h_1 \, p_{\mathrm{F},2h}^{h_1 (h_2+1) }
-(h_1\!-\!1) \, p_{\mathrm{B},2h}^{h_1 (h_2+1)}
\Big)
 \Phi^{(h_1+h_2-2h-2),ij}_{1}
\nonu\\
&+
&
\Big(
h_1 \, p_{\mathrm{F},2h}^{h_1 (h_2+1) }
+(h_1-1) \, p_{\mathrm{B},2h}^{h_1 (h_2+1)}
\Big)
\tilde{\Phi}^{(h_1+h_2-2h-2),ij}_{1}
\,
\Bigg)
\,,
\nonu\\
\Big[\Phi^{(h_1)}_{0},\tilde{\Phi}^{(h_2),i}_{\frac{3}{2}}\Big]
&=&
\sum_{h = 0}^{\frac{1}{2}(h_1+h_2-2)}
\!\!\!\!\!\!
\frac{-(2h_2-1)}{2 (h_1+h_2-2h)-5 }
\Big(
h_1\,q_{\mathrm{F},2h}^{h_1(h_2+\frac{3}{2}) }
-(h_1-1)\, q_{\mathrm{B}, 2h}^{h_1(h_2+\frac{3}{2})}
\Big)
\nonu \\
&\times & \tilde{\Phi}^{(h_1+h_2-2h-2),i}_{\frac{3}{2}}
+ \sum_{h =-1}^{\frac{1}{2}(h_1+h_2-3)}
\!\!\!\!\!\!
\frac{(2h_2-1)}{2 (h_1+h_2-2h)-5 }
\nonu \\
&\times & 
\Big(
h_1\,q_{\mathrm{F},2h+1}^{h_1(h_2+\frac{3}{2})}
-(h_1-1)\, q_{\mathrm{B},2h+1}^{h_1(h_2+\frac{1}{2})}
\Big)
\Phi^{(h_1+h_2-2h-2),i}_{\frac{1}{2}}
\,,
\nonu \\
\Big[\Phi^{(h_1)}_{0},\tilde{\Phi}^{(h_2)}_{2}\Big]
&=&
\sum_{h =0}^{\frac{1}{2}(h_1+h_2-1)}
\!\!
\frac{(2h_2-1)}{4 (h_1+h_2-2h)-2 }
\nonu \\
& \times &
\Bigg(
\Big(
h_1\,p_{\mathrm{F}, 2h}^{h_1 (h_2+2)}
+(h_1-1)\,p_{\mathrm{B}, 2h}^{h_1 (h_2+2)}
\Big)
\,\Phi^{(h_1+h_2-2h)}_{0}
\nonu\\
&
-&
\frac{1}{2(h_1+h_2+2h)-5}\Big(
h_1(h_1+h_2-2h-1)\,p_{\mathrm{F}, 2h}^{h_1 (h_2+2)}
\nonu\\
&
-&(h_1-1)(h_1+h_2-2h)\,p_{\mathrm{B}, 2h}^{h_1 (h_2+2)}
\Big)\,
\tilde{\Phi}^{(h_1+h_2-2h-2)}_{2}
\Bigg)
\,,
\label{completecomm}
\eea
where the mode dependent piece in (\ref{completecomm})
is given in (\ref{modedependence}) and for simplicity,
the third element appearing in the upper index of
$p_F, \cdots$ is located at
the lower index.
The remaining expressions for (anti)commutators are given in
(\ref{remaininganticomm}).
The relations (\ref{PhionePhione}) and (\ref{PhionePhitwo})
can be obtained from these complete results by substituting
$h_1$ and $h_2$.
The $m$ in the first relation is the mode of the
first element $\Phi_0^{(h_1)}$.
We also use the notations in the footnote \ref{BigJ}.
We can also reexpress these (anti)commutators
in terms of OPEs based on \cite{PRS} \footnote{
  By using the expressions in \cite{PRS}
  \bea
M^{h_1 h_2}_{h}(m,n)
&
=& \sum^{h+1}_{k=0}(-1)^k
\left(\begin{array}{c}
h+1 \\  k \\
\end{array}\right)
(2h_1-h-2)_k [2h_2-2-k]_{h+1-k}
\,m^{h+1-k}\,n^{k}\,,
\nonu\\
f^{h_1 h_2}_{\mathrm{B},h}(m,n)
&=& \frac{1}{2(h+1)!}\,\phi^{h_1,h_2}_{h}(0,0)
M^{h_1 h_2}_{h}(m,n)\,,
\,\,
f^{h_1 h_2}_{\mathrm{F},h}(m,n)
= \frac{1}{2(h+1)!}\,\phi^{h_1,h_2}_{h}(0,-\frac{1}{2})
M^{h_1 h_2}_{h}(m,n)\,,
\nonu
\eea
we can obtain the following OPE
corresponding to the first commutator in (\ref{completecomm})
as follows:
\bea
\Phi^{(h_1)}_0(z)  \Phi^{(h_2)}_0(w)
&
=&
\frac{1}{(z-w)^{h_1+h_2}}\,\delta^{h_1 h_2}\,N\,2^{2h_1-5}\,\frac{h_1\,(2h_1-1)!(h_1-1)!(h_1-1)!}{(2h_1-3)!!(2h_1-3)!!}
\nonu\\
&
+ & (-1) \, \sum_{h=0}^{\frac{1}{2}(h_1+h_2+1)}\!\!\!\!\!
\frac{1}{2(h_1+h_2-2h)-5}\,
\nonu\\
&
\times &
\Bigg(
\Big(\!
-h_1 \, h_2\,f_{\mathrm{F},2h}^{h_1 h_2 }(\pa_z, \pa_w)
+(h_1-1) \, (h_2-1)\,f_{\mathrm{B},2h}^{h_1 h_2 }(\pa_z, \pa_w)
\Big)\,\frac{\Phi^{(h_1+h_2-2h-2)}_{0}(w)}{(z-w)}
\nonu\\
&
+&
\frac{2}{2(h_1+h_2-2h)-9}
\Big(
h_1 \, h_2\,(h_1+h_2-2h-3)\,f_{\mathrm{F},2h}^{h_1 h_2 }(\pa_z,\pa_w)
\nonu\\
&+
&
(h_1-1) \, (h_2-1)\,(h_1+h_2-2h-2)\,f_{\mathrm{B},2h}^{h_1 h_2 }
(\pa_z,\pa_w)
\Big)\,
\frac{\tilde{\Phi}^{(h_1+h_2-2h-4)}_{2}(w)}{(z-w)}
\Bigg).
\nonu
\eea
Note that the central term is obtained from the one in the
corresponding commutator by considering further the relation between
the binomial coefficient and the product.
The mode dependent parts in next singular terms
$p_F$ and $p_B$ go to the differential operators
$f_F$ and $f_B$ respectively.
The fields are functions of  $w$ together with
the factor $\frac{1}{(z-w)}$.
The other numerical factors in the commutator remain as the same
with extra minus sign.
We can obtain the OPEs for
the other (anti)commutators similarly.}.
As before, the above relations (\ref{completecomm})
will be the fundamental ones because the other ones
can be obtained from (\ref{completecomm}) in the OPE language.
We can easily check that the coefficients of
the higher spin currents having negative spins in
(\ref{completecomm}) and (\ref{remaininganticomm})
are vanishing \footnote{
\label{vanishingF32}
  For example,
$\phi^{h_1,h_2}_{h_1+h_2-3}(0,0)$
is written as
$
 {}_4 F_3
 \Bigg[
\begin{array}{c}
-\frac{1}{2}, \frac{3}{2}, -\frac{(h_1+h_2-2)}{2},
 -\frac{(h_1+h_2-3)}{2} \\
 -h_1+\frac{3}{2},-h_2+\frac{3}{2},\frac{3}{2}
\end{array} ; 1
  \Bigg] 
=
 {}_3 F_2
 \Bigg[
\begin{array}{c}
a\equiv -\frac{1}{2},  b\equiv -\frac{(h_1+h_2-2)}{2},
 c \equiv -\frac{(h_1+h_2-3)}{2} \\
 d \equiv -h_1+\frac{3}{2}, e \equiv -h_2+\frac{3}{2}
\end{array} ; 1
  \Bigg]$ 
 because the upper second argument is equal to the lower
 third argument and they are cancelled each other in (\ref{defF43}).
 Therefore we are left with  ${}_3 F_2$. Using the definition of
 ${}_3 F_2$, we have the factor $\frac{1}{\sqrt{\Gamma(d-a)\,
     \Gamma(e-a)}} =
 \frac{1}{\sqrt{\Gamma(-h_1+2)\,
     \Gamma(-h_2+2)}}$ which goes to zero for $h_1, h_2 > 1$.
 Similarly, we have  vanishing $\phi^{h_1,h_2+2}_{h_1+h_2-1}(0,0)$
 by shifting $h_2 \rightarrow h_2+2$.}.

In the first and the last commutators of (\ref{completecomm}),
the right hande side contains the $SO(4)$ singlets with subscript
$0$ and $2$. In the second and fourth commutators,
the $SO(4)$ vectors
with subscript $\frac{1}{2}$ and $\frac{3}{2}$ appear.
In the third commutator, 
the $SO(4)$ adjoint with subscript $1$
appears in the right hand side
\footnote{
  In (\ref{PhionePhione}), the currents of
  the large ${\cal N}=4$ superconformal algebra
  occur in the right hand side.
  By introducing  the following
  notations $-\frac{1}{2}\,L  \rightarrow  \Phi^{(0)}_2$
and
$\frac{1}{4}\,G^i  \rightarrow  \Phi^{(0),i}_{\frac{3}{2}}$, we can
treat the currents and higher spin currents in (\ref{completecomm})
and (\ref{remaininganticomm}) 
simultaneously.}.

\section{$AdS_3$ higher spin theory with matrix generalization }

We consider the ``deformed''
oscillator construction \cite{AKP,CHR1211,AP1902}
corresponding to
the coset construction in sections, $2,3$, and $4$ associated with
the ${\cal N}=4$ higher spin multiplets under the large $(N,k)$
limit. The
Lie algebra $shs[\la]$ is generated by $\hat{y}_{\al}(\al=1,2)$
with defining relations \cite{Vasiliev89,Vasiliev91,Vasiliev1804} 
\bea
[\hat{y}_{\al}, \hat{y}_{\beta}]= 2 \, i\, \epsilon_{\al \beta} (1+ \nu \,
k), \qquad k \, \hat{y}_{\al} =-\hat{y}_{\al} \, k, \qquad k^2 =1.
\label{commutatordefinition}
\eea
The Chan-Paton factors are introduced and  
the generators of ${\cal N}=4$ higher spin algebra
denoted by $shs_2[\la]$ are given by the tensor product
between the generators of the $ {\cal N}=2$ higher spin
algebra $shs[\la]$ and
$U(2)$ generators.
There is a relation
between the parameter in the ${\cal N}=4$ higher spin algebra
and the one in the ${\cal N}=4$ coset model as follows:
\bea
\nu = 2 \la -1, \qquad \mbox{or} \qquad \la = \frac{(\nu+1)}{2}=
\mu.
\label{nula}
\eea
We construct the ${\cal N}=4$ (higher spin)generators at generic $\la$
from the
wedge subalgebra of the ${\cal W}_{\infty}^{{\cal N}=4}[\la]$ algebra
\footnote{The product in the oscillators is a Moyal product
  \cite{AKP}. On the other hand, the lone star product is introduced
  in \cite{PRS}. Recently, in \cite{BBB}, these two products
are equivalent to each other.}.

\subsection{ The $16$ generators and
  exceptional superalgebra $D(2,1|\frac{\la}{1-\la})$}

Spin-$\frac{3}{2}$ currents of the large ${\cal N}=4$ superconformal
algebra provide the eight fermionic
operators for ${\cal N}=4$ wedge algebra \cite{GG1305,Ferreira}
as follows:
\bea
G^{1}_{\pm \frac{1}{2}} & = &
-\frac{1}{2}\,e^{i\frac{\pi}{4}}\,\,\hat{y}_{\frac{3}{2}\mp \frac{1}{2}}\,k \otimes 
\left(\begin{array}{cc}
1 & 0 \\
0 & -1 \\
\end{array}\right),
\qquad
G^{2}_{\pm \frac{1}{2}}  = 
\frac{i}{2}\,e^{i\frac{\pi}{4}}\,\,\hat{y}_{\frac{3}{2}\mp \frac{1}{2}}\otimes 
\left(\begin{array}{cc}
1 & 0 \\
0 & 1 \\
\end{array}\right),
\nonu\\
G^{3}_{\pm \frac{1}{2}} & = &
\frac{i}{2}\,e^{i\frac{\pi}{4}}\,\,\hat{y}_{\frac{3}{2}\mp \frac{1}{2}}\,k \otimes 
\left(\begin{array}{cc}
0 & 1 \\
-1 & 0 \\
\end{array}\right),
\qquad
G^{4}_{\pm \frac{1}{2}}  = 
\frac{1}{2}\,e^{i\frac{\pi}{4}}\,\,\hat{y}_{\frac{3}{2}\mp \frac{1}{2}}\,k \otimes 
\left(\begin{array}{cc}
0 & 1 \\
1 & 0 \\
\end{array}\right).
\label{fourGexp}
\eea
Note that the second component of the spin-$\frac{3}{2}$ current
has a $2 \times 2$ identity matrix
in its expression.
After calculating the anticommutators between these operators
(\ref{fourGexp}), we obtain
the following result
\bea
\Big\{G^i_{r},G^j_{\rho}\Big\}
& 
=& \delta^{ij}\,2\, L_{r+\rho}
-i\,(r-\rho)\Big(
T^{ij}_{r+\rho}-(2\lambda-1)\,\widetilde{T}_{r+\rho}^{ij}
\Big)
\,.
\label{GGanti}
\eea
For the same index $i=j$,
the spin-$2$ current of
the large ${\cal N}=4$ superconformal algebra
gives the three bosonic operators of ${\cal N}=4$ wedge algebra
by using (\ref{commutatordefinition})
\bea
L_{+1} &=& \frac{1}{4i}\,\hat{y}_{1}\hat{y}_{1}
\otimes 
\left(\begin{array}{cc}
1 & 0 \\
0 & 1 \\
\end{array}\right),
\nonu\\
L_{0} & = & \frac{1}{8i}\Big(\,
\hat{y}_{1}\hat{y}_{2}
+\hat{y}_{2}\hat{y}_{1}
\,\Big)
\otimes 
\left(\begin{array}{cc}
1 & 0 \\
0 & 1 \\
\end{array}\right),
\nonu\\
L_{-1} & = & \frac{1}{4i}\,\hat{y}_{2}\hat{y}_{2}
\otimes 
\left(\begin{array}{cc}
1 & 0 \\
0 & 1 \\
\end{array}\right).
\label{Lexp}
\eea
Moreover, for different index
$i \neq j$,
we can write down the six spin-$1$ operators as follows: 
\bea
T^{12}_0 & = & \frac{1}{2}\,k \otimes 
\left(\begin{array}{cc}
1 & 0 \\
0 & -1 \\
\end{array}\right),
\qquad
T^{13}_0  =  \frac{1}{2}\, \otimes 
\left(\begin{array}{cc}
0 & 1 \\
1 & 0 \\
\end{array}\right),
\nonu\\
T^{14}_0 & =& -\frac{i}{2}\, \otimes 
\left(\begin{array}{cc}
0 & 1 \\
-1 & 0 \\
\end{array}\right),
\qquad
T^{23}_0  =  \frac{i}{2}\,k \otimes 
\left(\begin{array}{cc}
0 & 1 \\
-1 & 0 \\
\end{array}\right),
\nonu\\
T^{24}_0 & = & \frac{1}{2}\,k \otimes 
\left(\begin{array}{cc}
0 & 1 \\
1 & 0 \\
\end{array}\right),
\qquad
T^{34}_0  =  -\frac{1}{2}
\otimes \left(\begin{array}{cc}
1 & 0 \\
0 & -1 \\
\end{array}\right).
\label{Texp}
\eea
Note that for $\la =0$, the last term of (\ref{GGanti})
can combine the second term.
Then this leads to the similar relation as in (\ref{twoTT}).

We can check that the following commutators
are satisfied
\bea
\Big[L_{m},L_{n}\Big]
& 
=& (m-n)\,L_{m+n}
\,,
\qquad
\Big[L_{m},G^i_{r}\Big] 
= (\frac{m}{2}-r)\,G^i_{m+r}
\,,
\nonu\\
\Big[L_{m},T^{ij}_{n}\Big]
& 
=& 0
\,,
\qquad
\Big[G^i_{r},T^{jk}_{m}\Big]
=  \delta^{ij}\,
i\,G^k_{r+m}
-\delta^{ik}\,
i\,G^j_{r+m}
\,,
\nonu\\
\Big[T^{ij}_{m},T^{kl}_{n}\Big]
& 
=&
-  i\,\delta^{ik}\,T^{jl}
+i\,\delta^{il}\,T^{jk}
+i\,\delta^{jk}\,T^{il}
-i\,\delta^{jl}\,T^{ik}
\,.
\label{fiveremain}
\eea
The ${\cal N}=4$ wedge algebra which generates
nine bosonic and eight fermionic ones written in terms of
(\ref{fourGexp}), (\ref{Lexp}) and (\ref{Texp})
is characterized by
(\ref{GGanti}) and (\ref{fiveremain}).
We can easily see that the (anti)commutators in (\ref{16commanticomm})
will become the above ${\cal N}=4$ wedge algebra by restricting
the mode indices to the wedge indices. 

\subsection{The first ${\cal N}=4$ generators of ${\cal N}=4$
  higher spin algebra
  $shs_2[\la]$ }

Let us start with the following higher spin-$1$ operator
\cite{GG1305}
\bea
\Phi^{(1)}_{0,0} & = & (k+\nu)
\otimes 
\left(\begin{array}{cc}
1 & 0 \\
0 & 1 \\
\end{array}\right).
\label{spin1}
\eea
Here the parameter $\nu$ has a relation with $\la$
via (\ref{nula}).

We can calculate the following commutators
from (\ref{fourGexp}) and (\ref{spin1})
and read off the higher spin-$\frac{3}{2}$ operators
appearing in the right hand side
\bea
\Big[G^i_{r},\Phi^{(1)}_{0,0}\Big]
& 
=& -\Phi^{(1),i}_{\frac{1}{2},r}\,.
\label{relone}
\eea
This relation (\ref{relone})
is nothing but one of the ${\cal N}=4$ primary conditions.  
See also (\ref{N4primary}).
It turns out that the eight spin-$\frac{3}{2}$ operators
are given by
\bea
\Phi^{(1),1}_{\frac{1}{2},\pm \frac{1}{2}}& = &
e^{i\frac{\pi}{4}}\,\hat{y}_{\frac{3}{2}\mp\frac{1}{2}}
\otimes 
\left(\begin{array}{cc}
1 & 0 \\
0 & -1 \\
\end{array}\right),
\qquad
\Phi^{(1),2}_{\frac{1}{2},\pm \frac{1}{2}}  = 
-i\,e^{i\frac{\pi}{4}}\,\hat{y}_{\frac{3}{2}\mp\frac{1}{2}}\,k
\otimes 
\left(\begin{array}{cc}
1 & 0 \\
0 & 1 \\
\end{array}\right),
\nonu\\
\Phi^{(1),3}_{\frac{1}{2},\pm \frac{1}{2}} & = &
-i\,e^{i\frac{\pi}{4}}\,\hat{y}_{\frac{3}{2}\mp\frac{1}{2}}
\otimes 
\left(\begin{array}{cc}
0 & 1 \\
-1 & 0 \\
\end{array}\right),
\qquad
\Phi^{(1),4}_{\frac{1}{2},\pm \frac{1}{2}}  = 
-e^{i\frac{\pi}{4}}\,\hat{y}_{\frac{3}{2}\mp\frac{1}{2}}
\otimes 
\left(\begin{array}{cc}
0 & 1 \\
1 & 0 \\
\end{array}\right).
\label{3halfspin}
\eea
The second element has a $2 \times 2$ identity matrix.

In order to obtain eighteen spin-$2$ operators,
we should use the following anticommutators 
\bea
\Big\{G^i_{r},\Phi^{(1),j}_{\frac{1}{2},\rho}\Big\}
& 
= &
-\delta^{ij}\,(r-\rho)\,\Phi^{(1)}_{0,r+\rho}+
\widetilde{\Phi}^{(1),ij}_{1,r+\rho}\,.
\label{gphitwo}
\eea
Again, this is
one of the ${\cal N}=4$ primary conditions in component
approach. For equal index $i=j$, there is no nontrivial
relation.
This will provide the following eighteen spin-$2$ operators
for different $i$ and $j$ indices
as follows:
\bea
\Phi^{(1),12}_{1,-1} & = & 
-\hat{y}_{2}\hat{y}_{2}\,k
\otimes 
\left(\begin{array}{cc}
1 & 0 \\
0 & -1 \\
\end{array}\right),
\qquad
\Phi^{(1),12}_{1,0}  =  
-\frac{1}{2}\Big(\,
\hat{y}_{1}\hat{y}_{2}
+\hat{y}_{2}\hat{y}_{1}
\,\Big)\,k
\otimes 
\left(\begin{array}{cc}
1 & 0 \\
0 & -1 \\
\end{array}\right),
\nonu\\
\Phi^{(1),12}_{1,+1} & = & 
-\hat{y}_{1}\hat{y}_{1}\,k
\otimes 
\left(\begin{array}{cc}
1 & 0 \\
0 & -1 \\
\end{array}\right),
\qquad
\Phi^{(1),13}_{1,-1}  =  
-\hat{y}_{2}\hat{y}_{2}
\otimes 
\left(\begin{array}{cc}
0 & 1 \\
1 & 0 \\
\end{array}\right),
\nonu\\
\Phi^{(1),13}_{1,0} & = & 
-\frac{1}{2}\Big(\,
\hat{y}_{1}\hat{y}_{2}
+\hat{y}_{2}\hat{y}_{1}
\,\Big)
\otimes 
\left(\begin{array}{cc}
0 & 1 \\
1 & 0 \\
\end{array}\right),
\qquad
\Phi^{(1),13}_{1,+1}  =  
-\hat{y}_{1}\hat{y}_{1}
\otimes 
\left(\begin{array}{cc}
0 & 1 \\
1 &0 \\
\end{array}\right),
\nonu\\
\Phi^{(1),14}_{1,-1} & = & 
i\,\hat{y}_{2}\hat{y}_{2}
\otimes 
\left(\begin{array}{cc}
0 & 1 \\
-1 & 0 \\
\end{array}\right),
\qquad
\Phi^{(1),14}_{1,0}  =  
\,\frac{i}{2}\Big(\,
\hat{y}_{1}\hat{y}_{2}
+\hat{y}_{2}\hat{y}_{1}
\,\Big)
\otimes 
\left(\begin{array}{cc}
0 & 1 \\
-1 & 0 \\
\end{array}\right),
\nonu\\
\Phi^{(1),14}_{1,+1} & = & 
i\,\hat{y}_{1}\hat{y}_{1}
\otimes 
\left(\begin{array}{cc}
0 & 1 \\
-1 &0 \\
\end{array}\right),
\qquad
\Phi^{(1),23}_{1,-1}  =  
-i\,\hat{y}_{2}\hat{y}_{2}\,k
\otimes 
\left(\begin{array}{cc}
0 & 1 \\
-1 & 0 \\
\end{array}\right),
\nonu\\
\Phi^{(1),23}_{1,0} & = & 
-\,\frac{i}{2}\Big(\,
\hat{y}_{1}\hat{y}_{2}
+\hat{y}_{2}\hat{y}_{1}
\,\Big)\,k
\otimes 
\left(\begin{array}{cc}
0 & 1 \\
-1 & 0 \\
\end{array}\right),
\qquad
\Phi^{(1),23}_{1,+1}  =  
-i\,\hat{y}_{1}\hat{y}_{1}\,k
\otimes 
\left(\begin{array}{cc}
0 & 1 \\
-1 &0 \\
\end{array}\right),
\nonu\\
\Phi^{(1),24}_{1,-1} & = & 
-\hat{y}_{2}\hat{y}_{2}\,k
\otimes 
\left(\begin{array}{cc}
0 & 1 \\
1 & 0 \\
\end{array}\right),
\qquad
\Phi^{(1),24}_{1,0}  =  
-\,\frac{1}{2}\Big(\,
\hat{y}_{1}\hat{y}_{2}
+\hat{y}_{2}\hat{y}_{1}
\,\Big)\,k
\otimes 
\left(\begin{array}{cc}
0 & 1 \\
1 & 0 \\
\end{array}\right),
\nonu\\
\Phi^{(1),24}_{1,+1} & = & 
-\hat{y}_{1}\hat{y}_{1}\,k
\otimes 
\left(\begin{array}{cc}
0 & 1 \\
1 &0 \\
\end{array}\right),
\qquad
\Phi^{(1),34}_{1,-1}  =  
\hat{y}_{2}\hat{y}_{2}
\otimes 
\left(\begin{array}{cc}
1 & 0 \\
0 & -1 \\
\end{array}\right),
\nonu\\
\Phi^{(1),34}_{1,0} & = & 
\,\frac{1}{2}\Big(\,
\hat{y}_{1}\hat{y}_{2}
+\hat{y}_{2}\hat{y}_{1}
\,\Big)
\otimes 
\left(\begin{array}{cc}
1 & 0 \\
0 & -1 \\
\end{array}\right),
\qquad
\Phi^{(1),34}_{1,+1}  =  
\hat{y}_{1}\hat{y}_{1}
\otimes 
\left(\begin{array}{cc}
1 & 0 \\
0 & -1 \\
\end{array}\right).
\label{sspin2}
\eea
There are no elements having a $2 \times 2$ identity matrix. 

We can calculate the following commutators
\bea
\Big[G^i_{r},\Phi^{(1),jk}_{1,m}\Big]
& 
=& -\delta^{ij}\,(\tilde{\Phi}^{(1),k}_{\frac{3}{2},r+m}-
\frac{1}{3}\,(2\,r-m)
(2\lambda-1)\,\Phi^{(1),k}_{\frac{1}{2},r+m})
\nonu\\
& 
+& \delta^{ik}(\tilde{\Phi}^{(1),j}_{\frac{3}{2},r+m}-
\frac{1}{3}\,(2\,r-m)
(2\lambda-1)\Phi^{(1),j}_{\frac{1}{2},r+m})
+\varepsilon^{ijkl}(2\,r-m)\,\Phi^{(1),l}_{\frac{1}{2},r+m},
\label{gphi5half}
\eea
which is one of the ${\cal N}=4$ primary conditions
mentioned before.
This enables us to obtain
the following sixteen spin-$\frac{5}{2}$ operators
as follows:
\bea
\tilde{\Phi}^{(1),1}_{\frac{3}{2},-\frac{3}{2}} & = & 
-i\,e^{i\frac{\pi}{4}}\,\hat{y}_{2}\hat{y}_{2}\hat{y}_{2}\,k
\otimes 
\left(\begin{array}{cc}
1 & 0 \\
0 & -1 \\
\end{array}\right),
\nonu \\
\tilde{\Phi}^{(1),1}_{\frac{3}{2},-\frac{1}{2}} & = &  
-\frac{i}{3}\,e^{i\frac{\pi}{4}}\Big(\,
\hat{y}_{1}\hat{y}_{2}\hat{y}_{2}
+\hat{y}_{2}\hat{y}_{1}\hat{y}_{2}
+\hat{y}_{2}\hat{y}_{2}\hat{y}_{1}
\,\Big)\,k
\otimes 
\left(\begin{array}{cc}
1 & 0 \\
0 & -1 \\
\end{array}\right),
\nonu\\
\tilde{\Phi}^{(1),1}_{\frac{3}{2},+\frac{1}{2}} & = & 
-\frac{i}{3}\,e^{i\frac{\pi}{4}}\Big(\,
\hat{y}_{2}\hat{y}_{1}\hat{y}_{1}
+\hat{y}_{1}\hat{y}_{2}\hat{y}_{1}
+\hat{y}_{1}\hat{y}_{1}\hat{y}_{2}
\,\Big)\,k
\otimes 
\left(\begin{array}{cc}
1 & 0 \\
0 & -1 \\
\end{array}\right),
\nonu\\
\tilde{\Phi}^{(1),1}_{\frac{3}{2},+\frac{3}{2}} & = & 
-i\,e^{i\frac{\pi}{4}}\,\hat{y}_{1}\hat{y}_{1}\hat{y}_{1}\,k
\otimes 
\left(\begin{array}{cc}
1 & 0 \\
0 & -1 \\
\end{array}\right),
\qquad
\tilde{\Phi}^{(1),2}_{\frac{3}{2},-\frac{3}{2}}  =  
-\,e^{i\frac{\pi}{4}}\,\hat{y}_{2}\hat{y}_{2}\hat{y}_{2}
\otimes 
\left(\begin{array}{cc}
1 & 0 \\
0 & 1 \\
\end{array}\right),
\nonu\\
\tilde{\Phi}^{(1),2}_{\frac{3}{2},-\frac{1}{2}} & = & 
-\frac{1}{3}\,e^{i\frac{\pi}{4}}\Big(\,
\hat{y}_{1}\hat{y}_{2}\hat{y}_{2}
+\hat{y}_{2}\hat{y}_{1}\hat{y}_{2}
+\hat{y}_{2}\hat{y}_{2}\hat{y}_{1}
\,\Big)
\otimes 
\left(\begin{array}{cc}
1 & 0 \\
0 & 1 \\
\end{array}\right),
\nonu\\
\tilde{\Phi}^{(1),2}_{\frac{3}{2},+\frac{1}{2}} & = & 
-\frac{1}{3}\,e^{i\frac{\pi}{4}}\Big(\,
\hat{y}_{2}\hat{y}_{1}\hat{y}_{1}
+\hat{y}_{1}\hat{y}_{2}\hat{y}_{1}
+\hat{y}_{1}\hat{y}_{1}\hat{y}_{2}
\,\Big)
\otimes 
\left(\begin{array}{cc}
1 & 0 \\
0 & 1 \\
\end{array}\right),
\nonu\\
\tilde{\Phi}^{(1),2}_{\frac{3}{2},+\frac{3}{2}} & = & 
-\,e^{i\frac{\pi}{4}}\,\hat{y}_{1}\hat{y}_{1}\hat{y}_{1}
\otimes 
\left(\begin{array}{cc}
1 & 0 \\
0 & 1 \\
\end{array}\right),
\qquad
\tilde{\Phi}^{(1),3}_{\frac{3}{2},-\frac{3}{2}}  =  
-\,e^{i\frac{\pi}{4}}\,\hat{y}_{2}\hat{y}_{2}\hat{y}_{2}\,k
\otimes 
\left(\begin{array}{cc}
0 & 1 \\
-1 & 0 \\
\end{array}\right),
\nonu\\
\tilde{\Phi}^{(1),3}_{\frac{3}{2},-\frac{1}{2}} & = & 
-\frac{1}{3}\,e^{i\frac{\pi}{4}}\Big(\,
\hat{y}_{1}\hat{y}_{2}\hat{y}_{2}
+\hat{y}_{2}\hat{y}_{1}\hat{y}_{2}
+\hat{y}_{2}\hat{y}_{2}\hat{y}_{1}
\,\Big)\,k
\otimes 
\left(\begin{array}{cc}
0 & 1 \\
-1 & 0 \\
\end{array}\right),
\nonu\\
\tilde{\Phi}^{(1),3}_{\frac{3}{2},+\frac{1}{2}} & = & 
-\frac{1}{3}\,e^{i\frac{\pi}{4}}\Big(\,
\hat{y}_{2}\hat{y}_{1}\hat{y}_{1}
+\hat{y}_{1}\hat{y}_{2}\hat{y}_{1}
+\hat{y}_{1}\hat{y}_{1}\hat{y}_{2}
\,\Big)\,k
\otimes 
\left(\begin{array}{cc}
0 & 1 \\
-1 & 0 \\
\end{array}\right),
\nonu\\
\tilde{\Phi}^{(1),3}_{\frac{3}{2},+\frac{3}{2}} & = & 
-\,e^{i\frac{\pi}{4}}\,\hat{y}_{1}\hat{y}_{1}\hat{y}_{1}\,k
\otimes 
\left(\begin{array}{cc}
0 & 1 \\
-1 & 0 \\
\end{array}\right),
\qquad
\tilde{\Phi}^{(1),4}_{\frac{3}{2},-\frac{3}{2}}  =  
i\,\,e^{i\frac{\pi}{4}}\,\hat{y}_{2}\hat{y}_{2}\hat{y}_{2}\,k
\otimes 
\left(\begin{array}{cc}
0 & 1 \\
1 & 0 \\
\end{array}\right),
\nonu\\
\tilde{\Phi}^{(1),4}_{\frac{3}{2},-\frac{1}{2}} & = & 
\frac{i}{3}\,e^{i\frac{\pi}{4}}\Big(\,
\hat{y}_{1}\hat{y}_{2}\hat{y}_{2}
+\hat{y}_{2}\hat{y}_{1}\hat{y}_{2}
+\hat{y}_{2}\hat{y}_{2}\hat{y}_{1}
\,\Big)\,k
\otimes 
\left(\begin{array}{cc}
0 & 1 \\
1 & 0 \\
\end{array}\right),
\nonu\\
\tilde{\Phi}^{(1),4}_{\frac{3}{2},+\frac{1}{2}} & = & 
\frac{i}{3}\,e^{i\frac{\pi}{4}}\Big(\,
\hat{y}_{2}\hat{y}_{1}\hat{y}_{1}
+\hat{y}_{1}\hat{y}_{2}\hat{y}_{1}
+\hat{y}_{1}\hat{y}_{1}\hat{y}_{2}
\,\Big)\,k
\otimes 
\left(\begin{array}{cc}
0 & 1 \\
1 & 0 \\
\end{array}\right),
\nonu\\
\tilde{\Phi}^{(1),4}_{\frac{3}{2},+\frac{3}{2}} & = & 
i\,\,e^{i\frac{\pi}{4}}\,\hat{y}_{1}\hat{y}_{1}\hat{y}_{1}\,k
\otimes 
\left(\begin{array}{cc}
0 & 1 \\
1 & 0 \\
\end{array}\right).
\label{5halfspin}
\eea
Note that the $SO(4)$ index $i=2$ case
has a $2 \times 2$ identity matrix.

From the following anticommutators coming from ${\cal N}=4$
primary conditions with wedge conditions (see also (\ref{N4primary}))
\bea
\Big\{G^i_{r},\tilde{\Phi}^{(1),j}_{\frac{3}{2},\rho}\Big\}
& 
= & -\delta^{ij}\, \tilde{\Phi}^{(1)}_{2,r+\rho}
- (3r-\rho)\Big(
\Phi^{(1),ij}_{1,r+\rho}
-\frac{1}{3}(2\lambda-1)\,\tilde{\Phi}^{(1),ij}_{1,r+\rho}
\Big)
\,,
\label{gphithree}
\eea
we can determine 
the five spin-$3$ operators as follows:
\bea
\tilde{\Phi}^{(1)}_{2,-2} & = & 
-\hat{y}_{2}\hat{y}_{2}\hat{y}_{2}\hat{y}_{2}
\otimes 
\left(\begin{array}{cc}
1 & 0 \\
0 & 1 \\
\end{array}\right),
\nonu\\
\tilde{\Phi}^{(1)}_{2,-1} & = & 
-\frac{1}{4}\,\Big(\,
\hat{y}_{1}\hat{y}_{2}\hat{y}_{2}\hat{y}_{2}
+\hat{y}_{2}\hat{y}_{1}\hat{y}_{2}\hat{y}_{2}
+\hat{y}_{2}\hat{y}_{2}\hat{y}_{1}\hat{y}_{2}
+\hat{y}_{2}\hat{y}_{2}\hat{y}_{2}\hat{y}_{1}
\,\Big)
\otimes 
\left(\begin{array}{cc}
1 & 0 \\
0 & 1 \\
\end{array}\right),
\nonu\\
\tilde{\Phi}^{(1)}_{2,0} & = & 
-\frac{1}{6}\,\Big(\,
\hat{y}_{1}\hat{y}_{1}\hat{y}_{2}\hat{y}_{2}
+\hat{y}_{1}\hat{y}_{2}\hat{y}_{1}\hat{y}_{2}
+\hat{y}_{1}\hat{y}_{2}\hat{y}_{2}\hat{y}_{1}
+\hat{y}_{2}\hat{y}_{1}\hat{y}_{1}\hat{y}_{2}
+\hat{y}_{2}\hat{y}_{1}\hat{y}_{2}\hat{y}_{1}
+\hat{y}_{2}\hat{y}_{2}\hat{y}_{1}\hat{y}_{1}
\,\Big)
\nonu \\
& \otimes & 
\left(\begin{array}{cc}
1 & 0 \\
0 & 1 \\
\end{array}\right),
\nonu\\
\tilde{\Phi}^{(1)}_{2,+1} & = & 
-\frac{1}{4}\,\Big(\,
\hat{y}_{2}\hat{y}_{1}\hat{y}_{1}\hat{y}_{1}
+\hat{y}_{1}\hat{y}_{2}\hat{y}_{1}\hat{y}_{1}
+\hat{y}_{1}\hat{y}_{1}\hat{y}_{2}\hat{y}_{1}
+\hat{y}_{1}\hat{y}_{1}\hat{y}_{1}\hat{y}_{2}
\,\Big)
\otimes 
\left(\begin{array}{cc}
1 & 0 \\
0 & 1 \\
\end{array}\right),
\nonu\\
\tilde{\Phi}^{(1)}_{2,+2} & = & 
-\hat{y}_{1}\hat{y}_{1}\hat{y}_{1}\hat{y}_{1}
\otimes 
\left(\begin{array}{cc}
1 & 0 \\
0 & 1 \\
\end{array}\right).
\label{spin3}
\eea
These $SO(4)$ singlets
have a $2\times 2$ identity matrix in their expressions.
We can check that the following commutator satisfies
\bea
\Big[G^i_{r},\tilde{\Phi}^{(1)}_{2,m}\Big]
& 
=&
-(4r-m)\,\tilde{\Phi}^{(1),i}_{\frac{3}{2},r+m}
\,.
\label{reltwo}
\eea
The relation (\ref{reltwo}) can be checked from (\ref{fourGexp}),
(\ref{spin3}) and (\ref{5halfspin}).

Therefore, the oscillator realization for the first
${\cal N}=4$ 
higher spin operators
is summarized by
(\ref{spin1}), (\ref{3halfspin}), (\ref{sspin2}), (\ref{5halfspin}),
and (\ref{spin3}).
Note that the oscillators $\hat{y}_{\al}$ appear symmetrically.

\subsection{The second ${\cal N}=4$ generators of
 ${\cal N}=4$
  higher spin algebra
  $shs_2[\la]$}

Let us consider the second ${\cal N}=4$ higher spin generators.
For the higher spin-$2$ operator, we can find the higher
spin-$\frac{5}{2}$ operator $\Phi^{(2),i}_{\frac{1}{2}}$
  first. From the
  relation coming from the ${\cal N}=4$ wedge algebra
  in (\ref{PhionePhione})
\bea
\Big[\Phi^{(1)}_{0,m},\tilde{\Phi}^{(1),i}_{\frac{3}{2},r}\Big]
& 
=&
-\frac{1}{2}\,\Phi^{(2),i}_{\frac{1}{2},m+r},
\label{relthree}
\eea
we obtain sixteen higher spin-$\frac{5}{2}$ operators
by calculating the left hand side of (\ref{relthree}).
After that, we can determine the three
higher spin-$2$ operators by using
(\ref{gphitwo}) for upper index $s=2$ (or the fourth relation of
(\ref{N4primary})) for equal index $i=j$.
For the higher spin-$3$ operator, 
we can use (\ref{gphitwo}) for $s=2$ for different indices
$i\neq j$. Then thirty higher spin-$3$ operators can be determined.
The relation (\ref{gphi5half}) for upper index $s=2$
(or the fifth relation of (\ref{N4primary}))
is used for the twenty four
higher spin-$\frac{7}{2}$ operators.
Finally, the seven higher spin-$4$ operators can be obtained
from the relation (\ref{gphithree}) for upper index $s=2$
\footnote{
We present them explicitly
in Appendix $K$ where we can see a $2 \times 2 $ identity matrix
for the $SO(4)$ vector index $i=2$. 
Compared to the first ${\cal N}=4$ generators,
the number of oscillators is increased with different overall
factors. The form of $U(2)$ matrix elements
remain the same and the oscillators $\hat{y}_{\al}$ appear
symmetrically as before.}. 

\subsection{The s-th ${\cal N}=4$ generators of
 ${\cal N}=4$
  higher spin algebra
  $shs_2[\la]$}

From the results of the first and the second
${\cal N}=4$ higher spin generators (together with the third and
fourth ones)
in terms of oscillators,
we obtain the following expressions for the
$s$-th ${\cal N}=4$ higher spin generators
\bea
\Phi^{(s)}_{0,m} & = &
\Bigg[\frac{(s-1-m)!(s-1+m)!}{(2s-2)!}\Bigg]
  \,
\underbrace{
\hat{y}_{(1 }\,\cdots\,\hat{y}_{1}
}_{(s-1+m)}
\,
\underbrace{
\hat{y}_{2}\,\cdots \ \hat{y}_{2)}
}_{(s-1-m)}
\big((2s-1)k+\nu \big)
\otimes 
\left(\begin{array}{cc}
1 & 0 \\
0 & 1 \\
\end{array}\right),
\nonu \\
\Phi^{(s),1}_{\frac{1}{2},m}& = &
\Bigg[\frac{(s-\frac{1}{2}-m)!(s-\frac{1}{2}+m)!}{(2s-2)!}\Bigg]
\, e^{i\frac{\pi}{4}}
\,
\underbrace{
\hat{y}_{(1 }\,\cdots\,\hat{y}_{1}
}_{(s-\frac{1}{2}+m)}
\,
\underbrace{
\hat{y}_{2 }\,\cdots\ \hat{y}_{2)}
}_{(s-\frac{1}{2}-m)}
\otimes 
\left(\begin{array}{cc}
1 & 0 \\
0 & -1 \\
\end{array}\right),
\nonu\\
\Phi^{(s),2}_{\frac{1}{2},m} & = &
-i
\Bigg[\frac{(s-\frac{1}{2}-m)!(s-\frac{1}{2}+m)!}{(2s-2)!}\Bigg]
\, e^{i\frac{\pi}{4}}
\,
\underbrace{
\hat{y}_{(1 }\,\cdots\,\hat{y}_{1}
}_{(s-\frac{1}{2}+m)}
\,
\underbrace{
\hat{y}_{2 }\,\cdots\ \hat{y}_{2)}
}_{(s-\frac{1}{2}-m)}
\, k
\otimes 
\left(\begin{array}{cc}
1 & 0 \\
0 & 1 \\
\end{array}\right),
\nonu\\
\Phi^{(s),3}_{\frac{1}{2},m} & = &
-i
\Bigg[\frac{(s-\frac{1}{2}-m)!(s-\frac{1}{2}+m)!}{(2s-2)!}\Bigg]
\, e^{i\frac{\pi}{4}}
\,\underbrace{
\hat{y}_{(1 }\,\cdots\,\hat{y}_{1}
}_{(s-\frac{1}{2}+m)}
\,
\underbrace{
\hat{y}_{2 }\,\cdots\ \hat{y}_{2)}
}_{(s-\frac{1}{2}-m)}
\otimes 
\left(\begin{array}{cc}
0 & 1 \\
-1 & 0 \\
\end{array}\right),
\nonu\\
\Phi^{(s),4}_{\frac{1}{2},m} & = &
-
\Bigg[\frac{(s-\frac{1}{2}-m)!(s-\frac{1}{2}+m)!}{(2s-2)!}\Bigg]
\,e^{i\frac{\pi}{4}}
\,\underbrace{
\hat{y}_{(1 }\,\cdots\,\hat{y}_{1}
}_{(s-\frac{1}{2}+m)}
\,
\underbrace{
\hat{y}_{2 }\,\cdots\ \hat{y}_{2)}
}_{(s-\frac{1}{2}-m)}
\otimes 
\left(\begin{array}{cc}
0 & 1\\
1 & 0 \\
\end{array}\right),
\nonu
\\
\Phi^{(s),12}_{1,m} & = & -
\Bigg[\frac{(s-m)!(s+m)!}{2s(2s-2)!}\Bigg]
\,
\underbrace{
\hat{y}_{(1} \,\cdots\,\hat{y}_{1}
}_{(s+m)}
\,
\underbrace{
\hat{y}_{2 }\,\cdots\ \hat{y}_{2)}
}_{(s-m)}
\,k
\otimes 
\left(\begin{array}{cc}
1 & 0 \\
0 & -1 \\
\end{array}\right),
\nonu\\
\Phi^{(s),13}_{1,m} & = & -
\Bigg[\frac{(s-m)!(s+m)!}{2s(2s-2)!}\Bigg]
\,
\underbrace{
\hat{y}_{(1} \,\cdots\,\hat{y}_{1}
}_{(s+m)}
\,
\underbrace{
\hat{y}_{2 }\,\cdots\ \hat{y}_{2)}
}_{(s-m)}
\otimes 
\left(\begin{array}{cc}
0 & 1 \\
1 & 0 \\
\end{array}\right),
\nonu\\
\Phi^{(s),14}_{1,m} & = & i
\Bigg[\frac{(s-m)!(s+m)!}{2s(2s-2)!}\Bigg]
\,
\underbrace{
\hat{y}_{(1} \,\cdots\,\hat{y}_{1}
}_{(s+m)}
\,
\underbrace{
\hat{y}_{2 }\,\cdots\ \hat{y}_{2)}
}_{(s-m)}
\otimes 
\left(\begin{array}{cc}
0 & 1 \\
-1 & 0 \\
\end{array}\right),
\nonu\\
\Phi^{(s),23}_{1,m} & = & -i
\Bigg[\frac{(s-m)!(s+m)!}{2s(2s-2)!}\Bigg]
\,
\underbrace{
\hat{y}_{(1} \,\cdots\,\hat{y}_{1}
}_{(s+m)}
\,
\underbrace{
\hat{y}_{2 }\,\cdots\ \hat{y}_{2)}
}_{(s-m)}
\,k
\otimes 
\left(\begin{array}{cc}
0 & 1 \\
-1 & 0 \\
\end{array}\right),
\nonu\\
\Phi^{(s),24}_{1,m} &=& -
\Bigg[\frac{(s-m)!(s+m)!}{2s(2s-2)!}\Bigg]
\,
\underbrace{
\hat{y}_{(1} \,\cdots\,\hat{y}_{1}
}_{(s+m)}
\,
\underbrace{
\hat{y}_{2 }\,\cdots\ \hat{y}_{2)}
}_{(s-m)}
\,k
\otimes 
\left(\begin{array}{cc}
0 & 1 \\
1 & 0 \\
\end{array}\right),
\nonu\\
\Phi^{(s),34}_{1,m} & = &
\Bigg[\frac{(s-m)!(s+m)!}{2s(2s-2)!}\Bigg]
\,
\underbrace{
\hat{y}_{(1} \,\cdots\,\hat{y}_{1}
}_{(s+m)}
\,
\underbrace{
\hat{y}_{2 }\,\cdots\ \hat{y}_{2)}
}_{(s-m)}
\otimes 
\left(\begin{array}{cc}
1 & 0 \\
0 & -1 \\
\end{array}\right),
\nonu
\\
\tilde{\Phi}^{(s),1}_{\frac{3}{2},m}& = &
-i
\Bigg[\frac{(s+\frac{1}{2}-m)!(s+\frac{1}{2}+m)!}{2s(2s+1)(2s-2)!}\Bigg]
\,e^{i\frac{\pi}{4}}
\,
\underbrace{
\hat{y}_{(1 }\,\cdots\,\hat{y}_{1}
}_{(s+\frac{1}{2}+m)}
\,
\underbrace{
\hat{y}_{2 }\,\cdots\ \hat{y}_{2)}
}_{(s+\frac{1}{2}-m)}
\, k
\otimes 
\left(\begin{array}{cc}
1 & 0 \\
0 & -1 \\
\end{array}\right),
\nonu\\
\tilde{\Phi}^{(s),2}_{\frac{3}{2},m} & = &
-
\Bigg[\frac{(s+\frac{1}{2}-m)!(s+\frac{1}{2}+m)!}{2s(2s+1)(2s-2)!}\Bigg]
\,e^{i\frac{\pi}{4}}
\,
\underbrace{
\hat{y}_{(1 }\,\cdots\,\hat{y}_{1}
}_{(s+\frac{1}{2}+m)}
\,
\underbrace{
\hat{y}_{2 }\,\cdots\ \hat{y}_{2)}
}_{(s+\frac{1}{2}-m)}
\otimes 
\left(\begin{array}{cc}
1 & 0 \\
0 & 1 \\
\end{array}\right),
\nonu\\
\tilde{\Phi}^{(s),3}_{\frac{3}{2},m} & = &
-
\Bigg[\frac{(s+\frac{1}{2}-m)!(s+\frac{1}{2}+m)!}{2s(2s+1)(2s-2)!}\Bigg]
\,e^{i\frac{\pi}{4}}
\,
\underbrace{
\hat{y}_{(1 }\,\cdots\,\hat{y}_{1}
}_{(s+\frac{1}{2}+m)}
\,
\underbrace{
\hat{y}_{2 }\,\cdots\ \hat{y}_{2)}
}_{(s+\frac{1}{2}-m)}
k
\otimes 
\, 
\left(\begin{array}{cc}
0 & 1 \\
-1 & 0 \\
\end{array}\right),
\nonu\\
\tilde{\Phi}^{(s),4}_{\frac{3}{2},m} & = &
i
\Bigg[\frac{(s+\frac{1}{2}-m)!(s+\frac{1}{2}+m)!}{2s(2s+1)(2s-2)!}\Bigg]
\,e^{i\frac{\pi}{4}}
\,
\underbrace{
\hat{y}_{(1 }\,\cdots\,\hat{y}_{1}
}_{(s+\frac{1}{2}+m)}
\,
\underbrace{
\hat{y}_{2 }\,\cdots\ \hat{y}_{2)}
}_{(s+\frac{1}{2}-m)}
k
\otimes 
\, 
\left(\begin{array}{cc}
0 & 1\\
1 & 0 \\
\end{array}\right),
\nonu
\\
\tilde{\Phi}^{(s)}_{2,m} & = & -
\Bigg[\frac{(s+1-m)!(s+1+m)!}{2s(2s+1)(2s+2)(2s-2)!}\Bigg]
\,
\underbrace{
\hat{y}_{(1} \,\cdots\,\hat{y}_{1}
}_{(s+1+m)}
\,
\underbrace{
\hat{y}_{2} \,\cdots\ \hat{y}_{2)}
}_{(s+1-m)}
\otimes 
\left(\begin{array}{cc}
1 & 0 \\
0 & 1 \\
\end{array}\right).
\label{generalhigherspinoscillator}
\eea
The total number of
the $16$ higher spin operators
is given by $16(1+2s)$.
The difference between the number of $\hat{y}_1$
and the number of $\hat{y}_2$ is $2m$.
The number of $\hat{y}_{\alpha}$ for the higher spin
generator of spin $\hat{s}$ is given by
$2(\hat{s}-1)$.
We use the simplified notation for the symmetric product of
oscillators $\hat{y}_{\al}$. For example,
the expression $\hat{y}_{(1} \, \hat{y}_{1}
\,
\hat{y}_{2} \,\hat{y}_{2} \,\hat{y}_{2)}$
has ten terms as in $\tilde{\Phi}^{(2),1}_{\frac{3}{2},-\frac{1}{2}}$ of
Appendix $K$.
The higher spin generators of
$SO(4)$ singlet and $SO(4)$ vector index $2$ have
the $2 \times 2 $ identity matrix and they will
consist of the ${\cal N}=2$ higher spin algebra.
The question is what is the ${\cal N}=4$ higher spin algebra
generated by (\ref{generalhigherspinoscillator})
\footnote{There is an overall factor $\frac{1}{2}$ difference
  between the first
  ${\cal N}=4$ higher spin multiplet in the coset model and the one
  in the
  oscillator formalism. Similarly, the relative factor
  $-\frac{i}{3}(-\frac{6}{5})$ difference
  appears in the second(third)
  ${\cal N}=4$ higher spin multiplet of both  descriptions.}.

\subsection{The (anti)commutators in ${\cal N}=4$
  higher spin algebra $shs_2[\la]$}

In Appendix $G$, we present
the (anti)commutators between the ${\cal N}=4$
higher spin multiplets we have constructed in previous sections.
They originate from the results of OPEs in the coset construction.
Because we do not know any OPEs for the general spins $s_1, s_2$,
we cannot further obtain the corresponding (anti)commutators.
The next step we can consider is to use the known expressions
(\ref{generalhigherspinoscillator})
of the ${\cal N}=4$
higher spin generators in terms of oscillators. 
Is it possible to write down the $\la$ dependent structure
constants appearing in Appendix $G$ in terms of any closed form
using a special function like as the ones in section $6$ for the
$\la =0$ case?
According to the observation of \cite{BBB},
the ${\cal N}=2$ higher spin algebra $shs[\la]$
can be written in terms of closed form by generalizing
the mode dependent quantity in (\ref{Nphi}).
Moreover, the nontrivial dependences of
$s_1$, $s_2$, $s$ and $\la$ arise
in the structure constants.
Therefore, we expect that
the ${\cal N}=4$
higher spin algebra should be described in terms of
closed form.

\subsection{The ${\cal N}=2$ higher spin algebra $shs[\la]$}

As described in section $5$, 
among $16$ higher spin generators in
(\ref{generalhigherspinoscillator}),
there are four higher spin generators,
$\Phi^{(s)}_{0,m}$, $\Phi^{(s), 2}_{\frac{1}{2},m}$,
$\tilde{\Phi}^{(s),2}_{\frac{3}{2},m}$ and $\tilde{\Phi}^{(s)}_{2,m}$ which
have $U(2)$ identity matrix.
We expect that they consist of their own closed subalgebra.
In particular, from the identifications
(\ref{Spin2identification}) and (\ref{Spin3identification}),
we can calculate the corresponding (anti)commutators by using the
quantities appearing in the left hand sides with the help of
Appendix $G$.
In Appendix $L$, we collect the relevant (anti)commutators
from Appendix $G$. Moreover, the known (anti)commutators
from \cite{EGR,Romans} are presented in terms of the notations
of the right hand sides of (\ref{Spin2identification})
and (\ref{Spin3identification}).
Under the wedge condition, we explicitly check that
the ${\cal N}=2$ wedge algebra coming from
the first ${\cal N}=2$ higher spin multiplet, in addition to
the generators of ${\cal N}=2$ superconformal algebra,
can be realized by the ten higher spin operators
coming from the first, second and third ${\cal N}=4$ higher spin
generators as well as the two
operators from the large ${\cal N}=4$
linear superconformal algebra
\footnote{
Moreover, according to
the observation of \cite{EGR}, the wedge subalgebra
of ${\cal W}_{\infty}^{{\cal N}=2}[\la]$ algebra
matches with the corresponding ${\cal N}=2$ higher spin algebra
studied in \cite{FL}. Therefore, we can conclude that
the realization of (\ref{generalhigherspinoscillator})
with ${\cal N}=2$ supersymmetry
indeed leads to the
${\cal N}=2$ higher spin algebra.
That implies that we can
 write down the $\la$ dependent structure
 constants appearing in Appendix $G$
 in terms of closed form described in \cite{FL}.}.

 Let us consider the simplest case.
 We can rewrite the commutator between
 the first of ${\cal N}=2$ higher spin generators
 found by \cite{CG}.
 By using the ${\cal N}=2$ higher spin algebra
 and expressing the right hand side in terms of
 the elements of  ${\cal N}=2$ higher spin generators (and
 the generators of ${\cal N}=2$ superconformal algebra)
as follows:
 \bea
[W^{2\, 0}_{m}\, ,\, W^{2\, 0}_{n}]
& = & 
\frac{\sqrt{(1-m)!(1+m)!(1-n)!(1+n)!}}{6\sqrt{(1-m-n)!(1+m+n)!}}
\,C_{m\,n\,m+n}^{1\,1\,1}
\nonu\\
&
\times &
\Bigg(
\frac{1}{3}(\lambda+1)(\lambda-2)
\Big(
(\lambda+1)\,f_{TTT}^{1\,1\,1}
-(\lambda-2)\,f_{UUU}^{1\,1\,1}
\Big) \,L_{m+n}
\nonu\\
&+&
\frac{1}{\sqrt{3}}
\Big(
(\lambda+1)^2\,f_{TTT}^{1\,1\,1}
-(\lambda-2)^2\,f_{UUU}^{1\,1\,1}
\Big) \,W^{2\, 0}_{m+n}
\Bigg)\,,
 \label{22commutator}
 \eea
 where the structure constants $f_{AA'A''}^{jj'j''}$,
 which depend on the $\la$,
 are given in
 \cite{FL}  and $C_{mm'm''}^{jj'j''}$ are the $SL(2)$ Clebsch-Gordan
 coefficients \footnote{
\label{Cdef}   One representation of
 the Clebsch-Gordan
 coefficients   through 
   hypergeometric function is
   $
C_{m\,m'\,m''}^{j\,j'\,j''}
=
\delta_{m+m',m''} 
\frac{\sqrt{(2j''+1)(j+m)!(j'-m')!(j''+m'')!(j''-m'')!(j'+j''-j)!(j''+j-j')!}}
{\sqrt{(j-m)!(j'+m')!(j+j'+j''+1)!(j+j'-j'')!}(j''-j-m')!(j''-j'+m')!}
 {}_3 F_2
 \Bigg[
\begin{array}{c}
-j+m,-j'-m',-j-j'+j'' \\
1+j''-j-m',1+j''-j'+m
\end{array} ; 1
\Bigg]$
appearing in   \cite{PRS}.}.
 See also \cite{CG}.
We also have other (anti)commutators in Appendix $L$.
Therefore, we can
observe that the ${\cal N}=2$
higher spin generators of ${\cal N}=4$ higher spin generators
realized in (\ref{generalhigherspinoscillator})
satisfy the corresponding ${\cal N}=2$ higher spin algebra
studied in \cite{FL}.
We expect that the bosonic subalgebra should satisfy
the higher spin algebra.

\subsection{How to generate the ${\cal N}=4$
  higher spin algebra  $shs_2[\la]$}

For the general spins $s_1$ and $s_2$, we can think of
the (anti)commutators between two $16$ higher spin generators
and there are $256$ (anti)commutators.
Then how we can determine these nontrivial (anti)commutators?
For the vanishing 't Hooft-like coupling constant, 
there exist the previous relations in (\ref{completecomm})
and (\ref{remaininganticomm}).
The point is that we would like to determine the
structure constants for the nonzero $\la$. 
As an example, we take the simplest one where $h_1=1$ and $h_2=3$.
First of all, we do not know any OPEs from the coset construction.
All we have is the explicit form for the
$h$-th ${\cal N}=4$ higher spin generators
given by (\ref{generalhigherspinoscillator}). 
In principle, we can calculate the various (anti)commutators
from the first and the third ${\cal N}=4$ higher spin generators,
by using the relations (\ref{commutatordefinition}).
From the free field results in (\ref{completecomm}),
we can take the same $SO(4)$ index structure and mode dependence
and furthermore introduce
the undetermined structure constants in the right hand side.
Then we can compute the (anti)commutators
by using the explicit
forms in (\ref{generalhigherspinoscillator}).
We present the detailed computations in {\tt ancillary.nb} file
\footnote{
  Several comments are in order.
1) In order to simplify the computation, we intentionally
put the oscillator $\hat{y}_1$ to the left 
and the Klein operator $k$ to the right in the product
of oscillators along the line of \cite{Thielemans}.
2) Also the (anti)commutators of the
two tensor products are written in terms of
sum of the tensor product between
two product of oscillators and two $U(2)$ matrix product with
appropriate minus or plus sign depending on the commutators
or anticommutators. We can
insert the general $s$-th ${\cal N}=4$ higher spin multiplet
(\ref{generalhigherspinoscillator}) in the mathematica program
rather than fixed one.}. 

\section{ Conclusions and outlook}

As in the abstract,
the OPEs between the first and second
${\cal N}=4$ higher spin multiplets
are obtained in component and in ${\cal N}=4$ superspace.
By taking the large $(N,k)$ 't Hooft-like limit,
we obtain the ${\cal W}_{\infty}^{{\cal N}=4}[\la]$ algebra for low spins.
At $\la=0$, the free field construction is determined
and the corresponding
$SO(4)$ symmetric ${\cal W}_{\infty}^{{\cal N}=4}[0]$ algebra
is obtained for any spins $s_1$ and $s_2$.
At $\la \neq 0$, the ${\cal N}=4$ higher spin algebra
$shs_2[\la]$ is determined by using the ${\cal N}=4$ wedge
sublgebra of ${\cal W}_{\infty}^{{\cal N}=4}[\la]$ algebra
for low spins. 
We also
present how to determine the structure constants of
 the ${\cal N}=4$ higher spin algebra
$shs_2[\la]$
for fixed
spins $s_1$ and $s_2$ from the oscillator formalism.

We list the possible open problems  along the line of this paper
as follows:

\begin{itemize}
\item More OPEs

It is an open problem to construct
more OPEs (the OPE between the
first and the third ${\cal N}=4$ higher spin multiplets,
the OPE between the
second and the third ${\cal N}=4$ higher spin multiplets,
$\cdots$) and observe any new features beyond the corresponding
wedge subalgebra (for example, the three-point functions with
finite $N$ and $k$).
As the spin increases, the structure of the OPE
becomes more complicated and the problem in this direction can be
reduced if 
we can
manage to express the singular terms in the given OPE in terms of
known (higher spin) currents.

\item Application of free field construction to type IIB
string theory

For the infinity limit of the level $k$, the free field
construction was analyzed in the context of the higher spin
theory and the string theory \cite{GG1406}
where $(4N+4)$ free bosons and fermions are used.
It would be interesting
to analyze the free field construction with
$4N$ free bosons and fermions  obtained in this paper (under
the
large $(N,k)$ 't Hooft limit)
and see
how the higher spin symmetry can be embedded in the string theory.

\item Any `closed' form for the (anti)commutators of higher spin
algebra at generic $\la$

Maybe we can generalize the work of  \cite{Korybut,BBB}
to the matrix computations and obtain the ${\cal N}=4$ higher spin
algebra at the general $\la$ in closed form
for any spins $s_1$, $s_2$ (In this paper, we have found
them for 1) $s_1=1,s_2=1$, 2) $s_1=1,s_2=2$, 3) $s_1=2,s_2=2$
and 4) $s_1=1,s_2=3$).
Because we know the answer for $\la=0$,
we want to obtain the ${\cal N}=4$ higher spin algebra
$shs_2[\la]$ at nonzero $\la$ for general $s_1$ and $s_2$. 
Maybe we should examine the explicit cases for low spins
(See also the subsection $7.7$)
and see how to express those structure constants with the possible
spin
dependence. 

\item Orthogonal group

It is an open problem to describe the corresponding 
${\cal N}=4$ orthogonal coset model and to observe what the
corresponding ${\cal N}=4$ higher spin algebra, constructed
from the oscillators, is. See also the relevant paper \cite{AKP1904}.
Furthermore, according to \cite{EGR},
this case with fixed $K$ and $L$ which are different from
$2$ will be related to the explicit
application of type IIB string theory for the specific ratio of
two three spheres.

\item Nonsupersymmetric cases with general $(K,L)$ values

According to the observation of \cite{EGR}, the coset model
can be generalized by take the arbitrary values for $K$ and $L$.
It would be interesting to check whether there exists any nontrivial
extended conformal algebra, although there is no supersymmetry,
by calculating the OPEs for  several $K$ and $L$ values in the
coset model.  
One of the motivations in this direction is to consider
the application of type IIB string theory with orthogonal coset
model.

\item Construct the free fields with non $U(N)$ invariant

In this paper, we have restricted to focus on the $U(N)$ invariant
quantity for the
higher spin currents. However, we can construct the higher spin
currents which have the explicit $U(N)$ indices. In other words,
it is an open problem to understand the role of these
higher spin currents with nonsinglet $U(N)$. See also
the relevant papers on this direction \cite{BK,OS}.

\item Any relations with the rectangular $W$ algebras

  By allowing the group $G=SU(N+2)$ to generalize to $G=SU(N+K)$
  with $K \neq 2$,
  maybe we can see how the higher spin currents can be related to the
  ones
  of the
rectangular $W$ algebras studied recently in
\cite{CH1812,CH1906,CHU}.
For the case of $K \geq 3$, there is a symmetric tensor
of rank three in $SU(K)$.
Maybe we can construct the higher spin currents (like as
Sugawara construction) using these tensors.

\item The classical asymptotic symmetry algebra
${\cal W}_{\infty,\mbox{cl}}^{{\cal N}=4}[\la]$

According to the large ${\cal N}=4$ holography in \cite{GG1305},
there should be the relation between the
nonlinear ${\cal W}_{\infty}^{{\cal N}=4}[\la]$ algebra and the classical
asymptotic nonlinear
symmetry algebra (which can be obtained from
$shs_2[\la]$ by relaxing the wedge condition)
of the $AdS_3$ bulk theory.
It is an open problem to check whether the
classical asymptotic symmetry algebra of the
Vasiliev $AdS_3$ higher spin theory with matrix
generalization can be reproduced from
the  ${\cal W}_{\infty}^{{\cal N}=4}[\la]$ algebra in the coset model
by taking the large central charge limit or not. The nonlinear
terms in both sides should match with each other. 

\item Adding the bosonic spin-$1$ operators in the free field
construction

Although we have not analyzed for this possibility fully in this
paper, it would be interesting to observe whether we can add
the bosonic spin-$1$ current in the free field construction
described in section $6$.
Should we modify the stress energy tensor of spin-$2$?
Can we also consider for the general $K$ and $L$ in order to
have consistent closed algebra?
Maybe we should also consider the $U(N)$ nonsinglet cases.

\item Beyond the bilinear construction (cubic, quartic,
$\cdots$ ) in the free field
construction

In the construction of higher spin square \cite{GG1501,GG1512},
it is possible
to consider the free field construction beyond the bilinear terms.
The corresponding algebra coming from higher spin currents
will be, in general, nonlinear and it is interesting to obtain
the nontrivial structure behind this rather complicated algebras. 
See also the relevant paper \cite{AP1812} in the context of
horizontal algebra. 

\item The ${\cal N}=3$ example

In \cite{AK1607}, the ${\cal N}=2$
supersymmetry is enhanced by taking the
critical level which allows us to introduce the free fermionic
fields. We expect that by introducing the matrix generalization
in the $AdS_3$ higher spin theory, the corresponding higher spin
algebra can be obtained from the oscillator formalism. It will be
an open problem to study in detail. See also the relevant paper
\cite{AP1902}.  

\end{itemize}

\vspace{.7cm}

\centerline{\bf Acknowledgments}

We would like to thank  C. Peng for the free field
construction and how to read off the
$\la$ value from the free fields, S. Odake for
his algebra and how to obtain it from the mode expansions of
free fields and
Y. Hikida for the large level limit and its relation to
rectangular $W$ algebra. 
This research was supported by the Basic Science Research Program through the National Research Foundation of Korea   funded by the Ministry of Education   (grant. 2017R1D1A1A09079512).
CA acknowledges warm hospitality from 
the School of  Liberal Arts (and Institute of Convergence Fundamental
Studies), Seoul National University of Science and Technology.
MHK thanks the
Institut f$\ddot{u}$r Theoretische Physik
for the hospitality
during 
the completion of this work.
The activities of MHK
leading to the publication have been made possible by the support of
the Bilateral Science and Technology Cooperation Programme with Asia and the Korean-Swiss Science and Technology Programme
(Project no. EG-KR-05-092018).


\appendix

\renewcommand{\theequation}{\Alph{section}\mbox{.}\arabic{equation}}

\section{ Quasi (super)primary fields from section $2$ }

We present the various quasi primary super fields appearing in
(\ref{super1super1})
of the section
$2.3$ \footnote{All the coefficients which are not in the Appendices
are given explicitly in the  attached {\tt ancillary.nb} file.}
\bea
{\bf Q}^{(\frac{3}{2}),i}_{\frac{1}{2}} & = &
c_{1}^{\frac{1}{2},1}\,{\bf J}^{4-i}
\mathtt{+c_{2}^{\frac{1}{2},1}\,\partial {\bf J}^{i}
+c_{3}^{\frac{1}{2},1}\,{\bf J}^{ij}{\bf J}^{j}
+\varepsilon^{ijkl}\,c_{4}^{\frac{1}{2},1}\,{\bf J}^{j}{\bf J}^{k}{\bf J}^{l}}
\,,
\nonu\\
{\bf Q}^{(1),ij}_{1} & = &
\mathtt{c_{1}^{1,2}\,{\bf J}^{i}{\bf J}^{j}}
+c_{2}^{1,2}\,{\bf J}^{ij}
\mathtt{+\varepsilon^{ijkl}\,c_{3}^{1,2}\,{\bf J}^{k}{\bf J}^{l}}
+c_{4}^{1,2}\,{\bf J}^{4-ij}\,,
\nonu\\
{\bf Q}^{(2),ij}_{1} & = &
\mathtt{c_{1}^{1,1}\,({\bf J}^{4-i}{\bf J}^{j}-{\bf J}^{4-j}{\bf J}^{i})
+c_{2}^{1,1}\,\partial {\bf J}^{ij}
+c_{3}^{1,1}\,\partial ({\bf J}^{i} {\bf J}^{j})
+c_{4}^{1,1}\, \varepsilon^{ijkl}\, \partial ({\bf J}^{k} {\bf J}^{l})}
\nonu\\
&+&
\mathtt{
c_{5}^{1,1}\,({\bf J}^{ik}{\bf J}^{k}{\bf J}^{j}-{\bf J}^{jk}{\bf J}^{k}{\bf J}^{i})}\,,
\nonu\\
{\bf Q}^{(\frac{1}{2}),i}_{\frac{3}{2}} & = &
c_{1}^{\frac{3}{2},3}\,{\bf J}^{i}\,,
\nonu\\
{\bf Q}^{(\frac{3}{2}),i}_{\frac{3}{2}} & = &
c_{1}^{\frac{3}{2},2}\,{\bf J}^{4-i}
\mathtt{+c_{2}^{\frac{3}{2},2}\,\partial {\bf J}^{i}
+c_{3}^{\frac{3}{2},2}\, {\bf J}^{4-ij} {\bf J}^{j}
+c_{4}^{\frac{3}{2},2}\, \partial {\bf J} {\bf J}^{i}
+\varepsilon^{ijkl}\,
(
c_{5}^{\frac{3}{2},2}\,{\bf J}^{4-jk} {\bf J}^{l} 
+c_{6}^{\frac{3}{2},2}\,{\bf J}^{j} {\bf J}^{k} {\bf J}^{l} 
)}
\,,
\nonu\\
{\bf Q}^{(\frac{5}{2}),i}_{\frac{3}{2}} & = &
c_{1}^{\frac{3}{2},1}\, D^i {\bf \Phi}^{(2)}
\mathtt{+c_{2}^{\frac{3}{2},1}\, {\bf \Phi}^{(1)} D^i {\bf \Phi}^{(1)}
+c_{3}^{\frac{3}{2},1}\, {\bf J}^{ij} {\bf J}^{ij}  {\bf J}^{i} 
+c_{4}^{\frac{3}{2},1}\, \partial^2 {\bf J} {\bf J}^{i} 
+c_{5}^{\frac{3}{2},1}\, \partial^2 {\bf J}^{i} 
+c_{6}^{\frac{3}{2},1}\, {\bf J}^{ij} \partial {\bf J}^{j}  }
\nonu\\
& + &
\mathtt{c_{7}^{\frac{3}{2},1}\, {\bf J}^{i} \partial {\bf J}^{j} {\bf J}^{j}  
+c_{8}^{\frac{3}{2},1}\, {\bf J}^{ij} {\bf J}^{4-j}  
+c_{9}^{\frac{3}{2},1}\, \partial {\bf J}^{ij} {\bf J}^{j}
+c_{10}^{\frac{3}{2},1}\,  {\bf J}^{4-ij} {\bf J}^{4-j}
+c_{11}^{\frac{3}{2},1}\, {\bf J}^{ij} \partial {\bf J} {\bf J}^{j}}
\nonu\\
& + &
\mathtt{c_{12}^{\frac{3}{2},1}\,  {\bf J}^{i} {\bf J}^{j} {\bf J}^{4-j}
+c_{13}^{\frac{3}{2},1}\, \partial {\bf J}^{4-i}
+c_{14}^{\frac{3}{2},1}\, \partial {\bf J}  {\bf J}^{4-i}
+\varepsilon^{ijkl}
\Big(
c_{15}^{\frac{3}{2},1}\, {\bf J}^{j} {\bf J}^{k} {\bf J}^{4-l}
+c_{16}^{\frac{3}{2},1}\, {\bf J}^{4-ij} {\bf J}^{i k} {\bf J}^{l}}
\nonu\\
& + &
\mathtt{c_{17}^{\frac{3}{2},1}\, {\bf J}^{ij} {\bf J}^{kl} {\bf J}^{i}
+c_{18}^{\frac{3}{2},1}\,  \partial {\bf J}  {\bf J}^{jk} {\bf J}^{l}
+c_{19}^{\frac{3}{2},1}\,  {\bf J}^{ij}  {\bf J}^{i} {\bf J}^{k} {\bf J}^{l}
+c_{20}^{\frac{3}{2},1}\, {\bf J}^{jk} \partial {\bf J}^{l}
+c_{21}^{\frac{3}{2},1}\,  \partial {\bf J} {\bf J}^{j} {\bf J}^{k}  {\bf J}^{l}}
\nonu\\
& + &
\mathtt{c_{22}^{\frac{3}{2},1}\, \partial( {\bf J}^{j} {\bf J}^{k} {\bf J}^{l})
+c_{23}^{\frac{3}{2},1}\, \partial {\bf J}^{jk} {\bf J}^{l}
\Big)
+c_{24}^{\frac{3}{2},1}\,  \partial {\bf J} \partial {\bf J}^{i}}\,,
\nonu\\
{\bf Q}^{(1)}_{2} & = &
c_{1}^{2,3}\, \partial {\bf J}\,,
\nonu\\
{\bf Q}^{(2)}_{2} & = &
c_{1}^{2,2}\,  {\bf \Phi}^{(2)}
\mathtt{+c_{2}^{2,2}\, {\bf J}^{4-0}
+c_{3}^{2,2}\, {\bf J}^{4-i}{\bf J}^{i}
+\varepsilon^{ijkl}
(
c_{4}^{2,2}\, {\bf J}^{i} {\bf J}^{j} {\bf J}^{k} {\bf J}^{l}
+c_{5}^{2,2}\, {\bf J}^{ij} {\bf J}^{kl}
)
+c_{6}^{2,2}\, {\bf J}^{ij} {\bf J}^{ij}}
\nonu\\
& + &
\mathtt{
c_{7}^{2,2}\, {\bf J}^{4-ij} {\bf J}^{i} {\bf J}^{j}
+c_{8}^{2,2}\,  {\bf J}^{ij} {\bf J}^{i} {\bf J}^{j}
+c_{9}^{2,2}\, \partial^2 {\bf J}
+c_{10}^{2,2}\, \partial {\bf J}^{i}  {\bf J}^{i}
+c_{11}^{2,2}\, \partial{\bf J} \partial {\bf J}} \,,
\nonu\\
{\bf Q}^{(3)}_{2} & = &
\mathtt{
c_{1}^{2,1}\, {\bf J}^{4-i} \partial	{\bf J}^{i}
+c_{2}^{2,1}\,  \partial {\bf J}^{4-i} {\bf J}^{i}
+c_{3}^{2,1}\,  \partial^3 {\bf J}
+c_{4}^{2,1}\,  \partial^2 {\bf J}^i {\bf J}^i
+c_{5}^{2,1}\,  \partial ({\bf J}^{ij} {\bf J}^i {\bf J}^j )}
\nonu \\
& + & \mathtt{
  c_{6}^{2,1}\,    {\bf J}^i {\bf J}^j    \partial {\bf J}^{ij} 
+ 
c_{7}^{2,1}\,  \partial {\bf J} \partial {\bf J}^{i} {\bf J}^{i} }\,.
\label{quasisuper}
\eea
We have complete expressions of (\ref{twoquasiother}) in
(\ref{quasisuper}).

The fundamental five OPEs in component approach  corresponding to
(\ref{super1super1}) are
described as
\bea
\Phi_{0}^{(1)}(z)\,\Phi_{0}^{(1)}(w)
& = & \frac{1}{(z-w)^{2}}\,c^{0,2}_{0}+\cdots\,,
\nonu \\
\Phi_{\frac{1}{2}}^{(1),i}(z)\,\Phi_{0}^{(1)}(w)
& = & \frac{1}{(z-w)}\,Q^{(\frac{3}{2}),i}_{\frac{1}{2}}(w)+\cdots\,,
\nonu \\
\Phi_{1}^{(1),ij}(z)\,\Phi_{0}^{(1)}(w)
& = & 
\frac{1}{(z-w)^{2}}\,
 Q^{(1),ij}_{1}(w)
 + 
\frac{1}{(z-w)}\,\Bigg[\,
\partial Q^{(1),ij}_{1} 
+Q^{(2),ij}_{1}
\,
\Bigg](w)
+\cdots,
\nonu \\
\Phi_{\frac{3}{2}}^{(1),i}(z)\,\Phi_{0}^{(1)}(w)
& = & 
\frac{1}{(z-w)^3}\,Q^{(\frac{1}{2}),i}_{\frac{3}{2}}(w)
\nonu\\
&+ &
\frac{1}{(z-w)^2}\,
\Bigg[\,
2\,\partial Q^{(\frac{1}{2}),i}_{\frac{3}{2}}
+Q^{(\frac{3}{2}),i}_{\frac{3}{2}}
-\frac{(k-N)}{3(2+k+N)}\, Q^{(\frac{3}{2}),i}_{\frac{1}{2}}
\,\Bigg](w)
\nonu\\
&+ &
\frac{1}{(z-w)}\,
\Bigg[\,
\frac{3}{2}\,\partial^2 Q^{(\frac{1}{2}),i}_{\frac{3}{2}}
+\partial Q^{(\frac{3}{2}),i}_{\frac{3}{2}}
+Q^{(\frac{5}{2}),i}_{\frac{3}{2}}
\,\Bigg](w)
+\cdots\,,
\nonu \\
\Phi_{2}^{(1)}(z)\,\Phi_{0}^{(1)}(w)
& = & 
\frac{1}{(z-w)^{3}}\,
 Q^{(1)}_{2}(w)
 \nonu \\
 & + & 
\frac{1}{(z-w)^{2}}\,\Bigg[\,
\frac{3}{2}\,\partial Q^{(1)}_{2} 
+Q^{(2)}_{2}
\nonu \\
& - &
\frac{8(k-N)}{(5+4k+4N+3kN)}
\,(\Phi^{(1)}_0 \Phi^{(1)}_0+c^{0,2}_{0}L)
\,
\Bigg](w)
\nonu \\ 
& + & 
\frac{1}{(z-w)}\,\Bigg[\,
\partial^2 Q^{(1)}_{2} 
+\partial Q^{(2)}_{2}
+Q^{(3)}_{2}
\nonu \\
& - &
\frac{8(k-N)}{(5+4k+4N+3kN)}
\,\partial (\Phi^{(1)}_0 \Phi^{(1)}_0+c^{0,2}_{0}L)
\,
\Bigg](w)
+\cdots.
\label{fiveOPEs}
\eea
The quasi primary fields appearing in (\ref{fiveOPEs})
are summarized by \footnote{
  The expressions with typewriter font in Appendices will
  vanish under the large $(N,k)$ limit. Some of the linear terms also
vanish.}
\bea
Q^{(\frac{3}{2}),i}_{\frac{1}{2}}
& = &
w_{1,\frac{3}{2}}\,\tilde{G}^i
\mathtt{
+w_{2,\frac{3}{2}}\,\partial \Gamma^i
+w_{3,\frac{3}{2}}\,\tilde{T}^{ij}\Gamma^j
+w_{4,\frac{3}{2}}\,U\Gamma^i
+w_{5,\frac{3}{2}}\, \varepsilon^{ijkl}\, \Gamma^j \Gamma^k \Gamma^l}
\,,
\nonu \\
Q^{(1),ij}_{1}
& = &
\mathtt{w_{1,1}\,\Gamma^i \Gamma^j}
+w_{2,1}\,\tilde{T}^{ij}
+\mathtt{w_{3,1}\,\varepsilon^{ijkl}\, \Gamma^k \Gamma^l}
+w_{4,1}\,T^{ij}\,,
\nonu\\
Q^{(2),ij}_{1}
& = &
\mathtt{w_{1,2}\,(\tilde{G}^i \Gamma^j-\tilde{G}^j \Gamma^i)
+w_{2,2}\,\partial \tilde{T}^{ij} ,
+w_{3,2}\,\partial (\Gamma^i \Gamma^j)
+w_{4,4}\,\varepsilon^{ijkl}\, \partial (\Gamma^k \Gamma^l)}
\nonu \\ 
& + &
\mathtt{w_{5,3}\,(\tilde{T}^{ik} \Gamma^k \Gamma^j  -   \tilde{T}^{jk} \Gamma^k \Gamma^i)
+w_{6,4}\,U \Gamma^i \Gamma^j}
\,,
\nonu\\
Q^{(\frac{1}{2}),i}_{\frac{3}{2}}
& = &
w_{1,\frac{1}{2}}\,\Gamma^i,
\nonu\\
Q^{(\frac{3}{2}),i}_{\frac{3}{2}}
& = &
w_{1,\frac{3}{2}}\,\tilde{G}^i
+w_{2,\frac{3}{2}}\,\partial \Gamma^i
\mathtt{+w_{3,\frac{3}{2}}\,T^{ij} \Gamma^j
+w_{4,\frac{3}{2}}\,U \Gamma^i
+\varepsilon^{ijkl}
(
w_{5,\frac{3}{2}}\,T^{jk}\Gamma^l
+w_{6,\frac{3}{2}}\,\Gamma^j\Gamma^k\Gamma^l
)},
\nonu\\
Q^{(\frac{5}{2}),i}_{\frac{3}{2}}
& = &
w_{1,\frac{5}{2}}\,\Phi^{(2),i}_{\frac{1}{2}}
\mathtt{+w_{2,\frac{5}{2}}\,\Phi^{(1)}_0 \Phi^{(1),i}_{\frac{1}{2}}
+w_{3,\frac{5}{2}}\,\tilde{T}^{ij}\tilde{T}^{ij}\Gamma^i
+w_{4,\frac{5}{2}}\,U U \Gamma^i
+w_{5,\frac{5}{2}}\, \partial U \Gamma^i
+w_{6,\frac{5}{2}}\, \partial^2 \Gamma^i}
\nonu\\
&+ & 
\mathtt{w_{7,\frac{5}{2}}\,\tilde{T}^{ij} \partial \Gamma^j
+w_{8,\frac{5}{2}}\, \Gamma^{i} \partial \Gamma^{j}\Gamma^{j}
+w_{9,\frac{5}{2}}\,\tilde{T}^{ij} \tilde{G}^{j}
+w_{10,\frac{5}{2}}\, \partial \tilde{T}^{ij} \Gamma^j
+w_{11,\frac{5}{2}}\, T^{ij} \tilde{G}^j}
\nonu\\
&+ &
\mathtt{w_{12,\frac{5}{2}}\, \tilde{T}^{ij} U \Gamma^j
+w_{13,\frac{5}{2}}\, \Gamma^i \Gamma^j \tilde{G}^j
+w_{14,\frac{5}{2}}\, \partial \tilde{G}^i
+w_{15,\frac{5}{2}}\, U \tilde{G}^i
+
\varepsilon^{ijkl}
\Big(
w_{16,\frac{5}{2}}\, \Gamma^j\Gamma^k \tilde{G}^l}
\nonu\\
& + &
\mathtt{+w_{17,\frac{5}{2}}\, T^{ij} \tilde{T}^{ik} \Gamma^l
+w_{18,\frac{5}{2}}\, \tilde{T}^{ij} \tilde{T}^{kl}  \Gamma^i
+w_{19,\frac{5}{2}}\, U  \tilde{T}^{jk} \Gamma^l
+w_{20,\frac{5}{2}}\,\tilde{T}^{ij} \Gamma^i \Gamma^k \Gamma^l
+w_{21,\frac{5}{2}}\,\tilde{T}^{jk} \partial \Gamma^l }
\nonu\\
& + &
\mathtt{+w_{22,\frac{5}{2}}\,U \Gamma^j \Gamma^k \Gamma^l
+w_{23,\frac{5}{2}}\, \partial (\Gamma^j \Gamma^k \Gamma^l)
+w_{24,\frac{5}{2}}\, \partial \tilde{T}^{jk} \Gamma^l
\Big)
+w_{25,\frac{5}{2}}\,U \partial \Gamma^i } \,,
\nonu \\
Q^{(1)}_{2}
& = &
w_{1,1}\,U \,,
\nonu\\
Q^{(2)}_{2}
& = &
w_{1,2}\,\Phi^{(2)}_0
\mathtt{+w_{2,2}\, \tilde{L}
+w_{3,2}\, \tilde{G}^i \Gamma^i
+\varepsilon^{ijkl}
(
w_{4,2}\,\Gamma^i\Gamma^j\Gamma^k\Gamma^l
+w_{5,2}\,T^{ij} T^{kl}
)
+w_{6,2}\, T^{ij} T^{ij}}
\nonu\\
&+& 
\mathtt{w_{7,2}\, T^{ij} \Gamma^i\Gamma^j
+w_{8,2}\, \tilde{T}^{ij} \Gamma^i\Gamma^j
+w_{9,2}\, \partial U
+w_{10,2}\, \partial \Gamma^i\Gamma^i
+w_{11,2}\, U U} \,,
\nonu\\
Q^{(3)}_{2}
& = &
\mathtt{w_{1,3}\,\tilde{G}^{i} \partial \Gamma^i
+w_{2,3}\,\partial \tilde{G}^{i} \Gamma^i
+w_{3,3}\,\partial^2 U
+w_{4,3}\, \partial^2 \Gamma^i \Gamma^i
+w_{5,3}\, \partial(\tilde{T}^{ij} \Gamma^i \Gamma^j)}
\nonu\\
&+&
\mathtt{w_{6,3}\, \Gamma^i \Gamma^j \partial \tilde{T}^{ij}
+w_{7,3}\, U \partial \Gamma^i \Gamma^i  }\,.
\label{quasinonsuper}
\eea
We can obtain (\ref{quasinonsuper}) from (\ref{quasisuper})
by taking the appropriate procedures in (\ref{comptosuper}).
Note that the new higher spin-$2$ primary field
$\Phi_0^{(2)}$
appears in the quasi primary field $Q^{(2)}_{2}$. See also
(\ref{twoquasiother}) \footnote{
  All the other $120(=136-16)$ OPEs
  can be read off from the relations (\ref{comptosuper})
  and (\ref{super1super1})
  in the ${\cal N}=4$
  superspace. 
Then we obtain all the results in Appendix $H$ of \cite{AK1509}. 
}.
There are also linear terms having the typewriter font.
The fields with a tilde are defined in the footnote \ref{BigJ}.

\section{ Quasi (super)primary fields from section $3$}

The various quasi primary fields in  sections $3.1$, $3.2$,
and $3.3$
are presented
as follows:
\bea
Q^{(1)}_{0} & = & \mathtt{ d_{1}^{0,2}\,\Phi_{0}^{(1)}(w)}\,,
\nonu \\
Q^{(\frac{5}{2}),i}_{\frac{1}{2}}
& = &
w_{1,\frac{5}{2}}\,\Phi^{(1),i}_{\frac{3}{2}}
\mathtt{+w_{2,\frac{5}{2}}\,\partial \Phi^{(1),i}_{\frac{1}{2}}
+w_{3,\frac{5}{2}}\,\partial \Gamma^i \Phi^{(1)}_{0}
+w_{4,\frac{5}{2}}\, \Gamma^i \partial \Phi^{(1)}_{0}
+w_{5,\frac{5}{2}}\, U \Phi^{(1),i}_{\frac{1}{2}}
+w_{6,\frac{5}{2}}\,  \tilde{G}^i \Phi^{(1)}_{0}}
\nonu \\ 
& + &
\mathtt{w_{7,\frac{5}{2}}\, U \Gamma^i \Phi^{(1)}_{0}
+w_{8,\frac{5}{2}}\,  T^{ij}  \Phi^{(1),j}_{\frac{1}{2}}
+w_{9,\frac{5}{2}}\,   \Gamma^i   \Gamma^j  \Phi^{(1),j}_{\frac{1}{2}}
+\varepsilon^{ijkl}
\Big(
w_{10,\frac{5}{2}}\,  \Gamma^j \Phi^{(1),kl}_{1}
+w_{11,\frac{5}{2}}\, T^{jk} \Phi^{(1),l}_{\frac{1}{2}}}
\nonu \\ 
& + &
\mathtt{w_{12,\frac{5}{2}}\, T^{jk}  \Gamma^l  \Phi^{(1),l}_{0}
+w_{13,\frac{5}{2}}\,  \Gamma^j  \Gamma^k  \Phi^{(1),l}_{\frac{1}{2}}
+w_{14,\frac{5}{2}}\,  \Gamma^j  \Gamma^k \Gamma^l \Phi^{(1)}_{0}}
\Big)
\,,
\nonu \\
Q^{(2),ij}_{1}
& = &
w_{1,2}\,\Phi^{(1),ij}_1
\mathtt{+w_{2,2}\,T^{ij} \Phi^{(1)}_0
+w_{3,2}\,\tilde{T}^{ij} \Phi^{(1)}_0
+w_{4,2}\, \Gamma^i \Gamma^j \Phi^{(1)}_0
+w_{5,2}\,(\Gamma^i \Phi^{(1),j}_{\frac{1}{2}}-\Gamma^j \Phi^{(1),i}_{\frac{1}{2}})}
\nonu\\
& + &
w_{6,2}\, \tilde{\Phi}^{(1),ij}_{1}
\mathtt{+\varepsilon^{ijkl}
(
w_{7,2}\,\Gamma^k \Gamma^l \Phi^{(1)}_0
+w_{8,2}\,\Gamma^k \Phi^{(1),l}_{\frac{1}{2}}
)}
\,,
\nonu\\
Q^{(3),ij}_{1}
& = &
\mathtt{
w_{1,3}\,\partial \Phi^{(1),ij}_1
+w_{2,3}\,(\Gamma^i \Phi^{(1),j}_\frac{3}{2}-\Gamma^j \Phi^{(1),i}_\frac{3}{2})
+w_{3,3}\, (\Gamma^i \partial \Phi^{(1),j}_{\frac{1}{2}}-\Gamma^j \partial \Phi^{(1),i}_{\frac{1}{2}})}
\nonu\\
& + &
\mathtt{
w_{4,3}\, (\partial \Gamma^i \Phi^{(1),j}_{\frac{1}{2}}-\partial \Gamma^j \Phi^{(1),i}_{\frac{1}{2}})
+w_{5,3}\, (T^{ik} \Phi^{(1),jk}_1 -T^{jk} \Phi^{(1),ik}_1 )
+w_{6,3}\,  \partial T^{ij} \Phi^{(1)}_0}
\nonu\\
& + &
\mathtt{
w_{7,3}\,   T^{ij} \partial \Phi^{(1)}_0
+w_{8,3}\, (\Gamma^i \tilde{T}^{ij} \Phi^{(1),i}_{\frac{1}{2}}-\Gamma^j \tilde{T}^{ji} \Phi^{(1),j}_{\frac{1}{2}})
+w_{9,3}\, (\Gamma^i U \Phi^{(1),i}_{\frac{1}{2}}-\Gamma^j U\Phi^{(1),j}_{\frac{1}{2}})}
\nonu\\
& + & 
\mathtt{w_{10,3}\,  (\Gamma^i \tilde{G}^j
  \Phi^{(1)}_{0}-\Gamma^j \tilde{G}^i \Phi^{(1)}_{0})
+w_{11,3}\,  \Gamma^i \Gamma^j  \partial  \Phi^{(1)}_0
+w_{12,3}\, (\Gamma^i \partial \Gamma^j \Phi^{(1)}_0-\Gamma^j \partial \Gamma^i \Phi^{(1)}_0)}
\nonu\\
& + &
\mathtt{w_{13,3}\, \Gamma^i \Gamma^j U \Phi^{(1)}_0
+w_{14,3}\, (\Gamma^i \Gamma^k \Phi^{(1),jk}_1-  \Gamma^j \Gamma^k \Phi^{(1),ik}_1)
+w_{15,3}\,  (\Gamma^k T^{ik} \Phi^{(1),j}_{\frac{1}{2}} - \Gamma^k T^{jk} \Phi^{(1),i}_{\frac{1}{2}})}
\nonu\\
& + &
\mathtt{
+\varepsilon^{ijkl}
\Big(
w_{16,3}\,(\Gamma^i T^{ik} \Phi^{(1),l}_{\frac{1}{2}}+\Gamma^j T^{jk} \Phi^{(1),l}_{\frac{1}{2}})
+w_{17,3}\, \partial \Phi^{(1),kl}_1
+w_{18,3}\, \Gamma^k \partial \Phi^{(1),l}_\frac{1}{2}}
\nonu\\
& + &
\mathtt{+w_{19,3}\, \partial \Gamma^k \Phi^{(1),l}_\frac{1}{2}
+w_{20,3}\, (T^{ik} \Phi^{(1),il}_1 +T^{jk} \Phi^{(1),jl}_1 )
+w_{21,3}\, \partial T^{kl} \Phi^{(1)}_0
+w_{22,3}\, T^{kl}  \partial \Phi^{(1)}_0}
\nonu\\
& + &
\mathtt{
w_{23,3}\,  (\Gamma^i \Gamma^k \Phi^{(1),il}_1 + \Gamma^j \Gamma^k \Phi^{(1),jl}_1)
+w_{24,3}\,  (\Gamma^k T^{il} \Phi^{(1),i}_{\frac{1}{2}} +\Gamma^k T^{jl} \Phi^{(1),j}_{\frac{1}{2}})
+w_{25,3}\, \Gamma^k \Gamma^l \partial \Phi^{(1)}_{0}}
\nonu\\
& + &
\mathtt{
w_{26,3}\, \partial (\Gamma^k \Gamma^l )\Phi^{(1)}_{0}
+w_{27,3}\, (\Gamma^i \Gamma^k \Gamma^l  \Phi^{(1),i}_{\frac{1}{2}} +\Gamma^j \Gamma^k \Gamma^l  \Phi^{(1),j}_{\frac{1}{2}})}
\nonu\\
& + &
\mathtt{w_{28,3}\, (\Gamma^i \Gamma^k  T^{il} \Phi^{(1)}_0 +\Gamma^j \Gamma^k  T^{jl} \Phi^{(1)}_0 )
\Big)}
\,,
\nonu
\\
Q^{(\frac{3}{2}),i}_{\frac{3}{2}}
& = &
w_{1,\frac{3}{2}}\,\Phi^{(1),i}_{\frac{1}{2}}
\mathtt{+w_{2,\frac{3}{2}}\,\Gamma^i \Phi^{(1)}_0}\,,
\nonu\\
Q^{(\frac{5}{2}),i}_{\frac{3}{2}}
& = &
w_{1,\frac{5}{2}}\,\Phi^{(1),i}_{\frac{3}{2}}
\mathtt{+w_{2,\frac{5}{2}}\,\partial \Phi^{(1),i}_{\frac{1}{2}}
+w_{3,\frac{5}{2}}\, \Gamma^i \partial\Phi^{(1)}_0
+w_{4,\frac{5}{2}}\, \partial  \Gamma^i \Phi^{(1)}_0
+w_{5,\frac{5}{2}}\, U \Phi^{(1),i}_{\frac{1}{2}}
+w_{6,\frac{5}{2}}\,  \tilde{G}^i \Phi^{(1)}_0}
\nonu\\
& + &
\mathtt{w_{7,\frac{5}{2}}\,   \Gamma^i U  \Phi^{(1)}_0
+w_{8,\frac{5}{2}}\,  \Gamma^k \Phi^{(1),ik}_1
+w_{9,\frac{5}{2}}\,  T^{ik} \Phi^{(1),k}_{\frac{1}{2}}
+w_{10,\frac{5}{2}}\,  \Gamma^i \Gamma^k  \Phi^{(1),k}_{\frac{1}{2}}
+w_{11,\frac{5}{2}}\,  \Gamma^k  T^{ik}  \Phi^{(1)}_0}
\nonu\\
& + &
\mathtt{
\varepsilon^{ijkl}
\Big(
w_{12,\frac{5}{2}}\,  \Gamma^j \Phi^{(1),kl}_{1}
+w_{13,\frac{5}{2}}\,  T^{jk}  \Phi^{(1),l}_{\frac{1}{2}}
+w_{14,\frac{5}{2}}\,  \Gamma^j \Gamma^k  \Phi^{(1),l}_{\frac{1}{2}}
+w_{15,\frac{5}{2}}\,   \Gamma^j  T^{kl}  \Phi^{(1)}_0}
\nonu\\
& + &
\mathtt{w_{16,\frac{5}{2}}\,   \Gamma^j\Gamma^k\Gamma^l \Phi^{(1)}_0 
\Big)}\,,
\nonu\\
Q^{(\frac{7}{2}),i}_{\frac{3}{2}}
& = &
w_{1,\frac{7}{2}}\, \Phi^{(3),i}_{\frac{1}{2}}
\mathtt{+w_{2,\frac{7}{2}}\, \partial \Phi^{(1),i}_{\frac{3}{2}}
+w_{3,\frac{7}{2}}\,  \partial^2 \Phi^{(1),i}_{\frac{1}{2}}
+w_{4,\frac{7}{2}}\,  \Phi^{(1)}_0 \Phi^{(2),i}_{\frac{1}{2}}
+w_{5,\frac{7}{2}}\,  \Phi^{(1),i}_{\frac{1}{2}}  \Phi^{(2)}_0}
\nonu\\
& + &
\mathtt{w_{6,\frac{7}{2}}\,  \Phi^{(1)}_0 \Phi^{(1)}_0  \Phi^{(1),i}_{\frac{1}{2}}
+w_{7,\frac{7}{2}}\, \tilde{L} \Phi^{(1),i}_{\frac{1}{2}} 
+w_{8,\frac{7}{2}}\, U \Phi^{(1),i}_{\frac{3}{2}}
+w_{9,\frac{7}{2}}\,  U U \Phi^{(1),i}_{\frac{1}{2}}
+w_{10,\frac{7}{2}}\,  U \tilde{G}^i \Phi^{(1)}_{0}}
\nonu\\
& + &
\mathtt{w_{11,\frac{7}{2}}\, U  \partial \Phi^{(1),i}_{\frac{1}{2}}
+w_{12,\frac{7}{2}}\,  \partial U \Phi^{(1),i}_{\frac{1}{2}}
+w_{13,\frac{7}{2}}\,  \tilde{G}^i  \partial \Phi^{(1)}_{0}
+w_{14,\frac{7}{2}}\,  \partial  \tilde{G}^i  \Phi^{(1)}_{0}
+w_{15,\frac{7}{2}}\,  \Gamma^i  U U \Phi^{(1)}_{0}}
\nonu\\
& + &
\mathtt{w_{16,\frac{7}{2}}\,  \Gamma^i  \partial^2 \Phi^{(1)}_{0}
+w_{17,\frac{7}{2}}\,  \partial \Gamma^i  \partial \Phi^{(1)}_{0}
+w_{18,\frac{7}{2}}\,   \partial^2 \Gamma^i   \Phi^{(1)}_{0}
+w_{19,\frac{7}{2}}\,   \Gamma^i U  \partial\Phi^{(1)}_{0}
+w_{20,\frac{7}{2}}\,  \Gamma^i \partial U  \Phi^{(1)}_{0}}
\nonu\\
& + &
\mathtt{w_{21,\frac{7}{2}}\,    \partial  \Gamma^i U  \Phi^{(1)}_{0}
+w_{22,\frac{7}{2}}\,    T^{ij} \Phi^{(1),j}_{\frac{3}{2}}
+w_{23,\frac{7}{2}}\,   \partial T^{ij} \Phi^{(1),j}_{\frac{1}{2}}
+w_{24,\frac{7}{2}}\, T^{ij} \partial \Phi^{(1),j}_{\frac{1}{2}}}
\nonu\\
& + &
\mathtt{w_{25,\frac{7}{2}}\,  \Gamma^i  \tilde{G}^j  \Phi^{(1),j}_{\frac{1}{2}}
+w_{26,\frac{7}{2}}\,   \Gamma^j  \tilde{G}^j  \Phi^{(1),i}_{\frac{1}{2}}
+w_{27,\frac{7}{2}}\, \tilde{G}^j   \Phi^{(1),ij}_{1}
+w_{28,\frac{7}{2}}\,  \Gamma^j  \partial \Phi^{(1),ij}_{1}}
\nonu\\
& + &
\mathtt{w_{29,\frac{7}{2}}\,  \partial \Gamma^j   \Phi^{(1),ij}_{1}
+w_{30,\frac{7}{2}}\,  \Gamma^j  U \Phi^{(1),ij}_{1}
+w_{31,\frac{7}{2}}\,   \Gamma^i  \Gamma^j   \Phi^{(1),j}_{\frac{3}{2}}
+w_{32,\frac{7}{2}}\, \Gamma^i  \Gamma^j \partial \Gamma^j \Phi^{(1)}_{0}}
\nonu\\
& + &
\mathtt{w_{33,\frac{7}{2}}\,  \partial \Gamma^i  \Gamma^j  \Phi^{(1),j}_{\frac{1}{2}}
+w_{34,\frac{7}{2}}\,  \Gamma^i  \partial \Gamma^j  \Phi^{(1),j}_{\frac{1}{2}}
+w_{35,\frac{7}{2}}\,  \Gamma^i \Gamma^j   \partial  \Phi^{(1),j}_{\frac{1}{2}}
+w_{36,\frac{7}{2}}\,  \Gamma^j  \partial \Gamma^j  \Phi^{(1),i}_{\frac{1}{2}}}
\nonu\\
& + &
\mathtt{w_{37,\frac{7}{2}}\, \partial \Gamma^j  T^{ij} \Phi^{(1)}_{0}
+w_{38,\frac{7}{2}}\,  \Gamma^j  \partial T^{ij} \Phi^{(1)}_{0}
+w_{39,\frac{7}{2}}\,  \Gamma^j   T^{ij} \partial \Phi^{(1)}_{0}
+w_{40,\frac{7}{2}}\,  \Gamma^j   T^{ij} U \Phi^{(1)}_{0}}
\nonu\\
& + &
\mathtt{w_{41,\frac{7}{2}}\,  \Gamma^i  \tilde{T}^{ij} \tilde{T}^{ij}  \Phi^{(1)}_{0}  
+w_{42,\frac{7}{2}}\, \Gamma^i  T^{ij}  \Phi^{(1),ij}_{1}  
+w_{43,\frac{7}{2}}\,  \Gamma^i \Gamma^j  T^{ij}   \Phi^{(1),i}_{\frac{1}{2}}  
+w_{44,\frac{7}{2}}\,  T^{ij} \tilde{G}^j  \Phi^{(1)}_{0}}
\nonu\\
& + &
\mathtt{w_{45,\frac{7}{2}}\, \Gamma^i \Gamma^j  \tilde{G}^j \Phi^{(1)}_{0}
+w_{46,\frac{7}{2}}\,   T^{ij} U \Phi^{(1),j}_{\frac{1}{2}}  
+w_{47,\frac{7}{2}}\,  \Gamma^i T^{ij} T^{ij}  \Phi^{(1)}_{0}
+w_{48,\frac{7}{2}}\,  \Gamma^j T^{ik} T^{jk}  \Phi^{(1)}_{0}}
\nonu\\
& + &
\mathtt{w_{49,\frac{7}{2}}\,  \Gamma^i T^{jk}  \Phi^{(1),jk}_{1}
+w_{50,\frac{7}{2}}\,  \Gamma^j T^{ik}  \Phi^{(1),jk}_{1}
+w_{51,\frac{7}{2}}\, \Gamma^i \Gamma^j \Gamma^k T^{jk} \Phi^{(1)}_{0}
+w_{52,\frac{7}{2}}\,  \Gamma^j \Gamma^k T^{jk} \Phi^{(1),i}_{\frac{1}{2}} } 
\nonu\\
& + &
\mathtt{w_{53,\frac{7}{2}}\,  \Gamma^i  \Gamma^j  \Gamma^k \Phi^{(1),jk}_{1}  
+w_{54,\frac{7}{2}}\,  T^{jk} T^{jk} \Phi^{(1),i}_{\frac{1}{2}}  
+w_{55,\frac{7}{2}}\, T^{ij} T^{jk} \Phi^{(1),k}_{\frac{1}{2}}  
+
\varepsilon^{ijkl}
\Big(
w_{56,\frac{7}{2}}\, T^{jk} \Phi^{(1),l}_{\frac{3}{2}}  }
\nonu\\
& + &
\mathtt{w_{57,\frac{7}{2}}\,  \pa \,T^{jk} \Phi^{(1),l}_{\frac{1}{2}}  
+w_{58,\frac{7}{2}}\,  T^{jk} \partial \Phi^{(1),l}_{\frac{1}{2}}  
+w_{59,\frac{7}{2}}\, \Gamma^j \Gamma^k \Phi^{(1),l}_{\frac{1}{2}}  
+w_{60,\frac{7}{2}}\, \tilde{G}^j  \Phi^{(1),kl}_{1}}
\nonu\\
& + &
\mathtt{w_{61,\frac{7}{2}}\,  \Gamma^j  \partial \Phi^{(1),kl}_{1}
+w_{62,\frac{7}{2}}\,  \partial  \Gamma^j \Phi^{(1),kl}_{1}
+w_{63,\frac{7}{2}}\,   \Gamma^j U  \Phi^{(1),kl}_{1}
+w_{64,\frac{7}{2}}\,  \Gamma^j \Gamma^k  \Phi^{(1),l}_{\frac{3}{2}}  }
\nonu\\
& + &
\mathtt{w_{65,\frac{7}{2}}\, \Gamma^j \Gamma^k\Gamma^l \partial \Phi^{(1)}_{0}  
+w_{66,\frac{7}{2}}\,  \partial (\Gamma^j \Gamma^k\Gamma^l  \Phi^{(1)}_{0})
+w_{67,\frac{7}{2}}\,  \Gamma^j \Gamma^k  \partial  \Phi^{(1),l}_{\frac{1}{2}}  
+w_{68,\frac{7}{2}}\, \partial  (\Gamma^j \Gamma^k  \Phi^{(1),l}_{\frac{1}{2}}  )}
\nonu\\
& + &
\mathtt{w_{69,\frac{7}{2}}\,  \Gamma^j T^{kl} \Phi^{(1)}_{0}  
+w_{70,\frac{7}{2}}\, \Gamma^j \partial  T^{kl} \Phi^{(1)}_{0}  
+w_{71,\frac{7}{2}}\,  \Gamma^j  T^{kl} \partial  \Phi^{(1)}_{0} 
+w_{72,\frac{7}{2}}\, \Gamma^j  T^{kl} U  \Phi^{(1)}_{0} }
\nonu\\
& + &
\mathtt{w_{73,\frac{7}{2}}\,  \Gamma^j\Gamma^k U  \Phi^{(1),l}_{\frac{1}{2}} 
+w_{74,\frac{7}{2}}\,  \Gamma^j T^{ij} T^{kl} \Phi^{(1)}_{0} 
+w_{75,\frac{7}{2}}\, \Gamma^j \tilde{T}^{ik} \Phi^{(1),il}_{1}
+w_{76,\frac{7}{2}}\,  \Gamma^i T^{ij} \Phi^{(1),kl}_{1}}
\nonu\\
& + &
\mathtt{w_{77,\frac{7}{2}}\,  \Gamma^i T^{jk} \Phi^{(1),il}_{1}
+w_{78,\frac{7}{2}}\, \Gamma^j \tilde{T}^{ik} \tilde{\Phi}^{(1),il}_{1}
+w_{79,\frac{7}{2}}\,  \Gamma^i \Gamma^j \tilde{T}^{ik} \Phi^{(1),l}_{\frac{1}{2}}
+w_{80,\frac{7}{2}}\, \Gamma^i \Gamma^j T^{kl} \Phi^{(1),i}_{\frac{1}{2}}}
\nonu\\
& + &
\mathtt{w_{81,\frac{7}{2}}\,  \Gamma^j \Gamma^k T^{li} \Phi^{(1),i}_{\frac{1}{2}}
+w_{82,\frac{7}{2}}\,  T^{jk} \tilde{G}^{l} \Phi^{(1)}_{0} 
+w_{83,\frac{7}{2}}\, \Gamma^j \Gamma^k  \tilde{G}^{l} \Phi^{(1)}_{0} 
+w_{84,\frac{7}{2}}\, \Gamma^i \Gamma^j \Gamma^k \Gamma^l  \Phi^{(1),i}_{\frac{1}{2}}}
\nonu\\
& + &
\mathtt{w_{85,\frac{7}{2}}\, \Gamma^j \Gamma^k \Gamma^l  U \Phi^{(1)}_{0} 
+w_{86,\frac{7}{2}}\,  T^{jk} U \Phi^{(1),l}_{\frac{1}{2}}
+w_{87,\frac{7}{2}}\,  T^{ij} T^{kl} \Phi^{(1),i}_{\frac{1}{2}}
\Big)}
 \,,
\nonu \\
Q^{(1)}_{2}
& = &
w_{1,1}\,\Phi^{(1)}_0 \,,
\nonu\\
Q^{(2)}_{2}
& = &
\mathtt{w_{1,2}\, U \Phi^{(1)}_0}
\,,
\nonu\\
Q^{(3)}_{2}
& = &
w_{1,3}\,  \Phi^{(3)}_0
+w_{2,3}\,  \Phi^{(1)}_2
\mathtt{+w_{3,3}\,  \partial^2 \Phi^{(1)}_0
+w_{4,3}\,  \Phi^{(1)}_0 \Phi^{(1)}_0 \Phi^{(1)}_0
+w_{5,3}\,  \tilde{L} \Phi^{(1)}_0
+w_{6,3}\, U \partial  \Phi^{(1)}_0}
\nonu\\
& + &
\mathtt{w_{7,3}\, \partial U   \Phi^{(1)}_0
+w_{8,3}\,  U  U \Phi^{(1)}_0
+w_{9,3}\, \tilde{G}^i \Phi^{(1),i}_{\frac{1}{2}}
+w_{10,3}\,  \Gamma^i  \Phi^{(1),i}_{\frac{3}{2}}
+w_{11,3}\,  \Gamma^i \partial \Phi^{(1),i}_{\frac{1}{2}}}
\nonu\\
& + &
\mathtt{w_{12,3}\,  U  \Gamma^i \Phi^{(1),i}_{\frac{1}{2}}
+w_{13,3}\,  \Gamma^i \tilde{G}^i \Phi^{(1)}_{0}
+w_{14,3}\,  \partial \Gamma^i  \Phi^{(1),i}_{\frac{1}{2}}
+w_{15,3}\, \partial \Gamma^i \Gamma^i \Phi^{(1)}_{0}
+w_{16,3}\, T^{ij} \Phi^{(1),ij}_{1}}
\nonu\\
& + &
\mathtt{w_{17,3}\,  T^{ij} T^{ij} \Phi^{(1)}_{0}
+w_{18,3}\, \Gamma^i T^{ij} \Phi^{(1),j}_{\frac{1}{2}}
+w_{19,3}\, \Gamma^i \Gamma^j T^{ij} \Phi^{(1)}_{0} 
+w_{20,3}\,  \Gamma^i \Gamma^j \Phi^{(1),ij}_{1}}
\nonu\\
& + &
\mathtt{\varepsilon^{ijkl}
\Big(
w_{21,3}\, T^{ij} \Phi^{(1),kl}_{1}
+w_{22,3}\, T^{ij} T^{kl} \Phi^{(1)}_{0} 
+w_{23,3}\, T^{ij} \Gamma^k \Phi^{(1),l}_{\frac{1}{2}}
+w_{24,3}\,\Gamma^i \Gamma^j T^{kl} \Phi^{(1)}_{0} }
\nonu\\
& + &
\mathtt{w_{25,3}\,  \Gamma^i \Gamma^j \Phi^{(1),kl}_{1}
+w_{26,3}\,  \Gamma^i \Gamma^j \Gamma^k \Gamma^l \Phi^{(1)}_{0} 
\Big)}
\,,
\nonu\\
Q^{(4)}_{2}
& = &
\mathtt{w_{1,4}\,  \Phi^{(1)}_0 \partial \Phi^{(2)}_0
+w_{2,4}\, \partial \Phi^{(1)}_0 \Phi^{(2)}_0
+w_{3,4}\,  \partial^3 \Phi^{(1)}_0
+w_{4,4}\,  \tilde{L} \partial \Phi^{(1)}_0
+w_{5,4}\,    \partial \tilde{L} \Phi^{(1)}_0}
\nonu\\
&+&
\mathtt{w_{6,4}\,  U U  \partial \Phi^{(1)}_0
+w_{7,4}\,   U \partial U   \Phi^{(1)}_0
+w_{8,4}\,   U  \partial^2 \Phi^{(1)}_0
+w_{9,4}\, \partial  U  \partial \Phi^{(1)}_0
+w_{10,4}\,  \partial^2  U  \Phi^{(1)}_0}
\nonu\\
&+&
\mathtt{w_{11,4}\,    \partial \Gamma^i \Phi^{(1),i}_{\frac{3}{2}}
+w_{12,4}\,  \Gamma^i  \partial \Phi^{(1),i}_{\frac{3}{2}}
+w_{13,4}\,  \partial \Gamma^i  \tilde{G}^i   \Phi^{(1)}_0
+w_{14,4}\,   \Gamma^i  \partial  \tilde{G}^i   \Phi^{(1)}_0
+w_{15,4}\,   \Gamma^i   \tilde{G}^i  \partial  \Phi^{(1)}_0}
\nonu\\
&+&
\mathtt{w_{16,4}\,   \partial (\Gamma^i  U  \Phi^{(1),i}_{\frac{1}{2}})
+w_{17,4}\, \partial \Gamma^i  U  \Phi^{(1),i}_{\frac{1}{2}}
+w_{18,4}\, \partial \Gamma^i   \Gamma^i   U   \Phi^{(1)}_0
+w_{19,4}\, \Gamma^i   \Gamma^i  \partial U   \Phi^{(1)}_0}
\nonu\\
&+&
\mathtt{w_{20,4}\,  \partial \Gamma^i \Phi^{(1),i}_{\frac{3}{2}}
+w_{21,4}\, \partial^2 \Gamma^i \Phi^{(1),i}_{\frac{1}{2}}
+w_{22,4}\,  \partial \Gamma^i \partial \Phi^{(1),i}_{\frac{1}{2}}
+w_{23,4}\,  \Gamma^i \partial^2 \Phi^{(1),i}_{\frac{1}{2}}
+w_{24,4}\,  \partial^2 \Gamma^i \Gamma^i \Phi^{(1)}_0}
\nonu\\
&+&
\mathtt{w_{25,4}\, \Gamma^i \Gamma^i \partial^2 \Phi^{(1)}_0
+w_{26,4}\,  \partial \Gamma^i \Gamma^i \partial \Phi^{(1)}_0
+w_{27,4}\, \partial T^{ij}  \Phi^{(1),ij}_1
+w_{28,4}\,  T^{ij}  \partial \Phi^{(1),ij}_1}
\nonu\\
&+&
\mathtt{w_{29,4}\,  T^{ij} T^{ij}  \partial \Phi^{(1)}_0
+w_{30,4}\, \partial  T^{ij} T^{ij}  \Phi^{(1)}_0
+w_{31,4}\,    \Gamma^i  \Gamma^j T^{ij}  \partial \Phi^{(1)}_0
+w_{32,4}\,  \Gamma^i  \Gamma^j  \partial T^{ij}  \Phi^{(1)}_0}
\nonu\\
&+&
\mathtt{w_{33,4}\, \partial \Gamma^i  \Gamma^j   T^{ij}  \Phi^{(1)}_0
+w_{34,4}\,  \Gamma^i T^{ij}  \partial \Phi^{(1),j}_{\frac{1}{2}}
+w_{35,4}\,  \partial (\Gamma^i T^{ij}  \Phi^{(1),j}_{\frac{1}{2}})
+w_{36,4}\,   \Gamma^i  \Gamma^j  \partial \Phi^{(1),ij}_1}
\nonu\\
&+&
\mathtt{w_{37,4}\,  \partial \Gamma^i  \Gamma^j  \Phi^{(1),ij}_1
+\varepsilon^{ijkl}
\Big(
w_{38,4}\, \partial T^{ij}  \Phi^{(1),kl}_1
+w_{39,4}\,  T^{ij} \partial \Phi^{(1),kl}_1
+w_{40,4}\, T^{ij} T^{kl} \partial \Phi^{(1)}_0}
\nonu\\
&+&
\mathtt{w_{41,4}\, \partial (T^{ij} T^{kl} \Phi^{(1)}_0)
+w_{42,4}\,  \Gamma^i  \Gamma^j T^{kl} \partial \Phi^{(1)}_0
+w_{43,4}\,   \Gamma^i  \Gamma^j \partial T^{kl} \Phi^{(1)}_0
+w_{44,4}\, \partial \Gamma^i  \Gamma^j  T^{kl} \Phi^{(1)}_0}
\nonu\\
&+&
\mathtt{w_{45,4}\,  \pa \, \Gamma^i  T^{jk} \Phi^{(1),l}_{\frac{1}{2}}
+w_{46,4}\,  \Gamma^i  \partial T^{jk} \Phi^{(1),l}_{\frac{1}{2}}
+w_{47,4}\,   \Gamma^i   T^{jk} \partial \Phi^{(1),l}_{\frac{1}{2}}
+w_{48,4}\, \Gamma^i \Gamma^j   \pa \, \Phi^{(1),kl}_{1}}
\nonu\\
&+&
\mathtt{w_{49,4}\,  \partial \Gamma^i \Gamma^j   \Phi^{(1),kl}_{1}
+w_{50,4}\,   \partial( \Gamma^i \Gamma^j  \Gamma^k \Phi^{(1),l}_{\frac{1}{2}})
+w_{51,4}\,  \Gamma^i \Gamma^j  \Gamma^k \partial \Phi^{(1),l}_{\frac{1}{2}}
+w_{52,4}\, \Gamma^i \Gamma^j  \Gamma^k \Gamma^l  \partial \Phi^{(1)}_0}
\nonu\\
&+&
\mathtt{w_{53,4}\,\partial (\Gamma^i \Gamma^j  \Gamma^k \Gamma^l   \Phi^{(1)}_0)
  \Big)}\,.
\label{morequasi1}
\eea
The full expressions of (\ref{spin3ofthird}) and
(\ref{spin7halfquasi}) are given in (\ref{morequasi1}).

The quasi primary fields appearing in 
(\ref{singleOPE}) of section $3.4$ are presented
as follows:
\bea
{\bf Q}^{(1)}_0 & = &
\mathtt{d_{1}^{0,2}\,{\bf \Phi}^{(1)}}
\,,
\nonu\\
{\bf Q}^{(\frac{5}{2}),i}_{\frac{1}{2}} & = &
d_{1}^{\frac{1}{2},1}\,D^{4-i}{\bf \Phi}^{(1)}
\mathtt{
+d_{2}^{\frac{1}{2},1}\,  \partial D^i  {\bf \Phi}^{(1)}
+d_{3}^{\frac{1}{2},1}\, \partial {\bf J}^{i} {\bf \Phi}^{(1)}
+d_{4}^{\frac{1}{2},1}\, {\bf J}^{i} \partial {\bf \Phi}^{(1)}
+d_{5}^{\frac{1}{2},1}\, \partial {\bf J} D^i {\bf \Phi}^{(1)}
}
\nonu\\
&+&
\mathtt{
d_{6}^{\frac{1}{2},1}\,  {\bf J}^{4-i} {\bf \Phi}^{(1)}
+d_{7}^{\frac{1}{2},1}\, \partial {\bf J} {\bf J}^{i} {\bf \Phi}^{(1)}
+d_{8}^{\frac{1}{2},1}\, {\bf J}^{4-ij} D^j {\bf \Phi}^{(1)}
+d_{9}^{\frac{1}{2},1}\, {\bf J}^{i} {\bf J}^{j} D^j {\bf \Phi}^{(1)} }
\nonu\\
&+&
\mathtt{
\varepsilon^{ijkl}
\Big(
d_{10}^{\frac{1}{2},1}\,  {\bf J}^{j} D^{4-kl}{\bf \Phi}^{(1)}
+d_{11}^{\frac{1}{2},1}\, {\bf J}^{4-jk} D^l {\bf \Phi}^{(1)} 
+d_{12}^{\frac{1}{2},1}\,  {\bf J}^{4-jk} {\bf J}^{l} {\bf \Phi}^{(1)}
+d_{13}^{\frac{1}{2},1}\,  {\bf J}^{j} {\bf J}^{k} D^l{\bf \Phi}^{(1)}}
\nonu\\
&+&
\mathtt{
d_{14}^{\frac{1}{2},1}\, {\bf J}^{j}  {\bf J}^{k} {\bf J}^{l} {\bf \Phi}^{(1)}
\Big)}
\,,
\nonu\\
{\bf Q}^{(1),ij}_{1} & = &
d_{1}^{1,2}\, D^{4-ij}{\bf \Phi}^{(1)}
\mathtt{
+d_{2}^{1,2}\,  {\bf J}^{4-ij}  {\bf \Phi}^{(1)}
+d_{3}^{1,2}\,  {\bf J}^{ij}  {\bf \Phi}^{(1)}
+d_{4}^{1,2}\,  {\bf J}^{i}{\bf J}^{j}  {\bf \Phi}^{(1)}}
\nonu\\
&+&
\mathtt{d_{5}^{1,2}\,  ({\bf J}^{j}  D^i{\bf \Phi}^{(1)}-{\bf J}^{i}  D^j{\bf \Phi}^{(1)})}
+d_{6}^{1,2}\,   D^{ij}  {\bf \Phi}^{(1)}
\mathtt{+\varepsilon^{ijkl}\,
(
d_{7}^{1,2}\,{\bf J}^{k}{\bf J}^{l} {\bf \Phi}^{(1)}
+d_{8}^{1,2}\,{\bf J}^{k} D^l {\bf \Phi}^{(1)}}
)\,,
\nonu\\
{\bf Q}^{(2),ij}_{1} & = &
\mathtt{
d_{1}^{1,1}\,  \partial D^{4-ij} {\bf \Phi}^{(1)}
+d_{2}^{1,1}\, ({\bf J}^i D^{4-j}{\bf \Phi}^{(1)}-{\bf J}^j D^{4-i}{\bf \Phi}^{(1)})
+d_{3}^{1,1}\, ({\bf J}^i \partial D^{j}{\bf \Phi}^{(1)}-{\bf J}^j \partial D^{i}{\bf \Phi}^{(1)})
}
\nonu\\
&+&
\mathtt{
d_{4}^{1,1}\, (\partial {\bf J}^i D^{j}{\bf \Phi}^{(1)}-\partial {\bf J}^j D^{i}{\bf \Phi}^{(1)})
+d_{5}^{1,1}\, ( {\bf J}^{4-ik} D^{4-jk}{\bf \Phi}^{(1)} -{\bf J}^{4-jk} D^{4-ik}{\bf \Phi}^{(1)} )
}
\nonu\\
&+&
\mathtt{
d_{6}^{1,1}\, \partial {\bf J}^{4-ij} {\bf \Phi}^{(1)}
+d_{7}^{1,1}\, {\bf J}^{4-ij} \partial {\bf \Phi}^{(1)}
+d_{8}^{1,1}\, ({\bf J}^i {\bf J}^{ij} D^i {\bf \Phi}^{(1)} -{\bf J}^j {\bf J}^{ji} D^j {\bf \Phi}^{(1)})
}
\nonu\\
&+&
\mathtt{
d_{9}^{1,1}\, ( {\bf J}^i \partial {\bf J} D^j{\bf \Phi}^{(1)} -{\bf J}^j \partial {\bf J} D^i{\bf \Phi}^{(1)})
+d_{10}^{1,1}\, ( {\bf J}^i {\bf J}^{4-j} {\bf \Phi}^{(1)} -{\bf J}^j {\bf J}^{4-i} {\bf \Phi}^{(1)} )}
\nonu \\
&
+& \mathtt{ d_{11}^{1,1}\, {\bf J}^i {\bf J}^j \partial {\bf \Phi}^{(1)}}
+ \cdots,
\nonu\\
{\bf Q}^{(\frac{3}{2}),ij}_{\frac{3}{2}} & = &
d_{1}^{\frac{3}{2},3}\, D^i {\bf \Phi}^{(1)}
\mathtt{+d_{2}^{\frac{3}{2},3}\,  {\bf J}^i  {\bf \Phi}^{(1)}}\,,
\nonu\\
{\bf Q}^{(\frac{5}{2}),ij}_{\frac{3}{2}} & = &
d_{1}^{\frac{3}{2},2}\,  D^{4-i}{\bf \Phi}^{(1)}
\mathtt{
+d_{2}^{\frac{3}{2},2}\, \partial D^i{\bf \Phi}^{(1)}
+d_{3}^{\frac{3}{2},2}\, {\bf J}^i \partial {\bf \Phi}^{(1)}
+d_{4}^{\frac{3}{2},2}\, \partial {\bf J}^i {\bf \Phi}^{(1)}
+d_{5}^{\frac{3}{2},2}\, \partial {\bf J} D^i {\bf \Phi}^{(1)}}
\nonu\\
&+&
\mathtt{
d_{6}^{\frac{3}{2},2}\, {\bf J}^{4-i} {\bf \Phi}^{(1)}
+d_{7}^{\frac{3}{2},2}\, {\bf J}^i \partial {\bf J} {\bf \Phi}^{(1)}
+d_{8}^{\frac{3}{2},2}\,  {\bf J}^j D^{4-ij}{\bf \Phi}^{(1)}
+d_{9}^{\frac{3}{2},2}\,  {\bf J}^{4-ij} D^j {\bf \Phi}^{(1)}
}
\nonu\\
&+&
\mathtt{
d_{10}^{\frac{3}{2},2}\, {\bf J}^i {\bf J}^j D^j  {\bf \Phi}^{(1)}
+d_{11}^{\frac{3}{2},2}\,  {\bf J}^j {\bf J}^{4-ij} {\bf \Phi}^{(1)}}
+ \cdots,
\nonu\\
{\bf Q}^{(\frac{7}{2}),i}_{\frac{3}{2}} & = &
d_{1}^{\frac{3}{2},1}\,  D^i {\bf \Phi}^{(3)}
\mathtt{+d_{2}^{\frac{3}{2},1}\,  \partial D^{4-i} {\bf \Phi}^{(1)}
+d_{3}^{\frac{3}{2},1}\,  \partial^2 D^i {\bf \Phi}^{(1)}
+d_{4}^{\frac{3}{2},1}\,  {\bf \Phi}^{(1)} D^i {\bf \Phi}^{(2)}
+d_{5}^{\frac{3}{2},1}\,   D^i {\bf \Phi}^{(1)} {\bf \Phi}^{(2)}}
\nonu\\
&+&
\mathtt{
d_{6}^{\frac{3}{2},1}\,  {\bf \Phi}^{(1)}{\bf \Phi}^{(1)}D^i{\bf \Phi}^{(1)}
+d_{7}^{\frac{3}{2},1}\,  {\bf J}^{4-0} D^i{\bf \Phi}^{(1)}
+d_{8}^{\frac{3}{2},1}\, \partial {\bf J} D^{4-i}{\bf \Phi}^{(1)}
+d_{9}^{\frac{3}{2},1}\,  \partial {\bf J} \partial {\bf J} D^i{\bf \Phi}^{(1)}
}
\nonu\\
&+&
\mathtt{
d_{10}^{\frac{3}{2},1}\,  \partial {\bf J} {\bf J}^{4-i}  {\bf \Phi}^{(1)}
+d_{11}^{\frac{3}{2},1}\, \partial {\bf J} \partial D^i {\bf \Phi}^{(1)}
+d_{12}^{\frac{3}{2},1}\, \partial^2 {\bf J} D^i {\bf \Phi}^{(1)}
+d_{13}^{\frac{3}{2},1}\, {\bf J}^{4-i} \partial {\bf \Phi}^{(1)}
} + \cdots,
\nonu\\
{\bf Q}^{(1)}_{2} & = &
d_{1}^{2,4}\, {\bf \Phi}^{(1)}
 \,,
\nonu\\
{\bf Q}^{(2)}_{2} & = &
\mathtt{
d_{1}^{2,3}\, \partial{\bf J} {\bf \Phi}^{(1)}
}
 \,,
 \nonu\\
{\bf Q}^{(3)}_{2} & = &
d_{1}^{2,2}\, {\bf \Phi}^{(3)}
+d_{2}^{2,2}\, D^{4-0}{\bf \Phi}^{(1)}
\mathtt{
+d_{3}^{2,2}\, \partial^2 {\bf \Phi}^{(1)}
+d_{4}^{2,2}\, {\bf \Phi}^{(1)}{\bf \Phi}^{(1)}{\bf \Phi}^{(1)}
+d_{5}^{2,2}\, {\bf J}^{4-0}{\bf \Phi}^{(1)}}
\nonu\\
&+&
\mathtt{
d_{6}^{2,2}\, \partial {\bf J} \partial {\bf \Phi}^{(1)}
+d_{7}^{2,2}\, \partial^2 {\bf J}{\bf \Phi}^{(1)}
+d_{8}^{2,2}\, \partial {\bf J} \partial {\bf J} {\bf \Phi}^{(1)}
+d_{9}^{2,2}\, {\bf J}^{4-i} D^i {\bf \Phi}^{(1)}
+d_{10}^{2,2}\, {\bf J}^i D^{4-i} {\bf \Phi}^{(1)}
}
\nonu\\
&+&
\mathtt{
d_{11}^{2,2}\, {\bf J}^i \partial D^i {\bf \Phi}^{(1)}
+d_{12}^{2,2}\, \partial {\bf J} {\bf J}^i D^i{\bf \Phi}^{(1)}
+d_{13}^{2,2}\, {\bf J}^i {\bf J}^{4-i} {\bf \Phi}^{(1)}
+d_{14}^{2,2}\, \partial {\bf J}^i D^i{\bf \Phi}^{(1)}}
\nonu \\
&
+ & \mathtt{
  d_{15}^{2,2}\, \partial {\bf J}^i {\bf J}^i {\bf \Phi}^{(1)}
} + \cdots,
\nonu\\
{\bf Q}^{(4)}_{2} & = &
\mathtt{
d_{1}^{2,1}\, {\bf \Phi}^{(1)} \partial {\bf \Phi}^{(2)}
+d_{2}^{2,1}\, \partial {\bf \Phi}^{(1)}  {\bf \Phi}^{(2)}
+d_{3}^{2,1}\, \partial^3 {\bf \Phi}^{(1)}
+d_{4}^{2,1}\, {\bf J}^{4-0} \partial {\bf \Phi}^{(1)}
+d_{5}^{2,1}\, \partial {\bf J}^{4-0} {\bf \Phi}^{(1)}
}
\nonu\\
&+&
\mathtt{
d_{6}^{2,1}\, \partial {\bf J}\partial {\bf J}\partial {\bf \Phi}^{(1)}
+d_{7}^{2,1}\,\partial^2 {\bf J} \partial {\bf J} {\bf \Phi}^{(1)}
+d_{8}^{2,1}\,\partial {\bf J} \partial^2 {\bf \Phi}^{(1)}
+d_{9}^{2,1}\, \partial^2 {\bf J} \partial{\bf \Phi}^{(1)}
+d_{10}^{2,1}\, \partial^3 {\bf J} {\bf \Phi}^{(1)}
}
\nonu\\
&+&
\mathtt{
d_{11}^{2,1}\, \partial {\bf J}^i D^{4-i} {\bf \Phi}^{(1)}
+d_{12}^{2,1}\,{\bf J}^{4-i} \partial D^i {\bf \Phi}^{(1)}
+d_{13}^{2,1}\, \partial {\bf J}^i {\bf J}^{4-i} {\bf \Phi}^{(1)}
+d_{14}^{2,1}\, {\bf J}^i \partial {\bf J}^{4-i} {\bf \Phi}^{(1)}
} \nonu \\
& + & \cdots.
\label{morequasi}
\eea
The complete expressions of (\ref{Spin3quasi}) and
(\ref{Spin7halfquasi}) are given in (\ref{morequasi}).
Note that all the terms in these quasi primary fields in
(\ref{morequasi})
have the dependence of ${\cal N}=4$ higher spin multiplets ${
  \bf \Phi^{(s)}}$ where $s=1,2,3$ as well as the stress energy tensor
${\bf J}$.
We obtain (\ref{morequasi1}) from (\ref{morequasi})
as before.

\section{ Quasi primary fields from section $4$}

The previous OPE in (\ref{section4ope}) can be 
described further as
\bea
\Phi_{2}^{(2)}(z)\,\Phi_{0}^{(2)}(w)
& = & \frac{1}{(z-w)^{6}}\,
8 \, \alpha \, e_{0}^{0,4}
+\frac{1}{(z-w)^{5}}\,
Q^{(1)}_{2}(w)
\nonu\\
& + & \frac{1}{(z-w)^{4}}\,\Bigg[\,
\frac{3}{2}\,\partial Q^{(1)}_{2}
+Q^{(2)}_{2}-6\,p_{1}\,Q^{(2)}_{0}
-p_2 \, E^{(2)}_{2}
\,\Bigg](w)
\nonu\\
& + &
\frac{1}{(z-w)^{3}}\,\Bigg[\,
\partial^{2} Q^{(1)}_{2}
+\partial Q^{(2)}_{2}
+Q^{(3)}_{2} -p_1 \, \pa \, Q_0^{(2)}
- p_2 \, E^{(3)}_{2}
\,\Bigg](w)
\nonu\\
& + &
\frac{1}{(z-w)^{2}}\,\Bigg[\,
\frac{5}{12}\,\partial^{3} Q^{(1)}_{2}
+\frac{1}{2}\, \partial^{2} Q^{(2)}_{2}
+\frac{5}{6}\,\partial Q^{(3)}_{2}
+Q^{(4)}_{2}
-p_2 \, E^{(4)}_{2}
\,\Bigg](w)
\nonu\\
& + &
\frac{1}{(z-w)}\Bigg[
\frac{1}{8}\partial^{4} Q^{(1)}_{2}
+\frac{1}{6}\partial^{3} Q^{(2)}_{2}
+\frac{5}{14}\partial^{2} Q^{(3)}_{2}
+\frac{3}{4}\partial Q^{(4)}_{2}
+ Q^{(5)}_{2}
- p_2  E^{(5)}_{2}
\,\Bigg](w)
\nonu\\
&+& \cdots,
\label{ope4}
\eea
where we introduce the last
singular $p_2$ terms in (\ref{section4ope})
as
\bea
 E^{(4-n)}_{2} \equiv \Big\{
(L \Phi_{0}^{(2)})\,\Phi_{0}^{(2)}
\Big\}_{n+2}, \qquad n=-1,0,1,2,
\label{Edef}
\eea
and $p_1$ and $p_2$ are given in (\ref{p1p2sect4}). 
Then we obtain (\ref{Edef}) as follows:
\bea
E^{(2)}_{2}(w)& = &
(\,
4 Q^{(2)}_{0}+e^{0,4}_{0}\,L
\,)(w),
\nonu\\
E^{(3)}_{2}(w)& = &
(\,
\frac{5}{2} \, \pa \, Q^{(2)}_{0}+e^{0,4}_{0} \, \pa \, L
\,)(w),
\nonu\\
E^{(4)}_{2}(w)& = &
(\,
\frac{1}{2}\,\partial^{2} Q^{(2)}_{0}
+ L  Q^{(2)}_{0}
+\frac{1}{2}\,e^{0,4}_{0}\,\partial^{2} L
+2\,\Phi^{(2)}_{0}\Phi^{(2)}_{0}
\,)(w),
\nonu\\
E^{(5)}_{2}(w)& = &
(\,
-\frac{1}{24}\,\partial^{3} Q^{(2)}_{0}
+\frac{1}{2}\, L \partial Q^{(2)}_{0}
+\partial L  Q^{(2)}_{0}
+\frac{1}{6}\,e^{0,4}_{0}\,\,\partial^{3} L
+\frac{3}{2}\,\partial (\Phi^{(2)}_{0}\Phi^{(2)}_{0})
\,)(w).
\label{Eexp}
\eea

Moreover, the remaining four kinds of fundamental OPEs
can be described as 
\bea
\Phi_{0}^{(2)}(z)\,\Phi_{0}^{(2)}(w)
& = & 
\frac{1}{(z-w)^{4}}\,e^{0,4}_{0}+
\frac{1}{(z-w)^{2}}\,
 Q^{(2)}_{0}(w)
 + 
\frac{1}{(z-w)}\,\frac{1}{2}\,\partial  Q^{(2)}_{0}(w)
+\cdots,
\nonu \\
\Phi_{\frac{1}{2}}^{(2),i}(z)\,\Phi_{0}^{(2)}(w)
& = & \frac{1}{(z-w)^{3}}\,
Q^{(\frac{3}{2}),i}_{\frac{1}{2}}(w)
+\frac{1}{(z-w)^{2}}\,\Bigg[\,
\frac{2}{3}\,\partial  
Q^{(\frac{3}{2}),i}_{\frac{1}{2}}
+ Q^{(\frac{5}{2}),i}_{\frac{1}{2}}
\,\Bigg](w)
\nonu \\ 
& + &
\frac{1}{(z-w)}\,\Bigg[\,
\frac{1}{4}\,\partial^{2}  Q^{(\frac{3}{2}),i}_{\frac{1}{2}} +
\frac{3}{5}\,\partial  Q^{(\frac{5}{2}),i}_{\frac{1}{2}}
+ Q^{(\frac{7}{2}),i}_{\frac{1}{2}}\,
\Bigg](w)+\cdots,
\nonu \\
\Phi_{1}^{(2),ij}(z)\,\Phi_{0}^{(2)}(w)
& = & \frac{1}{(z-w)^{4}}\,
Q^{(1),ij}_{1}(w)
+\frac{1}{(z-w)^{3}}\,\Bigg[\,
\partial  Q^{(1),ij}_{1}+ Q^{(2),ij}_{1}
\,\Bigg](w)
\nonu \\ 
& + &
\frac{1}{(z-w)^{2}}\,\Bigg[\,
\frac{1}{2}\,\partial^{2}  Q^{(1),ij}_{1}
+\frac{3}{4}\,\partial  Q^{(2),ij}_{1}
+ Q^{(3),ij}_{1}
\,\Bigg](w)
\nonu \\ 
& + &
\frac{1}{(z-w)}\,\Bigg[\,
\frac{1}{6}\,\partial^{3}  Q^{(1),ij}_{1} 
+\frac{3}{10}\,\partial^{2} Q^{(2),ij}_{1}
+\frac{2}{3}\,\partial  Q^{(3),ij}_{1}
+ Q^{(4),ij}_{1}\,
\Bigg](w)
\nonu \\ 
& - &
\, (i\leftrightarrow j)
+\cdots,
\nonu \\
\Phi_{\frac{3}{2}}^{(2),i}(z)\,\Phi_{0}^{(2)}(w)
& = & \frac{1}{(z-w)^{5}}\,
Q^{(\frac{1}{2}),i}_{\frac{3}{2}}(w)
+\frac{1}{(z-w)^{4}}\,\Bigg[\,
2\,\partial Q^{(\frac{1}{2}),i}_{\frac{3}{2}}+Q^{(\frac{3}{2}),i}_{\frac{3}{2}}
\,\Bigg](w)
\nonu \\ 
&+&\frac{1}{(z-w)^{3}}\,\Bigg[\,
\frac{3}{2}\,\partial^{2} Q^{(\frac{1}{2}),i}_{\frac{3}{2}}
+\partial Q^{(\frac{3}{2}),i}_{\frac{3}{2}}
+Q^{(\frac{5}{2}),i}_{\frac{3}{2}}
\,\Bigg](w)
\nonu \\ 
& + &
\frac{1}{(z-w)^{2}}\,\Bigg[\,
\frac{2}{3}\,\partial^{3} Q^{(\frac{1}{2}),i}_{\frac{3}{2}}
+\frac{1}{2}\,\partial^{2} Q^{(\frac{3}{2}),i}_{\frac{3}{2}}
+\frac{4}{5}\,\partial  Q^{(\frac{5}{2}),i}_{\frac{3}{2}}
+ Q^{(\frac{7}{2}),i}_{\frac{3}{2}}
\,\Bigg](w)
\nonu \\ 
& + &
\frac{1}{(z-w)}\,\Bigg[\,
\frac{5}{24}\,\partial^{4} Q^{(\frac{1}{2}),i}_{\frac{3}{2}}
+\frac{1}{6}\,\partial^{3} Q^{(\frac{3}{2}),i}_{\frac{3}{2}}
+\frac{1}{3}\,\partial^{2} Q^{(\frac{5}{2}),i}_{\frac{3}{2}}
+\frac{5}{7}\,\partial Q^{(\frac{7}{2}),i}_{\frac{3}{2}}
+Q^{(\frac{9}{2}),i}_{\frac{3}{2}}
\,
\Bigg](w)
\nonu \\ 
& + &
\frac{2\,\alpha}{5}\,\sum\limits _{n=2}^{4}
\frac{1}{(z-w)^{n}} \, \Big\{
\partial \Phi_{\frac{1}{2}}^{(2),i} \,\Phi_{0}^{(2)}
\Big\}_{n}(w)
+\cdots.
\label{operem}
\eea
We have the final quasi primary fields appearing in
(\ref{ope4}) with (\ref{Eexp}) and (\ref{operem}) as 
follows:
\bea
 Q^{(2)}_{0}  & = & 
w_{0,2}\,\Phi^{(2)}_{0}
+w_{1,2}\,\Phi^{(1)}_{0}\Phi^{(1)}_{0}
+w_{2,2}\,\tilde{L}
+w_{3,2}\,\partial U
+w_{4,2}\,U U
+w_{5,2}\,T^{ij}T^{ij}
+w_{6,2}\,T^{ij}\tilde{T}^{ij}
\nonu\\
& + & 
w_{7,2}\,\Gamma^{i}\Gamma^{j}T^{ij}
+w_{8,2}\,\Gamma^{i}\Gamma^{j} \tilde{T}^{ij}
+w_{9,2}\,\Gamma^{i}  \partial \Gamma^{i}
+\varepsilon^{ijkl}\,
w_{10,2}\,\Gamma^{i}\Gamma^{j}\Gamma^{k}\Gamma^{l},
\nonu \\
Q^{(\frac{3}{2}),i}_{\frac{1}{2}} & = &
w_{1,\frac{3}{2}}\,\tilde{G}^{i}
+w_{2,\frac{3}{2}}\,\partial\Gamma^{i}
+w_{3,\frac{3}{2}}\,\Gamma^{i}U+\varepsilon^{ijkl}(\,
w_{4,\frac{3}{2}}\,\Gamma^{j}T^{kl}
+w_{5,\frac{3}{2}}\,\Gamma^{j}\Gamma^{k}\Gamma^{l}),
\nonu \\ 
Q^{(\frac{5}{2}),i}_{\frac{1}{2}} & = &
w_{1,\frac{5}{2}}\,\Phi_{\frac{1}{2}}^{(2),i}
+w_{2,\frac{5}{2}}\,\Phi_{0}^{(1)}\Phi_{\frac{1}{2}}^{(1),i}
+w_{3,\frac{5}{2}}\,\partial\tilde{G}^{i}
+w_{4,\frac{5}{2}}\,\partial^{2}\Gamma^{i}
+w_{5,\frac{5}{2}}\,\Gamma^{i}\partial U
+w_{5,\frac{5}{2}}\,\partial\Gamma^{i}U
\nonu \\ 
& + &w_{6,\frac{5}{2}}\,T^{ij}\tilde{G}^{i}
\cdots
+  \varepsilon^{ijkl}(\,
w_{16,\frac{5}{2}}\,T^{jk}\tilde{G^{l}}
+w_{17,\frac{5}{2}}\,\Gamma^{j}\Gamma^{k}\tilde{G}^{l}
\cdots
+w_{21,\frac{5}{2}}\,\Gamma^{j}\Gamma^{k}\Gamma^{l}U\,),
\nonu\\
 Q^{(\frac{7}{2}),i}_{\frac{1}{2}} & = &
w_{1,\frac{7}{2}}\,\Phi^{(2),j}_{\frac{3}{2}}
+w_{2,\frac{7}{2}}\,\partial \Phi^{(2),j}_{\frac{1}{2}}
+w_{3,\frac{7}{2}}\, \Phi^{(1)}_{0}\Phi^{(1),j}_{\frac{3}{2}}
+w_{4,\frac{7}{2}}\, \Phi^{(1)}_{0} \partial \Phi^{(1),i}_{\frac{1}{2}}
+\cdots
\nonu \\
& + & w_{20,\frac{7}{2}}\, U \Gamma^{i} \Phi^{(1)}_{0} \Phi^{(1)}_{0}
\nonu\\
& +& w_{21,\frac{7}{2}}\, \Gamma^{i} \partial \Phi^{(1)}_{0} \Phi^{(1)}_{0}
+w_{22,\frac{7}{2}}\, \partial \tilde{L} \Gamma^{i}
+w_{23,\frac{7}{2}}\,\tilde{L}  \partial \Gamma^{i}
+\cdots
+w_{67,\frac{7}{2}}\,\tilde{T}^{ij} \tilde{T}^{ij}  \partial \Gamma^{i}
\nonu\\
& +&
 \varepsilon^{ijkl}(\,
w_{68,\frac{7}{2}}\, \Gamma^{j} \Phi^{(2),kl}_{1}
+w_{69,\frac{7}{2}}\, T^{jk} \Phi^{(2),l}_{\frac{1}{2}}
+\cdots
+w_{117,\frac{7}{2}}\,\Gamma^{j} T^{ij} T^{ij} T^{kl} 
\nonu \\
& + & w_{118,\frac{7}{2}}\,\Gamma^{j} T^{ij} T^{ik} T^{lj} 
\,)
\nonu\\
& +&
(\varepsilon^{ijkl})^{2}\,w_{118,\frac{7}{2}}\,\Gamma^{j}\Gamma^{k}\Gamma^{l} T^{ij} T^{kl}  ,
\nonu \\
Q^{(1),ij}_{1} & = &
w_{1,1}\, T^{ij}
+w_{2,1}\,\Gamma^{i}\Gamma^{j}
+\varepsilon^{ijkl}(\,
w_{3,1}\, T^{kl}
+w_{4,1}\,\Gamma^{k}\Gamma^{l}\,)
-(i\leftrightarrow j),
\nonu \\ 
Q^{(2),ij}_{1}  & = &
w_{1,2}\,\tilde{G}^{i}\Gamma^{j}
+w_{2,2}\,\partial(\Gamma^{i}\Gamma^{j})
+w_{3,2}\,U \Gamma^{i}\Gamma^{j}
+w_{4,2}\,\tilde{T}^{jik}\Gamma^{j}\Gamma^{k}
+w_{5,2}\,\tilde{T}^{ij}
\nonu \\
& + & \varepsilon^{ijkl}\,w_{6,2}\,\partial(\Gamma^{k}\Gamma^{l})
\nonu\\
&-&(i\leftrightarrow j),
\nonu \\ 
Q^{(3),ij}_{1} & = &
w_{1,3}\,\Phi_{1}^{(2),ij}
+w_{2,3}\,\tilde{\Phi}_{1}^{(2),ij}
+w_{3,3}\,\Phi_{0}^{(1)j} \Phi_{1}^{(1),ij}
+w_{4,3}\,\Phi_{0}^{(1)j} \tilde{\Phi}_{1}^{(1),ij}
+\cdots
+ w_{42,3}\,T^{ki}T^{jl} T^{kl}
\nonu \\ 
& + &  w_{43,3}\,T^{ij}T^{kl} T^{kl}
+
\varepsilon^{ijkl}(\,
w_{44,3}\,\Phi_{\frac{1}{2}}^{(1),k}\Phi_{\frac{1}{2}}^{(1),l}
+w_{45,3}\,\Gamma^{k} \Phi_{0}^{(1)} \Phi_{\frac{1}{2}}^{(1),l}
+\cdots
\nonu\\
& + & 
w_{86,3}\,\partial \Gamma^{i}\Gamma^{i}\Gamma^{k}\Gamma^{l}
+w_{87,3}\,\Gamma^{i} \Gamma^{j}\Gamma^{k}\Gamma^{l} T^{kl}
\,)
+w_{88,3}\,\varepsilon^{ijkl}\varepsilon^{r \rho \delta \sigma} \Gamma^{r} \Gamma^{\rho} T^{\delta \sigma} T^{kl}
\nonu\\
& + & 
(\varepsilon^{ijkl})^{2}
(\,
w_{89,3}\,\Gamma^{i}\Gamma^{k}T^{jl}T^{kl}
+\cdots
+w_{92,3}\,\Gamma^{i}\Gamma^{j}\Gamma^{k}\Gamma^{l}T^{ij}
\,)
-(i\leftrightarrow j),
\nonu \\ 
Q^{(4),ij}_{1} & = &
w_{1,4}\,\partial \Phi_{1}^{(2),ij}
+w_{2,4}\, \Phi_{0}^{(1)} \partial \Phi_{1}^{(1),ij}
+w_{3,4}\, \Phi_{0}^{(1)} \partial \tilde{\Phi}_{1}^{(1),ij}
+\cdots
+w_{126,4}\, \tilde{G}^{l} T^{ik} T^{kl} \Gamma^{j}
\nonu \\ 
&+& w_{127,4}\, \tilde{G}^{l} T^{ik}  \Gamma^{j}\Gamma^{k}\Gamma^{l}
+w_{128,4}\, \tilde{G}^{l} T^{ik}  T^{jl}\Gamma^{k}
+\cdots
+w_{153,4}\,\partial \tilde{T}^{il} T^{kl} \Gamma^{j} \Gamma^{k}
\nonu\\ 
& + & 
\varepsilon^{ijkl}(\,
w_{154,4}\,U \Gamma^k \Phi_{\frac{1}{2}}^{(2),l}
+w_{155,4}\,\tilde{G}^k \Phi_{\frac{1}{2}}^{(2),l}
+\cdots
+w_{296,4}\,\partial^2 T^{ik} T^{il} \,)
\nonu \\ 
&+&
w_{297,4}\,\varepsilon^{ijkl} \varepsilon^{r \rho \delta \sigma}\,T^{r \rho}T^{\delta \sigma}U \Gamma^{k}\Gamma^{l}
-(i\leftrightarrow j),
\nonu \\
Q^{(\frac{1}{2}),i}_{\frac{3}{2}}  & = & 
w_{1,\frac{1}{2}}\,\Gamma^{i},
\nonu\\
Q^{(\frac{3}{2}),i}_{\frac{3}{2}}  & = & 
w_{1,\frac{3}{2}}\,\tilde{G}^{i}
+w_{2,\frac{3}{2}}\,\partial \Gamma^{i}
+w_{3,\frac{3}{2}}\,\Gamma^{i} U
+w_{4,\frac{3}{2}}\,\Gamma^{j} T^{ij}
+\varepsilon^{ijkl}(\,
w_{5,\frac{3}{2}}\,\Gamma^{j} T^{kl}
+w_{6,\frac{3}{2}}\,\Gamma^{j}\Gamma^{k}\Gamma^{l}
\,),
\nonu\\
Q^{(\frac{5}{2}),i}_{\frac{3}{2}}   & = & 
w_{1,\frac{5}{2}}\, \Phi_{\frac{1}{2}}^{(2),i}
+w_{2,\frac{5}{2}}\,\Gamma^{i}  \Phi_{0}^{(2)}
+w_{3,\frac{5}{2}}\,\Phi_{0}^{(1)} \Phi_{\frac{1}{2}}^{(1),i}
+\cdots
+w_{29,\frac{5}{2}}\,\varepsilon^{ijkl}\,T^{ji}\Gamma^{k}\Gamma^{i}
\Gamma^{l},
\nonu\\
Q^{(\frac{7}{2}),i}_{\frac{3}{2}} & = & 
w_{1,\frac{7}{2}}\,\Phi^{(2),i}_{\frac{3}{2}}
+w_{2,\frac{7}{2}}\,\partial \Phi^{(2),i}_{\frac{1}{2}}
+w_{3,\frac{7}{2}}\,U \Phi^{(2),i}_{\frac{1}{2}}
+\cdots
+\varepsilon^{ijkl}(\,
w_{34,\frac{7}{2}}\,\Gamma^{j} \Phi^{(1),k}_{\frac{1}{2}}\Phi^{(1),l}_{\frac{1}{2}}
\nonu\\
&+&
w_{35,\frac{7}{2}}\,\Gamma^{j}\Gamma^{k} \Phi^{(1)}_{0}\Phi^{(1),l}_{\frac{1}{2}}
+w_{36,\frac{7}{2}}\,\Gamma^{j}\Gamma^{k} \Gamma^{l} \Phi^{(1)}_{0} \Phi^{(1)}_{0}
\,)
+\cdots
+w_{93,\frac{7}{2}}\,T^{ij} \Gamma^{j} \partial \Gamma^{j} \Gamma^{j}
\nonu\\
&+&
\varepsilon^{ijkl}(\,
w_{94,\frac{7}{2}}\,\Gamma^{j} T^{kl} \tilde{L}
+w_{95,\frac{7}{2}}\,\Gamma^{j} T^{kl} \partial U
+\cdots
+w_{141,\frac{7}{2}}\,T^{ij} U \Gamma^{j} \Gamma^{k} \Gamma^{l}\,)
+\cdots
\nonu\\
&+&
w_{143,\frac{7}{2}}\,(\varepsilon^{ijkl})^2 T^{ij} \tilde{T}^{kl} \Gamma^{j} \Gamma^{k} \Gamma^{l}
\,,
\nonu\\
Q^{(1)}_{2}  & = &  w_{1,1}\,U,
\nonu\\
Q^{(2)}_{2}  & = &  
w_{1,2}\,\Phi_{0}^{(2)}
+w_{2,2}\, \Phi_{0}^{(1)}\Phi_{0}^{(1)}
+w_{3,2}\,\tilde{L}
+w_{4,2}\,\partial U+w_{5}\, UU
+w_{6,2}\, T^{ij}T^{ij}
+w_{7,2}\, T^{ij}\tilde{T}^{ij}
\nonu\\
&
+& w_{8,2}\,\Gamma^{i}\Gamma^{j}T^{ij}
+w_{9,2}\,\Gamma^{i}\Gamma^{j}\tilde{T}^{ij}
+w_{10,2}\,\Gamma^{i}\tilde{G}^{i}
+w_{11,2}\,\Gamma^{i}\partial\Gamma^{i}
+\varepsilon^{ijkl}\, 
w_{12,2}\,\Gamma^{i}\Gamma^{j}\Gamma^{k}\Gamma^{l},
\nonu\\
Q^{(3)}_{2}  & = &  
w_{1,3}\, U\Phi_{0}^{(2)}
+w_{2,3}\,U \Phi_{0}^{(1)} \Phi_{0}^{(1)}
+w_{3,3}\,\Gamma^{i} \Phi_{\frac{1}{2}}^{(2),i}
+w_{4,3}\,\Gamma^{i} \Phi_{0}^{(1)} \Phi_{\frac{1}{2}}^{(1),i}
+w_{5,3}\, U\tilde{L}
+w_{6,3}\, U\partial U
\nonu\\
&+&
w_{7,3}\, UUU
+w_{8,3}\,\Gamma^{i}\partial\tilde{G}^{i}
+w_{9,3}\,\partial\Gamma^{i}\tilde{G}^{i}
+w_{10,3}\,\Gamma^{i}\partial\Gamma^{i}U
+w_{11,3}\,\Gamma^{i}\partial^{2}\Gamma^{i}
+w_{10,3}\,\partial\Gamma^{i}\partial\Gamma^{i}
\nonu\\
&+&
w_{12,3}\,\Gamma^{i}T^{ij}\tilde{G}^{j}
+w_{13,3}\,\Gamma^{i}T^{ij}\tilde{G}^{j}
+w_{14,3}\,\Gamma^{i}\Gamma^{j}T^{ij}U
+ w_{15,3}\, T^{ij}T^{ij}U
+w_{16,3}\,\Gamma^{i}\Gamma^{j}\partial T^{ij}
\nonu\\
&+&
w_{17,3}(\partial\Gamma^{i}\Gamma^{j}T^{ij}
+\partial^{2}T^{ij})
+\varepsilon^{ijkl}(\, 
w_{18,3}\,\Gamma^{i}T^{jk}\tilde{G}^{l}
+w_{19,3}\,\Gamma^{i}\Gamma^{j}\Gamma^{k}\tilde{G}^{l}
+w_{20,3}\,\Gamma^{i}\Gamma^{j}T^{kl}U
\nonu\\
&+&
w_{21,3}\, T^{ij}T^{kl}U
+w_{22,3}\,\Gamma^{i}\Gamma^{j}\partial T^{kl}
+w_{23,3}\,\partial\Gamma^{i}\Gamma^{j}T^{kl}
+w_{24,3}\,\partial(\Gamma^{i}\Gamma^{j}\Gamma^{k}\Gamma^{l})\,),
\nonu\\
Q^{(4)}_{2}  & = &  
w_{0,4}\,\Phi_{0}^{(4)}
+w_{1,4}\,\Phi_{0}^{(2)}\Phi_{0}^{(2)}
+w_{2,4}\,\Phi_{2}^{(2)}
+w_{3,4}\,\tilde{L}\Phi_{0}^{(2)}
+\cdots
+w_{26,4}\,\Phi_{1}^{(1),ij} \Phi_{1}^{(1),ij}
\nonu\\
&+&
w_{27,4}\,\Phi_{1}^{(1),ij} \tilde{\Phi}_{1}^{(1),ij}
\cdots
+w_{131,4}\,T^{ij}T^{ij}T^{kl} \Gamma^{k}\Gamma^{l}
+w_{132,4}\,T^{ij}T^{kl} \tilde{T}^{ij}\tilde{T}^{kl}
\cdots
\nonu\\
&
+& 
\varepsilon^{ijkl}(\,
w_{133,4}\, \tilde{L} \Gamma^{i}\Gamma^{j}\Gamma^{k}\Gamma^{l}
+w_{134,4}\, \tilde{L} T^{ij} \Gamma^{k}\Gamma^{l}
+\cdots
+w_{168,4}\,\Gamma^{i}\Gamma^{j}\partial\Gamma^{k}\partial\Gamma^{l}
\,),
\label{quasiappendixc}
\eea
where the central term in (\ref{ope4}) contains
\bea
\alpha & = & \frac{(k-N)}{2(k+N+2)},
\nonu \\
e^{0,4}_{0}
&=&\frac{1}{(2 + k + N)^3 (4 + 3 k + 3 N + 2 k N) (5 + 4 k + 
     4 N + 3 k N)^2}\, 64 k N
\nonu\\
&\times&
     (-100 - 285 k - 358 k^2 - 246 k^3 - 88 k^4 - 12 k^5 - 285 N - 
     510 k N - 275 k^2 N + 8 k^3 N 
\nonu\\
     &+& 56 k^4 N + 16 k^5 N - 358 N^2 - 
     275 k N^2 + 406 k^2 N^2 + 573 k^3 N^2 + 242 k^4 N^2 + 
     34 k^5 N^2 
\nonu\\
&-& 246 N^3 + 8 k N^3 + 573 k^2 N^3 + 510 k^3 N^3 + 
     155 k^4 N^3 + 12 k^5 N^3 - 88 N^4 + 56 k N^4 
\nonu\\
     &+& 242 k^2 N^4 + 
     155 k^3 N^4 + 30 k^4 N^4 - 12 N^5 + 16 k N^5 + 34 k^2 N^5 + 
     12 k^3 N^5)\,.
\label{coeffexp}
     \eea     
Although the full expressions for
the $Q^{(\frac{9}{2}),i}_{\frac{3}{2}}$ of spin-$\frac{9}{2}$ and 
the $Q^{(5)}_{2}$ of spin-$5$ are not given in this Appendix,
they are presented in ${\tt ancillary.nb}$ we attach
and the complete expressions for the quasi primary fields
can be found also. The ${\cal N}=4$ version of quasi primary
fields will appear in
Appendix $F$.

\section{ The OPE between the first ${\cal N}=4$ higher spin
  multiplet
in the component approach under the large $(N,k)$ limit}

From the OPEs in \cite{AK1509} (or in section $2.3$), we write down
the following $15$ kinds of
OPEs \footnote{Although there are nonlinear
  terms with the overall factors $\frac{1}{N},
  \frac{1}{N^2}, \cdots$ in the OPEs,
  the infinity limit of $N$ leads to
  the fact that there will be no contributions from these nonlinear
  terms. We also take the infinity limit of $N$ for the central terms.
  For the classical asymptotical symmetry algebra of $AdS_3$ higher
spin theory, we should keep those nonlinear terms.  }
under the large $(N,k)$ limit
\bea
\Phi^{(1)}_0(z)\,\Phi^{(1)}_0(w)
&=&
-\frac{1}{(z-w)^2}\,
2\,N(\lambda-1)+\cdots
\,,
\nonu\\
\Phi^{(1)}_0(z)\,\Phi^{(1),i}_{\frac{1}{2}}(w)
&=&
\frac{1}{(z-w)}\,
G^i(w)+\cdots
\,,
\nonu\\
\Phi^{(1)}_0(z)\,\Phi^{(1),ij}_{1}(w)
&=&
-\frac{1}{(z-w)^2}\,2i
\,\bigg[\,(2\lambda-1)T^{ij}-\widetilde{T}^{ij} \,\bigg](w)+\cdots
\,,
\nonu\\
\Phi^{(1)}_0(z)\,\tilde{\Phi}^{(1),i}_{\frac{3}{2}}(w)
&=&
\frac{1}{(z-w)^3}
\,8i\lambda(\lambda-1)\, \Gamma^i(w)
\nonu \\
& - & \frac{1}{(z-w)^2}
8\,\bigg[\,
\frac{1}{3}\,(2\lambda-1)\,G^i+i \lambda(\lambda-1)\,\partial\Gamma^i
\,\bigg](w)
\nonu\\
&
-&\frac{1}{(z-w)}\,\frac{1}{2}\,\Phi^{(2),i}_{\frac{1}{2}}(w)+\cdots
\,,
\nonu\\
\Phi^{(1)}_0(z)\,\tilde{\Phi}^{(1)}_{2}(w)
&=&
\frac{1}{(z-w)^3}
\,16\,\lambda(\lambda-1)\,U(w)
\nonu \\
& + & \frac{1}{(z-w)^2}
\,\bigg[\,
2\,\Phi^{(2)}_0
-8\lambda(\lambda-1)\,\partial U
\,\bigg](w)+\cdots
\,,
\nonu \\
\Phi^{(1),i}_{\frac{1}{2}}(z)\,\Phi^{(1),j}_{\frac{1}{2}}(w)
&=&
\frac{\delta^{ij}}{(z-w)^3}\,
4\,N(\lambda-1)
+\frac{1}{(z-w)^2}\,
2i
\,\bigg[\,T^{ij}-(2\lambda-1)\widetilde{T}^{ij} \,\bigg](w)
\nonu\\
&
+ &\frac{1}{(z-w)}\,\bigg[
-2\,\delta^{ij}\,L
+i\,\partial T^{ij}-i\,(2\lambda-1)\,\partial \widetilde{T}^{ij}
\,\bigg](w)+\cdots
\,,
\nonu\\
\Phi^{(1),i}_{\frac{1}{2}}(z)\,\Phi^{(1),jk}_{1}(w)
&=&
-\frac{1}{(z-w)^3}\,8i\,\lambda(\lambda-1)(\delta^{ij}\,\Gamma^k-\delta^{ik}\,\Gamma^j)(w)
\nonu\\
&+&\frac{1}{(z-w)^2}
\,\bigg[\,
(2\lambda-1)(\delta^{ij}\,G^k-\delta^{ik}\,G^j)
+3\,\varepsilon^{ijkl}\,G^l
\,\bigg](w)
\nonu\\
&+&\frac{1}{(z-w)}
\,\bigg[\,
\Bigg(\delta^{ij}\Big(
\,\frac{1}{2}\,\Phi^{(2),k}_{\frac{1}{2}}+\frac{1}{3}(2\lambda-1)\,\partial G^k
\Big) -(j \leftrightarrow k)\Bigg)
\nonu \\
& + &
  \varepsilon^{ijkl}\,\partial G^l
\,\bigg](w)+\cdots,
\nonu\\
\Phi^{(1),i}_{\frac{1}{2}}(z)\,\tilde{\Phi}^{(1),j}_{\frac{3}{2}}(w)
&=&
-\frac{1}{(z-w)^3}\,8
\bigg[\,
\delta^{ij}\,\lambda(\lambda-1)\,U
-\frac{i}{3}\Big(
2(2\lambda-1)\,T^{ij} \nonu \\
& + & (\lambda+1)(\lambda-2)\,\widetilde{T}^{ij}
\Big)
\,\bigg](w)
\nonu\\
&
-&\frac{1}{(z-w)^2}\,2\,\delta^{ij}\,\Phi^{(2)}_{0}(w)
-\frac{1}{(z-w)}\,\frac{1}{2}\,
\bigg[\,
\delta^{ij}\,\partial \Phi^{(2)}_{0}
-\widetilde{\Phi}^{(2),ij}_{1}
\,\bigg](w)+\cdots,
\nonu\\
\Phi^{(1),i}_{\frac{1}{2}}(z)\,\tilde{\Phi}^{(1)}_{2}(w)
&=&
-\frac{1}{(z-w)^4}\,24i\,\lambda(\lambda-1)\,\Gamma^i(w)
\nonu\\
&
+&\frac{1}{(z-w)^3}\,\frac{8}{3}\,\bigg[\,
2(2\lambda-1)\,G^i+9i\,\lambda(\lambda-1)\,\partial \Gamma^i
\,\bigg](w)
\nonu \\
& + & \frac{1}{(z-w)^2}\,\frac{5}{2}\,\Phi^{(2),i}_{\frac{1}{2}}(w)
\nonu\\
&
+&\frac{1}{(z-w)}\,\frac{1}{2}\,\partial \Phi^{(2),i}_{\frac{1}{2}}(w)+
\cdots,
\nonu \\
\Phi^{(1),ij}_{1}(z)\,\Phi^{(1),kl}_{1}(w)
&=&
\frac{1}{(z-w)^4}\,\bigg[\,
12\,N(\lambda-1)\,
(\delta^{ik}\delta^{jl}-\delta^{il}\delta^{jk})
+4N(\lambda-1)(2\lambda-1)\,\varepsilon^{ijkl}
\bigg]
\nonu\\
&
-&\frac{1}{(z-w)^3}\,4i\,\bigg[ \Bigg(
\,
\delta^{ik}   \Big(
 (2\lambda-1)\,\widetilde{T}^{jl}+(2\lambda^2-2\lambda-1)\,T^{jl}
\Big) \nonu \\
&- & (k \leftrightarrow l) -( i \leftrightarrow j) + (i \leftrightarrow
j, k \leftrightarrow l)\Bigg)
\,\bigg](w)
\nonu\\
&
-&\frac{1}{(z-w)^2}\,\bigg[\,
( \delta^{ik}\delta^{jl}-\delta^{il}\delta^{jk})\,8\,L
-\varepsilon^{ijkl}\Big(\,
2\,\Phi^{(2)}_{0}
-\frac{8}{3}\,(2\lambda-1)\,L
\,\Big)
\nonu\\
&
+ & \Bigg(
2i\,\delta^{ik}   \Big(
 (2\lambda-1)\,\partial \widetilde{T}^{jl}+(2\lambda^2-2\lambda-1)\,\partial T^{jl}
\,\Big)
\nonu\\
&- & (k \leftrightarrow l) -( i \leftrightarrow j) + (i \leftrightarrow
j, k \leftrightarrow l)\Bigg)
\,\bigg](w)
\nonu\\
&
-&\frac{1}{(z-w)}\,\bigg[\,
( \delta^{ik}\delta^{jl}-\delta^{il}\delta^{jk})\,4\,\partial L
-\varepsilon^{ijkl}\Big(\,
\Phi^{(2)}_{0}
-\frac{4}{3}\,(2\lambda-1)\,L
\,\Big)
\nonu\\
&
+ & \Bigg( \delta^{ik}   \Big(\,
 \frac{1}{2}\,\Phi^{(2),jl}_{1}
+\frac{2i}{3}\,(2\lambda-1)\,\partial \widetilde{T}^{jl}+\frac{2i}{3}\,(2\lambda^2-2\lambda-1)\,\partial T^{jl}
\,\Big)
\nonu\\
&- & (k \leftrightarrow l) -( i \leftrightarrow j) + (i \leftrightarrow
j, k \leftrightarrow l)\Bigg)
\,\bigg](w)+  \cdots\,,
\nonu \\
\Phi^{(1),ij}_{1}(z)\,\tilde{\Phi}^{(1),k}_{\frac{3}{2}}(w)
&=&
\frac{1}{(z-w)^4}\,\bigg[\,
8i\,\lambda(\lambda-1)(2\lambda-1)\,(\delta^{ik}\,\Gamma^j-\delta^{jk}\,\Gamma^i)
\nonu \\
& + & \varepsilon^{ijkl}\,24i\,\lambda(\lambda-1)\,\Gamma^l
\,\bigg](w)
\nonu\\
&
-&\frac{1}{(z-w)^3}\,
\bigg[\,
\frac{16}{3}\,(\lambda+1)(\lambda-2)\,(\delta^{ik}\,G^j-\delta^{jk}\,G^i)
\,\bigg](w)
\nonu\\
&
-&\frac{1}{(z-w)^2}\,
\bigg[\,
\Bigg( \delta^{ik}\,\Big(\,
\frac{1}{6}\,(2\lambda-1)\,\Phi^{(2),j}_{\frac{1}{2}}
+\frac{16}{9}\,(\lambda+1)(\lambda-2)\,\partial G^j
\,\Big)
\nonu\\
&
- & (i \leftrightarrow j) \Bigg)
+\varepsilon^{ijkl}\,\frac{5}{2}\,\Phi^{(2),l}_{\frac{1}{2}}
\,\bigg](w)
\nonu\\
&
-&\frac{1}{(z-w)}\,
\bigg[
\Bigg( \delta^{ik}\,\Big(\,
\frac{1}{2}\,\tilde{\Phi}^{(2),j}_{\frac{3}{2}}
+\frac{1}{15}\,(2\lambda-1)\,\Phi^{(2),j}_{\frac{1}{2}}
+\frac{4}{9}\,(\lambda+1)(\lambda-2)\,\partial^2 G^j
\,\Big)
\nonu\\
&
- &
(i \leftrightarrow j) \Bigg)
+  \varepsilon^{ijkl}\,\partial \Phi^{(2),l}_{\frac{1}{2}}
\,\bigg](w)+\cdots\,,
\nonu\\
\Phi^{(1),ij}_{1}(z)\,\tilde{\Phi}^{(1)}_{2}(w)
&=&
\frac{1}{(z-w)^4}\,\bigg[\,
16i\,\Big(
(2\lambda^2-2\lambda-1)\,T^{ij}
+(2\lambda-1)\,\widetilde{T}^{ij}
\Big)
\,\bigg](w)
\nonu\\
&
+ & \frac{1}{(z-w)^2}\,3\,\Phi^{(2),ij}_{1}(w)
+\frac{1}{(z-w)}\,\partial \Phi^{(2),ij}_{1}(w)+\cdots\,,
\nonu \\
\tilde{\Phi}^{(1),i}_{\frac{3}{2}}(z)\,\tilde{\Phi}^{(1),j}_{\frac{3}{2}}(w)
&=&
\frac{1}{(z-w)^5}\,\delta^{ij}\,\frac{64}{3}\,N(\lambda-1)(\lambda+1)(\lambda-2)
\nonu\\
&
+&\frac{1}{(z-w)^4}\,\frac{32i}{3}\,\bigg[\,
(\lambda+1)(\lambda-2)\,T^{ij}
  \nonu \\
  & - & (\lambda+1)(\lambda-2)(2\lambda-1)\,\widetilde{T}^{ij}
\,\bigg](w)
\nonu\\
&
+& \frac{1}{(z-w)^3}\,\bigg[\,
\delta^{ij}\,\frac{8}{9}\Big(
3(2\lambda-1)\,\Phi^{(2)}_{0}-20(\lambda+1)(\lambda-2)\,L
\Big)
\nonu\\
&
+ &
\frac{16i}{3}\,\Big(
(\lambda+1)(\lambda-2)\,\partial T^{ij}
-(\lambda+1)(\lambda-2)(2\lambda-1)\,\partial \widetilde{T}^{ij}
\Big)
\,\bigg](w)
\nonu\\
&
+& \frac{1}{(z-w)^2}\,\bigg[\,
\delta^{ij}\,\frac{4}{9}\Big(
3(2\lambda-1)\,\partial \Phi^{(2)}_{0}-20(\lambda+1)(\lambda-2)\,\partial L
\Big)
\nonu\\
&
+ & 3\,\Phi^{(2),ij}_{1}
-\frac{1}{3}\,(2\lambda-1)\,\widetilde{\Phi}^{(2),ij}_{1}
\nonu\\
&
+&
\frac{16i}{9}\,\Big(
(\lambda+1)(\lambda-2)\,\partial^2 T^{ij}
-(\lambda+1)(\lambda-2)(2\lambda-1)\,\partial^2 \widetilde{T}^{ij}
\Big)
\,\bigg](w)
\nonu\\
&
+&\frac{1}{(z-w)}\,\bigg[\,
\delta^{ij}\,\Big(
\frac{1}{2}\,\Phi^{(2)}_{2}
+\frac{2}{5}(2\lambda-1)\,\partial^2 \Phi^{(2)}_{0}-\frac{8}{3}(\lambda+1)(\lambda-2)\,\partial^2 L
\Big)
\nonu\\
&
+& \frac{3}{2}\,\partial \Phi^{(2),ij}_{1}
+\frac{1}{6}\,(2\lambda-1)\,\partial \widetilde{\Phi}^{(2),ij}_{1}
\nonu\\
&
+&
\frac{4i}{9}\,\Big(
(\lambda+1)(\lambda-2)\,\partial^3 T^{ij}
-(\lambda+1)(\lambda-2)(2\lambda-1)\,\partial^3 \widetilde{T}^{ij}
\Big)
\,\bigg](w)
\nonu \\
& + & \cdots \,,
\nonu\\
\tilde{\Phi}^{(1),i}_{\frac{3}{2}}(z)\,\tilde{\Phi}^{(1)}_{2}(w)
&=&
\frac{1}{(z-w)^4}\,\frac{80}{3}\,(\lambda+1)(\lambda-2)\,G^i
\nonu\\
&
-&\frac{1}{(z-w)^3}\,\bigg[\,
\frac{4}{3}\,(2\lambda-1)\,\Phi^{(2),i}_{\frac{1}{2}}
-\frac{80}{9}\,(\lambda+1)(\lambda-2)\,\partial G^i
\,\bigg](w)
\nonu\\
&
+&\frac{1}{(z-w)^2}\,\bigg[\,
\frac{7}{2}\,\tilde{\Phi}^{(2),i}_{\frac{3}{2}}
-\frac{8}{15}\,(2\lambda-1)\,\partial \Phi^{(2),i}_{\frac{1}{2}}
+\frac{20}{9}\,(\lambda+1)(\lambda-2)\,\partial^2 G^i
\,\bigg]
\nonu\\
&
+&\frac{1}{(z-w)}\,\bigg[\,
\frac{3}{2}\,\partial \tilde{\Phi}^{(2),i}_{\frac{3}{2}}
-\frac{2}{15}\,(2\lambda-1)\,\partial^2 \Phi^{(2),i}_{\frac{1}{2}}
+\frac{4}{9}\,(\lambda+1)(\lambda-2)\,\partial^3 G^i
\,\bigg]\nonu \\
& + & \cdots \,,
\nonu \\
\tilde{\Phi}^{(1)}_{2}(z)\,\tilde{\Phi}^{(1)}_{2}(w)
&=&
\frac{1}{(z-w)^6}\,
\frac{320}{3}\,N (\lambda+1) (\lambda-1) (\lambda-2)
\nonu\\
&
+&\frac{1}{(z-w)^4}\,
\bigg[\,
8(2\lambda-1)\,\Phi^{(2)}_{0}
-\frac{320}{3}\,(\lambda+1)(\lambda-2)\,L
\,\bigg](w)
\nonu\\
&
+&\frac{1}{(z-w)^3}\,
\bigg[\,
4(2\lambda-1)\,\partial \Phi^{(2)}_{0}
-\frac{160}{3}\,(\lambda+1)(\lambda-2)\,\partial L
\,\bigg](w)
\nonu\\
&
+&\frac{1}{(z-w)^2}\,
\bigg[\,
4\,\tilde{\Phi}^{(2)}_{2}
+\frac{6}{5}(2\lambda-1)\,\partial^2 \Phi^{(2)}_{0}
-16\,(\lambda+1)(\lambda-2)\,\partial^2 L
\,\bigg](w)
\nonu\\
&
+&\frac{1}{(z-w)}\,
\bigg[\,
2\,\partial \tilde{\Phi}^{(2)}_{2}
+\frac{4}{15}(2\lambda-1)\,\partial^3 \Phi^{(2)}_{0}
-\frac{32}{9}\,(\lambda+1)(\lambda-2)\,\partial^3 L
\,\bigg](w)\nonu \\
& + & \cdots \,.
\label{ope1}
\eea
We keep the leading terms in the central terms after
the infinity limit of $N$ in (\ref{ope1}).
The corresponding (anti)commutators will be given in
(\ref{PhionePhione})
later.
One of the reasons why we present these OPEs is that
we need to know
the structure constants appearing in the quasi primary fields in
the right hand
sides explicitly for the (anti)commutators.
The relative coefficients appearing in all
the descendant terms of the quasi primary fields
are determined automatically.
For example, from the fourth relation of (\ref{ope1})
to the fourth relation of (\ref{PhionePhione}), we
need to have the explicit structure constants
of $\Gamma^i, G^i$, and
$\Phi_{\frac{1}{2}}^{(2),i}$ of the former:$8\,i\,
\la(\la-1)$, $-\frac{8}{3}\, (2\la-1)$, and
$-\frac{1}{2}$.

\section{ The OPE between the first and the second
  ${\cal N}=4$ higher spin multiplets
in the component approach under the large $(N,k)$ limit}

From the description of section $3$ (in particular, (\ref{singleOPE})),
we summarize the complete
$25$ kinds of OPEs under the large $(N,k)$ limit as follows:
\bea
\Phi^{(1)}_{0}(z)\,\Phi^{(2)}_{0}(w)
&=& + \cdots,
\nonu\\
\Phi^{(1)}_{0}(z)\,\Phi^{(2),i}_{\frac{1}{2}}(w)
&=&
-\frac{1}{(z-w)}\,2\,\tilde{\Phi}^{(1),i}_{\frac{3}{2}}(w)
+\cdots  \,,
\nonu\\
\Phi^{(1)}_{0}(z)\,\Phi^{(2),ij}_{1}(w)
&=&
\frac{1}{(z-w)^2}\,
\bigg[\,
\frac{8}{3}\,(2\lambda-1)\,\Phi^{(1),ij}_{1}
-8\,\widetilde{\Phi}^{(1),ij}_{1}
\,\bigg](w)+\cdots 
\,,
\nonu\\
\Phi^{(1)}_{0}(z)\,\tilde{\Phi}^{(2),i}_{\frac{3}{2}}(w)
&=&
\frac{1}{(z-w)^3}\,
\frac{32}{3}\,(\lambda+1)(\lambda-2)\,\Phi^{(1),i}_{\frac{1}{2}}(w)
\nonu\\
&
+& \frac{1}{(z-w)^2}\,
\bigg[\,
\frac{64}{15}\,(2\lambda-1)\,\tilde{\Phi}^{(1),i}_{\frac{3}{2}}
-\frac{32}{9}\,(\lambda+1)(\lambda-2)\,\partial \Phi^{(1),i}_{\frac{1}{2}}
\,\bigg](w)
\nonu\\
&
-&\frac{1}{(z-w)}\,\frac{1}{6}\,\Phi^{(3),i}_{\frac{1}{2}}(w)
+\cdots \,,
\nonu\\
\Phi^{(1)}_{0}(z)\,\tilde{\Phi}^{(2)}_{2}(w)
&=&
-\frac{1}{(z-w)^4}\,\frac{64}{3}\,(\lambda+1)(\lambda-2)\, \Phi^{(1)}_{0}(w)
\nonu \\
& + &
\frac{1}{(z-w)^3}\,\frac{64}{3}\,(\lambda+1)(\lambda-2)\, \partial \Phi^{(1)}_{0}(w)
\nonu\\
&
+&\frac{1}{(z-w)^2}\,
\bigg[\,
\Phi^{(3)}_{0}+\frac{64}{15}(2\lambda-1)\,\tilde{\Phi}^{(1)}_{2}
-\frac{32}{9}\,(\lambda+1)(\lambda-2)\,\partial^2 \Phi^{(1)}_{0}
\,\bigg](w)\nonu \\
& + & \cdots \,,
\nonu \\
\Phi^{(1),i}_{\frac{1}{2}}(z)\,\Phi^{(2)}_{0}(w)
&=&
\frac{1}{(z-w)}\,2\,\tilde{\Phi}^{(1),i}_{\frac{3}{2}}(w)+\cdots \,,
\nonu\\
\Phi^{(1),i}_{\frac{1}{2}}(z)\,\Phi^{(2),j}_{\frac{1}{2}}(w)
&=&
-\frac{1}{(z-w)^2}\,\bigg[\,
8\,\Phi^{(1),ij}_{1}-\frac{8}{3}\,(2\lambda-1)\,\widetilde{\Phi}^{(1),ij}_{1}
\,\bigg](w)
\nonu\\
&
-&\frac{1}{(z-w)}\,\bigg[\,
2\,\delta^{ij}\,\tilde{\Phi}^{(1)}_{2}
+2\,\partial \Phi^{(1),ij}_{1}-\frac{2}{3}\,(2\lambda-1)\,\partial \widetilde{\Phi}^{(1),ij}_{1}
\,\bigg](w)
+\cdots\,,
\nonu\\
\Phi^{(1),i}_{\frac{1}{2}}(z)\,\Phi^{(2),jk}_{1}(w)
&=&
-\frac{1}{(z-w)^3}\,
\bigg[\,
\frac{32}{3}\,(\lambda+1)(\lambda-2)\,(\delta^{ij}\,\Phi^{(1),k}_{\frac{1}{2}}
-\delta^{ik}\,\Phi^{(1),j}_{\frac{1}{2}})
\,\bigg](w)
\nonu\\
&
-&\frac{1}{(z-w)^2}\,
\bigg[\,
2(2\lambda-1)(\delta^{ij}\,\tilde{\Phi}^{(1),k}_{\frac{3}{2}}
-\delta^{ik}\,\tilde{\Phi}^{(1),j}_{\frac{3}{2}})
+\varepsilon^{ijkl}\,10\,\tilde{\Phi}^{(1),l}_{\frac{3}{2}}
\,\bigg](w)
\nonu\\
&
+&\frac{1}{(z-w)}\,
\frac{1}{6}\,\bigg[
\Bigg( \delta^{ij}\Big(
\Phi^{(3),k}_{\frac{1}{2}}
-\frac{12}{5}(2\lambda-1) \partial \tilde{\Phi}^{(1),k}_{\frac{3}{2}}
\Big) -(j \leftrightarrow k) \Bigg)
\nonu \\
& - &
\varepsilon^{ijkl}\,12\,\partial \tilde{\Phi}^{(1),l}_{\frac{3}{2}}
\,\bigg](w)
+\cdots\,,
\nonu\\
\Phi^{(1),i}_{\frac{1}{2}}(z)\,\tilde{\Phi}^{(2),j}_{\frac{3}{2}}(w)
&=&
\frac{1}{(z-w)^4}\,\delta^{ij}\,\frac{64}{3}\,(\lambda+1)(\lambda-2)\,\Phi^{(1)}_0(w)
\nonu \\
& - & \frac{1}{(z-w)^3}\,\bigg[
\delta^{ij}  \frac{32}{3}(\lambda+1)(\lambda-2)\,\partial \Phi^{(1)}_0
\nonu\\
&
+&\frac{128}{15}\,(2\lambda-1)\,\Phi^{(1),ij}_{1}
+\frac{32}{5}\,(\lambda+2)(\lambda-3)\,\widetilde{\Phi}^{(1),ij}_{1}
\,\bigg]
\nonu\\
&
-&\frac{1}{(z-w)^2}\,
\delta^{ij}\,\Phi^{(3)}_0
+\frac{1}{(z-w)}\,\frac{1}{6}\,\widetilde{\Phi}^{(3),ij}_1(w)+\cdots \,,
\nonu\\
\Phi^{(1),i}_{\frac{1}{2}}(z)\,\tilde{\Phi}^{(2)}_{2}(w)
&=&
-\frac{1}{(z-w)^4}\,\frac{160}{3}\,(\lambda+1)(\lambda-2)\,\Phi^{(1),i}_{\frac{1}{2}}(w)
\nonu\\
&
-&\frac{1}{(z-w)^3}\,
\bigg[\,
\frac{128}{15}\,(2\lambda-1)\,\tilde{\Phi}^{(1),i}_{\frac{3}{2}}
-\frac{160}{9}\,(\lambda+1)(\lambda-2)\,\partial \Phi^{(1),i}_{\frac{1}{2}}
\,\bigg](w)
\nonu\\
&
+&\frac{1}{(z-w)^2}\,\frac{7}{6}\,\Phi^{(3),i}_{\frac{1}{2}}(w)
+\frac{1}{(z-w)}\,\frac{1}{6}\,\partial \Phi^{(3),i}_{\frac{1}{2}}(w)+\cdots\,,
\nonu \\
\Phi^{(1),ij}_{1}(z)\,\Phi^{(2)}_{0}(w)
&=&
\frac{1}{(z-w)^2}\,\bigg[\,
\frac{8}{3}\,(2\lambda-1)\,\Phi^{(1),ij}_{1}
-8\,\widetilde{\Phi}^{(1),ij}_{1}
\,\bigg](w)
\nonu\\
&
+&\frac{1}{(z-w)}\,\bigg[\,
\frac{4}{3}\,(2\lambda-1)\,\partial \Phi^{(1),ij}_{1}
-4\,\partial \widetilde{\Phi}^{(1),ij}_{1}
\,\bigg](w)+\cdots\,,
\nonu\\
\Phi^{(1),ij}_{1}(z)\,\Phi^{(2),k}_{\frac{1}{2}}(w)
&=&
\frac{1}{(z-w)^3}\,
\bigg[\,
\frac{32}{3}\,(\lambda+1)(\lambda-2)\,(\delta^{ik}\,\Phi^{(1),j}_{\frac{1}{2}}
-\delta^{jk}\,\Phi^{(1),i}_{\frac{1}{2}})
\,\bigg](w)
\nonu\\
&
-&\frac{1}{(z-w)^2}\,\bigg[\,
\Bigg(\delta^{ik}\,\Big(
\frac{2}{3}\,(2\lambda-1)\,\tilde{\Phi}^{(1),j}_{\frac{3}{2}}
-\frac{32}{9}\,(\lambda+1)(\lambda-2)\,\partial\Phi^{(1),j}_{\frac{1}{2}}
\Big)\nonu \\
& - & (i \leftrightarrow j) \Bigg)
+
\varepsilon^{ijkl}\,10\,\tilde{\Phi}^{(1),l}_{\frac{3}{2}}
\,\bigg](w)
\nonu\\
&
-&\frac{1}{(z-w)}\,\bigg[\,
\Bigg( \delta^{ik}\,\Big(\,
\frac{1}{6}\,\Phi^{(3),j}_{\frac{1}{2}}
+\frac{4}{15}\,(2\lambda-1)\,\partial \tilde{\Phi}^{(1),j}_{\frac{3}{2}}
\nonu \\
& - &
\frac{8}{9}\,(\lambda+1)(\lambda-2)\,\partial^2\Phi^{(1),j}_{\frac{1}{2}}
\,\Big) -(i \leftrightarrow j ) \Bigg)
+ \varepsilon^{ijkl}\,4\,\partial \tilde{\Phi}^{(1),l}_{\frac{3}{2}}
\,\bigg](w)+\cdots \,,
\nonu \\
\Phi^{(1),ij}_{1}(z)\,\Phi^{(2),kl}_{1}(w)
&=&
-\frac{1}{(z-w)^4}\,\varepsilon^{ijkl}\,\frac{64}{3}\,(\lambda+1)(\lambda-2)\,\Phi^{(1)}_{0}(w)
\nonu\\
&
+&\frac{1}{(z-w)^3}\,
\frac{16}{3}\,
\bigg[\,
\Bigg( \delta^{ik}\,
\Big(
(2\lambda^2-2\lambda-7)\,\Phi^{(1),jl}_{1}+(2\lambda-1)\,\widetilde{\Phi}^{(1),jl}_{1}
\,\Big)
\nonu\\
&- & (k \leftrightarrow l) -( i \leftrightarrow j) + (i \leftrightarrow
j, k \leftrightarrow l)
\Bigg)
\,\bigg](w)
\nonu\\
&
-& \frac{1}{(z-w)^2}\,
\bigg[\,
  (\delta^{ik}\delta^{jl}-\delta^{il}\delta^{jk})\,12\,
  \tilde{\Phi}^{(1)}_{2}
\nonu\\
&
-& \Bigg( \delta^{ik}\,\frac{4}{3}\,
\Big(
(2\lambda^2-2\lambda-7)\,\partial \Phi^{(1),jl}_{1}+(2\lambda-1)\,\partial \widetilde{\Phi}^{(1),jl}_{1}
\,\Big)
\nonu\\
&- & (k \leftrightarrow l) -( i \leftrightarrow j) + (i \leftrightarrow
j, k \leftrightarrow l)
\Bigg)
\nonu \\
&
-& \varepsilon^{ijkl}\,\Big(
\Phi^{(3)}_{0}
-\frac{12}{5}(2\lambda-1)\,\tilde{\Phi}^{(1)}_2
\Big)
\,\bigg](w)
\nonu\\
&
-&\frac{1}{(z-w)}\,
\bigg[\,
  (\delta^{ik}\delta^{jl}-\delta^{il}\delta^{jk})\,4\,\partial
  \tilde{\Phi}^{(1)}_{2}
\nonu\\
&
+& \Bigg(
\delta^{ik}\,
\Big(\,
\frac{1}{6}\,\Phi^{(3),jl}_1
-\frac{4}{15}(2\lambda^2-2\lambda-7)\,\partial \Phi^{(1),jl}_{1}-\frac{4}{15}(2\lambda-1)\,\partial \widetilde{\Phi}^{(1),jl}_{1}
\,\Big)
\nonu\\
&- & (k \leftrightarrow l) -( i \leftrightarrow j) + (i \leftrightarrow
j, k \leftrightarrow l)
\Bigg)
\nonu \\
&
-&\varepsilon^{ijkl}\,\Big(\,
\frac{1}{3}\,\partial \Phi^{(3)}_{0}
-\frac{4}{5}\,(2\lambda-1)\,\partial \tilde{\Phi}^{(1)}_2
\,\Big)
\,\bigg](w)+\cdots\,,
\nonu \\
\Phi^{(1),ij}_{1}(z)\,\tilde{\Phi}^{(2),k}_{\frac{3}{2}}(w)
&=&
\frac{1}{(z-w)^4}\,\bigg[\,
\frac{32}{5}\,(\lambda+1)(\lambda-2)(2\lambda-1)\,
(
\delta^{ik}\,\Phi^{(1),j}_{\frac{1}{2}}
-\delta^{jk}\,\Phi^{(1),i}_{\frac{1}{2}}
)
\nonu\\
&
+&\varepsilon^{ijkl}\,\frac{160}{3}(\lambda+1)(\lambda-2)\,\Phi^{(1),l}_{\frac{1}{2}}
\,\bigg](w)
\nonu\\
&
+&\frac{1}{(z-w)^3}\,
\frac{48}{5}\,(\lambda+2)(\lambda-3)
(
\delta^{ik}\,\tilde{\Phi}^{(1),j}_{\frac{3}{2}}
-\delta^{jk}\,\tilde{\Phi}^{(1),i}_{\frac{3}{2}}
)(w)
\nonu\\
&
-&\frac{1}{(z-w)^2}\,
\bigg[\,
\Bigg( \delta^{ik}\,
\Big(\,
\frac{1}{30}\,(2\lambda-1)\,\Phi^{(3),j}_{\frac{1}{2}}
-\frac{48}{25}\,(\lambda+2)(\lambda-3)\,\partial
\tilde{\Phi}^{(1),j}_{\frac{3}{2}}
\,\Big)
\nonu\\
&
-&
(i \leftrightarrow j) \Bigg)
+\varepsilon^{ijkl}\,\frac{7}{6}\,\Phi^{(3),i}_{\frac{1}{2}}
\,\bigg]
\nonu\\
&
-&\frac{1}{(z-w)}\,
\bigg[\,
\Bigg( \delta^{ik}\,
\Big(\,
\frac{1}{6}\,\tilde{\Phi}^{(3),j}_{\frac{3}{2}}
+\frac{1}{105}\,(2\lambda-1)\,\partial \Phi^{(3),j}_{\frac{1}{2}}
\nonu \\
& - &
\frac{8}{25}\,(\lambda+2)(\lambda-3)\,\partial^2
\tilde{\Phi}^{(1),j}_{\frac{3}{2}}
\,\Big) -(i \leftrightarrow j) \Bigg)
+
\varepsilon^{ijkl}\,\frac{1}{3}\,\partial \Phi^{(3),l}_{\frac{1}{2}}
\,\bigg](w)+\cdots\,,
\nonu \\
\Phi^{(1),ij}_{1}(z)\,\tilde{\Phi}^{(2)}_{2}(w)
&=&
-\frac{1}{(z-w)^4}\,
\bigg[\,
\frac{64}{15}\,(17\lambda^2-17\lambda-52)\,\Phi^{(1),ij}_{1}
+\frac{128}{5}(2\lambda-1)\,\widetilde{\Phi}^{(1),ij}_{1}
\,\bigg](w)
\nonu\\
&
+&\frac{1}{(z-w)^2}\,
\frac{4}{3}\,\Phi^{(3),ij}_{1}(w)
+\frac{1}{(z-w)}\,
\frac{1}{3}\,\partial \Phi^{(3),ij}_{1}(w)+\cdots\,,
\nonu \\
\tilde{\Phi}^{(1),i}_{\frac{3}{2}}(z)\,\Phi^{(2)}_{0}(w)
&
=&
-\frac{1}{(z-w)^3}\,\frac{32}{3}\,(\lambda+1)(\lambda-2)\,\Phi^{(1),i}_{\frac{1}{2}}(w)
\nonu\\
&
+&\frac{1}{(z-w)^2}\,
\bigg[\,
\frac{8}{3}\,(2\lambda-1)\,\tilde{\Phi}^{(1),i}_{\frac{3}{2}}
-\frac{64}{9}\,(\lambda+1)(\lambda-2)\,\partial \Phi^{(1),i}_{\frac{1}{2}}
\,\bigg](w)
\nonu\\
&
+&\frac{1}{(z-w)}\,
\bigg[\,
\frac{1}{6}\,\Phi^{(3),i}_{\frac{1}{2}}
+\frac{8}{5}\,(2\lambda-1)\,\partial \tilde{\Phi}^{(1),i}_{\frac{3}{2}}
\nonu \\
& - &
\frac{8}{3}\,(\lambda+1)(\lambda-2)\,\partial^2 \Phi^{(1),i}_{\frac{1}{2}}
\,\bigg](w) +  \cdots\,,
\nonu\\
\tilde{\Phi}^{(1),i}_{\frac{3}{2}}(z)\,\Phi^{(2),j}_{\frac{1}{2}}(w)
&
=&
-\frac{1}{(z-w)^4}\,\delta^{ij}\,\frac{64}{3}\,(\lambda+1)(\lambda-2)\,\Phi^{(1)}_{0}(w)
\nonu \\
& - & \frac{1}{(z-w)^3}\,\frac{32}{3}\,\bigg[\,
\delta^{ij}\,(\lambda+1)(\lambda-2)\,\partial \Phi^{(1)}_{0}
\nonu\\
&
-&(2\lambda+1)\,\Phi^{(1),ij}_{1}
-\frac{1}{3}\,(\lambda^2-\lambda-11)\,\widetilde{\Phi}^{(1),ij}_{1}
\,\bigg](w)
\nonu\\
&
+&\frac{1}{(z-w)^2}\,
\bigg[\,
\delta^{ij}\,\Big(
\Phi^{(3)}_{0}-\frac{16}{15}\,(2\lambda-1)\,\tilde{\Phi}^{(1)}_{2}
-\frac{32}{9}\,(\lambda+1)(\lambda-2)\,\partial^2 \Phi^{(1)}_{0}
\Big)
\nonu\\
&
+&\frac{16}{3}\Big(
(2\lambda-1)\,\partial \Phi^{(1),ij}_{1}
+\frac{1}{3}\,(\lambda^2-\lambda-11)\,\partial \widetilde{\Phi}^{(1),ij}_{1}
\Big)
\,\bigg](w)
\nonu\\
&
+&\frac{1}{(z-w)}\,
\bigg[\,
\delta^{ij}\,\Big(\,
\frac{1}{2}\,\partial \Phi^{(3)}_{0}-\frac{8}{15}\,(2\lambda-1)\,\partial \tilde{\Phi}^{(1)}_{2}
\nonu \\
& - & \frac{8}{9}\,(\lambda+1)(\lambda-2)\,\partial^3 \Phi^{(1)}_{0}
\,\Big)
\nonu\\
&
-&\frac{1}{6}\,\widetilde{\Phi}^{(3),ij}_{1}
+\frac{8}{5}\Big(
(2\lambda-1)\,\partial^2 \Phi^{(1),ij}_{1}
+\frac{1}{3}\,(\lambda^2-\lambda-11)\,\partial^2 \widetilde{\Phi}^{(1),ij}_{1}
\Big)
\,\bigg](w)
\nonu\\
&
+&\cdots\,,
\nonu \\
\tilde{\Phi}^{(1),i}_{\frac{3}{2}}(z)\,\Phi^{(2),jk}_{1}(w)
&
=&
\frac{1}{(z-w)^4}\,\bigg[\,
\frac{32}{3}\,(\lambda+1)(\lambda-2)(2\lambda-1)\,(\delta^{ij}\,\Phi^{(1),k}_{\frac{1}{2}}
-\delta^{ik}\,\Phi^{(1),j}_{\frac{1}{2}})
\nonu\\
&
+&
\frac{160}{3}(\lambda+1)(\lambda-2)\,\varepsilon^{ijkl}\,\Phi^{(1),l}_{\frac{1}{2}}
\,\bigg](w)
\nonu\\
&
-&\frac{1}{(z-w)^3}\,
\bigg[\,
\Bigg( \delta^{ij}\,\Big(
\frac{16}{3}\,(\lambda^2-\lambda-11)\,\tilde{\Phi}^{(1),k}_{\frac{3}{2}}
\nonu \\
& - &
\frac{32}{9}\,(\lambda+1)(\lambda-2)(2\lambda-1)\,\partial \Phi^{(1),k}_{\frac{1}{2}}
\Big) -(j \leftrightarrow k) \Bigg)
\nonu\\
&
-&\varepsilon^{ijkl}\,\Big(\,
\frac{16}{3}(2\lambda-1)\,\tilde{\Phi}^{(1),l}_{\frac{3}{2}}
+\frac{160}{9}\,(\lambda+1)(\lambda-2)\,\partial \Phi^{(1),l}_{\frac{1}{2}}
\,\Big)
\,\bigg](w)
\nonu\\
&
-&\frac{1}{(z-w)^2}\,
\bigg[\,
\Bigg( \delta^{ij}\,\Big(
\frac{1}{18}\,(2\lambda-1)\,\Phi^{(3),k}_{\frac{1}{2}}
+\frac{32}{15}\,(\lambda^2-\lambda-11)\,\partial
\tilde{\Phi}^{(1),k}_{\frac{3}{2}}
\nonu\\
&
-&\frac{8}{9}\,(\lambda+1)(\lambda-2)(2\lambda-1)\,\partial^2 \Phi^{(1),k}_{\frac{1}{2}}
\Big) -(j \leftrightarrow k) \Bigg)
\nonu\\
&
+&\varepsilon^{ijkl}\,\Big(
\frac{7}{6}\Phi^{(3),l}_{\frac{1}{2}}
-\frac{32}{15}(2\lambda-1)\partial \tilde{\Phi}^{(1),l}_{\frac{3}{2}}
-\frac{40}{9}(\lambda+1)(\lambda-2)\partial^2 \Phi^{(1),l}_{\frac{1}{2}}
\Big)
\bigg](w)
\nonu\\
&
+&\frac{1}{(z-w)}\,
\bigg[\,
\Bigg( \delta^{ij}\,\Big(\,
\frac{1}{6}\,\tilde{\Phi}^{(3),k}_{\frac{3}{2}}
-\frac{1}{42}\,(2\lambda-1)\,\partial \Phi^{(3),k}_{\frac{1}{2}}
\nonu \\
& - &
\frac{8}{15}\,(\lambda^2-\lambda-11)\,\partial^2
\tilde{\Phi}^{(1),k}_{\frac{3}{2}}
\nonu\\
&
+&\frac{8}{45}\,(\lambda+1)(\lambda-2)(2\lambda-1)\,\partial^3 \Phi^{(1),k}_{\frac{1}{2}}
\,\Big) -(j \leftrightarrow k) \Bigg)
\nonu \\
& - & \varepsilon^{ijkl}\,\Big(\,
\frac{1}{2}\,\partial \Phi^{(3),l}_{\frac{1}{2}}
-\frac{8}{15}\,(2\lambda-1)\,\partial^2
\tilde{\Phi}^{(1),l}_{\frac{3}{2}}
\nonu\\
&
-&\frac{8}{9}\,(\lambda+1)(\lambda-2)\,\partial^3 \Phi^{(1),l}_{\frac{1}{2}}
\,\Big)
\,\bigg](w)+\cdots\,,
\nonu \\
\tilde{\Phi}^{(1),i}_{\frac{3}{2}}(z)\,\tilde{\Phi}^{(2),j}_{\frac{3}{2}}(w)
&
=&
-\frac{1}{(z-w)^5}\,\delta^{ij}\,\frac{2048}{45}\,(\lambda+1)(\lambda-2)(2\lambda-1)\,\Phi^{(1)}_{0}(w)
\nonu\\
&
-&\frac{1}{(z-w)^4}\,
\bigg[\,
\frac{192}{5}\,(\lambda+2)(\lambda-3)\,\Phi^{(1),ij}_{1}
\nonu \\
& - &
\frac{64}{5}\,(\lambda+2)(\lambda-3)(2\lambda-1)\,\widetilde{\Phi}^{(1),ij}_{1}
\,\bigg](w)
\nonu\\
&
+&\frac{1}{(z-w)^3}\,
\bigg[\,
\delta^{ij}\,\Big(
\frac{16}{15}\,(2\lambda-1)\,\Phi^{(3)}_{0}
-\frac{336}{25}\,(\lambda+2)(\lambda-3)\,\tilde{\Phi}^{(1)}_{2}
\Big)
\nonu\\
&
-&
\frac{48}{5}\,(\lambda+2)(\lambda-3)\,\partial \Phi^{(1),ij}_{1}
+\frac{16}{5}\,(\lambda+2)(\lambda-3)(2\lambda-1)\,\partial \widetilde{\Phi}^{(1),ij}_{1}
\,\bigg](w)
\nonu\\
&
+&\frac{1}{(z-w)^2}\,
\bigg[\,
\delta^{ij}\,\Big(
\frac{16}{45}\,(2\lambda-1)\,\partial \Phi^{(3)}_{0}
-\frac{112}{25}\,(\lambda+2)(\lambda-3)\,\partial
\tilde{\Phi}^{(1)}_{2}
\Big)
\nonu\\
&
+&
\frac{4}{3}\,\Phi^{(3),ij}_{1}
-\frac{4}{45}\,(2\lambda-1)\,\widetilde{\Phi}^{(3),ij}_{1}
-\frac{48}{25}\,(\lambda+2)(\lambda-3)\,\partial^2 \Phi^{(1),ij}_{1}
\nonu\\
&
+&
\frac{16}{25}\,(\lambda+2)(\lambda-3)(2\lambda-1)\,\partial^2 \widetilde{\Phi}^{(1),ij}_{1}
\,\bigg](w)
\nonu\\
&
+&\frac{1}{(z-w)}\,
\bigg[\,
\delta^{ij}\,\Big(
\frac{1}{6}\,\tilde{\Phi}^{(3)}_{2}
+\frac{8}{105}\,(2\lambda-1)\,\partial^2 \Phi^{(3)}_{0}
\nonu \\
& - & \frac{24}{25}\,(\lambda+2)(\lambda-3)\,\partial^2
\tilde{\Phi}^{(1)}_{2}
\Big)
\nonu\\
&
+&\frac{1}{2}\,\partial \Phi^{(3),ij}_{1}
-\frac{1}{30}\,(2\lambda-1)\,\partial\widetilde{\Phi}^{(3),ij}_{1}
-\frac{8}{25}\,(\lambda+2)(\lambda-3)\,\partial^3 \Phi^{(1),ij}_{1}
\nonu\\
&
+&\frac{8}{75}\,(\lambda+2)(\lambda-3)(2\lambda-1)\,\partial^3 \widetilde{\Phi}^{(1),ij}_{1}
\,\bigg](w)+\cdots\,,
\nonu\\
\tilde{\Phi}^{(1),i}_{\frac{3}{2}}(z)\,\tilde{\Phi}^{(2)}_{2}(w)
&
=&\frac{1}{(z-w)^5}\,\frac{2048}{45}\,
(\lambda+1)(\lambda-2)(2\lambda-1)\,\Phi^{(1),i}_{\frac{1}{2}}(w)
\nonu\\
&
-&\frac{1}{(z-w)^4}\,\frac{336}{5}\,
(\lambda+2)(\lambda-3)\,\tilde{\Phi}^{(1),i}_{\frac{3}{2}}(w)
\nonu\\
&
-&\frac{1}{(z-w)^3}\,\bigg[\,
\frac{32}{45}\,(2\lambda-1)\,\Phi^{(3),i}_{\frac{1}{2}}
+\frac{336}{25}\,(\lambda+2)(\lambda-3)\,\partial
\tilde{\Phi}^{(1),i}_{\frac{3}{2}}
\,\bigg](w)
\nonu\\
&
+&\frac{1}{(z-w)^2}\,\bigg[\,
\frac{3}{2}\,\tilde{\Phi}^{(3),i}_{\frac{3}{2}}
-\frac{64}{315}\,(2\lambda-1)\,\partial \Phi^{(3),i}_{\frac{1}{2}}
\nonu \\
& - &
\frac{56}{25}\,(\lambda+2)(\lambda-3)\,\partial^2
\tilde{\Phi}^{(1),i}_{\frac{3}{2}}
\,\bigg](w)
\nonu\\
&
+&\frac{1}{(z-w)}\,\bigg[\,
\frac{1}{2}\,\partial \tilde{\Phi}^{(3),i}_{\frac{3}{2}}
-\frac{4}{105}\,(2\lambda-1)\,\partial^2 \Phi^{(3),i}_{\frac{1}{2}}
\nonu \\
& - &
\frac{8}{25}\,(\lambda+2)(\lambda-3)\,\partial^3
\tilde{\Phi}^{(1),i}_{\frac{3}{2}}
\,\bigg](w)
+\cdots\,,
\nonu \\
\tilde{\Phi}^{(1)}_{2}(z)\,\Phi^{(2)}_{0}(w)
&
=&
-\frac{1}{(z-w)^4}\,\frac{64}{3}\,(\lambda+1)(\lambda-2)\,\Phi^{(1)}_{0}(w)
\nonu \\
& - &
\frac{1}{(z-w)^3}\,\frac{64}{3}\,(\lambda+1)(\lambda-2)\,\partial \Phi^{(1)}_{0}(w)
\nonu\\
&
+&\frac{1}{(z-w)^2}\,
\bigg[\,
\Phi^{(3)}_{0}
+\frac{8}{5}\,(2\lambda-1)\,\tilde{\Phi}^{(1)}_{2}
-\frac{32}{3}\,(\lambda+1)(\lambda-2)\,\partial^2 \Phi^{(1)}_{0}
\,\bigg](w)\nonu\\
&
+&\frac{1}{(z-w)}\,
\bigg[\,
\frac{2}{3}\,\partial \Phi^{(3)}_{0}
+\frac{16}{15}\,(2\lambda-1)\,\partial \tilde{\Phi}^{(1)}_{2}
\nonu \\
& - & \frac{32}{9}\,(\lambda+1)(\lambda-2)\,\partial^3 \Phi^{(1)}_{0}
\,\bigg](w)
+  \cdots\,,
\nonu\\
\tilde{\Phi}^{(1)}_{2}(z)\,\Phi^{(2),i}_{\frac{1}{2}}(w)
&
=&
-\frac{1}{(z-w)^4}\,\frac{160}{3}\,(\lambda+1)(\lambda-2)\,\Phi^{(1),i}_{\frac{1}{2}}(w)
\nonu\\
&
+&\frac{1}{(z-w)^3}\,\bigg[\,
\frac{16}{3}\,(2\lambda-1)\, \tilde{\Phi}^{(1),i}_{\frac{3}{2}}
-\frac{320}{9}\,(\lambda+1)(\lambda-2)\,\partial \Phi^{(1),i}_{\frac{1}{2}}
\,\bigg](w)
\nonu\\
&
+&\frac{1}{(z-w)^2}\,
\bigg[\,
\frac{7}{6}\,\Phi^{(3),i}_{\frac{1}{2}}
+\frac{16}{5}\,(2\lambda-1)\,\partial \tilde{\Phi}^{(1),i}_{\frac{3}{2}}
\nonu \\
& - &
\frac{40}{3}\,(\lambda+1)(\lambda-2)\,\partial^2 \Phi^{(1),i}_{\frac{1}{2}}
\,\bigg](w)\nonu\\
&
+&\frac{1}{(z-w)}\,
\bigg[\,
\frac{2}{3}\,\partial \Phi^{(3),i}_{\frac{1}{2}}
+\frac{16}{15}\,(2\lambda-1)\,\partial^2 \tilde{\Phi}^{(1),i}_{\frac{3}{2}}
\nonu \\
& - &
\frac{32}{9}\,(\lambda+1)(\lambda-2)\,\partial^3 \Phi^{(1),i}_{\frac{1}{2}}
\,\bigg](w)
+  \cdots\,,
\nonu\\
\tilde{\Phi}^{(1)}_{2}(z)\,\Phi^{(2),ij}_{1}(w)
&
=&
-\frac{1}{(z-w)^4}\,
\bigg[\,
32\,(2\lambda^2-2\lambda-7)\,\Phi^{(1),ij}_{1}
+32\,(2\lambda-1)\,\widetilde{\Phi}^{(1),ij}_{1}
\,\bigg](w)
\nonu\\
&
-&\frac{1}{(z-w)^3}\,
\bigg[\,
16\,(2\lambda^2-2\lambda-7)\,\partial \Phi^{(1),ij}_{1}
+16\,(2\lambda-1)\,\partial \widetilde{\Phi}^{(1),ij}_{1}
\,\bigg](w)
\nonu\\
&
+&\frac{1}{(z-w)^2}\,
\bigg[\,
\frac{4}{3}\,\Phi^{(3),ij}_{1}
-\frac{24}{5}\,(2\lambda^2-2\lambda-7)\,\partial^2 \Phi^{(1),ij}_{1}
\nonu \\
& - & \frac{24}{5}\,(2\lambda-1)\,\partial^2 \widetilde{\Phi}^{(1),ij}_{1}
\,\bigg](w)
\nonu\\
&
+&\frac{1}{(z-w)}\,
\bigg[\,
\frac{2}{3}\,\partial \Phi^{(3),ij}_{1}
-\frac{16}{15}\,(2\lambda^2-2\lambda-7)\,\partial^3 \Phi^{(1),ij}_{1}
\nonu \\
& - &
\frac{16}{15}\,(2\lambda-1)\,\partial^3 \widetilde{\Phi}^{(1),ij}_{1}
\,\bigg](w)
+\cdots\,,
\nonu\\
\tilde{\Phi}^{(1)}_{2}(z)\,\tilde{\Phi}^{(2),i}_{\frac{3}{2}}(w)
&
=&
-\frac{1}{(z-w)^4}\,
\frac{336}{5}\,
(\lambda+2)(\lambda-3)\,\tilde{\Phi}^{(1),i}_{\frac{3}{2}}(w)
\nonu\\
&
+&\frac{1}{(z-w)^3}\,
\bigg[\,
\frac{16}{45}\,(2\lambda-1)\, \Phi^{(3),i}_{\frac{1}{2}}
-\frac{672}{25}\,(\lambda+2)(\lambda-3)\,\partial
\tilde{\Phi}^{(1),i}_{\frac{3}{2}}
\,\bigg](w)
\nonu\\
&
+&\frac{1}{(z-w)^2}\,
\bigg[\,
\frac{3}{2}\,\tilde{\Phi}^{(3),i}_{\frac{3}{2}}
+\frac{16}{105}\,(2\lambda-1)\,\partial \Phi^{(3),i}_{\frac{1}{2}}
\nonu \\
& - &
\frac{168}{25}\,(\lambda+2)(\lambda-3)\,\partial^2
\tilde{\Phi}^{(1),i}_{\frac{3}{2}}
\,\bigg](w)
\nonu\\
&
+&\frac{1}{(z-w)}\,
\bigg[\,
\frac{2}{3}\,\partial \tilde{\Phi}^{(3),i}_{\frac{3}{2}}
+\frac{4}{105}\,(2\lambda-1)\,\partial^2 \Phi^{(1),i}_{\frac{1}{2}}
\nonu \\
& - &
\frac{32}{25}\,(\lambda+2)(\lambda-3)\,\partial^3
\tilde{\Phi}^{(1),i}_{\frac{3}{2}}
\,\bigg](w)
+\cdots\,,
\nonu \\
\tilde{\Phi}^{(1)}_{2}(z)\,\tilde{\Phi}^{(2)}_{2}(w)
&
=&
-\frac{1}{(z-w)^6}\,
\frac{2048}{9}\,
(\lambda+1)(\lambda-2)(2\lambda-1)\,\Phi^{(1)}_{0}(w)
\nonu\\
&
+&\frac{1}{(z-w)^4}\,
\bigg[\,
\frac{16}{5}\,(2\lambda-1)\, \Phi^{(3)}_{0}
-\frac{2688}{25}\,(\lambda+2)(\lambda-3)\, \tilde{\Phi}^{(1)}_{2}
\,\bigg](w)
\nonu\\
&
+&\frac{1}{(z-w)^3}\,
\bigg[\,
\frac{16}{15}\,(2\lambda-1)\,\partial \Phi^{(3)}_{0}
-\frac{896}{25}\,(\lambda+2)(\lambda-3)\,\partial
\tilde{\Phi}^{(1)}_2
\,\bigg](w)
\nonu\\
&
+&\frac{1}{(z-w)^2}\,
\bigg[\,
\frac{5}{3}\,\tilde{\Phi}^{(3)}_{2}
+\frac{8}{35}\,(2\lambda-1)\,\partial^2 \Phi^{(3)}_{0}
\nonu \\
& - & \frac{192}{25}\,(\lambda+2)(\lambda-3)\,\partial^2
\tilde{\Phi}^{(1)}_2
\,\bigg](w)
\nonu\\
&
+&\frac{1}{(z-w)}\,
\bigg[\,
\frac{2}{3}\,\partial \tilde{\Phi}^{(3)}_{2}
+\frac{4}{105}\,(2\lambda-1)\,\partial^3 \Phi^{(3)}_{0}
\nonu \\
& - & \frac{32}{25}\,(\lambda+2)(\lambda-3)\,\partial^3
\tilde{\Phi}^{(1)}_2
\,\bigg](w)
+\cdots\,.
\label{ope2}
\eea 
Since we are considering the OPEs between the
first and second ${\cal N}=4$ higher spin multiplets in (\ref{ope2}),
we have more OPEs compared to the ones in Appendix $D$.
As before, the corresponding (anti)commutators are presented
in (\ref{PhionePhitwo}) later.
There is no $\la$ factor in the structure constants in the
right hand side of (\ref{ope2}), contrary to the case of (\ref{ope1}). 
Therefore, at $\la =0$, all the terms in (\ref{ope2}) survive.

\section{ The OPE between the second
  ${\cal N}=4$ higher spin multiplet
in the ${\cal N}=4$ superspace with $N=5$}

From the results of section $4$ and
Appendix $C$, we can summarize the complete
OPE with $N=5$ as follows:
\bea
&& {\bf \Phi}^{(2)}(Z_{1})\,{\bf \Phi}^{(2)}(Z_{2}) =
\frac{1}{z_{12}^{4}}\, e^{0,4}_{0}
+\frac{\theta_{12}^{4-0}}{z_{12}^{6}}\,
8 \, \alpha \, e_{0}^{0,4}
+\frac{\theta_{12}^{4-i}}{z_{12}^{5}}\,
{\bf Q}^{(\frac{1}{2}),i}_{\frac{3}{2}}(Z_{2})
\nonu \\
&& +\frac{\theta_{12}^{4-0}}{z_{12}^{5}}\,
{\bf Q}^{(1)}_{2}(Z_{2})
+
\frac{\theta_{12}^{4-ij}}{z_{12}^{4}}\,
{\bf Q}^{(1),ij}_{1}(Z_{2})
 +\frac{\theta_{12}^{4-i}}{z_{12}^{4}}\,\Bigg[\,
2\,\partial {\bf Q}^{(\frac{1}{2}),i}_{\frac{3}{2}}+{\bf Q}^{(\frac{3}{2}),i}_{\frac{3}{2}}
+{\bf R}^{(\frac{3}{2}),i}_{\frac{3}{2}}
\,\Bigg](Z_{2})
 \nonu\\
&&
+\frac{\theta_{12}^{4-0}}{z_{12}^{4}}\,\Bigg[\,
\frac{3}{2}\,\partial {\bf Q}^{(1)}_{2}
 +{\bf Q}^{(2)}_{2}
+{\bf R}^{(2)}_{2}\,\Bigg](Z_{2})
+
\frac{\theta_{12}^{i}}{z_{12}^{3}}\,
{\bf Q}^{(\frac{3}{2}),i}_{\frac{1}{2}}(Z_{2})
 +
\frac{\theta_{12}^{4-ij}}{z_{12}^{3}}\,\Bigg[\,
\partial{\bf Q}^{(1),ij}_{1}
+{\bf Q}^{(2),ij}_{1}
\,\Bigg](Z_{2})
\nonu \\
&& +\frac{\theta_{12}^{4-i}}{z_{12}^{3}}\,\Bigg[\,
\frac{3}{2}\,\partial^{2} {\bf Q}^{(\frac{1}{2}),i}_{\frac{3}{2}}
+\partial {\bf Q}^{(\frac{3}{2}),i}_{\frac{3}{2}}
+{\bf Q}^{(\frac{5}{2}),i}_{\frac{3}{2}}
+{\bf R}^{(\frac{5}{2}),i}_{\frac{3}{2}}
\,\Bigg](Z_{2})
\nonu\\
&&
+\frac{\theta_{12}^{4-0}}{z_{12}^{3}}\,\Bigg[\,
\partial^{2} {\bf Q}^{(1)}_{2}
+\partial {\bf Q}^{(2)}_{2}
+{\bf Q}^{(3)}_{2}
+{\bf R}^{(3)}_{2}\,\Bigg](Z_{2})
+\frac{1}{z_{12}^{2}}\,
{\bf Q}^{(2)}_{0}(Z_{2})
\nonu \\
&&
+
\frac{\theta_{12}^{i}}{z_{12}^{2}}\,\Bigg[\,
\frac{2}{3}\,\partial  {\bf Q}^{(\frac{3}{2}),i}_{\frac{1}{2}}
 + {\bf Q}^{(\frac{5}{2}),i}_{\frac{1}{2}}
\,\Bigg](Z_{2})
+
\frac{\theta_{12}^{4-ij}}{z_{12}^{2}}\,\Bigg[\,
\frac{1}{2}\,\partial^{2} {\bf Q}^{(1),ij}_{1}
+\frac{3}{4}\,\partial {\bf Q}^{(2),ij}_{1}
+{\bf Q}^{(3),ij}_{1}
\,\Bigg](Z_{2})
\nonu\\
&&
+
\frac{\theta_{12}^{4-i}}{z_{12}^{2}}\,\Bigg[\,
\frac{2}{3}\,\partial^{3} {\bf Q}^{(\frac{1}{2}),i}_{\frac{3}{2}}
+\frac{1}{2}\,\partial^{2} {\bf Q}^{(\frac{3}{2}),i}_{\frac{3}{2}}
+\frac{4}{5}\,\partial {\bf Q}^{(\frac{5}{2}),i}_{\frac{3}{2}}
+ {\bf Q}^{(\frac{7}{2}),i}_{\frac{3}{2}}
+{\bf R}^{(\frac{7}{2}),i}_{\frac{3}{2}}
\,\Bigg](Z_{2})
\nonu\\
&&
+\frac{\theta_{12}^{4-0}}{z_{12}^{2}}\,\Bigg[\,
\frac{5}{12}\,\partial^{3} {\bf Q}^{(1)}_{2}
+\frac{1}{2}\, \partial {\bf Q}^{(2)}_{2}
+\frac{5}{6}\,{\bf Q}^{(3)}_{2}
+{\bf Q}^{(4)}_{2}
+{\bf R}^{(4)}_{2}
\,\Bigg](Z_{2})
\nonu\\
&&
+\frac{1}{z_{12}}\,
\frac{1}{2}\partial {\bf Q}^{(2)}_{0}(Z_{2})
+
\frac{\theta_{12}^{i}}{z_{12}}\,\Bigg[\,
\frac{1}{4}\,\partial^{2}  {\bf Q}^{(\frac{3}{2}),i}_{\frac{1}{2}} +
\frac{3}{5}\,\partial {\bf Q}^{(\frac{5}{2}),i}_{\frac{1}{2}}
+ {\bf Q}^{(\frac{7}{2}),i}_{\frac{1}{2}}
\,\Bigg](Z_{2})
\nonu\\
&&
\nonu \\ 
&&
+
\frac{\theta_{12}^{4-ij}}{z_{12}}\,\Bigg[\,
\frac{1}{6}\,\partial^{3} {\bf Q}^{(1),ij}_{1} 
+\frac{3}{10}\,\partial^{2} {\bf Q}^{(2),ij}_{1}
+\frac{2}{3}\,\partial {\bf Q}^{(3),ij}_{1}
+{\bf Q}^{(4),ij}_{1}
\,\Bigg](Z_{2})
\nonu \\ 
&&
+\frac{\theta_{12}^{4-i}}{z_{12}}\,\Bigg[\,
\frac{5}{24}\,\partial^{4} {\bf Q}^{(\frac{1}{2}),i}_{\frac{3}{2}}
+\frac{1}{6}\,\partial^{3} {\bf Q}^{(\frac{3}{2}),i}_{\frac{3}{2}}
+\frac{1}{3}\,\partial^{2} {\bf Q}^{(\frac{5}{2}),i}_{\frac{3}{2}}
+\frac{5}{7}\,\partial {\bf Q}^{(\frac{7}{2}),i}_{\frac{3}{2}}
+ {\bf Q}^{(\frac{9}{2}),i}_{\frac{3}{2}}
\,\Bigg](Z_{2})
\nonu \\ 
&&
+\frac{\theta_{12}^{4-0}}{z_{12}}\,\Bigg[\,
\frac{1}{8}\,\partial^{4} {\bf Q}^{(1)}_{2}
+\frac{1}{6}\,\partial^{3} {\bf Q}^{(2)}_{2}
+\frac{5}{14}\,\partial^{2} {\bf Q}^{(3)}_{2}
+\frac{3}{4}\,\partial {\bf Q}^{(4)}_{2}
+ {\bf Q}^{(5)}_{2} + {\bf R}_{2}^{(5)}\,
\Bigg](Z_{2})
+\cdots.
\label{superopeappF}
\eea
The various quasi super primary fields
appearing in (\ref{superopeappF}) are given by
\bea
{\bf Q}^{(2)}_{0} & = &
e_{0}^{0,2}\,{\bf \Phi}^{(2)}
+e_{1}^{0,2}\,{\bf \Phi}^{(1)} {\bf \Phi}^{(1)}
+e_{2}^{0,2}\,{\bf J}^{4-0}
+e_{3}^{0,2}\,\partial^{2}{\bf J}
+e_{4}^{0,2}\,\partial {\bf J}\partial {\bf J}
+e_{5}^{0,2}\,{\bf J}^{ij}{\bf J}^{ij}
\nonu \\
& + & e_{6}^{0,2}\,{\bf J}^{ij}{\bf J}^{4-ij}
\nonu \\ 
& + & 
e_{7}^{0,2}\,{\bf J}^{i}{\bf J}^{j}{\bf J}^{ij}
+e_{8}^{0,2}\,{\bf J}^{i}{\bf J}^{j}{\bf J}^{4-ij}
+e_{9}^{0,2}\,{\bf J}^{i}\partial {\bf J}^{i}
+e_{10}^{0,2}\,\varepsilon^{ijkl}\,{\bf J}^{i}{\bf J}^{j}{\bf J}^{k}{\bf J}^{l},
\nonu\\
 {\bf Q}^{(\frac{3}{2}),i}_{\frac{1}{2}}      &  = &
e_{1}^{\frac{1}{2},3}\,{\bf J}^{4-i}
+e_{2}^{\frac{1}{2},3}\,\partial {\bf J}^{i}
+e_{3}^{\frac{1}{2},3}\,{\bf J}^{i}\partial {\bf J}
+\varepsilon^{ijkl}(
e_{4}^{\frac{1}{2},3}\,{\bf J}^{j}{\bf J}^{4-kl}
+e_{5}^{\frac{1}{2},3}\,{\bf J}^{j}{\bf J}^{k}{\bf J}^{l}),
\nonu \\ 
{\bf Q}^{(\frac{5}{2}),i}_{\frac{1}{2}}  & = &
e_{1}^{\frac{1}{2},2}\,D^{i}{\bf \Phi}^{(2)}
+e_{2}^{\frac{1}{2},2}\,{\bf \Phi}^{(1)} D^{i}{\bf \Phi}^{(1)}
+e_{3}^{\frac{1}{2},2}\,\partial {\bf J}^{4-i}
+e_{4}^{\frac{1}{2},2}\,\partial^{2}{\bf J}^{i}
+e_{5}^{\frac{1}{2},2}\,{\bf J}^{i}\partial^{2}{\bf J}
\nonu \\
& + & e_{6}^{\frac{1}{2},2}\,\partial {\bf J}^{i}\partial {\bf J}
\nonu \\ 
&
+& 
e_{7}^{\frac{1}{2},2}\,{\bf J}^{4-ij}{\bf J}^{4-j}
+e_{8}^{\frac{1}{2},2}\,{\bf J}^{i}{\bf J}^{j}{\bf J}^{4-j}
+e_{9}^{\frac{1}{2},2}\,\partial {\bf J}^{j}{\bf J}^{4-ij}
+e_{10}^{\frac{1}{2},2}\,{\bf J}^{j}\partial {\bf J}^{4-ij}
+e_{11}^{\frac{1}{2},2}\,{\bf J}^{j}{\bf J}^{4-ij}\partial {\bf J}
\nonu \\ 
&
+& 
e_{12}^{\frac{1}{2},2}\,{\bf J}^{i}{\bf J}^{j}\partial {\bf J}^{j}
+e_{13}^{\frac{1}{2},2}\,{\bf J}^{i}{\bf J}^{jk}{\bf J}^{jk}
+e_{14}^{\frac{1}{2},2}\,{\bf J}^{i}{\bf J}^{j}{\bf J}^{k}{\bf J}^{4-jk}
+e_{15}^{\frac{1}{2},2}\,{\bf J}^{j}{\bf J}^{4-ik}{\bf J}^{4-jk}
\nonu \\ 
&
+&
\varepsilon^{ijkl}(\,
e_{16}^{\frac{1}{2},2}\,{\bf J}^{4-jk}{\bf J}^{4-l}
+e_{17}^{\frac{1}{2},2}\,{\bf J}^{j}{\bf J}^{k}{\bf J}^{4-l}
+e_{18}^{\frac{1}{2},2}\,\partial {\bf J}^{j}{\bf J}^{4-kl}
+e_{19}^{\frac{1}{2},2}\, {\bf J}^{j}\partial{\bf J}^{4-kl}
\nonu \\
&
+ & e_{20}^{\frac{1}{2},2}\,{\bf J}^{j}{\bf J}^{4-kl}\partial {\bf J}
\nonu \\ 
&
+&
e_{21}^{\frac{1}{2},2}\,{\bf J}^{i}{\bf J}^{ij}{\bf J}^{kl}
+e_{22}^{\frac{1}{2},2}\,\partial({\bf J}^{j}{\bf J}^{k}{\bf J}^{l})
+e_{23}^{\frac{1}{2},2}\,{\bf J}^{j}{\bf J}^{k}{\bf J}^{l}\partial {\bf J}\,),
\nonu\\
    {\bf Q}^{(\frac{7}{2}),i}_{\frac{1}{2}} & = &
e_{1}^{\frac{1}{2},1}\,D^{4-i}{\bf \Phi}^{(2)}
+e_{2}^{\frac{1}{2},1}\,\partial D^{i}{\bf \Phi}^{(2)}
+e_{3}^{\frac{1}{2},1}\, {\bf \Phi}^{(1)} D^{4-i}{\bf \Phi}^{(1)}
+e_{4}^{\frac{1}{2},1}\, {\bf \Phi}^{(1)} \partial D^{i}{\bf \Phi}^{(1)}
+\cdots
\nonu \\ 
&+ & e_{20}^{\frac{1}{2},1}\,{\bf J}^{4-0}  \partial {\bf J}^i
+e_{21}^{\frac{1}{2},1}\,{\bf J}^{4-i} {\bf J}^{4-0}
+e_{22}^{\frac{1}{2},1}\, {\bf J}^{i} \partial {\bf J} {\bf J}^{4-0}
+e_{23}^{\frac{1}{2},1}\,\partial^2 {\bf J}^{4-i}
+e_{24}^{\frac{1}{2},1}\,\partial^3 {\bf J}^{i}
+\cdots
\nonu \\ 
&+ &
e_{67}^{\frac{1}{2},1}\,\varepsilon^{ijkl}\,{\bf J}^{j}{\bf J}^{k}D^l {\bf \Phi}^{(2)}
+e_{68}^{\frac{1}{2},1}\,\varepsilon^{ijkl}\,{\bf J}^{j} D^k {\bf \Phi}^{(1)} D^l {\bf \Phi}^{(1)}
+e_{69}^{\frac{1}{2},1}\,\varepsilon^{ijkl}\, {\bf J}^{j} {\bf J}^{k} {\bf \Phi}^{(1)} D^l {\bf \Phi}^{(1)}
+ \cdots
\nonu \\ 
&+ &
e_{115}^{\frac{1}{2},1}\,\varepsilon^{ijkl}\,{\bf J}^{j}{\bf J}^{4-ij}{\bf J}^{4-ik}{\bf J}^{4-lj}
+e_{116}^{\frac{1}{2},1}\,(\varepsilon^{ijkl})^2 \,{\bf J}^{j}{\bf J}^{k}{\bf J}^{l}{\bf J}^{4-ij}{\bf J}^{4-kl}, 
\nonu \\
{\bf Q}^{(1),ij}_{1} & = &
e_{1}^{1,4}\,{\bf J}^{4-ij}
+e_{2}^{1,4}\,{\bf J}^{i}{\bf J}^{j}
+e_{3}^{1,4}\,{\bf J}^{ij}
+\varepsilon^{ijkl}\,e_{4}^{1,4}\,{\bf J}^{k}{\bf J}^{l}
-(i\leftrightarrow j),
\nonu\\
 {\bf Q}^{(2),ij}_{1} & = &
e_{1}^{1,3}\,{\bf J}^{4-j}{\bf J}^{i}
+e_{2}^{1,3}\,\partial{\bf J}^{i}{\bf J}^{j}
+e_{3}^{1,3}\,{\bf J}^{ik}{\bf J}^{j}{\bf J}^k
+e_{4}^{1,3}\,\partial{\bf J}^{ij}
+e_{5}^{1,3}\,\varepsilon^{ijkl}\,\partial ({\bf J}^{k}{\bf J}^{l})
-(i\leftrightarrow j),
\nonu\\
 {\bf Q}^{(3),ij}_{1} & = &
e_{1}^{1,2}\, D^{4-ij}{\bf \Phi}^{(2)}
+e_{2}^{1,2}\, D^{ij}{\bf \Phi}^{(2)}
+e_{3}^{1,2}\, {\bf \Phi}^{(1)}D^{4-ij}{\bf \Phi}^{(1)}
+e_{4}^{1,2}\, {\bf \Phi}^{(1)}D^{ij}{\bf \Phi}^{(2)}
+\cdots
\nonu \\ 
&
+& e_{42}^{1,2}\, {\bf J}^{4-ik}{\bf J}^{4-jl}{\bf J}^{4-kl}
+e_{43}^{1,2}\, {\bf J}^{4-ij}{\bf J}^{4-kl}{\bf J}^{4-kl}
+\varepsilon^{ijkl}
(\,
e_{44}^{1,2}\, D^{k}{\bf \Phi}^{(1)} D^{l}{\bf \Phi}^{(1)}
\nonu \\
& + & e_{45}^{1,2}\, {\bf J}^{k} {\bf \Phi}^{(1)} D^{l}{\bf \Phi}^{(1)}
\nonu \\
& + &
e_{86}^{1,2}\, \partial {\bf J}^{i} {\bf J}^{i} {\bf J}^{k}  {\bf J}^{l} 
+e_{87}^{1,2}\, {\bf J}^{i} {\bf J}^{j} {\bf J}^{k} {\bf J}^{l}  {\bf J}^{4-kl} 
\,)
+e_{88}^{1,2}\, \varepsilon^{ijkl}\varepsilon^{r \rho \delta \sigma} 
{\bf J}^{r} {\bf J}^{\rho} {\bf J}^{4-\delta \sigma} {\bf J}^{4-kl}
\nonu \\
& + &
e_{89}^{1,2}\, (\varepsilon^{ijkl})^2 {\bf J}^{j}{\bf J}^{k}{\bf J}^{l} {\bf J}^{4-ij}
+\cdots
+e_{92}^{1,2}\,\varepsilon^{ijkl}\, {\bf J}^{i}{\bf J}^{j}{\bf J}^{j}{\bf J}^{k}{\bf J}^{l}{\bf J}^{4-ij}
-(i\leftrightarrow j),
\nonu\\
 {\bf Q}^{(4),ij}_{1} & = & 
e_{1}^{1,1}\,\partial D^{4-ij}{\bf \Phi}^{(2)}
+e_{2}^{1,1}\,{\bf \Phi}^{(1)} \partial D^{4-ij} {\bf \Phi}^{(1)})
+e_{3}^{1,1}\,{\bf \Phi}^{(1)} \partial D^{ij} {\bf \Phi}^{(1)})
\nonu \\
& + & e_{4}^{1,1}\,D^i {\bf \Phi}^{(1)} D^{4-j} {\bf \Phi}^{(1)})
\nonu\\
& + & 
\cdots
+e_{126}^{1,1}\,{\bf J}^{l}{\bf J}^{4-ik}{\bf J}^{4-kl}{\bf J}^{j}
+e_{127}^{1,1}\,{\bf J}^{l}{\bf J}^{4-ik}{\bf J}^{j}{\bf J}^{k}{\bf J}^{l}
+e_{128}^{1,1}\,{\bf J}^{l}{\bf J}^{4-ik}{\bf J}^{4-jl}{\bf J}^{k}
+\cdots
\nonu\\
& + & 
e_{153}^{1,1}\,{\bf J}^{l}{\bf J}^{4-ik}{\bf J}^{4-kl}{\bf J}^{j}
+\varepsilon^{ijkl}
(\,
e_{154}^{1,1}\,\partial {\bf J} {\bf J}^{k} D^l  {\bf \Phi}^{(2)}
+e_{155}^{1,1}\,{\bf J}^{4-k} D^l  {\bf \Phi}^{(2)}
+\cdots
\nonu\\
& + & 
e_{296}^{1,1}\,\partial^2 {\bf J}^{4-ik}{\bf J}^{4-il} 
\,)
+e_{297}^{1,1}\, \varepsilon^{ijkl}\varepsilon^{r \rho \delta \sigma} 
{\bf J}^{4-r \rho} {\bf J}^{4-\delta \sigma} \partial{\bf J} {\bf J}^{k} {\bf J}^{l}
- (i\leftrightarrow j),
\nonu \\
{\bf Q}^{(\frac{1}{2}),i}_{\frac{3}{2}} & = &
e_{1}^{\frac{3}{2},5}\,{\bf J}^{i},
\nonu\\
 {\bf Q}^{(\frac{3}{2}),i}_{\frac{3}{2}} & = &
e_{1}^{\frac{3}{2},4}\, {\bf J}^{4-i}
+e_{2}^{\frac{3}{2},4}\,\partial {\bf J}^{i}
+e_{3}^{\frac{3}{2},4}\, {\bf J}^{i}\partial {\bf J}
+e_{4}^{\frac{3}{2},4}\, {\bf J}^{j}{\bf J}^{4-ij}
+\varepsilon^{ijkl}\,
(\,
e_{5}^{\frac{3}{2},4}\, {\bf J}^{j}{\bf J}^{4-kl}
+e_{6}^{\frac{3}{2},4}\, {\bf J}^{j}{\bf J}^{k}{\bf J}^{l}\,
),
\nonu\\
 {\bf Q}^{(\frac{5}{2}),i}_{\frac{3}{2}} & = &
e_{1}^{\frac{3}{2},3}\,D^{i}{\bf \Phi}^{(2)}
+e_{2}^{\frac{3}{2},3}\, {\bf J}^{i}{\bf \Phi}^{(2)}
+e_{3}^{\frac{3}{2},3}\,{\bf \Phi}^{(1)}  D^{i}{\bf \Phi}^{(1)} 
+e_{4}^{\frac{3}{2},3}\,{\bf J}^{i} {\bf \Phi}^{(1)} {\bf \Phi}^{(1)} 
+\cdots
\nonu \\
& + &
e_{29}^{\frac{3}{2},3}\,\varepsilon^{ijkl}\, {\bf J}^{4-ji}{\bf J}^{k}{\bf J}^{i}{\bf J}^{l},
\nonu\\
 {\bf Q}^{(\frac{7}{2}),i}_{\frac{3}{2}} &  = &
e_{1}^{\frac{3}{2},2}\, D^{4-i}{\bf \Phi}^{(2)}
+e_{2}^{\frac{3}{2},2}\,\partial D^{i}{\bf \Phi}^{(2)}
+e_{3}^{\frac{3}{2},2}\,\partial {\bf J} D^{i}{\bf \Phi}^{(2)}
+\cdots 
+\varepsilon^{ijkl}\
(\,e_{34}^{\frac{3}{2},2}\, {\bf J}^{j} D^k {\bf \Phi}^{(1)} D^l {\bf \Phi}^{(1)}
\nonu\\
&  
+ &
e_{35}^{\frac{3}{2},2}\,{\bf J}^{j}{\bf J}^{k} {\bf \Phi}^{(1)} D^l {\bf \Phi}^{(1)}
+e_{36}^{\frac{3}{2},2}\,{\bf J}^{j} {\bf J}^{k}  {\bf J}^{l} {\bf \Phi}^{(1)}{\bf \Phi}^{(1)}
\,)
+e_{37}^{\frac{3}{2},2}\,\partial^{2}{\bf J}^{4-i}
+e_{38}^{\frac{3}{2},2}\,\partial^{3}{\bf J}^{i}
\nonu\\
&  
+ &e_{93}^{\frac{3}{2},2}\,{\bf J}^{4-ij}{\bf J}^j \partial {\bf J}^{k}  {\bf J}^{k} 
+ \varepsilon^{ijkl}(\,
e_{94}^{\frac{3}{2},2}\, {\bf J}^{j}{\bf J}^{4-kl}{\bf J}^{4-0}
+e_{95}^{\frac{3}{2},2}\, {\bf J}^{j}{\bf J}^{4-kl} \partial^2 {\bf J}
+\cdots
\nonu\\
&  
+ &
e_{141}^{\frac{3}{2},2}\, {\bf J}^{4-ij} \partial{\bf J} {\bf J}^{i} {\bf J}^{k} {\bf J}^{l} 
\,)
+\cdots
+e_{143}^{\frac{3}{2},2}\,\varepsilon^{ijkl}\, {\bf J}^{kl} {\bf J}^{kl} {\bf J}^j {\bf J}^{k} {\bf J}^{l} 
,\nonu\\
{\bf Q}^{(1)}_{2} & = &
e_{1}^{2,5}\,\partial {\bf J},
\nonu\\
{\bf Q}^{(2)}_{2} &  = &
e_{1}^{2,4}\,{\bf \Phi}^{(2)}
+e_{2}^{2,4}\,{\bf \Phi}^{(1)}{\bf \Phi}^{(1)}
+e_{3}^{2,4}\, {\bf J}^{4-0}
+e_{4}^{2,4}\,\partial^{2}{\bf J}
+e_{5}^{2,4}\,\partial {\bf J}\partial {\bf J}
+e_{6}^{2,4}\, {\bf J}^{ij}{\bf J}^{ij}
\nonu \\
& + & e_{7}^{2,4}\, {\bf J}^{ij}{\bf J}^{4-ij}
\nonu\\
&  
+ &
e_{8}^{2,4}\, {\bf J}^{i}{\bf J}^{j}{\bf J}^{4-ij}
+e_{9}^{2,4}\,  {\bf J}^{i}{\bf J}^{j}{\bf J}^{ij}
+e_{10}^{2,4}\, {\bf J}^{i}{\bf J}^{4-i}
+e_{11}^{2,4}\, {\bf J}^{i}\partial{\bf J}^{i}
+e_{12}^{2,4}\,\varepsilon^{ijkl}\, {\bf J}^{i}{\bf J}^{j}{\bf J}^{k}{\bf J}^{l},
\nonu\\
{\bf Q}^{(3)}_{2}  & = &
e_{1}^{2,3}\,\partial {\bf J} {\bf \Phi}^{(2)}
+e_{2}^{2,3}\, \partial {\bf J} {\bf \Phi}^{(1)}{\bf \Phi}^{(1)}
+e_{3}^{2,3}\, {\bf J}^{i}D^{i}{\bf \Phi}^{(2)}
+e_{4}^{2,3}\, {\bf J}^{i} {\bf \Phi}^{(1)} D^i{\bf \Phi}^{(1)}
+e_{5}^{2,3}\,\partial {\bf J}{\bf J}^{4-0}
\nonu\\
& 
+ &e_{6}^{2,3}\,\partial {\bf J}\partial^{2}{\bf J}
+e_{7}^{2,3}\,\partial {\bf J}\partial {\bf J}\partial {\bf J}
+e_{8}^{2,3}\, {\bf J}^{i}\partial {\bf J}^{4-i}
+e_{9}^{2,3}\,\partial {\bf J}^{i}{\bf J}^{4-i}
+e_{10}^{2,3}\, {\bf J}^{i}\partial {\bf J}^{i}\partial {\bf J}
+e_{11}^{2,3}\, {\bf J}^{i}\partial^{2}{\bf J}^{i}
\nonu\\
& 
+ &
e_{12}^{2,3}\,\partial {\bf J}^{i}\partial {\bf J}^{i}
+e_{13}^{2,3}\, {\bf J}^{i}{\bf J}^{4-ij}{\bf J}^{4-j}
+e_{14}^{2,3}\, {\bf J}^{i}{\bf J}^{j}{\bf J}^{4-ij}\partial {\bf J}
+e_{15}^{2,3}\, {\bf J}^{4-ij}{\bf J}^{4-ij}\partial {\bf J}
\nonu \\
& + & e_{16}^{2,3}\, {\bf J}^{i}{\bf J}^{j}\partial {\bf J}^{4-ij}
\nonu\\
& 
+ &
e_{17}^{2,3}(\partial {\bf J}^{i}{\bf J}^{j}{\bf J}^{4-ij}
+\partial^{2} {\bf J}^{ij})
+\varepsilon^{ijkl}(\, 
e_{18}^{2,3}\, {\bf J}^{i}{\bf J}^{4-jk}{\bf J}^{4-l}
+e_{19}^{2,3}\, {\bf J}^{i}{\bf J}^{j}{\bf J}^{k}{\bf J}^{4-l}
\nonu \\
& + & e_{20}^{2,3}\, {\bf J}^{i}{\bf J}^{j}{\bf J}^{4-kl}\partial {\bf J}
\nonu\\
& 
+ &
e_{21}^{2,3}\, {\bf J}^{4-ij}{\bf J}^{4-kl}\partial {\bf J}
+e_{22}^{2,3}\, {\bf J}^{i}{\bf J}^{j}\partial {\bf J}^{4-kl}
+e_{23}^{2,3}\,\partial {\bf J}^{i}{\bf J}^{j}{\bf J}^{4-kl}
+e_{24}^{2,3}\,\partial({\bf J}^{i}{\bf J}^{j}{\bf J}^{k}{\bf J}^{l})\,
),
\nonu\\
{\bf Q}^{(4)}_{2} & = &
e_{0}^{2,2}\,{\bf \Phi}^{(4)}
+e_{1}^{2,2}\,{\bf \Phi}^{(2)}{\bf \Phi}^{(2)}
+e_{2}^{2,2}\, D^{4-0}{\bf \Phi}^{(2)}
+e_{3}^{2,2}\, {\bf J}^{4-0}{\bf \Phi}^{(2)}+\cdots
+e_{26}^{2,2}\,D^{ij}{\bf \Phi}^{(1)}D^{ij}{\bf \Phi}^{(1)}
\nonu\\
& 
+ & 
e_{27}^{2,2}\, D^{ij}{\bf \Phi}^{(1)}D^{4-ij}{\bf \Phi}^{(1)}
+\cdots
+e_{131}^{2,2}\, {\bf J}^{ij}{\bf J}^{ij}{\bf J}^{kl}{\bf J}^{4-kl} {\bf J}^{i} {\bf J}^{j}
+e_{132}^{2,2}\, {\bf J}^{ij}{\bf J}^{kl} {\bf J}^{ij}{\bf J}^{kl}
\nonu\\
& 
+ & 
\varepsilon^{ijkl}
(\,
e_{133}^{2,2}\, {\bf J}^{4-0}{\bf J}^{i}{\bf J}^{j}{\bf J}^{k}{\bf J}^{l}
+e_{134}^{2,2}\,{\bf J}^{4-0}{\bf J}^{4-ij}{\bf J}^{k}{\bf J}^{l}
+\cdots
+e_{168}^{2,2}\,{\bf J}^{i}{\bf J}^{j}\partial {\bf J}^{k}\partial {\bf J}^{l}\,).
\label{Quasi}
\eea
As in the component approach in Appendix $C$,
the
${\bf Q}^{(\frac{9}{2}),i}_{\frac{3}{2}}$ of super spin-$\frac{9}{2}$ and 
the ${\bf Q}^{(5)}_{2}$ of super
spin-$5$ are presented in ${\tt ancillary.nb}$.
Similarly, the higher spin currents
${\bf R}^{(\frac{3}{2}),i}_{\frac{3}{2}}, \cdots,
{\bf R}^{(5)}_{2}$ can be found also
\bea
{\bf R}_{\frac{3}{2}}^{(\frac{7}{2}),i}(Z_2) & \equiv &
-\frac{2  \, \alpha}{5}\, (\frac{1}{4} \, \pa^2 \,
{\bf Q}_{\frac{1}{2}}^{(\frac{3}{2}),i} +\frac{3}{5} \, \pa \,
{\bf Q}_{\frac{1}{2}}^{(\frac{5}{2}),i} +
{\bf Q}_{\frac{1}{2}}^{(\frac{7}{2}),i})(Z_2),
\nonu \\
{\bf R}_{\frac{3}{2}}^{(\frac{5}{2}),i}(Z_2) & \equiv &
-\frac{4 \, \alpha}{5} \, (\frac{2}{3} \, \pa \,
{\bf Q}_{\frac{1}{2}}^{(\frac{3}{2}),i} +
{\bf Q}_{\frac{1}{2}}^{(\frac{5}{2}),i})(Z_2),
\nonu \\
{\bf R}_{\frac{3}{2}}^{(\frac{3}{2}),i}(Z_2) & \equiv &
-\frac{6 \, \alpha}{5} \, {\bf Q}_{\frac{1}{2}}^{(\frac{3}{2}),i}(Z_2),
\nonu \\
{\bf R}_{2}^{(4-n)}(Z_2) & \equiv &
-p_2 \, {\bf E}_2^{(4-n)}(Z_2) -p_1 \, n(n+1)\,
{\bf Q}_{0}^{(4-n)}(Z_2), \qquad n=-1,0,1,2.
\label{BigR}
\eea
We also have the following quantities corresponding to (\ref{Eexp})
\bea
{\bf E}^{(2)}_{2}(Z_2)& \equiv &
(\,
4 {\bf Q}^{(2)}_{0}+ \frac{e^{0,4}_{0}}{2}{\bf J}^{4-0}
\,)(Z_2),
\qquad
{\bf E}^{(3)}_{2}(Z_2) \equiv 
(\,
\frac{5}{2} \, \partial \, {\bf Q}^{(2)}_{0}+
\frac{e^{0,4}_{0}}{2} \, \partial \, {\bf J}^{4-0}
\,)(Z_2),
\nonu\\
{\bf E}^{(4)}_{2}(Z_2) & \equiv &
(\,
\frac{1}{2}\,\partial^{2} {\bf Q}^{(2)}_{0}
+\frac{1}{2}\, {\bf J}^{4-0}  {\bf Q}^{(2)}_{0}
+\frac{e^{0,4}_{0}}{4}\,\partial^{2} {\bf J}^{4-0}
+2\,{\bf \Phi}^{(2)} {\bf\Phi}^{(2)}
\,)(Z_2),
\label{BigE}
\\
{\bf E}^{(5)}_{2}(Z_2) & \equiv &
(\,
-\frac{1}{24}\,\partial^{3} {\bf Q}^{(2)}_{0}
+\frac{1}{4}\, {\bf J}^{4-0} \partial {\bf Q}^{(2)}_{0}
+\frac{1}{2}\, \partial {\bf J}^{4-0}  {\bf Q}^{(2)}_{0}
+\frac{e^{0,4}_{0}}{12}\,\partial^{3} {\bf J}^{4-0}
+\frac{3}{2}\,\partial ({\bf \Phi}^{(2)} {\bf\Phi}^{(2)})
\,)(Z_2).
\nonu
\eea
As emphasized before, all the coefficients
appearing in (\ref{Quasi})((\ref{BigR}) and (\ref{BigE}))
depend on $k$ with fixed $N=5$.
It is an open problem to obtain them for generic $(N,k)$.

\section{ The (anti)commutators  from the coset construction of
sections $2,3$, and $4$}

In order to extract the ${\cal N}=4$ higher spin algebra from
the ${\cal W}_{\infty}^{{\cal N}=4}[\la]$ algebra in
two dimensional conformal field theory, we should
express the corresponding OPEs obtained in previous sections
$2,3$ and $4$
in terms of (anti)commutators by using the explicit formula in
\cite{Blumenhagenetal,Blumenhagenbook} \footnote{All the nonlinear
  terms appearing in this Appendix disappear
under the infinity limit of $N$ as before.}.
By starting with the (anti)commutators of $16$ currents and
those between the $16$ currents and $16$ higher spin currents,
we will present  the (anti)commutators corresponding to
Appendices $A$ and $B$.

\subsection{ The (anti)commutators between the $16$ currents}

From the standard OPEs of the large ${\cal N}=4$ superconformal
algebra \cite{AK1509},
we can write down them in terms of (anti)commutators
under the large $(N,k)$ 't Hooft limit
\footnote{The expressions with typewriter font will 
disappear after taking the wedge condition.}
\bea
\Big[L_{m},L_{n}\Big]
& 
=& \mathtt{-m(m^2-1)\,\frac{N(1-\lambda)}{2}\,\delta_{m+n}}
+(m-n)\,L_{m+n}
\,,
\nonu\\
\Big[L_{m},G^i_{r}\Big]
& 
=& (\frac{m}{2}-r)\,G^i_{m+r}
\,,
\nonu\\
\Big[L_{m},T^{ij}_{n}\Big]
& 
=& \mathtt{-n\,T^{ij}_{m+n}}
\,,
\nonu\\
\Big\{G^i_{r},G^j_{\rho}\Big\}
& 
=& -\delta^{ij}\,\Big(
\mathtt{(4r^2-1)\frac{N(1-\lambda)}{2}\,\delta_{r+\rho}}
-2\,L_{r+\rho}
\Big)
-i\,(r-\rho)\Big(
T^{ij}_{r+\rho}-(2\lambda-1)\,\widetilde{T}_{r+\rho}^{ij}
\Big)
\,,
\nonu\\
\Big[G^i_{r},T^{jk}_{m}\Big]
& 
= & \delta^{ij}\Big(
i\,G^k_{r+m}+
\mathtt{m(2\lambda-1)\,\Gamma^k_{r+m}}
\Big)
-\delta^{ik}\Big(
i\,G^j_{r+m}+
\mathtt{m(2\lambda-1)\,\Gamma^j_{r+m}}
\Big)
+\mathtt{\varepsilon^{ijkl}\,m\,\Gamma^l_{r+m}}
\,,
\nonu\\
\Big[T^{ij}_{m},T^{kl}_{n}\Big]
& 
= & \mathtt{
m\,\delta_{m+n}\Big((\delta^{ik}\delta^{jl}-\delta^{il}\delta^{jk})\frac{N}{2\lambda}
+\varepsilon^{ijkl}\,N(1-\frac{1}{2\lambda})\Big)
}
\nonu\\
&
- & i\,\delta^{ik}\,T^{jl}
+i\,\delta^{il}\,T^{jk}
+i\,\delta^{jk}\,T^{il}
-i\,\delta^{jl}\,T^{ik}
\,.
\label{16commanticomm}
\eea
We obtain the ${\cal N}=4$ wedge algebra, (\ref{GGanti})
 and
(\ref{fiveremain}),
by removing the parts
having a typewriter
font as in section $7.1$.

\subsection{ The (anti)commutators between the $16$ currents
and the $s$-th ${\cal N}=4$ higher spin multiplet}

From the standard ${\cal N}=4$ primary condition for the ${\cal N}=4$
higher spin multiplet in component approach \cite{AK1509},
we can write down them in terms of (anti)commutators
under the large $(N,k)$ 't Hooft limit 
as follows:
\bea
\Big[L_{m},\Phi^{(s)}_{s_c,n}\Big]
& 
=& ((s+s_c-1)m-n)\,\Phi^{(s)}_{s_c,\,m+n}\,, \qquad s_c=0,\frac{1}{2},1
\, ,
\nonu\\
\Big[L_{m},\tilde{\Phi}^{(s)}_{s_c,n}\Big]
& 
=& ((s+s_c-1)m-n)\,\tilde{\Phi}^{(s)}_{s_c,\,m+n}\,, \qquad
s_c=\frac{3}{2},2\,,
\nonu\\
\Big[G^i_{r},\Phi^{(s)}_{0,m}\Big]
& 
=& -\Phi^{(s),i}_{\frac{1}{2},r+m}\,,
\nonu\\
\Big\{G^i_{r},\Phi^{(s),j}_{\frac{1}{2},\rho}\Big\}
& 
= &
-\delta^{ij}\,((2s-1)r-\rho)\,\Phi^{(s)}_{0,r+\rho}+\widetilde{\Phi}^{(s),ij}_{1,r+\rho}\,,
\nonu\\
\Big[G^i_{r},\Phi^{(s),jk}_{1,m}\Big]
& 
=& \Bigg(-\delta^{ij}\,\Big(\tilde{\Phi}^{(s),k}_{\frac{3}{2},r+m}-
\frac{1}{(2s+1)}\,(2s\,r-m)(2\lambda-1)\,\Phi^{(s),k}_{\frac{1}{2},r+m}
\Big) -(j \leftrightarrow k) \Bigg)
\nonu\\
& + & \varepsilon^{ijkl}\,(2s\,r-m)\,\Phi^{(s),l}_{\frac{1}{2},r+m}
\,,
\nonu\\
\Big\{G^i_{r},\tilde{\Phi}^{(s),j}_{\frac{3}{2},\rho}\Big\}
& 
= & -\delta^{ij}\,\Big( \tilde{\Phi}^{(s)}_{2,r+\rho}+
\mathtt{
\frac{(s^2+s)}{(2s+1)}\,(4m^2-1)(2\lambda-1)\,\Phi^{(s)}_{0,r+\rho}
}
\Big)
\nonu\\
&
-& ((2s+1)r-\rho)\Big(
\Phi^{(s),ij}_{1,r+\rho}
-\frac{1}{(2s+1)}(2\lambda-1)\,\tilde{\Phi}^{(s),ij}_{1,r+\rho}
\Big)
\,,
\nonu\\
\Big[G^i_{r},\tilde{\Phi}^{(s)}_{2,m}\Big]
& 
=&
-(2(s+1)r-m)\, \tilde{\Phi}^{(s),i}_{\frac{3}{2},r+m}
+
\mathtt{
\frac{(s^2+s)}{(2s+1)}(4r^2-1)(2\lambda-1)\,\Phi^{(s),i}_{\frac{1}{2},r+m}
}
\,,
\nonu
\\
\Big[T^{ij}_m,\Phi^{(s),k}_{\frac{1}{2},r}\Big]
& 
=&
-i\,\delta^{ik}\, \Phi^{(s),j}_{\frac{1}{2},m+r}
+i\, \delta^{jk}\,\Phi^{(s),i}_{\frac{1}{2},m+r}
\,,
\nonu\\
\Big[T^{ij}_m,\Phi^{(s),kl}_{1,n}\Big]
& 
=&
-i\,\delta^{ik}\,\Phi^{(s),jl}_{1,m+n}
+i\,\delta^{il}\,\Phi^{(s),jk}_{1,m+n}
+i\,\delta^{jk}\,\Phi^{(s),il}_{1,m+n}
-i\,\delta^{jl}\,\Phi^{(s),ik}_{1,m+n}
+\mathtt{
\varepsilon^{ijkl}\,2s\,i\,m\,\Phi^{(s)}_{0,m+n}
}
\,,
\nonu\\
\Big[T^{ij}_m,\tilde{\Phi}^{(s),k}_{\frac{3}{2},r}\Big]
& 
=&
\Bigg( -i\,\delta^{ik}\,\Big(
\tilde{\Phi}^{(s),j}_{\frac{3}{2},m+r}
+
\mathtt{\frac{1}{(2s+1)}\,m\,(2\lambda-1)\,
  \tilde{\Phi}^{(s),j}_{\frac{3}{2},m+r}
}
\Big) -(i \leftrightarrow j) \Bigg)
\nonu\\
&
-&
\mathtt{
\varepsilon^{ijkl}\,(2s+1)i\,\Phi^{(s),l}_{\frac{1}{2},m+r}
}\,,
\nonu\\
\Big[T^{ij}_m,\tilde{\Phi}^{(s)}_{2,n}\Big]
& 
=&
\mathtt{
2(s+1)i\,m\,\Phi^{(s),ij}_{1,m+n}}
\,.
\label{N4primary}
\eea
The $\la$ dependence appears as $(2\la-1)$.
We observe that there are terms having
typewriter font in the right hand side and they
will vanish under the wedge restriction.
In particular, we have only
vanishing commutator for the higher spin-$1$
current as follows:
\bea
\Big[L_{m},\Phi^{(1)}_{0,n}\Big]
& 
=& \mathtt{-n \,\Phi^{(1)}_{0,\,m+n}}\,.
\nonu
\eea
For the other higher spin currents, in general,
we have the first and second relations of (\ref{N4primary}).
The crucial point in (\ref{N4primary}) is that we can
determine all the remaining $15$
higher spin currents starting with the commutator between
the spin-$\frac{3}{2}$ current $G^i$ and
the lowest higher spin current $\Phi_0^{(s)}$.


\subsection{ The (anti)commutators between the first
  ${\cal N}=4$ higher spin multiplet}

The OPEs appearing in Appendix $D$
can be written in terms of (anti)commutators by using the works of
\cite{Blumenhagenetal,Blumenhagenbook} as follows:
\bea
\Big[\Phi^{(1)}_{0,m},\Phi^{(1)}_{0,n}\Big]
& 
=& \mathtt{-2\,m\,N(\lambda-1)\,\delta_{m+n}},
\nonu\\
\Big[\Phi^{(1)}_{0,m},\Phi^{(1),i}_{\frac{1}{2},r}\Big]
&
=& G^i_{m+r},
\nonu\\
\Big[\Phi^{(1)}_{0,m},\Phi^{(1),ij}_{1,n}\Big]
&
= &
\mathtt{2i\,m\Big((1-2\lambda)\, T^{ij}_{m+n}+\widetilde{T}^{ij}_{m+n}\Big)},
\nonu\\
\Big[\Phi^{(1)}_{0,m},\tilde{\Phi}^{(1),i}_{\frac{3}{2},r}\Big]
& 
=&
-\frac{1}{2}\,\Phi^{(2),i}_{\frac{1}{2},m+r}
-\mathtt{
\frac{8}{3}m(2\lambda-1)\,G^i_{m+r}
+4 i\, m(3m+2r)\lambda(\lambda-1)\, \Gamma^i_{m+r}},
\nonu\\
\Big[\Phi^{(1)}_{0,m},\tilde{\Phi}^{(1)}_{2,n}\Big]
&
=&
\tt{
2\,m\,\Phi^{(2)}_{0,m+n}
+
8m(2m+n)\lambda(\lambda-1)\,U_{m+n}
},
\nonu\\
\Big\{\Phi^{(1),i}_{\frac{1}{2},r},\Phi^{(1),j}_{\frac{1}{2},s}\Big\}
&
=& \mathtt{
2(r^2-\frac{1}{4})\,N(\lambda-1)
\,\delta^{ij}\,\delta_{r+s}}
-2\delta^{ij}\,L_{r+s}
+i(r-s)\Big(T^{ij}_{r+s}-(2\lambda-1)\widetilde{T}^{ij}_{r+s}\Big),
\nonu\\
\Big[\Phi^{(1),i}_{\frac{1}{2},r},\Phi^{(1),jk}_{1,_m}\Big]
&
=& \Bigg(
\delta^{ij}\bigg(\,\frac{1}{2}\,\Phi^{(2),k}_{\frac{1}{2},r+m}+\frac{1}{3}(2r-m)(2\lambda-1)\,G^k_{r+m}-
\mathtt{
i(4r^2-1)\lambda(\lambda-1)\,\Gamma^k_{r+m}
}
\,\bigg)
\nonu\\
&-& ( j \leftrightarrow k ) \Bigg)
+ \varepsilon^{ijkl}(2r-m)\,G^l_{r+m},
\nonu\\
\Big\{\Phi^{(1),i}_{\frac{1}{2},r},\tilde{\Phi}^{(1),j}_{\frac{3}{2},s}\Big\}
&
=& \delta^{ij}\bigg(\,\frac{1}{2}(-3r+s)\,\Phi^{(2)}_{0,r+s}
-
\mathtt{
(4r^2-1)\lambda(\lambda-1)\,U_{r+s}\bigg)
}
\nonu\\
&
+& \frac{1}{2}\,\widetilde{\Phi}^{(2),ij}_{1,r+s}
+
\mathtt{
\frac{i}{3}(4r^2-1)\Big(2(2\lambda-1)\,T^{ij}_{r+s}+(\lambda+1)(\lambda-2)\,\widetilde{T}^{ij}_{r+s}
}
\Big),
\nonu\\
\Big[\Phi^{(1),i}_{\frac{1}{2},r},\tilde{\Phi}^{(1)}_{2,m}\Big]
&
=& (2r-\frac{m}{2})\,\Phi^{(2),i}_{\frac{1}{2},r+m}
\nonu \\
& + &
\mathtt{
(4r^2-1)\Big(\,
\frac{2}{3}(2\lambda-1)\,G^i_{r+m}-i(4r+3m)\lambda(\lambda-1)\,\Gamma^{i}_{r+m}
}
\,\Big),
\nonu\\
\Big[\Phi^{(1),ij}_{1,m},\Phi^{(1),kl}_{1,n}\Big]
&
=& \mathtt{
2m(m^2-1)\,N(\lambda-1)\,\delta_{m+n}
\,\Big(
(\delta^{ik}\delta^{jl}-\delta^{il}\delta^{jk})
-\varepsilon^{ijkl}\,
\frac{1}{3}\,(2\lambda-1)\,
\Big)
}
\nonu\\
&-& ( \delta^{ik}\delta^{jl}-\delta^{il}\delta^{jk})4(m-n)L_{m+n}
\nonu \\
& + &\varepsilon^{ijkl}(m-n)\Big(\,
\Phi^{(2)}_{0,m+n}
-\frac{4}{3}(2\lambda-1)L_{m+n}
\,\Big)
\nonu\\
&
+& \frac{1}{2}\,\Big(
-\delta^{ik}\,\Phi^{(2),jl}_{1,m+n}
+\delta^{il}\,\Phi^{(2),jk}_{1,m+n}
+\delta^{jk}\,\Phi^{(2),il}_{1,m+n}
-\delta^{jl}\,\Phi^{(2),ik}_{1,m+n}
\,\Big)
\nonu\\
&+& \Bigg(-\delta^{ik}   
\frac{2i}{3}(m^2-mn+n^2-1)\Big( (2\lambda-1)\widetilde{T}^{jl}_{m+n}+(2\lambda^2-2\lambda-1)T^{jl}_{m+n}  \Big)
\nonu\\
&-& ( k \leftrightarrow l ) -(i \leftrightarrow j) +
( i \leftrightarrow j, k \leftrightarrow l) \Bigg),
\nonu \\
\Big[\Phi^{(1),ij}_{1,m},\tilde{\Phi}^{(1),k}_{\frac{3}{2},r}\Big]
&
= & \Bigg( -\delta^{ik}\bigg(\,
\frac{1}{2}\,\tilde{\Phi}^{(2),j}_{\frac{3}{2},m+r}
+\frac{1}{30}(3m-2r)(2\lambda-1)\,\Phi^{(2),j}_{\frac{1}{2},m+r}
\nonu \\
& + & \frac{1}{9}(12m^2-8mr+4r^2-9)
(\lambda+1)(\lambda-2)G^{j}_{m+r}
\nonu \\
&- &
\mathtt{
\frac{4i}{3}m(m^2-1)\lambda(\lambda-1)(2\lambda-1)\Gamma^j_{m+r}
}
\,\bigg) -( i \leftrightarrow j ) \Bigg)
\nonu\\
&
-& \varepsilon^{ijkl}\bigg(
(\frac{3}{2}m-r)\,\Phi^{(2),l}_{\frac{1}{2},m+r}
-
\tt{
4i\,m(m^2-1)\lambda(\lambda-1)\Gamma^l_{m+r}
}
\bigg),
\nonu\\
\Big[\tilde{\Phi}^{(1),ij}_{1,m},\tilde{\Phi}^{(1)}_{2,n}\Big]
&
=& (2m-n)\,\Phi^{(2),ij}_{1,m+n}
+\mathtt{
\frac{8i}{3}m(m^2-1)\Big(
(2\lambda^2-2\lambda-1)T^{ij}_{m+n}
+(2\lambda-1)\widetilde{T}^{ij}_{m+n}
\Big)
},
\nonu\\
\Big\{\tilde{\Phi}^{(1),i}_{\frac{3}{2},r},
\tilde{\Phi}^{(1),j}_{\frac{3}{2},s}\Big\}
&
= & \mathtt{
\frac{8}{9}
\,(r^2-\frac{1}{4})(r^2-\frac{9}{4})\,N\,(\lambda+1)(\lambda-1)(\lambda-2)\,
\delta^{ij}\,\delta_{r+s}
}
\nonu\\
&  
+ & \delta^{ij}
\bigg(
\frac{1}{2}\,\tilde{\Phi}^{(2)}_{2,r+s}
+(6r^2-8rs+6s^2-9)\Big(
\frac{1}{15}(2\lambda-1)\,\Phi^{(2)}_{0,r+s}
\nonu \\
& - & \frac{4}{9}(\lambda+1)(\lambda-2)L_{r+s}
\Big)
\bigg)
\nonu\\
  &   
+ & (r-s)\Big(\,
\,\frac{3}{2}\,\Phi^{(2),ij}_{1,r+s}
-\frac{1}{6}(2\lambda-1)\widetilde{\Phi}^{(2),ij}_{1,r+s}
\,\Big)
\nonu\\
&
+&(r-s)(2r^2+2s^2-5)(\lambda+1)(\lambda-2)\frac{2i}{9}
\Big(\,
T^{ij}_{r+s}-(2\lambda-1)\widetilde{T}^{ij}_{r+s}
\,\Big),
\nonu\\
\Big[\tilde{\Phi}^{(1),i}_{\frac{3}{2},r},\tilde{\Phi}^{(1)}_{2,m}\Big]
&=&
\frac{(4r-3m)}{2}\,\tilde{\Phi}^{(2),i}_{\frac{3}{2},r+m}
-\frac{1}{15}(4r^2-4rm+2m^2-5)(2\lambda-1)\,\Phi^{(2),i}_{\frac{1}{2},r+m}
\nonu\\
&
+& \frac{1}{9}(16r^3-12r^2m-36r+19m+8rm^2-4m^3)
(\lambda+1)(\lambda-2)\,G^{i}_{r+m},
\nonu\\
\Big[\tilde{\Phi}^{(1)}_{2,m},\tilde{\Phi}^{(1)}_{2,n}\Big]
&
=&
\mathtt{
\frac{8}{9}
\,
m(m^2-4)(m^2-1)
\,
N\,(\lambda+1)(\lambda-1)(\lambda-2)
\,\delta_{m+n}}
+2(m-n)\,\tilde{\Phi}^{(2)}_{2,m+n}
\nonu\\
&
+& (m-n)(2m^2-mn+2n^2-8)\nonu \\
&\times & \Big(
\frac{2}{15}(2\lambda-1)\,\Phi^{(2)}_{0,m+n}
-\frac{16}{9}(\lambda+1)(\lambda-2)L_{m+n}
\Big).
\label{PhionePhione}
\eea
We observe the antisymmetric property of the $SO(4)$ indices.
The central terms vanish under the wedge condition.
Although the $\la$ (or $(1-\la)$)
factor appears in the structure constants,
those terms have typewriter font and they vanish by taking
the wedge condition.
The quadratic $\la$ terms appear under the wedge
condition in the right hand side.

\subsection{ The (anti)commutators between  the
  first and the second ${\cal N}=4$ higher spin multiplets}

The OPEs appearing in Appendix $E$
can be written in terms of (anti)commutators as follows:
\bea
\Big[\Phi^{(1)}_{0,m},\Phi^{(2)}_{0,n}\Big]
&
=&
0,
\nonu\\
\Big[\Phi^{(1)}_{0,m},\Phi^{(2),i}_{\frac{1}{2},r}\Big]
&
=& -2 \,\tilde{\Phi}^{(1),i}_{\frac{3}{2},m+r},
\nonu\\
\Big[\Phi^{(1)}_{0,m},\Phi^{(2),ij}_{1,n}\Big]
&
=& \mathtt{
8m\Big(\,
\frac{1}{3}(2\lambda-1)\,\Phi^{(1),ij}_{1,m+n}
-\widetilde{\Phi}^{(1),ij}_{1,m+n}
\,\Big)
},
\nonu\\
\Big[\Phi^{(1)}_{0,m},\tilde{\Phi}^{(2),i}_{\frac{3}{2},r}\Big]
&
=& -\frac{1}{6}\,\Phi^{(3),i}_{\frac{1}{2},m+r}
+
\mathtt{
\frac{64}{15}m(2\lambda-1)\,\tilde{\Phi}^{(1),i}_{\frac{3}{2},m+r}
+\frac{16}{9}m(5m+2n)(\lambda+1)(\lambda-2)\,\Phi^{(1),i}_{\frac{1}{2},m+r}
},
\nonu\\
\Big[\Phi^{(1)}_{0,m},\tilde{\Phi}^{(2)}_{2,n}\Big]
&
=&
\mathtt{
m\,\Phi^{(3)}_{0,m+n}
+\frac{64}{15}m(2\lambda-1)\,\tilde{\Phi}^{(1)}_{2,m+n}}
\nonu\\
& - &
\mathtt{
\frac{32}{9}m(5m^2+5mn+n^2+1)(\lambda+1)(\lambda-2)\,\Phi^{(1)}_{0,m+n}
},
\nonu\\
\Big[\Phi^{(1),i}_{\frac{1}{2},r},\Phi^{(2)}_{0,m}\Big]
&
=& 2 \,\tilde{\Phi}^{(1),i}_{\frac{3}{2},r+m},
\nonu\\
\Big\{\Phi^{(1),i}_{\frac{1}{2},r},\Phi^{(2),j}_{\frac{1}{2},s}\Big\}
&
=& -2\,\delta^{ij}\,\tilde{\Phi}^{(1)}_{2,r+s}
-2(3r-s)\Big(\,
\Phi^{(1),ij}_{1,r+s}-\frac{1}{3}\,(2\lambda-1)\,\widetilde{\Phi}^{(1),ij}_{1,r+s}
\,\Big),
\nonu\\
\Big[\Phi^{(1),i}_{\frac{1}{2},r},\Phi^{(2),jk}_{1,m}\Big]
&
=& \Bigg( \delta^{ij}\,\bigg(
\frac{1}{6}\,\,\Phi^{(3),k}_{\frac{1}{2},r+m}
-\frac{2}{5}(4r-m)(2\lambda-1)\,\tilde{\Phi}^{(1),k}_{\frac{3}{2},r+m}
\nonu \\
& - &
\mathtt{
\frac{4}{3}(4r^2-1)(\lambda+1)(\lambda-2)\,\Phi^{(1),k}_{\frac{1}{2},r+m}
}
\bigg) -( j \leftrightarrow k ) \Bigg)
\nonu\\
& - &  2\,\varepsilon^{ijkl}(4r-m)\,\tilde{\Phi}^{(1),l}_{\frac{3}{2},r+m},
\nonu\\
\Big\{\Phi^{(1),i}_{\frac{1}{2},r},\tilde{\Phi}^{(2),j}_{\frac{3}{2},s}\Big\}
&
=& -\delta^{ij}
\Big(\,
(\frac{1}{2}+r)\,\Phi^{(3)}_{0,r+s}-
\mathtt{
\frac{4}{9}(4r^2-1)(5r+3s)(\lambda+1)(\lambda-2)\,\Phi^{(1)}_{0,r+s}
}
\,\Big)
\nonu\\
&
+& \frac{1}{6}\,\widetilde{\Phi}^{(3),ij}_{1,r+s}
-
\mathtt{
(4r^2-1)\Big(\,
\frac{16}{15}(2\lambda-1)\,\Phi^{(1),ij}_{1,r+s}
+\frac{4}{5}(\lambda+2)(\lambda-3)\,\widetilde{\Phi}^{(1),ij}_{1,r+s}
\,\Big)
},
\nonu\\
\Big[\Phi^{(1),i}_{\frac{1}{2},r},\tilde{\Phi}^{(2)}_{2,m}\Big]
&
= & (r-\frac{m}{6})\,\Phi^{(3),i}_{\frac{1}{2},r+m}
-
\mathtt{
\frac{16}{15}(4r^2-1)(2\lambda-1)\,\tilde{\Phi}^{(1),i}_{\frac{3}{2},r+m}}
\nonu\\
&
- &
\mathtt{
\frac{20}{9}(4r^2-1)(2r+m)(\lambda+1)(\lambda-2)\,\Phi^{(1),i}_{\frac{1}{2},r+m}}
,
\nonu \\
\Big[\Phi^{(1),ij}_{1,m},\Phi^{(2)}_{0,n}\Big]
&
=& 4(m-n)\Big(\,
\frac{1}{3}(2\lambda-1)\,\Phi^{(1),ij}_{1,m+n}
-\widetilde{\Phi}^{(1),ij}_{1,m+n},
\,\Big),
\nonu\\
\Big[\Phi^{(1),ij}_{1,m},\Phi^{(2),k}_{\frac{1}{2},r}\Big]
&
=&
\Bigg(-\delta^{ik}\bigg(\,
\frac{1}{6}\,\Phi^{(3),j}_{\frac{1}{2},m+r}
+\frac{2}{15}(3m-2r)(2\lambda-1)\,\tilde{\Phi}^{(1),j}_{\frac{3}{2},m+r}
\nonu \\
& - & \frac{2}{9}(12m^2-8mr+4r^2-9)
  (\lambda+1)(\lambda-2)\,\Phi^{(1),j}_{\frac{1}{2},m+r}
\,\bigg) -( i \leftrightarrow j ) \Bigg)
\nonu \\
& - & \varepsilon^{ijkl}\,2(3m-2r)\,\tilde{\Phi}^{(1),l}_{\frac{3}{2},m+r},
\nonu\\
\Big[\Phi^{(1),ij}_{1,m},\Phi^{(2),kl}_{1,n}\Big]
&
=&
-4(2m-n)(\delta^{ik}\delta^{jl}-\delta^{il}\delta^{jk})\,
\tilde{\Phi}^{(1)}_{2,m+n}
\nonu\\
&
-& \frac{1}{6}\Big(\,
\delta^{ik}\,\Phi^{(3),jl}_{1,m+n}
-\delta^{il}\,\Phi^{(3),jk}_{1,m+n}
-\delta^{jk}\,\Phi^{(3),il}_{1,m+n}
+\delta^{jl}\,\Phi^{(3),ik}_{1,m+n}
\,\Big)
\nonu\\
&
+& \Bigg( \delta^{ik}
\frac{4}{15}(6m^2-3mn+n^2-4)\Big(
(2\lambda^2-2\lambda-7)\,\Phi^{(1),jl}_{1,m+n}
+(2\lambda-1)\,\widetilde{\Phi}^{(1),jl}_{1,m+n}
\,\Big)
\nonu\\
&-& ( k \leftrightarrow l) -( i \leftrightarrow j) +
 (i \leftrightarrow j,  k \leftrightarrow l) \Bigg)
\nonu \\
&
+ & \varepsilon^{ijkl}\,\bigg(
(2m-n)
\Big(\,
\frac{1}{3}\,\Phi^{(3)}_{0,m+n}
-\frac{4}{5}(2\lambda-1)\,\tilde{\Phi}^{(1)}_{2,m+n}
\,\Big)
\nonu\\
&
- & \mathtt{
\frac{32}{9}m(m^2-1)
(\lambda+1)(\lambda-2)\,\Phi^{(1)}_{0,m+n}
}
\,\bigg),
\nonu\\
\Big[\Phi^{(1),ij}_{1,m},\tilde{\Phi}^{(2),k}_{\frac{3}{2},r}\Big]
&=& \Bigg(
-\delta^{ik}\bigg(\,
\frac{1}{6}\,\tilde{\Phi}^{(3),j}_{\frac{3}{2},m+r}
+\frac{1}{210}(5m-2r)(2\lambda-1)\,\Phi^{(3),j}_{\frac{1}{2},m+r}
\nonu \\
& - & \frac{2}{25}(40m^2-16mr+4r^2
-  25)
(\lambda+2)(\lambda-3)\,\tilde{\Phi}^{(1),j}_{\frac{3}{2},m+r}
\nonu \\
& - &
\mathtt{
\frac{16}{15}m(m^2-1)(\lambda+1)(\lambda-2)(2\lambda-1)\,\Phi^{(1),j}_{\frac{1}{2},m+r}
}
\,\bigg) -( i \leftrightarrow  j) \Bigg)
\nonu\\
&
-& \varepsilon^{ijkl}\bigg(\,
\frac{1}{6}(5m-2r)\,\Phi^{(3),l}_{\frac{1}{2},m+r}
-\frac{80}{9}m(m^2-1)(\lambda+1)(\lambda-2)\,\Phi^{(1),l}_{\frac{1}{2},m+r}
\,\bigg),
\nonu\\
\Big[\Phi^{(1),ij}_{1,m},\tilde{\Phi}^{(2)}_{2,n}\Big]
&
= & \frac{1}{3}(3m-n)\,\Phi^{(3),ij}_{1,m+n}
\nonu\\
&
-& \mathtt{
m(m^2-1)
\Big(\,
\frac{32}{45}(17\lambda^2-17\lambda-52)\,\Phi^{(1),ij}_{1,m+n}
+\frac{64}{15}(2\lambda-1)\,\widetilde{\Phi}^{(1),ij}_{1,m+n}
}
\,\Big),
\nonu\\
\Big[\tilde{\Phi}^{(1),i}_{\frac{3}{2},r},\Phi^{(2)}_{0,m}\Big]
&= &\frac{1}{6}\,\Phi^{(3),i}_{\frac{1}{2},r+m}
+\frac{8}{15}(2r-3m)(2\lambda-1)\,\tilde{\Phi}^{(1),i}_{\frac{3}{2},r+m}
\nonu\\
&
-&\frac{2}{9}(4r^2-8rm+12m^2-9)(\lambda+1)(\lambda-2)\,\Phi^{(1),i}_{\frac{1}{2},r+m},
\nonu\\
\Big\{\tilde{\Phi}^{(1),i}_{\frac{3}{2},r},\Phi^{(2),j}_{\frac{1}{2},s}\Big\}
&
=& \delta^{ij}\,(m-n)\bigg(\,
\frac{1}{2}\,\Phi^{(3)}_{0,r+s}
-\frac{8}{15}(2\lambda-1)\,\tilde{\Phi}^{(1)}_{2,r+s}
\nonu\\
&
-& \frac{4}{9}(2r^2+2s^2-5)(\lambda+1)(\lambda-2)\,\Phi^{(1)}_{0,r+s}
\,\bigg)
-\frac{1}{6}\,\widetilde{\Phi}^{(3),ij}_{1,r+s}
\nonu\\
&
+& 4(6r^2-8rs+6s^2-9)\Big(\,
\frac{1}{15}(2\lambda-1)\,\Phi^{(1),ij}_{1,r+s}
+\frac{1}{45}(\lambda^2-\lambda-11)\,\widetilde{\Phi}^{(1),ij}_{1,r+s}
\,\Big),
\nonu\\
\Big[\tilde{\Phi}^{(1),i}_{\frac{3}{2},r},\Phi^{(2),jk}_{1,m}\Big]
&
=& \Bigg( \delta^{ij}\bigg(\,
\frac{1}{6}\,\tilde{\Phi}^{(3),k}_{\frac{3}{2},r+m}
-\frac{1}{126}(4r-3m)(2\lambda-1)\,\Phi^{(3),k}_{\frac{1}{2},r+m}
\nonu\\
&
- &
\frac{4}{15}(4r^2-4rm+2m^2-5)(\lambda^2-\lambda-11)\,
\tilde{\Phi}^{(1),k}_{\frac{3}{2},r+m}
\nonu\\
&
+ &
\frac{2}{45}(16r^3-12r^2m+8rm^2-36r-4m^3+19m)
\nonu \\
& \times &
(\lambda+1)(\lambda-2)(2\lambda-1)\,\Phi^{(1),k}_{\frac{1}{2},r+m}
\,\bigg) -(j \leftrightarrow k) \Bigg)
\nonu\\
&
+ & \varepsilon^{ijkl}\bigg(
-\frac{1}{6}(4r-3m)\,\Phi^{(3),l}_{\frac{1}{2},r+m}
+\frac{4}{15}(4r^2-4rm+2m^2-5)(2\lambda-1)\,
\tilde{\Phi}^{(1),l}_{\frac{3}{2},r+m}
\nonu\\
&
+ & \frac{2}{9}(16r^3-12r^2m+8rm^2-36r-4m^3+19m)(\lambda+1)(\lambda-2)\,\Phi^{(1),l}_{\frac{1}{2},r+m}
\,\bigg),
\nonu\\
\Big\{\tilde{\Phi}^{(1),i}_{\frac{3}{2},r},\tilde{\Phi}^{(2),j}_{\frac{3}{2},s}\Big\}
&
=&
\delta^{ij}\bigg(\,
\frac{1}{6}\,\tilde{\Phi}^{(3)}_{2,r+s}
+2(40r^2-32rs+12s^2-45)\Big(
\frac{1}{315}(2\lambda-1)\,\Phi^{(3)}_{0,r+s}
\nonu\\
&
-& \frac{1}{25}(\lambda+2)(\lambda-3)\,\tilde{\Phi}^{(1)}_{2,r+s}
\Big)
\nonu \\
& - &
\mathtt{
\frac{256}{135}(r^2-\frac{9}{4})(r^2-\frac{1}{4})(\lambda+1)(\lambda-2)
(2\lambda-1)\,\Phi^{(1)}_{0,r+s}}
\bigg)
\nonu\\
&
+ & \frac{1}{6}(5r-3s)\Big(\,
\Phi^{(3),ij}_{1,r+s}
-\frac{1}{15}(2\lambda-1)\,\widetilde{\Phi}^{(3),ij}_{1,r+s}
\,\Big)
\nonu \\
& - & \frac{2}{25}(40r^3-24r^2s+12rs^2-85r
+  31s-4s^3) \nonu \\
& \times & 
\Big(\,
(\lambda+2)(\lambda-3)\,\Phi^{(1),ij}_{1,r+s}
-\frac{1}{3}(\lambda+2)(\lambda-3)(2\lambda-1)\,\widetilde{\Phi}^{(1),ij}_{1,r+s}
\,\Big),
\nonu\\
\Big[\tilde{\Phi}^{(1),i}_{\frac{3}{2},r},\tilde{\Phi}^{(2)}_{2,m}\Big]
&
=& \frac{1}{2}(2r-m)\,\Phi^{(3),i}_{\frac{1}{2},r+m}
-\frac{1}{315}(60r^2-40rm+12m^2-63)(2\lambda-1)\,\Phi^{(3),i}_{\frac{1}{2},r+m}
\nonu\\
&
-&
\frac{4}{25}(40r^3-20r^2m+8rm^2-82r+23m-2m^3)(\lambda-3)(\lambda+2)\,
\tilde{\Phi}^{(1),i}_{\frac{3}{2},r+m}
\nonu\\
&
+ &
\mathtt{
\frac{256}{135}(r^2-\frac{9}{4})(r^2-\frac{1}{4})(\lambda+1)(\lambda-2)(2\lambda-1)\,\Phi^{(1),i}_{\frac{1}{2},r+m}
},
\nonu\\
\Big[\tilde{\Phi}^{(1)}_{2,m},\Phi^{(2)}_{0,n}\Big]
&
= & \frac{1}{3}(m-2n)\,\Phi^{(3)}_{0,m+n}
+\frac{8}{15}(m-2n)(2\lambda-1)\,\tilde{\Phi}^{(1)}_{2,m+n}
\nonu\\
&
+&
\mathtt{
\frac{32}{9}n(n^2-1)(\lambda+1)(\lambda-2)\,\Phi^{(1)}_{0,m+n}
},
\nonu\\
\Big[\tilde{\Phi}^{(1)}_{2,m},\Phi^{(2),i}_{\frac{1}{2},r}\Big]
&
= & \frac{1}{6}(3m-4r)\,\Phi^{(3),i}_{\frac{1}{2},m+r}
+\frac{4}{15}(2m^2-4mr+4r^2-5)(2\lambda-1)\,
\tilde{\Phi}^{(1),i}_{\frac{3}{2},m+r}
\nonu\\
&
- &
\frac{2}{9}(4m^3-8m^2r+12mr^2-19m+36r-16r^3)(\lambda+1)(\lambda-2)\,\Phi^{(1),i}_{\frac{1}{2},m+r},
\nonu\\
\Big[\tilde{\Phi}^{(1)}_{2,m},\Phi^{(2),ij}_{1,n}\Big]
&
= & \frac{2}{3}(m-n)\,\Phi^{(3),ij}_{1,m+n}
\nonu\\
&
- & \frac{8}{15}(m-n)(2m^2-mn+2n^2-8)
\nonu \\
& \times & \Big(\,
(2\lambda^2-2\lambda-7)\,\Phi^{(1),ij}_{1,m+n}
+(2\lambda-1)\,\widetilde{\Phi}^{(1),ij}_{1,m+n}
\,\Big),
\nonu\\
\Big[\tilde{\Phi}^{(1)}_{2,m}, \tilde{\Phi}^{(2),i}_{\frac{3}{2},r}\Big]
&=&
\frac{1}{6}(5m-4r)\,\tilde{\Phi}^{(3),i}_{\frac{3}{2},m+r}
+\frac{1}{315}(20m^2-24mr+12r^2-35)(2\lambda-1)\,\Phi^{(3),i}_{\frac{1}{2},m+r}
\nonu\\
&
-&
\frac{2}{25}(40m^3-48m^2r+36mr^2-145m+100r-16r^3)
\nonu \\
& \times & (\lambda+2)(\lambda-3)\,\tilde{\Phi}^{(1),i}_{\frac{3}{2},m+r},
\nonu\\
\Big[\tilde{\Phi}^{(1)}_{2,m},\tilde{\Phi}^{(2)}_{2,n}\Big]
&
=& (m-\frac{2n}{3})\,\tilde{\Phi}^{(3)}_{2,m+n}
+(5m^3-5m^2n+3mn^2-17m+9n-n^3)
\nonu\\
&
\times & \Big(\,
\frac{4}{105}\,(2\lambda-1)
\Phi^{(3)}_{0,m+n}
-\frac{32}{25}\,(\lambda+2)(\lambda-3)\,\tilde{\Phi}^{(1)}_{2,m+n}
\,\Big)
\nonu\\
&
- &
\mathtt{
\frac{256}{135}m(m^2-1)(m^2-4)(\lambda+1)(\lambda-2)(2\lambda-1)\,\Phi^{(1)}_{0,m+n}
}.
\label{PhionePhitwo}
\eea
There is no second ${\cal N}=4$ higher spin multiplet in the right hand
side of (\ref{PhionePhitwo}).
Then, from the (anti)commutators
having the components of first ${\cal N}=4$ higher spin multiplet
in the right hand side, we can obtain the information of
the components of
second ${\cal N}=4$ higher spin multiplet
in the left hand side. For example, see the second relation of
(\ref{PhionePhitwo}).
The cubic $\la$ terms appear under the wedge
condition in the right hand side.

%

\section{ The second ${\cal N}=4$ higher spin
  multiplet in terms free fields
at $\la =0$}

In section $6.2$, we observed the first ${\cal N}=4$ higher spin
multiplet in terms of the free fields. For the second
${\cal N}=4$ higher spin multiplet, we present them as follows: 
\bea
\Phi^{(2)}_{0}
&=&
\frac{8}{3}\,(W^{11}_{\mathrm{B},2}+W^{22}_{\mathrm{B},2}-2\,W^{11}_{\mathrm{F},2}-2\,W^{22}_{\mathrm{F},2})\,,
\nonu \\
\Phi^{(2),1}_{\frac{1}{2}}
&
=&
-\frac{1}{q}\,(
Q^{11}_{\frac{5}{2}}
+i \sqrt{2}\,Q^{12}_{\frac{5}{2}}
+2i \sqrt{2}\,Q^{21}_{\frac{5}{2}}
-2Q^{22}_{\frac{5}{2}}
+2\,\bar{Q}^{11}_{\frac{5}{2}}
+2i\sqrt{2}\,\bar{Q}^{12}_{\frac{5}{2}}
+i \sqrt{2}\,\bar{Q}^{21}_{\frac{5}{2}}
-\bar{Q}^{22}_{\frac{5}{2}})\,,
\nonu\\
\Phi^{(2),2}_{\frac{1}{2}}
&
=&
\frac{1}{q}\,(
i\,Q^{11}_{\frac{3}{2}}
-2\sqrt{2}\,Q^{21}_{\frac{3}{2}}
-2i \,Q^{22}_{\frac{3}{2}}
+2i\,\bar{Q}^{11}_{\frac{3}{2}}
-2\sqrt{2}\,\bar{Q}^{12}_{\frac{3}{2}}
-i\,\bar{Q}^{22}_{\frac{3}{2}}
)
\,,
\nonu\\
\Phi^{(2),3}_{\frac{1}{2}}
&
=&
\frac{1}{q}\,(
i\,Q^{11}_{\frac{3}{2}}
-\sqrt{2}\,Q^{12}_{\frac{3}{2}}
-2i \,Q^{22}_{\frac{3}{2}}
+2i\,\bar{Q}^{11}_{\frac{3}{2}}
-\sqrt{2}\,\bar{Q}^{21}_{\frac{3}{2}}
-i\,\bar{Q}^{22}_{\frac{3}{2}}
)
\,,
\nonu\\
\Phi^{(2),4}_{\frac{1}{2}}
&
=&
\frac{1}{q}\,(
Q^{11}_{\frac{3}{2}}
+2\,Q^{22}_{\frac{3}{2}}
-2\,\bar{Q}^{11}_{\frac{3}{2}}
-\bar{Q}^{22}_{\frac{3}{2}}
)
\,,
\nonu \\
\Phi^{(2),12}_{1}
&
=&
\frac{2}{q}\,
(
-2i\,W^{11}_{\mathrm{B},3}
+\sqrt{2}\,W^{12}_{\mathrm{B},3}
+2i\,W^{22}_{\mathrm{B},3}
-2i\,W^{11}_{\mathrm{F},3}
+2\sqrt{2}\,W^{12}_{\mathrm{F},3}
+2i\,W^{22}_{\mathrm{F},3}
)\,,
\nonu\\
\Phi^{(2),13}_{1}
&
=&
\frac{2}{q}\,
(
2i\,W^{11}_{\mathrm{B},3}
-4\sqrt{2}\,W^{21}_{\mathrm{B},3}
-2i\,\,W^{22}_{\mathrm{B},3}
+2i\,W^{11}_{\mathrm{F},3}
-2\sqrt{2}\,W^{21}_{\mathrm{F},3}
-2i\,W^{22}_{\mathrm{F},3}
)\,,
\nonu\\
\Phi^{(2),14}_{1}
&
=&
\frac{2}{q}\,
(
-2\,W^{11}_{\mathrm{B},3}
-i\sqrt{2}\,W^{12}_{\mathrm{B},3}
-4i\sqrt{2}\,\,W^{21}_{\mathrm{B},3}
+2\,W^{22}_{\mathrm{B},3}
+2\,W^{11}_{\mathrm{F},3}
+2i\sqrt{2}\,W^{12}_{\mathrm{F},3}
+2i\sqrt{2}\,W^{21}_{\mathrm{F},3}
\nonu \\
& - & 2\,W^{22}_{\mathrm{F},3}
)\,,
\nonu\\
\Phi^{(2),23}_{1}
&
=&
\frac{2}{q}\,
(
2\,W^{11}_{\mathrm{B},3}
+i\sqrt{2}\,W^{12}_{\mathrm{B},3}
+4i\sqrt{2}\,\,W^{21}_{\mathrm{B},3}
-2\,W^{22}_{\mathrm{B},3}
+2\,W^{11}_{\mathrm{F},3}
+2i\sqrt{2}\,W^{12}_{\mathrm{F},3}
+2i\sqrt{2}\,W^{21}_{\mathrm{F},3}
\nonu \\
& - & 2\,W^{22}_{\mathrm{F},3}
)\,,
\nonu\\
\Phi^{(2),24}_{1}
&
=&
\frac{2}{q}\,
(
2i\,W^{11}_{\mathrm{B},3}
-4\sqrt{2}\,W^{21}_{\mathrm{B},3}
-2i\,\,W^{22}_{\mathrm{B},3}
-2i\,W^{11}_{\mathrm{F},3}
+2\sqrt{2}\,W^{21}_{\mathrm{F},3}
+2i\,W^{22}_{\mathrm{F},3}
)\,,
\nonu\\
\Phi^{(2),34}_{1}
&
=&
\frac{2}{q}\,
(
2i\,W^{11}_{\mathrm{B},3}
-\sqrt{2}\,W^{12}_{\mathrm{B},3}
-2i\,\,W^{22}_{\mathrm{B},3}
-2i\,W^{11}_{\mathrm{F},3}
+2\sqrt{2}\,W^{12}_{\mathrm{F},3}
+2i\,W^{22}_{\mathrm{F},3}
)\,,
\nonu \\
\tilde{\Phi}^{(2),1}_{\frac{3}{2}}
&
=&
\frac{1}{q^2}\,(
Q^{11}_{\frac{7}{2}}
+i\sqrt{2}\,Q^{12}_{\frac{7}{2}}
+2i\sqrt{2}\,Q^{21}_{\frac{7}{2}}
-2\,Q^{22}_{\frac{7}{2}}
-2\,\bar{Q}^{11}_{\frac{7}{2}}
-2i\sqrt{2}\,\bar{Q}^{12}_{\frac{7}{2}}
-i\sqrt{2}\,\bar{Q}^{21}_{\frac{7}{2}}
+\bar{Q}^{22}_{\frac{7}{2}}
)\,,
\nonu\\
\tilde{\Phi}^{(2),2}_{\frac{3}{2}}
&
=&
-\frac{i}{q^2}\,(\,
Q^{11}_{\frac{7}{2}}
+2i\sqrt{2}\,Q^{21}_{\frac{7}{2}}
-2\,Q^{22}_{\frac{7}{2}}
-2\,\bar{Q}^{11}_{\frac{7}{2}}
-2i\sqrt{2}\,\bar{Q}^{12}_{\frac{7}{2}}
+\bar{Q}^{22}_{\frac{7}{2}}
)\,,
\nonu\\
\tilde{\Phi}^{(2),3}_{\frac{3}{2}}
&
=&
-\frac{i}{q}\,(
Q^{11}_{\frac{7}{2}}
+i\sqrt{2}\,Q^{12}_{\frac{7}{2}}
-2\,Q^{22}_{\frac{7}{2}}
-2\,\bar{Q}^{11}_{\frac{7}{2}}
-i\sqrt{2}\,\bar{Q}^{21}_{\frac{7}{2}}
+\bar{Q}^{22}_{\frac{7}{2}}
)\,,
\nonu\\
\tilde{\Phi}^{(2),4}_{\frac{3}{2}}
&
=&
-\frac{1}{q^2}\,
(
Q^{11}_{\frac{7}{2}}
+2\,Q^{22}_{\frac{7}{2}}
+2\,\bar{Q}^{11}_{\frac{7}{2}}
+\bar{Q}^{22}_{\frac{7}{2}}
)\,,
\nonu \\
\tilde{\Phi}^{(2)}_{2}
&
=&
\frac{4}{q^2}\,(
W^{11}_{\mathrm{B},4}
+W^{22}_{\mathrm{B},4}
+W^{11}_{\mathrm{F},4}
+W^{22}_{\mathrm{F},4}
).
\label{higherstructure}
\eea
The structure of these higher spin currents
(\ref{higherstructure}) looks like
the one for the first ${\cal N}=4$ higher spin multiplet in
section $6.2$. There are only sign changes
in front of the fields  for the $SO(4)$ singlets.
For the $SO(4)$ nonsinglet case, the only overall factors differ.  
Once one of the components of
the second ${\cal N}=4$ higher spin multiplet is found
by using some of the relations in (\ref{PhionePhitwo}), then
the remaining $15$ higher spin currents can be fixed
with the help of spin-$\frac{3}{2}$ currents as described
before.

\section{ The first ${\cal N}=4$ higher spin
  multiplet in terms free fields
at $\la =1$}

For the $\la =1$, we can analyze the description in section
$6.1$,
$6.2$,
and $6.3$
similarly.
For some of the $16$ currents, we have the following expressions
\bea
L & = &
W^{11}_{\mathrm{B},2}+W^{22}_{\mathrm{B},2}+W^{11}_{\mathrm{F},2}+
W^{22}_{\mathrm{F},2},
\nonu \\
\frac{1}{4q}\,(T^{14}-T^{23})
&
=&
i\,W^{11}_{\mathrm{F},1}
-\sqrt{2}\,W^{12}_{\mathrm{F},1}
-\sqrt{2}\,W^{21}_{\mathrm{F},1}
-i\,W^{22}_{\mathrm{F},1}
\,,
\nonu\\
\frac{1}{4q}\,(T^{13}-T^{42})
&
=&
-W^{11}_{\mathrm{F},1}
-i\,\sqrt{2}\,W^{21}_{\mathrm{F},1}
+W^{22}_{\mathrm{F},1}
\,,
\nonu\\
\frac{1}{4q}\,(T^{12}-T^{34})
&
=&
-W^{11}_{\mathrm{F},1}
-i\,\sqrt{2}\,W^{12}_{\mathrm{F},1}
+W^{22}_{\mathrm{F},1}
\,,
\nonu \\
G^1
&
=&
-\frac{1}{2}\,(
Q^{11}_{\frac{3}{2}}
+i\sqrt{2}\,Q^{12}_{\frac{3}{2}}
+2i \sqrt{2}\,Q^{21}_{\frac{3}{2}}
-2\,Q^{22}_{\frac{3}{2}}
-2\,\bar{Q}^{11}_{\frac{3}{2}}
-2i \sqrt{2}\,Q^{12}_{\frac{3}{2}}
\nonu \\
& - & i\sqrt{2}\,\bar{Q}^{21}_{\frac{3}{2}}
+\bar{Q}^{22}_{\frac{3}{2}}
)\,,
\nonu\\
G^2
&
=&
-\frac{i}{2}\,(
Q^{11}_{\frac{3}{2}}
+2i\sqrt{2} \,Q^{21}_{\frac{3}{2}}
-2 \,Q^{22}_{\frac{3}{2}}
-2\, \bar{Q}^{11}_{\frac{3}{2}}
-2i\sqrt{2} \, \bar{Q}^{12}_{\frac{3}{2}}
+\bar{Q}^{22}_{\frac{3}{2}}
)\,,
\nonu\\
G^3
&
=&
\frac{i}{2}\,(
Q^{11}_{\frac{3}{2}}
+i\sqrt{2} \,Q^{12}_{\frac{3}{2}}
-2\,Q^{22}_{\frac{3}{2}}
-2\, \bar{Q}^{11}_{\frac{3}{2}}
-i \sqrt{2} \, \bar{Q}^{21}_{\frac{3}{2}}
+\bar{Q}^{22}_{\frac{3}{2}}
)\,,
\nonu\\
G^4
&
=&
\frac{1}{2}\,Q^{11}_{\frac{3}{2}}
+Q^{22}_{\frac{3}{2}}
+\bar{Q}^{11}_{\frac{3}{2}}
+\frac{1}{2} \, \bar{Q}^{22}_{\frac{3}{2}}
\,.
\label{Expression}
\eea
We observe that there are some extra minus signs
in the coefficients of (\ref{Expression}),
compared to the $\la=0$ case in subsection $6.1$.

For the first ${\cal N}=4$ higher spin multiplet, we obtain
\bea
\Phi^{(1)}_{0}
&=&
4q\,(W^{11}_{\mathrm{F},1}+W^{22}_{\mathrm{F},1})\,,
\nonu \\
\Phi^{(1),1}_{\frac{1}{2}}
&
=&
\frac{1}{2}\,(
Q^{11}_{\frac{3}{2}}
+i\sqrt{2}\,Q^{12}_{\frac{3}{2}}
+2i \sqrt{2}\,Q^{21}_{\frac{3}{2}}
-2\,Q^{22}_{\frac{3}{2}}
+2\,\bar{Q}^{11}_{\frac{3}{2}}
+2i \sqrt{2}\,Q^{12}_{\frac{3}{2}}
+i\sqrt{2}\,\bar{Q}^{21}_{\frac{3}{2}}
-\bar{Q}^{22}_{\frac{3}{2}}
)\,,
\nonu\\
\Phi^{(1),2}_{\frac{1}{2}}
&
=&
\frac{i}{2}\,(
Q^{11}_{\frac{3}{2}}
+2i\sqrt{2} \,Q^{21}_{\frac{3}{2}}
-2 \,Q^{22}_{\frac{3}{2}}
+2\, \bar{Q}^{11}_{\frac{3}{2}}
+2i\sqrt{2} \, \bar{Q}^{12}_{\frac{3}{2}}
-\bar{Q}^{22}_{\frac{3}{2}}
)\,,
\nonu\\
\Phi^{(1),3}_{\frac{1}{2}}
&
=&
-\frac{i}{2}\,(
Q^{11}_{\frac{3}{2}}
+i\sqrt{2} \,Q^{12}_{\frac{3}{2}}
-2\,Q^{22}_{\frac{3}{2}}
+2\, \bar{Q}^{11}_{\frac{3}{2}}
+i \sqrt{2} \, \bar{Q}^{21}_{\frac{3}{2}}
-\bar{Q}^{22}_{\frac{3}{2}}
)\,,
\nonu\\
\Phi^{(1),4}_{\frac{1}{2}}
&
=&
-\frac{1}{2}\,Q^{11}_{\frac{3}{2}}
-Q^{22}_{\frac{3}{2}}
+\bar{Q}^{11}_{\frac{3}{2}}
+\frac{1}{2}\,\bar{Q}^{22}_{\frac{3}{2}}
\,,
\nonu \\
\Phi^{(1),12}_{1}
&
=&
2i\,W^{11}_{\mathrm{B},2}
-\sqrt{2}\,W^{12}_{\mathrm{B},2}
-2i\,\,W^{22}_{\mathrm{B},2}
+2i\,W^{11}_{\mathrm{F},2}
-2\sqrt{2}\,W^{12}_{\mathrm{F},2}
-2i\,W^{22}_{\mathrm{F},2}\,,
\nonu\\
\Phi^{(1),13}_{1}
&
=&
2i\,W^{11}_{\mathrm{B},2}
-4\sqrt{2}\,W^{21}_{\mathrm{B},2}
-2i\,\,W^{22}_{\mathrm{B},2}
+2i\,W^{11}_{\mathrm{F},2}
-2\sqrt{2}\,W^{21}_{\mathrm{F},2}
-2i\,W^{22}_{\mathrm{F},2}\,,
\nonu\\
\Phi^{(1),14}_{1}
&
=&
-2\,W^{11}_{\mathrm{B},2}
-i\sqrt{2}\,W^{12}_{\mathrm{B},2}
-4i\sqrt{2}\,\,W^{21}_{\mathrm{B},2}
+2\,W^{22}_{\mathrm{B},2}
+2\,W^{11}_{\mathrm{F},2}
+2i\sqrt{2}\,W^{12}_{\mathrm{F},2}
\nonu \\
& + & 2i\sqrt{2}\,W^{21}_{\mathrm{F},2}
-2\,W^{22}_{\mathrm{F},2}\,,
\nonu\\
\Phi^{(1),23}_{1}
&
=&
-2\,W^{11}_{\mathrm{B},2}
-i\sqrt{2}\,W^{12}_{\mathrm{B},2}
-4i\sqrt{2}\,\,W^{21}_{\mathrm{B},2}
+2\,W^{22}_{\mathrm{B},2}
-2\,W^{11}_{\mathrm{F},2}
-2i\sqrt{2}\,W^{12}_{\mathrm{F},2}
\nonu \\
& - & 2i\sqrt{2}\,W^{21}_{\mathrm{F},2}
+2\,W^{22}_{\mathrm{F},2}\,,
\nonu\\
\Phi^{(1),24}_{1}
&
=&
-2i\,W^{11}_{\mathrm{B},2}
+4\sqrt{2}\,W^{21}_{\mathrm{B},2}
+2i\,\,W^{22}_{\mathrm{B},2}
+2i\,W^{11}_{\mathrm{F},2}
-2\sqrt{2}\,W^{21}_{\mathrm{F},2}
-2i\,W^{22}_{\mathrm{F},2}\,,
\nonu\\
\Phi^{(1),34}_{1}
&
=&
2i\,W^{11}_{\mathrm{B},2}
-\sqrt{2}\,W^{12}_{\mathrm{B},2}
-2i\,\,W^{22}_{\mathrm{B},2}
-2i\,W^{11}_{\mathrm{F},2}
+2\sqrt{2}\,W^{12}_{\mathrm{F},2}
+2i\,W^{22}_{\mathrm{F},2}\,,
\nonu \\
\tilde{\Phi}^{(1),1}_{\frac{3}{2}}
&
=&
\frac{1}{2q}\,(
Q^{11}_{\frac{5}{2}}
+i\sqrt{2}\,Q^{12}_{\frac{5}{2}}
+2i\sqrt{2}\,Q^{21}_{\frac{5}{2}}
-2\,Q^{22}_{\frac{5}{2}}
-2\,\bar{Q}^{11}_{\frac{5}{2}}
-2i\sqrt{2}\,\bar{Q}^{12}_{\frac{5}{2}}
-i\sqrt{2}\,\bar{Q}^{21}_{\frac{5}{2}}
+\bar{Q}^{22}_{\frac{5}{2}}
)\,,
\nonu\\
\tilde{\Phi}^{(1),2}_{\frac{3}{2}}
&
=&
\frac{i}{2q}\,(\,
Q^{11}_{\frac{5}{2}}
+2i\sqrt{2}\,Q^{21}_{\frac{5}{2}}
-2\,Q^{22}_{\frac{5}{2}}
-2\,\bar{Q}^{11}_{\frac{5}{2}}
-2i\sqrt{2}\,\bar{Q}^{12}_{\frac{5}{2}}
+\bar{Q}^{22}_{\frac{5}{2}}
)\,,
\nonu\\
\tilde{\Phi}^{(1),3}_{\frac{3}{2}}
&
=&
-\frac{i}{2q}\,(
Q^{11}_{\frac{5}{2}}
+i\sqrt{2}\,Q^{12}_{\frac{5}{2}}
-2\,Q^{22}_{\frac{5}{2}}
-2\,\bar{Q}^{11}_{\frac{5}{2}}
-i\sqrt{2}\,\bar{Q}^{21}_{\frac{5}{2}}
+\bar{Q}^{22}_{\frac{5}{2}}
)\,,
\nonu\\
\tilde{\Phi}^{(1),4}_{\frac{3}{2}}
&
=&
-\frac{1}{2q}\,
(
Q^{11}_{\frac{5}{2}}
+2\,Q^{22}_{\frac{5}{2}}
+2\,\bar{Q}^{11}_{\frac{5}{2}}
+\bar{Q}^{22}_{\frac{5}{2}}
)\,,
\nonu \\
\tilde{\Phi}^{(1)}_{2}
&
=&
\frac{2}{q}\,(
W^{11}_{\mathrm{B},3}
+W^{22}_{\mathrm{B},3}
+W^{11}_{\mathrm{F},3}
+W^{22}_{\mathrm{F},3}
).
\label{laonehigherspin}
\eea
The expression in (\ref{laonehigherspin})
looks similar to the ones in the subsection $6.2$.
For the next ${\cal N}=4$ higher spin multiplet,
we can obtain similar results. We do not present them
in this paper.

\section{ The remaining (anti)commutators with free fields
of section $6.4$}

We present the remaining (anti)commutators
described in section $6.4$ as follows:
\bea
\Big\{\Phi^{(h_1),i}_{\frac{1}{2}},\Phi^{(h_2),j}_{\frac{1}{2}}\Big\}
&=&
\delta^{ij}
\sum_{h = 0}^{\frac{1}{2}(h_1+h_2-1)}
\!\!\!\!\!\!
\frac{(2h_1-1)(2h_2-1)}{32(h_1+h_2-2h)\!-\!16}
\nonu \\
&\times & \Bigg(\!
\Big(
o_{\mathrm{F},2h}^{(h_1+\frac{1}{2})(h_2+\frac{1}{2}) }
\!-o_{\mathrm{B},2h}^{(h_1+\frac{1}{2})(h_2+\frac{1}{2}) } 
\Big)
\Phi^{(h_1+h_2-2h)}_{0}
\nonu\\
&
-&
\frac{1}{2(h_1+h_2-2h)-5}
\Big(
(h_1+h_2-2h-1)\,o_{\mathrm{F},2h}^{(h_1+\frac{1}{2})(h_2+\frac{1}{2}) }
\nonu\\
&
+&(h_1+h_2-2h)\,o_{\mathrm{B},2h}^{(h_1+\frac{1}{2})(h_2+\frac{1}{2}) } 
\Big)\,
 \tilde{\Phi}^{(h_1+h_2-2h-2)}_{2}
\Bigg)
\nonu\\
&
+&
\sum_{h = -1}^{\frac{1}{2}(h_1+h_2-2)}
\!\!\!\!\!
\frac{(2h_1-1)(2h_2-1)}{32(h_1+h_2-2h)-80 }
\nonu \\
& \times & \Bigg(\!
\Big(
o_{\mathrm{F},2h+1}^{(h_1+\frac{1}{2})(h_2+\frac{1}{2})} 
\!-\!o_{\mathrm{B},2h+1}^{(h_1+\frac{1}{2})(h_2+\frac{1}{2})} 
 \Big)
 \Phi^{(h_1+h_2-2h-2),ij}_{1}
\nonu\\
&
+& \Big(
o_{\mathrm{F},2h+1}^{(h_1+\frac{1}{2})(h_2+\frac{1}{2})} 
+o_{\mathrm{B},2h+1}^{(h_1+\frac{1}{2})(h_2+\frac{1}{2})} 
 \Big)
\, \tilde{\Phi}^{(h_1+h_2-2h-2),ij}_{1}
\,\Bigg)
\nonu\\
&
-& \delta^{h_1 h_2}\,\delta^{ij}
\,\frac{N\,2^{2h_1-5}h_1!(h_1-1)!}{(2h_1-3)!!(2h_1-3)!!}\prod_{j=-h_1}^{h_1-1}(r+j+\frac{1}{2})
\,,
\nonu \\
\Big[\Phi^{(h_1),i}_{\frac{1}{2}},\Phi^{(h_2),jk}_{1}\Big]
&=&
\Bigg[-\delta^{ik}\,
\sum_{h = -1}^{\frac{1}{2}(h_1+h_2-2)}\frac{(2h_1-1)(2h_2-1)}{4(h_1+h_2-2h)-10}\,
\nonu\\
&\times &
\Bigg(\!
\Big(
\,q_{\mathrm{F},2h}^{(h_2+1)(h_1+\frac{1}{2}) }
-q_{\mathrm{B},2h}^{(h_2+1)(h_1+\frac{1}{2}) }
\Big)\,
\tilde{\Phi}^{(h_1+h_2-2h-2),j}_{\frac{3}{2}}
\nonu \\
& - & \!
\Big(
\,q_{\mathrm{F},2h+1}^{(h_2+1)(h_1+\frac{1}{2}) }
-q_{\mathrm{B},2h+1}^{(h_2+1)(h_1+\frac{1}{2})}
\Big)\,
\Phi^{(h_1+h_2-2h-2),j}_{\frac{1}{2}}
\Bigg) -(j \leftrightarrow k) \Bigg]
\nonu\\
&+&
\varepsilon^{ijkl}\,
\sum_{h = -1}^{\frac{1}{2}(h_1+h_2-2)}
\frac{(2h_1-1)(2h_2-1)}{4 (h_1+h_2-2h)-10}\,
\nonu\\
&\times&
\Bigg(\!
\Big(
\,q_{\mathrm{F} ,2h}^{(h_2+1)(h_1+\frac{1}{2})}
+q_{\mathrm{B} ,2h}^{(h_2+1)(h_1+\frac{1}{2})}
\Big)
\tilde{\Phi}^{(h_1+h_2-2h-2),l}_{\frac{3}{2}}
\nonu \\
& - & \!
\Big(
\,q_{\mathrm{F},2h+1}^{(h_2+1)(h_1+\frac{1}{2}) }
+q_{\mathrm{B},2h+1}^{(h_2+1)(h_1+\frac{1}{2}) }
\Big)
\Phi^{(h_1+h_2-2h-2),l}_{\frac{1}{2}}
\Bigg)
\,,
\nonu\\
\Big\{\Phi^{(h_1),i}_{\frac{1}{2}},\tilde{\Phi}^{(h_2),j}_{\frac{3}{2}}\Big\}
&=&
\delta^{ij}\sum_{h = 0}^{\frac{1}{2}(h_1+h_2-1)}
\!\!\!\!\!
\frac{(2h_1-1)(2h_2-1)}{32 (h_1+h_2-2h)\!-\!16}
\nonu \\
&\times & \Bigg(\!
\Big(
o_{\mathrm{F},2h+1}^{(h_1+\frac{1}{2})(h_2+\frac{3}{2})}
-o_{\mathrm{B},2h+1}^{(h_1+\frac{1}{2})(h_2+\frac{3}{2})}
\Big)
\Phi^{(h_1+h_2-2h)}_{0}
\nonu\\
&
-&
\frac{2}{2(h_1+h_2-2h)-5}\,
\Big(
(h_1+h_2-2h-1)\,o_{\mathrm{F},2h+1}^{(h_1+\frac{1}{2})(h_2+\frac{3}{2})}
\nonu\\
&
+&
(h_1+h_2-2h)\,o_{\mathrm{B},2h+1}^{(h_1+\frac{1}{2})(h_2+\frac{3}{2})}
\,\Big)
\,\tilde{\Phi}^{(h_1+h_2-2h-2)}_{2}
\,\Bigg)
\nonu\\
&
+&
\sum_{h = 0}^{\frac{1}{2}(h_1+h_2)}
\frac{(2h_1-1)(2h_2-1)}{32(h_1+h_2-2h)-16}
\nonu \\
& \times & \Bigg(\!
\Big(
o_{\mathrm{F},2h}^{(s_1+\frac{1}{2})(s_2+\frac{3}{2})}
-o_{\mathrm{B},2h}^{(s_1+\frac{1}{2})(s_2+\frac{3}{2})}
\Big)
\, \Phi^{(h_1+h_2-2h),ij}_{1}
\nonu\\
&+&
\Big(
o_{\mathrm{F},2h}^{(s_1+\frac{1}{2})(s_2+\frac{3}{2})}
+o_{\mathrm{B},2h}^{(s_1+\frac{1}{2})(s_2+\frac{3}{2})}
\Big)
\, \tilde{\Phi}^{(h_1+h_2-2h),ij}_{1}
\Bigg)
\,,
\nonu \\
\Big[\Phi^{(h_1),i}_{\frac{1}{2}},\tilde{\Phi}^{(h_2)}_{2}\Big]
&=&
\sum_{h = -1}^{\frac{1}{2}(h_1+h_2-2)}
\!\!
\frac{(2h_1-1)(2h_2-1)}{4(h_1+h_2-2h)\!-\!10}\,
\nonu \\
&\times & \Big(
q_{\mathrm{F},2h+1}^{(h_2+2)(h_1+\frac{1}{2})}
+ q_{\mathrm{B},2h+1}^{(h_2+2)(h_1+\frac{1}{2}) }
\Big)\,
\tilde{\Phi}^{(h_1+h_2-2h-2),i}_{\frac{3}{2}}
\nonu\\
&
-& \sum_{h =-1}^{\frac{1}{2}(h_1+h_2-1)}
\!\!
\frac{(2h_1-1)(2h_2-1)}{4(h_1+h_2-2h)-2}
\nonu \\
& \times & \Big(
q_{\mathrm{F},2h}^{(h_2+2)(h_1+\frac{1}{2}) }
+q_{\mathrm{B},2h}^{(h_2+2)(h_1+\frac{1}{2})}
\Big)\,
\Phi^{(h_1+h_2-2h),i}_{\frac{1}{2}}
\,,
\nonu \\
\Big[\Phi^{(s_1),ij}_{1},\Phi^{(s_2),kl}_{1}\Big]
& = &
(\delta^{ik}\delta^{jl}-\delta^{il}\delta^{jk})
\sum_{r = 0}^{\frac{1}{2}(h_1+h_2-1)}
\frac{(2h_1-1)(2h_2-1)}{8(h_1+h_2-2h)-4}\,
\nonu \\
&\times & \Bigg(
\Big(
p_{\mathrm{F},2h}^{(h_1+1)(h_2+1) }
-p_{\mathrm{B},2h}^{(h_1+1)(h_2+1) } 
\Big)
\, \Phi^{(h_1+h_2-2h)}_{0}
\nonu\\
&
-&
\frac{2}{2(h_1+h_2-2h)-5}
\Big(
(h_1+h_2-2h-1)\,p_{\mathrm{F},2h}^{(h_1+1)(h_2+1) }
\nonu \\
&+ &(h_1+h_2-2h)\,p_{\mathrm{B},2h}^{(h_1+1)(h_2+1) }
\Big)
\, \tilde{\Phi}^{(h_1+h_2-2h-2)}_{2}
\Bigg)
\nonu\\
&
+& \varepsilon^{ijkl}
\sum_{r = 0}^{\frac{1}{2}(h_1+h_2-1)}
\frac{(2h_1-1)(2h_2-1)}{8(h_1+h_2-2h)-4}\,
\nonu \\
& \times & \Bigg( \Big(
p_{\mathrm{F},2h}^{(h_1+1)(h_2+1) }
+p_{\mathrm{B},2h}^{(h_1+1)(h_2+1) } 
\Big)
\, \Phi^{(h_1+h_2-2h)}_{0}
\nonu\\
&
-&
\frac{2}{2(h_1+h_2-2h)-5}
\Big(
(h_1+h_2-2h-1)\,p_{\mathrm{F},2h}^{(h_1+1)(h_2+1) }
\nonu \\
&-& (h_1+h_2-2h)\,p_{\mathrm{B},2h}^{(h_1+1)(h_2+1) }
\Big)
\, \tilde{\Phi}^{(h_1+h_2-2h-2)}_{2}
\Bigg)
\,
\nonu\\
&+& \Bigg[-\delta^{ik}
\,\sum_{r= -1}^{\frac{1}{2}(h_1+h_2-2)}
\,\frac{(2h_1-1)(2h_2-1)}{8(h_1+h_2-2h)-20}
\nonu\\
&
\times &
\Bigg(\!
\Big(
p_{\mathrm{F},2h+1}^{(h_1+1)(h_2+1)} 
+p_{\mathrm{B},2h+1}^{(h_1+1)(h_2+1)} 
 \Big)
\Phi^{(h_1+h_2-2h-2),jl}_{1}
\nonu \\
& + &
\Big(
p_{\mathrm{F},2h+1}^{(h_1+1)(h_2+1)} 
-p_{\mathrm{B},2h+1}^{(h_1+1)(h_2+1)} 
 \Big)
\tilde{\Phi}^{(h_1+h_2-2h-2),jl}_{1}
\Bigg) \nonu \\
&-& (k \leftrightarrow l) -(i \leftrightarrow j) +
(i \leftrightarrow j, k \leftrightarrow l) \Bigg]
\nonu\\
&
-&
\delta^{h_1h_2}\Bigg(\!
(\delta^{ik}\delta^{jl}-\delta^{jk}\delta^{il})
\frac{N\,2^{2h_1-5}h_1!(h_1-1)!}{(2h_1-3)!!^2}
\nonu \\
& - & \varepsilon^{ijkl}
\frac{N\,2^{2h_1-5}h_1!(h_1-1)!(2h_1-1)}{(2h_1+1)!!(2h_1-3)!!}
\Bigg)\!
\prod_{j=-h_1}^{h_1}(m+j)
\,,
\nonu \\
\Big[\Phi^{(h_1),ij}_{1},\tilde{\Phi}^{(h_2),k}_{\frac{3}{2}}\Big]
&=&
\Bigg[ \delta^{ik}\,\frac{1}{2}\,(2h_1-1)(2h_2-1)
\nonu\\
&\times &
\Bigg(\!
\sum_{h = -1}^{\frac{1}{2}(h_1+h_2-2)}
\!\!
\frac{1}{2 (h_1+h_2-2h)-5}
\Big(
q_{\mathrm{F},2h+1}^{(h_1+1)(s_2+\frac{3}{2}) }
- q_{\mathrm{B},2h+1}^{(h_1+1)(s_2+\frac{3}{2}) }
\Big)\,
\nonu \\
&\times & \tilde{\Phi}^{(h_1+h_2-2h-2),j}_{\frac{3}{2}}
+  \sum_{h = -1}^{\frac{1}{2}(h_1+h_2-1)}
\!\!
\frac{1}{2(h_1+h_2-2h)-1}
\nonu \\
& \times & \Big(
q_{\mathrm{F},2h}^{(h_1+1)(h_2+\frac{3}{2}) }
- q_{\mathrm{B},2h}^{(h_1+1)(h_2+\frac{3}{2})}
\Big)\,
\Phi^{(h_1+h_2-2h),j}_{\frac{1}{2}}
\,\Bigg) -(i \leftrightarrow j)\Bigg]
\nonu\\
&+&
\varepsilon^{ijkl}\,
\frac{1}{2}\,(2h_1-1)(2h_2-1)
\Bigg(\!
\sum_{h = -1}^{\frac{1}{2}(h_1+h_2-2)}
\!\!
\frac{1}{2(h_1+h_2-2h)-5}
\nonu \\
& \times & \Big(
q_{\mathrm{F},2h+1}^{(h_1+1)(s_2+\frac{3}{2}) }
+ q_{\mathrm{B},2h+1}^{(h_1+1)(s_2+\frac{3}{2}) }
\Big)\,
\tilde{\Phi}^{(h_1+h_2-2h-2),l}_{\frac{3}{2}}
\nonu\\
&
+ & \sum_{h = -1}^{\frac{1}{2}(h_1+h_2-1)}
\!\!
\frac{1}{2(h_1+h_2-2h)-1}
\Big(
q_{\mathrm{F},2h}^{(h_1+1)(h_2+\frac{3}{2}) }
+q_{\mathrm{B},2h}^{(h_1+1)(h_2+\frac{3}{2})}
\Big)\,
\nonu \\
& \times & \Phi^{(h_1+h_2-2h),l}_{\frac{1}{2}}
\,\Bigg)
\,,
\nonu \\
\Big[\Phi^{(h_1),ij}_{1},\tilde{\Phi}^{(h_2)}_{2}\Big]
&=&
-\sum_{h = 0}^{\frac{1}{2}(h_1+h_2)}
\frac{(2h_1-1)(2h_2-1)}{8 (h_1+h_2-2h)-4}
\nonu \\
&\times & 
\Bigg(
\Big(
p_{\mathrm{F},2h}^{(h_2+2)(h_2+1)}
+p_{\mathrm{B},2h}^{(h_2+2)(h_2+1) }
\Big)\,
 \Phi^{(h_1+h_2-2h),ij}_{1}
\nonu\\
&
+&
\Big(
p_{\mathrm{F},2h}^{(h_2+2)(h_2+1)}
-p_{\mathrm{B},2h}^{(h_2+2)(h_2+1) }
\Big)\,
\tilde{\Phi}^{(h_1+h_2-2h),ij}_{1}
\Bigg)
\,,
\nonu \\
\Big\{\tilde{\Phi}^{(h_1),i}_{\frac{3}{2}},
\tilde{\Phi}^{(h_2),j}_{\frac{3}{2}}\Big\}
&=&
-\delta^{ij}
\!\!\!\! \,\,\,\,\,\,
\sum_{h = 0}^{\frac{1}{2}(h_1+h_2+1)}
\!\!\!\!\!\!
\frac{(2h_1-1)(2h_2-1)}{32 (h_1+h_2-2h)\!+\!48}
\nonu \\
& \times & \Bigg(\!
\Big(
o_{\mathrm{F},2h}^{(h_1+\frac{3}{2})(h_2+\frac{3}{2}) }
\!-\!o_{\mathrm{B},2h}^{(h_1+\frac{3}{2})(h_2+\frac{3}{2}) }
\Big)
\Phi^{(h_1+h_2-2h+2)}_{0}
\nonu\\
&
-& \frac{2}{2(h_1+h_2+2h)-1}
\Big(
(h_1+h_2-2h+1)\,o_{\mathrm{F},2h}^{(h_1+\frac{3}{2})(h_2+\frac{3}{2}) }
\nonu\\
&
+&
(h_1+h_2-2h+2)\,o_{\mathrm{B},2h}^{(h_1+\frac{3}{2})(h_2+\frac{3}{2}) }
\Big)
\, \tilde{\Phi}^{(h_1+h_2-2h)}_{2}
\,\Bigg)
\nonu\\
&
-&
\sum_{h = 0}^{\frac{1}{2}(h_1+h_2)}\,
\frac{(2h_1-1)(2h_2-1)}{32(h_1+h_2-2h)-16}\,
\nonu \\
& \times & \Bigg(\!
\Big(
o_{\mathrm{F},2h+1}^{(h_1+\frac{3}{2})(h_2+\frac{3}{2}) }
-o_{\mathrm{B},2h+1}^{(h_1+\frac{3}{2})(h_2+\frac{3}{2}) }
\Big)\,
\Phi^{(h_1+h_2-2h),ij}_{1}
\nonu\\
&+&
\Big(
o_{\mathrm{F},2h+1}^{(h_1+\frac{3}{2})(h_2+\frac{3}{2}) }
+o_{\mathrm{B},2h+1}^{(h_1+\frac{3}{2})(h_2+\frac{3}{2}) }
\Big)\,
\tilde{\Phi}^{(h_1+h_2-2h),ij}_{1}
\Bigg)
\nonu\\
&
+&
\delta^{h_1 h_2}\,\delta^{ij}
\,\frac{N\,2^{2h_1-3}h_1!(h_1+1)!(2h_1-1)^2}
       {(2h_1+1)!!(2h_1+1)!!}\prod_{j=-h_1-1}^{h_1}(r+j+\frac{1}{2})
\,,
\nonu \\
\Big[\tilde{\Phi}^{(h_1),i}_{\frac{3}{2}},\tilde{\Phi}^{(h_2)}_{2}\Big]
&=&
-
\!\!
\sum_{h = -1}^{\frac{1}{2}(h_1+h_2)}\!
\frac{(2h_1-1)(2h_2-1)}{4(h_1+h_2-2h)-2}\,
\nonu \\
& \times & \Bigg(
\Big(
q_{\mathrm{F}, 2h}^{(h_2+2)(h_1+\frac{3}{2})}
+ q_{\mathrm{B}, 2h}^{(h_2+2)(h_1+\frac{3}{2})}
\Big)\,
\tilde{\Phi}^{(h_1+h_2-2h),i}_{\frac{3}{2}}
\nonu\\
&
-&
\Big(
q_{\mathrm{F}, 2h+1}^{(h_2+2)(h_1+\frac{3}{2})}
+ q_{\mathrm{B}, 2h+1}^{(h_2+2)(h_1+\frac{3}{2})}
\Big)\,
\Phi^{(h_1+h_2-2h),i}_{\frac{1}{2}}
\,\Bigg)
\,,
\nonu \\
\Big[\tilde{\Phi}^{(h_1)}_{2}, \tilde{\Phi}^{(h_2)}_{2}\Big]
&=&
-\!\!\!\!
\sum_{h= 0}^{\frac{1}{2}(h_1+h_2+1)}
\!\!
\frac{(2h_1-1)(2h_2-1)}{8(h_1+h_2-2h)+12}
\nonu \\
& \times & \Bigg(
\Big(
p_{\mathrm{F}, 2h}^{(h_1+2)(h_2+2)}
-p_{\mathrm{B}, 2h}^{(h_1+2)(h_2+2)}
\Big)\,
\Phi^{(h_1+h_2-2h+2)}_{0}
\nonu\\
&
-&
\frac{2}{2(h_1+h_2-2h)-1}
\Big(
(h_1+h_2-2h+1)\,p_{\mathrm{F}, 2h}^{(h_1+2)(h_2+2)}
\nonu\\
&
+ & (h_1+h_2-2h+2)\,p_{\mathrm{B}, 2h}^{(h_1+2)(h_2+2)}
\Big)\,
\tilde{\Phi}^{(h_1+h_2-2h)}_{2}
\,\Bigg)
\nonu\\
&
+&
\delta^{h_1 h_2}\,\frac{N\,2^{2h_1-3}h!(h_1+1)!(2h_1-1)^2}{
  (2h_1+1)!!  (2h_1+1)!!}\prod_{j=-h_1-1}^{h_1+1}(m+j)
\,.
\label{remaininganticomm}
\eea
Note that
in the second,  fourth, seventh and ninth
of (\ref{remaininganticomm}), the order of two modes
appearing in $q_F, q_B,p_F, p_B$ in the right hand side 
is reversed, compared to the ones in the left hand side.
For example, the indices of $(r, m)$ appearing
in the left hand side of
the second of
(\ref{remaininganticomm})
occur in the form of $q_F(m,r)$ and $q_B(m,r)$
in the right hand side. 
We can obtain $\Big[\Phi^{(h_1),i}_{\frac{1}{2}},
  \Phi^{(h_2)}_{0}\Big]$ from the second expression of
(\ref{completecomm}) by substituting $h_1 \leftrightarrow h_2$
with minus sign. Furthermore,
the other nine (anti)commutators we do not present
explicitly can be red off from the known ones similarly.
As usual, the mode parameter appearing in the central terms
refers to the one of the first higher spin current
of the left hand side
\footnote{
In the first, third and eigth anticommutators,
the $SO(4)$ singlets
with subscript $0$ and $2$ and $SO(4)$ adjoints appear
in the right hand side of (\ref{remaininganticomm}).
The second and sixth  commutators
contain the $SO(4)$ vectors with subscript $\frac{1}{2}$ and
$\frac{3}{2}$.
In the fourth and ninth commutators,
the $SO(4)$ vectors
with subscript $\frac{1}{2}$ and $\frac{3}{2}$ appear.
The fifth  commutator
contains
the $SO(4)$ singlets
with subscript $0$ and $2$ and $SO(4)$ adjoints.
In the seventh commutator,
the $SO(4)$ adjoints with subscript $1$ appear.
Finally, the last commutator
contains the $SO(4)$ singlets
with subscript $0$ and $2$.
The coefficients of
the higher spin currents having negative spins in
(\ref{remaininganticomm})
are vanishing.
As in the footnote \ref{vanishingF32}, we have 
$\phi^{h_1+1,h_2+1}_{h_1+h_2-1}(0,0)=0=\phi^{h_1+2,h_2+2}_{h_1+h_2+1}(0,0)$.
Moreover,
$\phi^{h_1+\frac{1}{2},h_2+\frac{1}{2}}_{h_1+h_2-1}(\frac{1}{2},-\frac{1}{4})$
is written as
$
 {}_3 F_2
 \Bigg[
\begin{array}{c}
 -\frac{1}{2},   -\frac{(h_1+h_2)}{2},
  -\frac{(h_1+h_2-1)}{2} \\
  -h_1+\frac{1}{2},  -h_2+\frac{1}{2}
\end{array} ; 1
\Bigg]$ which contains $\frac{1}{\sqrt{\Gamma(-h_1+1) \,
     \Gamma(-h_2+1)}}$ and is zero.
 There are also vanishing
 $\phi^{h_1+\frac{1}{2},h_2+\frac{3}{2}}_{h_1+h_2}(\frac{1}{2},-\frac{1}{4})$
 and $\phi^{h_1+\frac{3}{2},h_2+\frac{3}{2}}_{h_1+h_2+1}(\frac{1}{2},-\frac{1}{4})$.
}.

For convenience, we present the quasi primary fields in (\ref{WWQQ})
in terms of the components of ${\cal N}=4$ higher spin multiplets 
\bea
W^{11}_{\mathrm{F},h}
&=&
q^{h-2} 
\Big(
\frac{1}{2(1-2h)}\,\Phi^{(h)}_0
+\frac{(h-1)}{(2 h-5) (2 h-1)}\,\tilde{\Phi}^{(h-2)}_2
\Big)
+
q^{h-2} \,\frac{1}{2(2h-3)}
\Big(
i\,\Phi^{(h-1),12}_1
\nonu \\
& - & i\,\Phi^{(h-1),13}_1
-  \Phi^{(h-1),14}_1
-\Phi^{(h-1),23}_1
+i\,\Phi^{(h-1),24}_1
+i\,\Phi^{(h-1),34}_1
\Big)\,,
\nonu\\
W^{12}_{\mathrm{F},h}
&=&
q^{h-2}\,\frac{1}{\sqrt{2}(2h-3)} 
\Big(
\Phi^{(h-1),13}_1
-i\,\Phi^{(h-1),14}_1
-i\,\Phi^{(h-1),23}_1
-\Phi^{(h-1),24}_1
\Big)\,,
\nonu\\
W^{21}_{\mathrm{F},h}
&=&
-q^{h-2}\,\frac{1}{\sqrt{2}(2h-3)} 
\Big(
\Phi^{(h-1),12}_1
+i\,\Phi^{(h-1),14}_1
+i\,\Phi^{(h-1),23}_1
+\Phi^{(h-1),34}_1
\Big)\,,
\nonu\\
W^{22}_{\mathrm{F},h}
&=&
q^{h-2} 
\Big(
\frac{1}{2(1-2h)}\,\Phi^{(h)}_0
+\frac{(h-1)}{(2 h-5) (2 h-1)}\,\tilde{\Phi}^{(h-2)}_2
\Big)
-
q^{h-2} \,\frac{1}{2(2h-3)}
\Big(
i\,\Phi^{(h-1),12}_1
\nonu \\
& - & i\,\Phi^{(h-1),13}_1
-\Phi^{(h-1),14}_1
-\Phi^{(h-1),23}_1
+i\,\Phi^{(h-1),24}_1
+i\,\Phi^{(h-1),34}_1
\Big)\,,
\nonu \\
W^{11}_{\mathrm{B},h}
&=&
-q^{h-2} 
\Big(
\frac{1}{2(1-2h)}\,\Phi^{(h)}_0
-\frac{h}{(2 h-5) (2 h-1)}\, \tilde{\Phi}^{(h-2)}_2
\Big)
+
q^{h-2} \,\frac{1}{2(2h-3)}
\Big(
i\,\Phi^{(h-1),12}_1
\nonu \\
& - & i\,\Phi^{(h-1),13}_1
+\Phi^{(h-1),14}_1
-\Phi^{(h-1),23}_1
-i\,\Phi^{(h-1),24}_1
-i\,\Phi^{(h-1),34}_1
\Big)\,,
\nonu\\
W^{12}_{\mathrm{B},h}
&=&
q^{h-2}\,\frac{\sqrt{2}}{(2h-3)} 
\Big(
\Phi^{(h-1),13}_1
+i\,\Phi^{(h-1),14}_1
-i\,\Phi^{(h-1),23}_1
+\Phi^{(h-1),24}_1
\Big)\,,
\nonu\\
W^{21}_{\mathrm{B},h}
&=&
-q^{h-2}\,\frac{1}{2\sqrt{2}(2h-3)} 
\Big(
\Phi^{(h-1),12}_1
-i\,\Phi^{(h-1),14}_1
+i\,\Phi^{(h-1),23}_1
-\Phi^{(h-1),34}_1
\Big)\,,
\nonu\\
W^{22}_{\mathrm{B},h}
&=&
-q^{h-2} 
\Big(
\frac{1}{2(1-2h)}\,\Phi^{(h)}_0
-\frac{h}{(2 h-5) (2 h-1)}\,\tilde{\Phi}^{(h-2)}_2
\Big)
-
q^{h-2} \,\frac{1}{2(2h-3)}
\Big(
i\,\Phi^{(h-1),12}_1
\nonu \\
& - & i\,\Phi^{(h-1),13}_1
+\Phi^{(h-1),14}_1
-\Phi^{(h-1),23}_1
-i\,\Phi^{(h-1),24}_1
-i\,\Phi^{(h-1),34}_1
\Big)\,,
\nonu \\
Q^{11}_{h+\frac{1}{2}}
&=&
q^{h-1} \,\frac{2}{(2h-1)}
\Big(
\Phi^{(h),1}_{\frac{1}{2}}
-i\,\Phi^{(h),2}_{\frac{1}{2}}
-i\,\Phi^{(h),3}_{\frac{1}{2}}
+\Phi^{(h),4}_{\frac{1}{2}}
\Big)
\nonu\\
&-&
q^{h-1} \,\frac{2}{(2h-3)}
\Big(
\tilde{\Phi}^{(h-1),1}_{\frac{3}{2}}
-i\,\tilde{\Phi}^{(h-1),2}_{\frac{3}{2}}
-i\,\tilde{\Phi}^{(h-1),3}_{\frac{3}{2}}
+\tilde{\Phi}^{(h-1),4}_{\frac{3}{2}}
\Big)\,,
\nonu\\
Q^{12}_{h+\frac{1}{2}}
&=&
q^{h-1} \,\frac{2\sqrt{2}}{(2h-1)}
\Big(
i\,\Phi^{(h),1}_{\frac{1}{2}}
+\Phi^{(h),2}_{\frac{1}{2}}
\Big)
-q^{h-1} \,\frac{2\sqrt{2}}{(2h-3)}
\Big(
i\,\tilde{\Phi}^{(h-1),1}_{\frac{3}{2}}
+\tilde{\Phi}^{(h-1),2}_{\frac{3}{2}}
\Big)\,,
\nonu\\
Q^{21}_{h+\frac{1}{2}}
&=&
q^{h-1} \,\frac{\sqrt{2}}{(2h-1)}
\Big(
i\,\Phi^{(h),1}_{\frac{1}{2}}
+\Phi^{(h),3}_{\frac{1}{2}}
\Big)
-q^{h-1} \,\frac{\sqrt{2}}{(2h-3)}
\Big(
i\,\tilde{\Phi}^{(h-1),1}_{\frac{3}{2}}
+\tilde{\Phi}^{(h-1),3}_{\frac{3}{2}}
\Big)\,,
\nonu\\
Q^{22}_{h+\frac{1}{2}}
&=&
-q^{h-1} \,\frac{1}{(2h-1)}
\Big(
\Phi^{(h),1}_{\frac{1}{2}}
-i\,\Phi^{(h),2}_{\frac{1}{2}}
-i\,\Phi^{(h),3}_{\frac{1}{2}}
-\Phi^{(h),4}_{\frac{1}{2}}
\Big)
\nonu\\
&+&
q^{h-1} \,\frac{1}{(2h-3)}
\Big(
\tilde{\Phi}^{(h-1),1}_{\frac{3}{2}}
-i\,\tilde{\Phi}^{(h-1),2}_{\frac{3}{2}}
-i\,\tilde{\Phi}^{(h-1),3}_{\frac{3}{2}}
-\tilde{\Phi}^{(h-1),4}_{\frac{3}{2}}
\Big)\,,
\nonu \\
\bar{Q}^{11}_{h+\frac{1}{2}}
&=&
q^{h-1} \,\frac{1}{(2h-1)}
\Big(
\Phi^{(h),1}_{\frac{1}{2}}
-i\,\Phi^{(h),2}_{\frac{1}{2}}
-i\,\Phi^{(h),3}_{\frac{1}{2}}
-\Phi^{(h),4}_{\frac{1}{2}}
\Big)
\nonu\\
&+&
q^{h-1} \,\frac{1}{(2h-3)}
\Big(
\tilde{\Phi}^{(h-1),1}_{\frac{3}{2}}
-i\,\tilde{\Phi}^{(h-1),2}_{\frac{3}{2}}
-i\,\tilde{\Phi}^{(h-1),3}_{\frac{3}{2}}
-\tilde{\Phi}^{(h-1),4}_{\frac{3}{2}}
\Big)\,,
\nonu\\
\bar{Q}^{12}_{h+\frac{1}{2}}
&=&
q^{h-1} \,\frac{\sqrt{2}}{(2h-1)}
\Big(
i\,\Phi^{(h),1}_{\frac{1}{2}}
+\Phi^{(h),3}_{\frac{1}{2}}
\Big)
+q^{h-1} \,\frac{\sqrt{2}}{(2h-3)}
\Big(
i\,\tilde{\Phi}^{(h-1),1}_{\frac{3}{2}}
+\tilde{\Phi}^{(h-1),3}_{\frac{3}{2}}
\Big)\,,
\nonu\\
\bar{Q}^{21}_{h+\frac{1}{2}}
&=&
q^{h-1} \,\frac{2\sqrt{2}}{(2h-1)}
\Big(
i\,\Phi^{(h),1}_{\frac{1}{2}}
+\Phi^{(h),2}_{\frac{1}{2}}
\Big)
+q^{h-1} \,\frac{2\sqrt{2}}{(2h-3)}
\Big(
i\,\tilde{\Phi}^{(h-1),1}_{\frac{3}{2}}
+\tilde{\Phi}^{(h-1),2}_{\frac{3}{2}}
\Big)\,,
\nonu\\
\bar{Q}^{22}_{h+\frac{1}{2}}
&=&
-q^{h-1} \,\frac{2}{(2h-1)}
\Big(
\Phi^{(h),1}_{\frac{1}{2}}
-i\,\Phi^{(h),2}_{\frac{1}{2}}
-i\,\Phi^{(h),3}_{\frac{1}{2}}
+\Phi^{(h),4}_{\frac{1}{2}}
\Big)
\nonu\\
&-&
q^{h-1} \,\frac{2}{(2h-3)}
\Big(
\tilde{\Phi}^{(h-1),1}_{\frac{3}{2}}
-i\,\tilde{\Phi}^{(h-1),2}_{\frac{3}{2}}
-i\,\tilde{\Phi}^{(h-1),3}_{\frac{3}{2}}
+\tilde{\Phi}^{(h-1),4}_{\frac{3}{2}}
\Big)\,.
\nonu
\eea
Note that the mixture of $h$-th, $(h-1)$-th and $(h-2)$-th
of the ${\cal N}=4$ higher spin multiplets occurs.

\section{ The second ${\cal N}=4$ higher spin
  generators in terms of
  oscillators  }

We continue to calculate the
second ${\cal N}=4$ higher spin multiplet in terms of
oscillators by using the method in section $7$
and we summarize them as follows:
\bea
\Phi^{(2)}_{0,-1} & = & 
\hat{y}_{2}\hat{y}_{2}\,(3k+\nu)
\otimes 
\left(\begin{array}{cc}
1 & 0 \\
0 & 1 \\
\end{array}\right),
\qquad
\Phi^{(2)}_{0,0}  =  
\frac{1}{2}\Big(\,
\hat{y}_{1}\hat{y}_{2}
+\hat{y}_{2}\hat{y}_{1}
\,\Big)(3k+\nu)
\otimes 
\left(\begin{array}{cc}
1 & 0 \\
0 & 1 \\
\end{array}\right),
\nonu\\
\Phi^{(2)}_{0,+1} & = & 
\hat{y}_{1}\hat{y}_{1}\,(3k+\nu)
\otimes 
\left(\begin{array}{cc}
1 & 0 \\
0 & 1 \\
\end{array}\right),
\qquad
\Phi^{(2),1}_{\frac{1}{2},-\frac{3}{2}}  =  
3\,e^{i\frac{\pi}{4}}\,\hat{y}_{2}\hat{y}_{2}\hat{y}_{2}
\otimes 
\left(\begin{array}{cc}
1 & 0 \\
0 & -1 \\
\end{array}\right),
\nonu\\
\Phi^{(2),1}_{\frac{1}{2},-\frac{1}{2}} & = & 
e^{i\frac{\pi}{4}}\Big(\,
\hat{y}_{1}\hat{y}_{2}\hat{y}_{2}
+\hat{y}_{2}\hat{y}_{1}\hat{y}_{2}
+\hat{y}_{2}\hat{y}_{2}\hat{y}_{1}
\,\Big)
\otimes 
\left(\begin{array}{cc}
1 & 0 \\
0 & -1 \\
\end{array}\right),
\nonu\\
\Phi^{(2),1}_{\frac{1}{2},+\frac{1}{2}} & =& 
e^{i\frac{\pi}{4}}\Big(\,
\hat{y}_{2}\hat{y}_{1}\hat{y}_{1}
+\hat{y}_{1}\hat{y}_{2}\hat{y}_{1}
+\hat{y}_{1}\hat{y}_{1}\hat{y}_{2}
\,\Big)
\otimes 
\left(\begin{array}{cc}
1 & 0 \\
0 & -1 \\
\end{array}\right),
\nonu\\
\Phi^{(2),1}_{\frac{1}{2},+\frac{3}{2}} & = & 
3\,e^{i\frac{\pi}{4}}\,\hat{y}_{2}\hat{y}_{2}\hat{y}_{2}
\otimes 
\left(\begin{array}{cc}
1 & 0 \\
0 & -1 \\
\end{array}\right),
\qquad
\Phi^{(2),2}_{\frac{1}{2},-\frac{3}{2}}  =  
-3i\,e^{i\frac{\pi}{4}}\,\hat{y}_{2}\hat{y}_{2}\hat{y}_{2}\,k
\otimes 
\left(\begin{array}{cc}
1 & 0 \\
0 & 1 \\
\end{array}\right),
\nonu\\
\Phi^{(2),2}_{\frac{1}{2},-\frac{1}{2}} & = & 
-i\,e^{i\frac{\pi}{4}}\Big(\,
\hat{y}_{1}\hat{y}_{2}\hat{y}_{2}
+\hat{y}_{2}\hat{y}_{1}\hat{y}_{2}
+\hat{y}_{2}\hat{y}_{2}\hat{y}_{1}
\,\Big)\,k
\otimes 
\left(\begin{array}{cc}
1 & 0 \\
0 & 1 \\
\end{array}\right),
\nonu\\
\Phi^{(2),2}_{\frac{1}{2},+\frac{1}{2}} & = & 
-i\,e^{i\frac{\pi}{4}}\Big(\,
\hat{y}_{2}\hat{y}_{1}\hat{y}_{1}
+\hat{y}_{1}\hat{y}_{2}\hat{y}_{1}
+\hat{y}_{1}\hat{y}_{1}\hat{y}_{2}
\,\Big)\,k
\otimes 
\left(\begin{array}{cc}
1 & 0 \\
0 & 1 \\
\end{array}\right),
\nonu\\
\Phi^{(2),2}_{\frac{1}{2},+\frac{3}{2}} & = & 
-3i\,e^{i\frac{\pi}{4}}\,\hat{y}_{2}\hat{y}_{2}\hat{y}_{2}\,k
\otimes 
\left(\begin{array}{cc}
1 & 0 \\
0 & 1 \\
\end{array}\right),
\qquad
\Phi^{(2),3}_{\frac{1}{2},-\frac{3}{2}}  =  
-3i\,e^{i\frac{\pi}{4}}\,\hat{y}_{2}\hat{y}_{2}\hat{y}_{2}
\otimes 
\left(\begin{array}{cc}
0 & 1 \\
-1 & 0 \\
\end{array}\right),
\nonu\\
\Phi^{(2),3}_{\frac{1}{2},-\frac{1}{2}} &= & 
-i\,e^{i\frac{\pi}{4}}\Big(\,
\hat{y}_{1}\hat{y}_{2}\hat{y}_{2}
+\hat{y}_{2}\hat{y}_{1}\hat{y}_{2}
+\hat{y}_{2}\hat{y}_{2}\hat{y}_{1}
\,\Big)
\otimes 
\left(\begin{array}{cc}
0 & 1 \\
-1 & 0 \\
\end{array}\right),
\nonu\\
\Phi^{(2),3}_{\frac{1}{2},+\frac{1}{2}} &=& 
-i\,e^{i\frac{\pi}{4}}\Big(\,
\hat{y}_{2}\hat{y}_{1}\hat{y}_{1}
+\hat{y}_{1}\hat{y}_{2}\hat{y}_{1}
+\hat{y}_{1}\hat{y}_{1}\hat{y}_{2}
\,\Big)
\otimes 
\left(\begin{array}{cc}
0 & 1 \\
-1 & 0 \\
\end{array}\right),
\nonu\\
\Phi^{(2),3}_{\frac{1}{2},+\frac{3}{2}} &=& 
-3i\,e^{i\frac{\pi}{4}}\,\hat{y}_{2}\hat{y}_{2}\hat{y}_{2}
\otimes 
\left(\begin{array}{cc}
0 & 1 \\
-1 & 0 \\
\end{array}\right),
\qquad
\Phi^{(2),4}_{\frac{1}{2},-\frac{3}{2}} = 
-3\,e^{i\frac{\pi}{4}}\,\hat{y}_{2}\hat{y}_{2}\hat{y}_{2}
\otimes 
\left(\begin{array}{cc}
0 & 1 \\
1 & 0 \\
\end{array}\right),
\nonu\\
\Phi^{(2),4}_{\frac{1}{2},-\frac{1}{2}} &=& 
-e^{i\frac{\pi}{4}}\Big(\,
\hat{y}_{1}\hat{y}_{2}\hat{y}_{2}
+\hat{y}_{2}\hat{y}_{1}\hat{y}_{2}
+\hat{y}_{2}\hat{y}_{2}\hat{y}_{1}
\,\Big)
\otimes 
\left(\begin{array}{cc}
0 & 1 \\
1 & 0 \\
\end{array}\right),
\nonu\\
\Phi^{(2),4}_{\frac{1}{2},+\frac{1}{2}} &=& 
-e^{i\frac{\pi}{4}}\Big(\,
\hat{y}_{2}\hat{y}_{1}\hat{y}_{1}
+\hat{y}_{1}\hat{y}_{2}\hat{y}_{1}
+\hat{y}_{1}\hat{y}_{1}\hat{y}_{2}
\,\Big)
\otimes 
\left(\begin{array}{cc}
0 & 1 \\
1 & 0 \\
\end{array}\right),
\nonu\\
\Phi^{(2),4}_{\frac{1}{2},+\frac{3}{2}} &=& 
-3\,e^{i\frac{\pi}{4}}\,\hat{y}_{2}\hat{y}_{2}\hat{y}_{2}
\otimes 
\left(\begin{array}{cc}
0 & 1 \\
1 & 0 \\
\end{array}\right),
\qquad
\Phi^{(2),12}_{1,-2} = 
-3\,\hat{y}_{2}\hat{y}_{2}\hat{y}_{2}\hat{y}_{2}\,k
\otimes 
\left(\begin{array}{cc}
1 & 0 \\
0 & -1 \\
\end{array}\right),
\nonu\\
\Phi^{(2),12}_{1,-1} &=& 
-\frac{3}{4}\,\Big(\,
\hat{y}_{1}\hat{y}_{2}\hat{y}_{2}\hat{y}_{2}
+\hat{y}_{2}\hat{y}_{1}\hat{y}_{2}\hat{y}_{2}
+\hat{y}_{2}\hat{y}_{2}\hat{y}_{1}\hat{y}_{2}
+\hat{y}_{2}\hat{y}_{2}\hat{y}_{2}\hat{y}_{1}
\,\Big)\,k
\otimes 
\left(\begin{array}{cc}
1 & 0 \\
0 & -1 \\
\end{array}\right),
\nonu\\
\Phi^{(2),12}_{1,0} &=& 
-\frac{1}{2}\,\Big(\,
\hat{y}_{1}\hat{y}_{1}\hat{y}_{2}\hat{y}_{2}
+\hat{y}_{1}\hat{y}_{2}\hat{y}_{1}\hat{y}_{2}
+\hat{y}_{1}\hat{y}_{2}\hat{y}_{2}\hat{y}_{1}
+\hat{y}_{2}\hat{y}_{1}\hat{y}_{1}\hat{y}_{2}
+\hat{y}_{2}\hat{y}_{1}\hat{y}_{2}\hat{y}_{1}
+\hat{y}_{2}\hat{y}_{2}\hat{y}_{1}\hat{y}_{1}
\,\Big)\,k
\nonu \\
& \otimes & 
\left(\begin{array}{cc}
1 & 0 \\
0 & -1 \\
\end{array}\right),
\nonu\\
\Phi^{(2),12}_{1,+1} &=& 
-\frac{3}{4}\,\Big(\,
\hat{y}_{2}\hat{y}_{1}\hat{y}_{1}\hat{y}_{1}
+\hat{y}_{1}\hat{y}_{2}\hat{y}_{1}\hat{y}_{1}
+\hat{y}_{1}\hat{y}_{1}\hat{y}_{2}\hat{y}_{1}
+\hat{y}_{1}\hat{y}_{1}\hat{y}_{1}\hat{y}_{2}
\,\Big)\,k
\otimes 
\left(\begin{array}{cc}
1 & 0 \\
0 & -1 \\
\end{array}\right),
\nonu\\
\Phi^{(2),12}_{1,+2} &=& 
-3\,\hat{y}_{1}\hat{y}_{1}\hat{y}_{1}\hat{y}_{1}\,k
\otimes 
\left(\begin{array}{cc}
1 & 0 \\
0 & -1 \\
\end{array}\right),
\qquad
\Phi^{(2),13}_{1,-2} = 
-3\,\hat{y}_{2}\hat{y}_{2}\hat{y}_{2}\hat{y}_{2}
\otimes 
\left(\begin{array}{cc}
0 & 1 \\
1 & 0 \\
\end{array}\right),
\nonu\\
\Phi^{(2),13}_{1,-1} &=& 
-\frac{3}{4}\,\Big(\,
\hat{y}_{1}\hat{y}_{2}\hat{y}_{2}\hat{y}_{2}
+\hat{y}_{2}\hat{y}_{1}\hat{y}_{2}\hat{y}_{2}
+\hat{y}_{2}\hat{y}_{2}\hat{y}_{1}\hat{y}_{2}
+\hat{y}_{2}\hat{y}_{2}\hat{y}_{2}\hat{y}_{1}
\,\Big)
\otimes 
\left(\begin{array}{cc}
0 & 1 \\
1 & 0 \\
\end{array}\right),
\nonu\\
\Phi^{(2),13}_{1,0} &=& 
-\frac{1}{2}\,\Big(\,
\hat{y}_{1}\hat{y}_{1}\hat{y}_{2}\hat{y}_{2}
+\hat{y}_{1}\hat{y}_{2}\hat{y}_{1}\hat{y}_{2}
+\hat{y}_{1}\hat{y}_{2}\hat{y}_{2}\hat{y}_{1}
+\hat{y}_{2}\hat{y}_{1}\hat{y}_{1}\hat{y}_{2}
+\hat{y}_{2}\hat{y}_{1}\hat{y}_{2}\hat{y}_{1}
+\hat{y}_{2}\hat{y}_{2}\hat{y}_{1}\hat{y}_{1}
\,\Big)
\nonu \\
& \otimes & 
\left(\begin{array}{cc}
0 & 1 \\
1 & 0 \\
\end{array}\right),
\nonu\\
\Phi^{(2),13}_{1,+1} &=& 
-\frac{3}{4}\,\Big(\,
\hat{y}_{2}\hat{y}_{1}\hat{y}_{1}\hat{y}_{1}
+\hat{y}_{1}\hat{y}_{2}\hat{y}_{1}\hat{y}_{1}
+\hat{y}_{1}\hat{y}_{1}\hat{y}_{2}\hat{y}_{1}
+\hat{y}_{1}\hat{y}_{1}\hat{y}_{1}\hat{y}_{2}
\,\Big)
\otimes 
\left(\begin{array}{cc}
0 & 1 \\
1 & 0 \\
\end{array}\right),
\nonu\\
\Phi^{(2),13}_{1,+2} &=& 
-3\,\hat{y}_{1}\hat{y}_{1}\hat{y}_{1}\hat{y}_{1}
\otimes 
\left(\begin{array}{cc}
0 & 1 \\
1 & 0 \\
\end{array}\right),
\qquad
\Phi^{(2),14}_{1,-2} = 
3i\,\hat{y}_{2}\hat{y}_{2}\hat{y}_{2}\hat{y}_{2}
\otimes 
\left(\begin{array}{cc}
0 & 1 \\
-1 & 0 \\
\end{array}\right),
\nonu\\
\Phi^{(2),14}_{1,-1} &=& 
\frac{3i}{4}\,\Big(\,
\hat{y}_{1}\hat{y}_{2}\hat{y}_{2}\hat{y}_{2}
+\hat{y}_{2}\hat{y}_{1}\hat{y}_{2}\hat{y}_{2}
+\hat{y}_{2}\hat{y}_{2}\hat{y}_{1}\hat{y}_{2}
+\hat{y}_{2}\hat{y}_{2}\hat{y}_{2}\hat{y}_{1}
\,\Big)
\otimes 
\left(\begin{array}{cc}
0 & 1 \\
-1 & 0 \\
\end{array}\right),
\nonu\\
\Phi^{(2),14}_{1,0} &=& 
\frac{i}{2}\,\Big(\,
\hat{y}_{1}\hat{y}_{1}\hat{y}_{2}\hat{y}_{2}
+\hat{y}_{1}\hat{y}_{2}\hat{y}_{1}\hat{y}_{2}
+\hat{y}_{1}\hat{y}_{2}\hat{y}_{2}\hat{y}_{1}
+\hat{y}_{2}\hat{y}_{1}\hat{y}_{1}\hat{y}_{2}
+\hat{y}_{2}\hat{y}_{1}\hat{y}_{2}\hat{y}_{1}
+\hat{y}_{2}\hat{y}_{2}\hat{y}_{1}\hat{y}_{1}
\,\Big)
\nonu \\
& \otimes & 
\left(\begin{array}{cc}
0 & 1 \\
-1 & 0 \\
\end{array}\right),
\nonu\\
\Phi^{(2),14}_{1,+1} &=& 
\frac{3i}{4}\,\Big(\,
\hat{y}_{2}\hat{y}_{1}\hat{y}_{1}\hat{y}_{1}
+\hat{y}_{1}\hat{y}_{2}\hat{y}_{1}\hat{y}_{1}
+\hat{y}_{1}\hat{y}_{1}\hat{y}_{2}\hat{y}_{1}
+\hat{y}_{1}\hat{y}_{1}\hat{y}_{1}\hat{y}_{2}
\,\Big)
\otimes 
\left(\begin{array}{cc}
0 & 1 \\
-1 & 0 \\
\end{array}\right),
\nonu\\
\Phi^{(2),14}_{1,+2} &=& 
3i\,\hat{y}_{1}\hat{y}_{1}\hat{y}_{1}\hat{y}_{1}
\otimes 
\left(\begin{array}{cc}
0 & 1 \\
-1 & 0 \\
\end{array}\right),
\qquad
\Phi^{(2),23}_{1,-2} = 
-3i\,\hat{y}_{2}\hat{y}_{2}\hat{y}_{2}\hat{y}_{2}\,k
\otimes 
\left(\begin{array}{cc}
0 & 1 \\
-1 & 0 \\
\end{array}\right),
\nonu\\
\Phi^{(2),23}_{1,-1} &=& 
-\frac{3i}{4}\,\Big(\,
\hat{y}_{1}\hat{y}_{2}\hat{y}_{2}\hat{y}_{2}
+\hat{y}_{2}\hat{y}_{1}\hat{y}_{2}\hat{y}_{2}
+\hat{y}_{2}\hat{y}_{2}\hat{y}_{1}\hat{y}_{2}
+\hat{y}_{2}\hat{y}_{2}\hat{y}_{2}\hat{y}_{1}
\,\Big) \,k
\otimes
\left(\begin{array}{cc}
0 & 1 \\
-1 & 0 \\
\end{array}\right),
\nonu\\
\Phi^{(2),23}_{1,0} &=& 
-\frac{i}{2}\,\Big(\,
\hat{y}_{1}\hat{y}_{1}\hat{y}_{2}\hat{y}_{2}
+\hat{y}_{1}\hat{y}_{2}\hat{y}_{1}\hat{y}_{2}
+\hat{y}_{1}\hat{y}_{2}\hat{y}_{2}\hat{y}_{1}
+\hat{y}_{2}\hat{y}_{1}\hat{y}_{1}\hat{y}_{2}
+\hat{y}_{2}\hat{y}_{1}\hat{y}_{2}\hat{y}_{1}
+\hat{y}_{2}\hat{y}_{2}\hat{y}_{1}\hat{y}_{1}
\,\Big)\,k
\nonu \\
&\otimes& 
\left(\begin{array}{cc}
0 & 1 \\
-1 & 0 \\
\end{array}\right),
\nonu\\
\Phi^{(2),23}_{1,+1} &=& 
-\frac{3i}{4}\,\Big(\,
\hat{y}_{2}\hat{y}_{1}\hat{y}_{1}\hat{y}_{1}
+\hat{y}_{1}\hat{y}_{2}\hat{y}_{1}\hat{y}_{1}
+\hat{y}_{1}\hat{y}_{1}\hat{y}_{2}\hat{y}_{1}
+\hat{y}_{1}\hat{y}_{1}\hat{y}_{1}\hat{y}_{2}
\,\Big) \,k
\otimes 
\left(\begin{array}{cc}
0 & 1 \\
-1 & 0 \\
\end{array}\right),
\nonu\\
\Phi^{(2),23}_{1,+2} &=& 
-3i\,\hat{y}_{1}\hat{y}_{1}\hat{y}_{1}\hat{y}_{1} \,k
\otimes 
\left(\begin{array}{cc}
0 & 1 \\
-1 & 0 \\
\end{array}\right),
\qquad
\Phi^{(2),24}_{1,-2} = 
-3\,\hat{y}_{2}\hat{y}_{2}\hat{y}_{2}\hat{y}_{2}\,k
\otimes 
\left(\begin{array}{cc}
0 & 1 \\
1 & 0 \\
\end{array}\right),
\nonu\\
\Phi^{(2),24}_{1,-1} &=& 
-\frac{3}{4}\,\Big(\,
\hat{y}_{1}\hat{y}_{2}\hat{y}_{2}\hat{y}_{2}
+\hat{y}_{2}\hat{y}_{1}\hat{y}_{2}\hat{y}_{2}
+\hat{y}_{2}\hat{y}_{2}\hat{y}_{1}\hat{y}_{2}
+\hat{y}_{2}\hat{y}_{2}\hat{y}_{2}\hat{y}_{1}
\,\Big)\,k
\otimes 
\left(\begin{array}{cc}
0 & 1 \\
1 & 0 \\
\end{array}\right),
\nonu\\
\Phi^{(2),24}_{1,0} &=& 
-\frac{1}{2}\,\Big(\,
\hat{y}_{1}\hat{y}_{1}\hat{y}_{2}\hat{y}_{2}
+\hat{y}_{1}\hat{y}_{2}\hat{y}_{1}\hat{y}_{2}
+\hat{y}_{1}\hat{y}_{2}\hat{y}_{2}\hat{y}_{1}
+\hat{y}_{2}\hat{y}_{1}\hat{y}_{1}\hat{y}_{2}
+\hat{y}_{2}\hat{y}_{1}\hat{y}_{2}\hat{y}_{1}
+\hat{y}_{2}\hat{y}_{2}\hat{y}_{1}\hat{y}_{1}
\,\Big)\,k
\nonu \\
&\otimes& 
\left(\begin{array}{cc}
0 & 1 \\
1 & 0 \\
\end{array}\right),
\nonu\\
\Phi^{(2),24}_{1,+1} &=& 
-\frac{3}{4}\,\Big(\,
\hat{y}_{2}\hat{y}_{1}\hat{y}_{1}\hat{y}_{1}
+\hat{y}_{1}\hat{y}_{2}\hat{y}_{1}\hat{y}_{1}
+\hat{y}_{1}\hat{y}_{1}\hat{y}_{2}\hat{y}_{1}
+\hat{y}_{1}\hat{y}_{1}\hat{y}_{1}\hat{y}_{2}
\,\Big)\,k
\otimes 
\left(\begin{array}{cc}
0 & 1 \\
1 & 0 \\
\end{array}\right),
\nonu\\
\Phi^{(2),24}_{1,+2} &=& 
-3\,\hat{y}_{1}\hat{y}_{1}\hat{y}_{1}\hat{y}_{1}\,k
\otimes 
\left(\begin{array}{cc}
0 & 1 \\
1 & 0 \\
\end{array}\right),
\qquad
\Phi^{(2),34}_{1,-2} = 
3\,\hat{y}_{2}\hat{y}_{2}\hat{y}_{2}\hat{y}_{2}
\otimes 
\left(\begin{array}{cc}
1 & 0 \\
0 & -1 \\
\end{array}\right),
\nonu\\
\Phi^{(2),34}_{1,-1} &=& 
\frac{3}{4}\,\Big(\,
\hat{y}_{1}\hat{y}_{2}\hat{y}_{2}\hat{y}_{2}
+\hat{y}_{2}\hat{y}_{1}\hat{y}_{2}\hat{y}_{2}
+\hat{y}_{2}\hat{y}_{2}\hat{y}_{1}\hat{y}_{2}
+\hat{y}_{2}\hat{y}_{2}\hat{y}_{2}\hat{y}_{1}
\,\Big)
\otimes 
\left(\begin{array}{cc}
1 & 0 \\
0 & -1 \\
\end{array}\right),
\nonu\\
\Phi^{(2),34}_{1,0} &=& 
\frac{1}{2}\,\Big(\,
\hat{y}_{1}\hat{y}_{1}\hat{y}_{2}\hat{y}_{2}
+\hat{y}_{1}\hat{y}_{2}\hat{y}_{1}\hat{y}_{2}
+\hat{y}_{1}\hat{y}_{2}\hat{y}_{2}\hat{y}_{1}
+\hat{y}_{2}\hat{y}_{1}\hat{y}_{1}\hat{y}_{2}
+\hat{y}_{2}\hat{y}_{1}\hat{y}_{2}\hat{y}_{1}
+\hat{y}_{2}\hat{y}_{2}\hat{y}_{1}\hat{y}_{1}
\,\Big)
\nonu \\
& \otimes & 
\left(\begin{array}{cc}
1 & 0 \\
0 & -1 \\
\end{array}\right),
\nonu\\
\Phi^{(2),34}_{1,+1} &=& 
\frac{3}{4}\,\Big(\,
\hat{y}_{2}\hat{y}_{1}\hat{y}_{1}\hat{y}_{1}
+\hat{y}_{1}\hat{y}_{2}\hat{y}_{1}\hat{y}_{1}
+\hat{y}_{1}\hat{y}_{1}\hat{y}_{2}\hat{y}_{1}
+\hat{y}_{1}\hat{y}_{1}\hat{y}_{1}\hat{y}_{2}
\,\Big)
\otimes 
\left(\begin{array}{cc}
1 & 0 \\
0 & -1 \\
\end{array}\right),
\nonu\\
\Phi^{(2),34}_{1,+2} &=& 
3\,\hat{y}_{1}\hat{y}_{1}\hat{y}_{1}\hat{y}_{1}
\otimes 
\left(\begin{array}{cc}
1 & 0 \\
0 & -1 \\
\end{array}\right),
\qquad
\tilde{\Phi}^{(2),1}_{\frac{3}{2},-\frac{5}{2}}
=
-3i\,e^{i\frac{\pi}{4}}\,\hat{y}_{2}\hat{y}_{2}\hat{y}_{2}\hat{y}_{2}\hat{y}_{2}\,k
\otimes 
\left(\begin{array}{cc}
1 & 0 \\
0 & -1 \\
\end{array}\right),
\nonu \\
\tilde{\Phi}^{(2),1}_{\frac{3}{2},-\frac{3}{2}} &
= &
-\frac{3i}{5}\,e^{i\frac{\pi}{4}}\,
\hat{y}_{(1}\hat{y}_{2}\hat{y}_{2}\hat{y}_{2}\hat{y}_{2)}
\,k
 \otimes  
\left(\begin{array}{cc}
1 & 0 \\
0 & -1 \\
\end{array}\right),
\nonu \\
\tilde{\Phi}^{(2),1}_{\frac{3}{2},-\frac{1}{2}}
& = & 
-\frac{3i}{10}\,e^{i\frac{\pi}{4}}\,
\hat{y}_{(1}\hat{y}_{1}\hat{y}_{2}\hat{y}_{2}\hat{y}_{2)}
\,k
\otimes 
\left(\begin{array}{cc}
1 & 0 \\
0 & -1 \\
\end{array}\right),
\nonu\\
\tilde{\Phi}^{(2),1}_{\frac{3}{2},+\frac{1}{2}}
&
=& 
-\frac{3i}{10}\,e^{i\frac{\pi}{4}}\,
\hat{y}_{(2}\hat{y}_{2}\hat{y}_{1}\hat{y}_{1}\hat{y}_{1)}
\,k
\otimes 
\left(\begin{array}{cc}
1 & 0 \\
0 & -1 \\
\end{array}\right),
\nonu\\
\tilde{\Phi}^{(2),1}_{\frac{3}{2},+\frac{3}{2}}
&
=& 
-\frac{3i}{5}\,e^{i\frac{\pi}{4}}\,
\hat{y}_{(2}\hat{y}_{1}\hat{y}_{1}\hat{y}_{1}\hat{y}_{1)}
\,k
 \otimes  
\left(\begin{array}{cc}
1 & 0 \\
0 & -1 \\
\end{array}\right),
\tilde{\Phi}^{(2),1}_{\frac{3}{2},+\frac{5}{2}}
=
-3i\,e^{i\frac{\pi}{4}}\,\hat{y}_{1}\hat{y}_{1}\hat{y}_{1}\hat{y}_{1}\hat{y}_{1}\,k
\otimes 
\left(\begin{array}{cc}
1 & 0 \\
0 & -1 \\
\end{array}\right),
\nonu\\
\tilde{\Phi}^{(2),2}_{\frac{3}{2},-\frac{5}{2}}
&
=& 
-3\,e^{i\frac{\pi}{4}}\,\hat{y}_{2}\hat{y}_{2}\hat{y}_{2}\hat{y}_{2}\hat{y}_{2}
\otimes 
\left(\begin{array}{cc}
1 & 0 \\
0 & 1 \\
\end{array}\right),
\tilde{\Phi}^{(2),2}_{\frac{3}{2},-\frac{3}{2}}
=
-\frac{3}{5}\,e^{i\frac{\pi}{4}}\,
\hat{y}_{(1}\hat{y}_{2}\hat{y}_{2}\hat{y}_{2}\hat{y}_{2)}
\otimes
\left(\begin{array}{cc}
1 & 0 \\
0 & 1 \\
\end{array}\right),
\nonu\\
\tilde{\Phi}^{(2),2}_{\frac{3}{2},-\frac{1}{2}}
&
=& 
-\frac{3}{10}\,e^{i\frac{\pi}{4}}\,
\hat{y}_{(1}\hat{y}_{1}\hat{y}_{2}\hat{y}_{2}\hat{y}_{2)}
\otimes 
\left(\begin{array}{cc}
1 & 0 \\
0 & 1 \\
\end{array}\right),
\tilde{\Phi}^{(2),2}_{\frac{3}{2},+\frac{1}{2}}
=
-\frac{3}{10}\,e^{i\frac{\pi}{4}}\,
\hat{y}_{(2}\hat{y}_{2}\hat{y}_{1}\hat{y}_{1}\hat{y}_{1)}
\otimes 
\left(\begin{array}{cc}
1 & 0 \\
0 & 1 \\
\end{array}\right),
\nonu\\
\tilde{\Phi}^{(2),2}_{\frac{3}{2},+\frac{3}{2}}
&
=& 
-\frac{3}{5}\,e^{i\frac{\pi}{4}}\,
\hat{y}_{(2}\hat{y}_{1}\hat{y}_{1}\hat{y}_{1}\hat{y}_{1)}
\otimes
\left(\begin{array}{cc}
1 & 0 \\
0 & 1 \\
\end{array}\right),
\tilde{\Phi}^{(2),2}_{\frac{3}{2},+\frac{5}{2}}
= 
-3\,e^{i\frac{\pi}{4}}\,\hat{y}_{1}\hat{y}_{1}\hat{y}_{1}\hat{y}_{1}\hat{y}_{1}
\otimes 
\left(\begin{array}{cc}
1 & 0 \\
0 & 1 \\
\end{array}\right),
\nonu \\
\tilde{\Phi}^{(2),3}_{\frac{3}{2},-\frac{5}{2}}
& = &
-3\,e^{i\frac{\pi}{4}}\,\hat{y}_{2}\hat{y}_{2}\hat{y}_{2}\hat{y}_{2}\hat{y}_{2}\,k
\otimes 
\left(\begin{array}{cc}
0 & 1 \\
-1 &0 \\
\end{array}\right),
\tilde{\Phi}^{(2),3}_{\frac{3}{2},-\frac{3}{2}}
=
-\frac{3}{5}\,e^{i\frac{\pi}{4}}\,
\hat{y}_{(1}\hat{y}_{2}\hat{y}_{2}\hat{y}_{2}\hat{y}_{2)}
\,k
\otimes 
\left(\begin{array}{cc}
0 & 1 \\
-1 &0 \\
\end{array}\right),
\nonu\\
\tilde{\Phi}^{(2),3}_{\frac{3}{2},-\frac{1}{2}}
&
=& 
-\frac{3}{10}\,e^{i\frac{\pi}{4}}\,
\hat{y}_{(1}\hat{y}_{1}\hat{y}_{2}\hat{y}_{2}\hat{y}_{2)}
\,k
\otimes 
\left(\begin{array}{cc}
0 & 1 \\
-1 &0 \\
\end{array}\right),
\nonu\\
\tilde{\Phi}^{(2),3}_{\frac{3}{2},+\frac{1}{2}}
&
=& 
-\frac{3}{10}\,e^{i\frac{\pi}{4}}\,
\hat{y}_{(2}\hat{y}_{2}\hat{y}_{1}\hat{y}_{1}\hat{y}_{1)}
\,k
\otimes 
\left(\begin{array}{cc}
0 & 1 \\
-1 &0 \\
\end{array}\right),
\nonu\\
\tilde{\Phi}^{(2),3}_{\frac{3}{2},+\frac{3}{2}}
&
=& 
-\frac{3}{5}\,e^{i\frac{\pi}{4}}\,
\hat{y}_{(2}\hat{y}_{1}\hat{y}_{1}\hat{y}_{1}\hat{y}_{1)}
\otimes
\left(\begin{array}{cc}
0 & 1 \\
-1 &0 \\
\end{array}\right),
\nonu\\
\tilde{\Phi}^{(2),3}_{\frac{3}{2},+\frac{5}{2}}
&
=& 
-3\,e^{i\frac{\pi}{4}}\,\hat{y}_{1}\hat{y}_{1}\hat{y}_{1}\hat{y}_{1}\hat{y}_{1}\,k
\otimes 
\left(\begin{array}{cc}
0 & 1 \\
-1 &0 \\
\end{array}\right),
\tilde{\Phi}^{(2),4}_{\frac{3}{2},-\frac{5}{2}}
=
3i\,e^{i\frac{\pi}{4}}\,\hat{y}_{2}\hat{y}_{2}\hat{y}_{2}\hat{y}_{2}\hat{y}_{2}\,k
\otimes 
\left(\begin{array}{cc}
0 & 1 \\
1 &0 \\
\end{array}\right),
\nonu\\
\tilde{\Phi}^{(2),4}_{\frac{3}{2},-\frac{3}{2}}
&
=& 
\frac{3i}{5}\,e^{i\frac{\pi}{4}}\,
\hat{y}_{(1}\hat{y}_{2}\hat{y}_{2}\hat{y}_{2}\hat{y}_{2)}
\,k
\otimes 
\left(\begin{array}{cc}
0 & 1 \\
1 &0 \\
\end{array}\right),
\tilde{\Phi}^{(2),4}_{\frac{3}{2},-\frac{1}{2}}
=
\frac{3i}{10}\,e^{i\frac{\pi}{4}}\,
\hat{y}_{(1}\hat{y}_{1}\hat{y}_{2}\hat{y}_{2}\hat{y}_{2)}
\,k
\otimes 
\left(\begin{array}{cc}
0 & 1 \\
1 &0 \\
\end{array}\right),
\nonu\\
\tilde{\Phi}^{(2),4}_{\frac{3}{2},+\frac{1}{2}}
&
=& 
\frac{3i}{10}\,e^{i\frac{\pi}{4}}\,
\hat{y}_{(2}\hat{y}_{2}\hat{y}_{1}\hat{y}_{1}\hat{y}_{1)}
\,k
\otimes 
\left(\begin{array}{cc}
0 & 1 \\
1 &0 \\
\end{array}\right),
\tilde{\Phi}^{(2),4}_{\frac{3}{2},+\frac{3}{2}}
= 
\frac{3i}{5}\,e^{i\frac{\pi}{4}}\,
\hat{y}_{(2}\hat{y}_{1}\hat{y}_{1}\hat{y}_{1}\hat{y}_{1)}
\,k
\otimes 
\left(\begin{array}{cc}
0 & 1 \\
1 &0 \\
\end{array}\right),
\nonu\\
\tilde{\Phi}^{(2),4}_{\frac{3}{2},+\frac{5}{2}}
&
=& 
3i\,e^{i\frac{\pi}{4}}\,\hat{y}_{1}\hat{y}_{1}\hat{y}_{1}\hat{y}_{1}\hat{y}_{1}\,k
\otimes 
\left(\begin{array}{cc}
0 & 1 \\
1 &0 \\
\end{array}\right),
\qquad
\tilde{\Phi}^{(2)}_{2,-3}
= 
-3\,\hat{y}_{2}\hat{y}_{2}\hat{y}_{2}\hat{y}_{2}\hat{y}_{2}\hat{y}_{2}
\otimes 
\left(\begin{array}{cc}
1 & 0 \\
0 & 1 \\
\end{array}\right),
\nonu\\
\tilde{\Phi}^{(2)}_{2,-2}
&
=& 
-\frac{1}{2}\,
\hat{y}_{(1}
\hat{y}_{2}\hat{y}_{2}\hat{y}_{2}\hat{y}_{2}\hat{y}_{2)}
\otimes 
\left(\begin{array}{cc}
1 & 0 \\
0 & 1 \\
\end{array}\right),
\qquad
\tilde{\Phi}^{(2)}_{2,-1}
= 
-\frac{1}{5}\,
\hat{y}_{(1}\hat{y}_{1}\hat{y}_{2}\hat{y}_{2}\hat{y}_{2}
\hat{y}_{2)}
\otimes 
\left(\begin{array}{cc}
1 & 0 \\
0 & 1 \\
\end{array}\right),
\nonu\\
\tilde{\Phi}^{(2)}_{2,0}
&
=& 
-\frac{3}{20}\,
\hat{y}_{(1}\hat{y}_{1}\hat{y}_{1}\hat{y}_{2}\hat{y}_{2}
\hat{y}_{2)}
\otimes 
\left(\begin{array}{cc}
1 & 0 \\
0 & 1 \\
\end{array}\right),
\qquad
\tilde{\Phi}^{(2)}_{2,+1}
=
-\frac{1}{5}\,
\hat{y}_{(2}\hat{y}_{2}
\hat{y}_{1}\hat{y}_{1}\hat{y}_{1}\hat{y}_{1)}
\otimes 
\left(\begin{array}{cc}
1 & 0 \\
0 & 1 \\
\end{array}\right),
\nonu\\
\tilde{\Phi}^{(2)}_{2,+2}
&
=& 
-\frac{1}{2}\Big(\,
\hat{y}_{(2}\hat{y}_{1}\hat{y}_{1}\hat{y}_{1}\hat{y}_{1}
\hat{y}_{1)}
\otimes 
\left(\begin{array}{cc}
1 & 0 \\
0 & 1 \\
\end{array}\right),
\qquad
\tilde{\Phi}^{(2)}_{2,+3}
= 
-3\,\hat{y}_{1}\hat{y}_{1}\hat{y}_{1}\hat{y}_{1}\hat{y}_{1}\hat{y}_{1}
\otimes 
\left(\begin{array}{cc}
1 & 0 \\
0 & 1 \\
\end{array}\right).
\label{sequalto2}
\eea 
The symmetrized product for the oscillators
in these expressions (\ref{sequalto2})
is used. 
For the third and more ${\cal N}=4$ higher spin generators,
we can use (\ref{generalhigherspinoscillator}).

\section{The ${\cal N}=2$
  wedge subalgebra of ${\cal W}_{\infty}^{{\cal N}=4}[\la]$
  algebra }

After we consider the subalgebras of
${\cal W}_{\infty}^{{\cal N}=4}[\la]$, we will compare it with
${\cal N}=2$ higher spin algebra.

\subsection{The ${\cal N}=2$ wedge subalgebra of
 ${\cal W}_{\infty}^{{\cal N}=4}[\la]$ algebra}

In this Appendix, we select the relevant (anti)commutators
appearing in Appendix $G$ with ${\cal N}=2$ supersymmetry. 
The ${\cal N}=2$ superconformal algebra, by recalling
the description of sections $5$ and $7$, is described by
\bea
\Big[L_m,L_n\Big]
&
=&
(m-n)\,L_{m+n}
-
\mathtt{
m(m^2-1)\,\frac{N}{2}\,(\lambda-1)\,\delta_{m+n}}
\,,
\nonu\\
\Big[L_m,\Phi^{(1)}_{0,n}\Big]
&
=&
-\mathtt{n\,\Phi^{(1)}_{0,m+n}}\,,
\qquad
\Big[L_m,\Phi^{(1),2}_{\frac{1}{2},r}\Big]
=
(\frac{m}{2}-r)\,\Phi^{(1),2}_{\frac{1}{2},m+r}\,,
\nonu\\
\Big[L_m,G^{2}_{r}\Big]
&
=&
(\frac{m}{2}-r)\,G^{2}_{m+r}\,,
\qquad
\Big[\Phi^{(1)}_{0,m},\Phi^{(1)}_{0,n}\Big]
=
\mathtt{
-2\,m\,N (\lambda-1)\,\delta_{m+n}
}\,,
\nonu\\
\Big[\Phi^{(1)}_{0,m},\Phi^{(1),2}_{\frac{1}{2},r}\Big]
&
=&
G^2_{m+r}\,,
\qquad
\Big[\Phi^{(1)}_{0,m},G^{2}_{\frac{1}{2},r}\Big]
=
\Phi^{(1),2}_{\frac{1}{2},m+r}\,,
\nonu\\
\Big\{\Phi^{(1),2}_{\frac{1}{2},r},\Phi^{(1),2}_{\frac{1}{2},s}\Big\}
&
=&
-2L_{r+s}-
\mathtt{
(4m^2-1)\,\frac{N}{2}\,(\lambda-1)\,\delta_{r+s}
}\,,\nonu\\
\Big\{G^2_r,G^2_s\Big\}
&
=&
2L_{r+s}+
\mathtt{
(4m^2-1)\,
\frac{N}{2}\,(\lambda-1)}
\,\delta_{r+s}\,,
\nonu\\
\Big\{\Phi^{(1),2}_{\frac{1}{2},r},G^2_s\Big\}
&
=&
(r-s)\,\Phi^{(1)}_{0,r+s}\,.
\label{n2superconformal}
\eea

The primary condition for the higher spin currents
under the stress energy tensor can be summarized by
\bea
\Big[L_m,\Phi^{(2)}_{0,n}\Big]
&
=&
(m-n)\,\Phi^{(2)}_{0,m+n}\,,
\qquad
\Big[L_m,\Phi^{(2),2}_{\frac{1}{2},r}\Big]
=
(\frac{3m}{2}-r)\,\Phi^{(2),2}_{\frac{1}{2},m+r}\,,
\nonu\\
\Big[L_m,\tilde{\Phi}^{(1),2}_{\frac{3}{2},r}\Big]
&
=&
(\frac{3m}{2}-r)\,\tilde{\Phi}^{(1),2}_{\frac{3}{2},m+r}\,,
\qquad
\Big[L_m,\tilde{\Phi}^{(1)}_{2,n}\Big]
=
(2m-n)\,\tilde{\Phi}^{(1)}_{2,m+n}\,.
\label{bosonicprimary}
\eea

The remaining ${\cal N}=2$ primary condition can be
described as 
\bea
\Big[\Phi^{(1)}_{0,m},\Phi^{(2)}_{0,n}\Big]
&
=&
0\,,
\qquad
\Big[\Phi^{(1)}_{0,m},\Phi^{(2),2}_{\frac{1}{2},r}\Big]
=
-2\, \tilde{\Phi}^{(1),2}_{\frac{3}{2},m+r}\,,
\nonu\\
\Big[\Phi^{(1)}_{0,m},\tilde{\Phi}^{(1),2}_{\frac{3}{2},r}\Big]
&
=&
-\frac{1}{2}\, \Phi^{(2),2}_{\frac{1}{2},m+r}
-\mathtt{\frac{8}{3}\,m\,(2\lambda-1)\,G^2}
+\mathtt{4i\,m(3m+2r)\,\lambda(\lambda-1)\,\Gamma^2_{m+r}}\,,
\nonu\\
\Big[\Phi^{(1)}_{0,m},\tilde{\Phi}^{(1)}_{2,n}\Big]
&
=&
\mathtt{2m\, \Phi^{(2)}_{0,m+n}}
+\mathtt{8\,m(2m+n)\,\lambda(\lambda-1)\,U_{m+n}}\,,
\nonu\\
\Big[\Phi^{(1),2}_{\frac{1}{2},r},\Phi^{(2)}_{0,m}\Big]
&
=&
2\, \tilde{\Phi}^{(1),2}_{\frac{3}{2},r+m}\,,
\qquad
\Big\{\Phi^{(1),2}_{\frac{1}{2},r},\Phi^{(2),2}_{\frac{1}{2},s}\Big\}
=
-2\,\tilde{\Phi}^{(1)}_{2,r+s}\,,
\nonu\\
\Big\{\Phi^{(1),2}_{\frac{1}{2},r},\tilde{\Phi}^{(1),2}_{\frac{3}{2},s}\Big\}
&
=&
-\frac{1}{2}\,(3r-s)\,\Phi^{(2)}_{0,r+s}
-\mathtt{4(m^2-\frac{1}{4})\,\lambda(\lambda-1)\,U_{r+s}\,},
\nonu\\
\Big[\Phi^{(1),2}_{\frac{1}{2},r}, \tilde{\Phi}^{(1)}_{2,m}\Big]
&
=&
\frac{1}{2}\,(4r-m)\,\Phi^{(2),2}_{\frac{1}{2},r+m}
+\mathtt{\frac{1}{6}\,(r^2-\frac{1}{4})\,(2\lambda-1)\,G^2_{r+m}}
  \nonu \\
  & - &
  \mathtt{
    \frac{i}{4}\,(r^2-1)(4r+3m)\lambda(\lambda-1)\,\Gamma^2_{r+m}\,},
\nonu\\
\Big[G^{2}_{r},\Phi^{(2)}_{0,m}\Big]
&
=&
- \Phi^{(2),2}_{\frac{1}{2},r+m}\,,
\qquad
\Big\{G^{2}_{r},\Phi^{(2),2}_{\frac{1}{2},s}\Big\}
=
-(3r-s)\,\Phi^{(2)}_{0,r+s}\,,
\nonu\\
\Big\{G^{2}_{r},\tilde{\Phi}^{(1),2}_{\frac{3}{2},s}\Big\}
&
=&
-\,\tilde{\Phi}^{(1)}_{2,r+s}
-\frac{8}{3}\,(r^2-\frac{1}{4})(2\lambda-1)\,\Phi^{(1)}_{0,r+s}
\,,
\nonu\\
\Big[G^{2}_{r},\tilde{\Phi}^{(1)}_{2,m}\Big]
&
=&
-(4r-m)\,\tilde{\Phi}^{(1),2}_{\frac{3}{2},r+m}
+\frac{8}{3}\,(r^2-\frac{1}{4})\,(2\lambda-1)\,\Phi^{(1),2}_{\frac{1}{2},r+m}
\,.
\label{remainingn2primary}
\eea
The whole ${\cal N}=2$ primary conditions are given by
(\ref{bosonicprimary}) and (\ref{remainingn2primary}).

The (anti)commutators between the ${\cal N}=2$
higher spin currents with the descriptions of sections $5$
and $7$
are summarized by
\bea
\Big[\Phi^{(2)}_{0,m},\Phi^{(2)}_{0,n}\Big]
&
=&
\frac{8}{3}\,(m-n)(2\lambda-1)\,\Phi^{(2)}_{0,m+n}
-\frac{64}{9}\,(m-n)(\lambda+1)(\lambda-2)\,L_{m+n}
\nonu\\
&
+&
\mathtt{
m(m^2-1)\,
\frac{32\,N}{9}\,(\lambda+1)(\lambda-1)(\lambda-2)
\,\delta_{m+n}
}
\,,
\nonu\\
\Big[\Phi^{(2)}_{0,m},\tilde{\Phi}^{(1),2}_{\frac{3}{2},r}\Big]
&
=&
-\frac{1}{6}\,\Phi^{(3),2}_{\frac{1}{2},m+r}
+\frac{8}{15}\,(3m-2r)(2\lambda-1)\,\tilde{\Phi}^{(1),2}_{\frac{3}{2},m+r}
\nonu\\
&
+&\frac{2}{9}\,(12m^2-8mr+4r^2-9)(\lambda+1)(\lambda-2)\,\Phi^{(1),2}_{\frac{1}{2},m+r}
\,,
\nonu\\
\Big[\Phi^{(2)}_{0,m},\tilde{\Phi}^{(1)}_{2,n}\Big]
&
=&
\frac{1}{3}\,(2m-n)\,\Phi^{(3)}_{0,m+n}
+\frac{8}{15}\,(2m-n)(2\lambda-1)\,\tilde{\Phi}^{(1)}_{2,m+n}
\nonu\\
&
-&\frac{32}{9}\,m(m^2-1)(\lambda+1)(\lambda-2)\,\Phi^{(1)}_{0,m+n}
\,,
\nonu\\
\Big\{\Phi^{(2),2}_{\frac{1}{2},r},\tilde{\Phi}^{(1),2}_{\frac{3}{2},s}\Big\}
&
=&
-\frac{1}{2}\,(r-s)\,\Phi^{(3)}_{0,r+s}
+\frac{4}{9}\,(r-s)(2r^2+2s^2-5)(\lambda+1)(\lambda-2)\,\Phi^{(1)}_{0,r+s}
\nonu\\
&
+&\frac{8}{15}\,(r-s)(2\lambda-1)\,\tilde{\Phi}^{(1)}_{2,r+s}
\,,
\nonu\\
\Big[\Phi^{(2),2}_{\frac{1}{2},r},\tilde{\Phi}^{(1)}_{2,m}\Big]
&
=&
\frac{1}{6}\,(4r-3m)\,\Phi^{(3),2}_{\frac{1}{2},r+m}
-\frac{4}{15}\,(4r^2-4rm+2m^2-5)(2\lambda-1)\,
\tilde{\Phi}^{(1),2}_{\frac{3}{2},r+m}
\nonu\\
&
-&\frac{2}{9}\,(16r^3-12r^2m+8rm^2-36r+19m-4m^3)(\lambda+1)(\lambda-2)\,\Phi^{(1),2}_{\frac{1}{2},r+m}
\,,
\nonu\\
\Big\{\tilde{\Phi}^{(1),2}_{\frac{3}{2},r},
\tilde{\Phi}^{(1),2}_{\frac{3}{2},s}\Big\}
&
=&
\frac{1}{2}\,\tilde{\Phi}^{(2)}_{2,r+s}
\nonu \\
& + & (6r^2-8rs+6s^2-9)
\Big(
\frac{1}{15}\,(2\lambda-1)\,\Phi^{(2)}_{0,r+s}
-\frac{4}{9}\,(\lambda+1)(\lambda-2)\,L_{r+s}
\Big)
\nonu\\
&+&
\mathtt{
(m^2-\frac{1}{4})(m^2-\frac{9}{4})\frac{8\,N}{9}\,(\lambda+1)(\lambda-1)(\lambda-2)
\,\delta_{r+s}}\,,
\nonu\\
\Big[\tilde{\Phi}^{(1),2}_{\frac{3}{2},r},\tilde{\Phi}^{(1)}_{2,m}\Big]
&
=&
\frac{1}{2}\,(4r-3m)\,\tilde{\Phi}^{(2),2}_{\frac{3}{2},r+m}
-\frac{1}{15}\,(4r^2-4rm+2m^2-5)(2\lambda-1)\,\Phi^{(2),2}_{\frac{1}{2},r+m}
\nonu\\
&
+&
\frac{1}{9}\,(16r^3-12r^2m+8rm^2-36r+19m-4m^3)(\lambda+1)(\lambda-2)\,G^{2}_{r+m}
\,,
\nonu\\
\Big[\tilde{\Phi}^{(1)}_{2,m},\tilde{\Phi}^{(1)}_{2,n}\Big]
&
=&
2\,(m-n)\,\tilde{\Phi}^{(2)}_{2,m+n}
\nonu\\
&
+&(m-n)(2m^2-mn+2n^2-8)\nonu \\
&\times & \Big(
\frac{2}{15}\,(2\lambda-1)\,\Phi^{(2)}_{0,m+n}
-\frac{16}{9}\,(\lambda+1)(\lambda-2)\,L_{m+n}
\Big)
\nonu\\
&
+&
\mathtt{
m(m^2-1)(m^2-4)
\,
\frac{8\,N}{9}\,(\lambda+1)(\lambda-1)(\lambda-2)\,\delta_{m+n}
}
\,.
\label{higherhigher1}
\eea


There are also two relations
\bea
\Big[\Phi^{(2)}_{0,m},\Phi^{(2),2}_{\frac{1}{2},r}\Big]
&=&
-2\,\tilde{\Phi}^{(2),2}_{\frac{3}{2},m+r}
+\frac{8}{15}(3m-2r)(2\lambda-1)\, \Phi^{(2),2}_{\frac{1}{2},m+r}
\nonu\\
&- &
\frac{9}{4}(12m^2-8m r +4r^2-9) (\lambda+1)(\lambda-2)\, G^{2}_{m+r},
\nonu\\
\Big\{\Phi^{(2),2}_{\frac{1}{2},r},\Phi^{(2),2}_{\frac{1}{2},\rho}\Big\}
&=&
-2\,\tilde{\Phi}^{(2)}_{2,r+\rho}
\label{higherhigher2}
\\
& + &
(6r^2-8 r \rho +6 \rho^2-9)\Big(
\frac{16}{9} (\lambda+1)(\lambda-2)\, L_{r+\rho}
+\frac{4}{15} (2\lambda-1)\, \Phi^{(2)}_{0,r+\rho}
\Big),
\nonu
\eea
where we do not see the contributions from the outside of the
wedge.

\subsection{The ${\cal N}=2$ wedge subalgebra of
  ${\cal W}_{\infty}^{{\cal N}=2}[\la]$ algebra and the
  ${\cal N}=2$ higher spin algebra $shs[\la]$}

Let us return to the work of \cite{CG}.
We can collect the relevant (anti)commutators. 
By linear combinations of the above (anti)commutators
in (\ref{n2superconformal})-(\ref{higherhigher2})
with the help of (\ref{n2sca}), (\ref{Spin2identification}),
and (\ref{Spin3identification}), we can check
the corresponding ones written in terms of the notations in
\cite{CG}
\bea
\big[L_m,L_n\big] & = &
(m-n)L_{m+n} +\frac{c}{12}m(m^2-1)\delta_{m+n,0}\ ,
\nonu\\
\big[L_m,J_n\big] & = & \mathtt{-n J_{m+n}}\ , \qquad [L_m,G^\pm_r] =
\left(\frac{m}{2}-r\right) G^\pm_{m+r}\ , 
\nonu\\
\big[J_m,J_n\big] & = &
\mathtt{ m\, \frac{c}{3}\,\delta_{m+n,0}}\ ,\qquad [J_m,G^\pm_r] = \pm G^\pm_{m+r}\ ,
\nonu\\
\big\{G^+_r,G^-_s\big\} & = & 2L_{r+s}+(r-s)J_{r+s} +
\mathtt{
\frac{c}{3}\left(r^2-\frac{1}{4}\right)\delta_{r+s,0}
} \ ,
\qquad  \{G^\pm_r,G^\pm_s\}=0\, ,
\nonu
\\
\big[L_m,W^{s\, 0}_n\big] & = & [(s-1)m-n]W^{s\, 0}_{m+n}\ ,\quad 
\big[L_m,W^{s\, \pm}_r\big]=\left[\left(s-\frac{1}{2}\right)m-r\right] W^{s\, \pm}_{m+r}\ , 
\nonu\\
\big[L_m,W^{s\, 1}_n\big] & = &
    [s m-n]W^{s\, 1}_{m+n}\ ,\quad  [J_m,W^{s\, 0}_n]=0 \ ,\quad 
\big[J_m,W^{s\, \pm}_r]=\pm W^{s\, \pm}_{m+r}\ , 
\nonu\\
\big[J_m,W^{s \, 1}_n\big] & = & \mathtt{ s \,m\, W^{s\, 0}_{m+n}}\ , \quad
\big[G^\pm_r,W^{s\, 0}_n\big]= \mp W^{s\, \pm}_{r+n}\ ,
\nonu\\
\big\{G^\pm_r,W^{s\, \mp}_t\big\} & = &
\pm [(2s-1)r-t] W^{s\, 0}_{r+t}+2 W^{s\, 1}_{r+t}\ , \quad
\big\{G^\pm_r,W^{s\, \pm}_t\big\}= 0\ , \nonu \\
\big [G^\pm_r, W^{s\, 1}_m\big] & = &
\frac{1}{2}[(2r+1)s-2-m ] W^{s\, \pm}_{r+m}\ , 
\nonu
\\
\big[W^{2\, 0}_{m}\, ,\, W^{2\, 0}_{n}\big] & = & 
 (m-n)\mathscr{A}^{[2]}_{m+n}+\mathtt{\frac{c}{12} m(m^2-1)\delta_{m+n,0}} \,,
\nonu\\
\big[W^{2\, 1}_{m}\, ,\, W^{2\, 1}_{n}\big] & = &
\mathtt{ \frac{c}{48} m(m^2-1)(m^2-4)\delta_{m+n,0}}
+(m-n)(2m^2-mn+2n^2-8)\mathscr{B}^{[2]}_{m+n}
\nonu\\
 & + &
(m-n)\left(\mathtt{ \mathscr{B}^{[4]}_{m+n}} -2 c_{22}^3 W^{3\, 1}_{m+n}\right) \,,
\nonu\\
\big[W^{2\, 0}_{m}\, ,\, W^{2\, 1}_{n}\big] & = &
\mathtt{  \mathscr{C}^{[4]}_{m+n}}+
(2m-n)\left( \mathscr{C}^{[3]}_{m+n}-c_{22}^3 W^{3\, 0}_{m+n}\right)
\nonu\\
& + &
\mathtt{ (6m^2-3mn+n^2-4)\mathscr{C}^{[2]}_{m+n}}+\mathtt{ m(m^2-1)\mathscr{C}^{[1]}_{m+n}} \,,
\nonu\\
\big\{W^{2\, +}_{r}\, ,\, W^{2\, -}_{s}\big\} & = &
\left(\mathtt{\mathscr{D}^{[4]}_{r+s}} -2c_{22}^3 W^{3\, 1}_{r+s}\right)
+(r-s)\left(\mathscr{D}^{[3]}_{r+s} - 3c_{22}^3 W^{3\, 0}_{r+s}\right)
\nonu\\
 &
+ & \left(3r^2-4rs+3s^2-\frac{9}{2}\right)\mathscr{D}^{[2]}_{r+s}
+(r-s)\left(r^2+s^2-\frac{5}{2}\right)\mathscr{D}^{[1]}_{r+s}
\nonu\\
&
+ & \mathtt{\frac{c}{12}\left( r^2-\frac{1}{4}\right)\left( r^2-\frac{9}{4}\right)\delta_{r+s,0}} \,,
\nonu\\
\big\{W^{2\, +}_{r}\, ,\, W^{2\, +}_{s}\big\} & = &
\mathtt{\mathscr{E}^{[4]}_{r+s}}\,,
\nonu\\
\big[W^{2\, 0}_{m}\, ,\, W^{2\, +}_{r}\big] & = &
\left(\mathtt{\Phi^{[7/2]}_{m+r}}-c_{22}^3 W^{3\, +}_{m+r}\right)
+\left( \frac{3}{2}m - r\right) \Phi^{[5/2]}_{m+r}
\nonu\\
&
+& \left( 3m^2 - 2mr +r^2-\frac{9}{4}\right) \Phi^{[3/2]}_{m+r} \,,
\nonu\\
\big[W^{2\, +}_{r}\, ,\, W^{2\, 1}_{m}\big] & = &
\mathtt{ \Psi^{[9/2]}_{r+m}} + \left( 2r-\frac{3}{2}m \right)
\left(\mathtt{ \Psi^{[7/2]}_{r+m}} -c_{22}^3 W^{3\, +}_{r+m}\right)\,
\nonu\\
& + & \left( 2r^2 - 2rm +m^2 -\frac{5}{2}\right) \Psi^{[5/2]}_{r+m}
\nonu\\
&+ &
\left( 4r^3 -3r^2m +2rm^2 - m^3 - 9r +\frac{19}{4}m  \right)\Psi^{[3/2]}_{r+m}\ .
\label{finalcomm}
\eea
The first four lines correspond to (\ref{n2superconformal}),
the next five lines correspond to (\ref{bosonicprimary}) and
(\ref{remainingn2primary}), and the last
lines correspond to (\ref{higherhigher1}) and (\ref{higherhigher2}).
Note that the quantities
$\mathscr{A}^{[s]}$,$\mathscr{B}^{[s]}$,$\mathscr{C}^{[s]}$,
$\mathscr{D}^{[s]}$, $\mathscr{E}^{[s]}$, $\Phi^{[s]}$ and $\Psi^{[s]}$
in (\ref{finalcomm})
are found in \cite{Romans}.
Under the large $(N,k)$ 't Hooft-like limit, we have the
following nonzero contributions with explicit $\la$-dependent terms 
\bea
\mathscr{A}^{[2]}
& \rightarrow &
-\frac{1}{3}(\la+1)(\la-2)\,L-\frac{1}{\sqrt{3}}(2\lambda-1)\,W^{2\,0}\,,
\nonu\\
\mathscr{B}^{[2]}
& \rightarrow &
-\frac{1}{12}(\la+1)(\la-2)\,L-\frac{1}{20\sqrt{3}}(2\lambda-1)\,W^{2\,0}\,,
\nonu\\
\mathscr{C}^{[3]}
& \rightarrow & -\frac{1}{5\sqrt{3}}(2\lambda-1)\,W^{2\,1}\,,
\qquad
\mathscr{D}^{[3]}
 \rightarrow \frac{1}{5\sqrt{3}}(2\lambda-1)\,W^{2\,1}\,,
\nonu\\
\mathscr{D}^{[2]}
& \rightarrow &
-\frac{1}{6}(\la+1)(\la-2)\,L-\frac{1}{5\sqrt{3}}(2\lambda-1)\,W^{2\,0}\,,
\nonu\\
\mathscr{D}^{[1]}
& \rightarrow &-\frac{1}{12}(\la+1)(\la-2)\,J \,,
\qquad
\Phi^{[5/2]}
 \rightarrow -\frac{2}{5\sqrt{3}}(2\lambda-1)\,W^{2\,+}\,,
\nonu\\
\Phi^{[3/2]}
& \rightarrow & -\frac{1}{12}(\la+1)(\la-2)\,G^{+}\,,
\qquad
\Psi^{[5/2]} 
 \rightarrow -\frac{1}{10\sqrt{3}}(2\lambda-1)\,W^{2\,+}\,,
\nonu\\
\Psi^{[3/2]} 
& \rightarrow &-\frac{1}{24}(\la+1)(\la-2)\,G^{+}\,.
\label{finallimit}
\eea

Therefore, by substituting
(\ref{finallimit}) into (\ref{finalcomm}),
our ${\cal N}=2$ wedge subalgebra of
${\cal W}_{\infty}^{{\cal N}=4}[\la]$ algebra reproduces
the ${\cal N}=2$ higher spin algebra \cite{FL}, which is equal to
 ${\cal N}=2$ wedge subalgebra of
${\cal W}_{\infty}^{{\cal N}=2}[\la]$ algebra from the observation of
\cite{CG}.

In (\ref{22commutator}), we explicitly
check that the structure constants in the right hand side
can be written in terms of the ones in \cite{FL}.
There are also other (anti)commutators in addition to
(\ref{22commutator}) as follows:
\bea
&&[W^{2\, 0}_{m}\, ,\, W^{2\, +}_{r}]
 = 
-\frac{1}{6}
\sqrt{(1-m)!(1+m)!(\frac{3}{2}-r)!(\frac{3}{2}+r)!}
\nonu\\
&&
\times
\Bigg(
\sum_{m''=-\frac{1}{2}}^{\frac{1}{2}}
\sqrt{\frac{1}{2(\frac{1}{2}-m-r)!(\frac{1}{2}+m+r)!}}\,
\Big(
(\lambda+1)\,f_{T\Psi\Psi}^{1\,\frac{3}{2}\,\frac{1}{2}}
+(\lambda-2)\,f_{U\Psi\Psi}^{1\,\frac{3}{2}\,\frac{1}{2}}
\Big)\,C_{m\,r\,m''}^{1\,\frac{3}{2}\,\frac{1}{2}} \,G^{+}_{m''}
\nonu\\
&&
-
\sum_{m''=-\frac{3}{2}}^{\frac{3}{2}}
\sqrt{\frac{3}{(\frac{3}{2}-m-r)!(\frac{3}{2}+m+r)!}}\,
\Big(
(\lambda+1)\,f_{T\Psi\Psi}^{1\,\frac{3}{2}\,\frac{3}{2}}
+(\lambda-2)\,f_{U\Psi\Psi}^{1\,\frac{3}{2}\,\frac{3}{2}}
\Big)\,C_{m\,r\,m''}^{1\,\frac{3}{2}\,\frac{3}{2}} \,W^{2\,+}_{m''}
\nonu\\
&&
-
\sum_{m''=-\frac{5}{2}}^{\frac{5}{2}}
10\,c_{22,3}\sqrt{\frac{2}{3(\frac{5}{2}-m-r)(\frac{5}{2}+m+r)}}
\Big(
(\lambda+1)f_{T\Psi\Psi}^{1\,\frac{3}{2}\,\frac{5}{2}}
+(\lambda-2)f_{U\Psi\Psi}^{1\,\frac{3}{2}\,\frac{5}{2}}
\Big)C_{m\,r\,m''}^{1\,\frac{3}{2}\,\frac{5}{2}} W^{3\,+}_{m''}
\Bigg),
\nonu \\
&& [W^{2\, 0}_{m}\, ,\, W^{2\, -}_{r}]
 = 
-\frac{1}{6}
\sqrt{(1-m)!(1+m)!(\frac{3}{2}-r)!(\frac{3}{2}+r)!}
\nonu\\
&&
\times
\Bigg(
\sum_{m''=-\frac{1}{2}}^{\frac{1}{2}}
\sqrt{\frac{1}{2(\frac{1}{2}-m-r)!(\frac{1}{2}+m+r)!}}\,
\Big(
(\lambda+1)\,f_{T \bar{\Psi} \bar{\Psi}}^{1\,\frac{3}{2}\,\frac{1}{2}}
+(\lambda-2)\,f_{U \bar{\Psi} \bar{\Psi}}^{1\,\frac{3}{2}\,\frac{1}{2}}
\Big)\,C_{m\,r\,m''}^{1\,\frac{3}{2}\,\frac{1}{2}} \,G^{-}_{m''}
\nonu\\
&&
-
\sum_{m''=-\frac{3}{2}}^{\frac{3}{2}}
\sqrt{\frac{3}{(\frac{3}{2}-m-r)!(\frac{3}{2}+m+r)!}}\,
\Big(
(\lambda+1)\,f_{T \bar{\Psi} \bar{\Psi}}^{1\,\frac{3}{2}\,\frac{3}{2}}
+(\lambda-2)\,f_{U \bar{\Psi} \bar{\Psi}}^{1\,\frac{3}{2}\,\frac{3}{2}}
\Big)\,C_{m\,r\,m''}^{1\,\frac{3}{2}\,\frac{3}{2}} \,W^{2\,-}_{m''}
\nonu\\
&&
-
\sum_{m''=-\frac{5}{2}}^{\frac{5}{2}}
10c_{22,3}\sqrt{\frac{2}{3(\frac{5}{2}-m-r)(\frac{5}{2}+m+r)}}
\Big(
(\lambda+1)f_{T \bar{\Psi} \bar{\Psi}}^{1\,\frac{3}{2}\,\frac{5}{2}}
+(\lambda-2)f_{U \bar{\Psi} \bar{\Psi}}^{1\,\frac{3}{2}\,\frac{5}{2}}
\Big)C_{m\,r\,m''}^{1\,\frac{3}{2}\,\frac{5}{2}} W^{3\,-}_{m''}
\Bigg),
\nonu \\
&& [W^{2\, 0}_{m}\, ,\, W^{2\, 1}_{n}]
 = 
\frac{1}{3\sqrt{30}}\sqrt{\frac{(1-m)!(1-m)!(2-n)!(2+n)!}{(2-m-n)!(2+m+n)!}}
\nonu\\
&&
\times
\Bigg(
5\sqrt{3}\,c_{22,3}\,
\Big(
(\lambda+1)\,f_{T T T}^{1\,2\,2}
-(\lambda-2)\,f_{U U U}^{1\,2\,2}\,
\Big) \,
C_{m\,n\,m+n}^{1\,2\,2}
 \,W^{3\,0}_{m+n}
\nonu\\
&&
-
\Big(
3(\lambda+1)(\lambda-3)\,f_{T T T}^{1\,2\,2}
-3(\lambda+2)(\lambda-2)\,f_{U U U}^{1\,2\,2}
\Big)
\,C_{m\,n\,m+n}^{1\,2\,2}
 \,W^{2\,1}_{m+n}
\Bigg),
\nonu \\
&& [W^{2\, +}_{r}\, ,\, W^{2\, +}_{\rho}]
= 0\,,
\nonu \\
&& [W^{2\, +}_{r}\, ,\, W^{2\, -}_{\rho}]
 =
\frac{1}{3}
\sqrt{(\frac{3}{2}-r)!(\frac{3}{2}+r)!(\frac{3}{2}-\rho)!(\frac{3}{2}+\rho)!}
\nonu\\
&&
\times
\Bigg(
\frac{1}{(2\lambda-1) \sqrt{(-r-\rho)!(r+\rho)!}}
\Big(
-\lambda\,
f_{\Psi\bar{\Psi}T}^{\frac{3}{2}\,\frac{3}{2}\,0}
+(\lambda-1)\,f_{\Psi\bar{\Psi}U}^{\frac{3}{2}
\,\frac{3}{2}\,0}\Big)\,C_{r\,\rho\,0}^{\frac{3}{2}\,\frac{3}{2}\,0} \, J_{0}
\nonu\\
&&
+
\sum_{m''=-1}^{1}
\frac{2}{3\sqrt{(1-r-\rho)!(1+r+\rho)!}}\Big((\lambda-2)\, f_{\Psi\bar{\Psi}T}^{\frac{3}{2}\,\frac{3}{2}\,1}-(\lambda+1)\, f_{\Psi\bar{\Psi}U}^{\frac{3}{2}\,\frac{3}{2}\,1}\Big)\,C_{r\,\rho\,m''}^{\frac{3}{2}\,\frac{3}{2},1}\, L_{m''}
\nonu\\
&&
+
\sum_{m''=-1}^{1}
\frac{2}{\sqrt{3(1-r-\rho)!(1+r+\rho)!}}\Big(f_{\Psi\bar{\Psi}T}^{\frac{3}{2}\,\frac{3}{2}\,1}-f_{\Psi\bar{\Psi}U}^{\frac{3}{2}\,\frac{3}{2}\,1}\Big)\,C_{r\,\rho\,m''}^{\frac{3}{2}\,\frac{3}{2}\,1}\, W_{m''}^{2\,0}
\nonu\\
&&
+
\sum_{m''=-2}^{2}
\frac{4\, c_{22,3}}{\sqrt{(2-r-\rho)!(2+r+\rho)!}}\Big(f_{\Psi\bar{\Psi}T}^{\frac{3}{2}\,\frac{3}{2}\,2}-f_{\Psi\bar{\Psi}U}^{\frac{3}{2}\,\frac{3}{2},2}\Big)\,C_{r\,\rho\,m''}^{\frac{3}{2}\,\frac{3}{2}\,2}\, W_{m''}^{3\,0}
\nonu\\
&&
-
\sum_{m''=-2}^{2}
\frac{4\sqrt{3}}{5\sqrt{(2-r-\rho)!(2+r+\rho)!}}\Big((\lambda-3)\, f_{\Psi\bar{\Psi}T}^{\frac{3}{2}\,\frac{3}{2}\,2}-(\lambda+2)\, f_{\Psi\bar{\Psi}U}^{\frac{3}{2}\,\frac{3}{2}\,2}\Big)\,C_{r\,\rho\,m''}^{\frac{3}{2}\,\frac{3}{2}\,2}\, W_{m''}^{2\,1}
\nonu\\
&&
+
\sum_{m''=-3}^{3}
\frac{40\, c_{22,3}\, c_{23,4}}{\sqrt{3(3-r-\rho)!(3+r+\rho)!}}\Big(f_{\Psi\bar{\Psi}T}^{\frac{3}{2}\,\frac{3}{2}\,3}-f_{\Psi\bar{\Psi}U}^{\frac{3}{2}\,\frac{3}{2}\,3}\Big)\,C_{r\,\rho\,m''}^{\frac{3}{2}\,\frac{3}{2}\,3}\, W_{m''}^{4\,0}
\nonu\\
&&
-
\sum_{m''=-3}^{3}
\frac{40\, c_{22,3}}{7\sqrt{(3-r-\rho)!(3+r+\rho)!}}\Big((\lambda-4)\, f_{\Psi\bar{\Psi}T}^{\frac{3}{2}\,\frac{3}{2}\,3}-(\lambda+3)\, f_{\Psi\bar{\Psi}U}^{\frac{3}{2}\,\frac{3}{2}\,3}\Big)\,C_{r\,\rho\,m''}^{\frac{3}{2}\,\frac{3}{2}\,3}\, W_{m''}^{3\,1}
\Bigg),
\nonu \\
&& [W^{2\, +}_{r}\, ,\, W^{2\, 1}_{m}]
 = \frac{1}{12}\sqrt{(\frac{3}{2}-r)!(\frac{3}{2}+r)!(2-m)!(2+m)!}
 \nonu\\
&&
\times
\Bigg(
\sum_{m''=-\frac{1}{2}}^{\frac{1}{2}}
\frac{1}{\sqrt{2(\frac{1}{2}-r-m)!(\frac{1}{2}+r+m)!}}
\Big(
f_{T \Psi \Psi}^{2\,\frac{3}{2}\,\frac{1}{2}}
+f_{U \Psi \Psi}^{2\,\frac{3}{2}\,\frac{1}{2}}
\Big)\,C_{r\,m\,m''}^{2\,\frac{3}{2}\,\frac{1}{2}}\,G^{+}_{m''}
\nonu\\
&&
-
\sum_{m''=-\frac{3}{2}}^{\frac{3}{2}}
\frac{\sqrt{3}}{\sqrt{(\frac{3}{2}-r-m)!(\frac{3}{2}+r+m)!}}
\Big(
f_{T \Psi \Psi}^{2\,\frac{3}{2}\,\frac{3}{2}}
+f_{U \Psi \Psi}^{2\,\frac{3}{2}\,\frac{3}{2}}
\Big)\,C_{r\,m\,m''}^{2\,\frac{3}{2}\,\frac{3}{2}}\,W^{2\,+}_{m''}
\nonu\\
&&
-
\sum_{m''=-\frac{5}{2}}^{\frac{5}{2}}
\frac{10\,\sqrt{2}\,c_{22,3}}{\sqrt{3(\frac{5}{2}-r-m)!(\frac{5}{2}+r+m)!}}\,
\Big(
f_{T \Psi \Psi}^{2\,\frac{3}{2}\,\frac{5}{2}}
+f_{U \Psi \Psi}^{2\,\frac{3}{2}\,\frac{5}{2}}
\Big)\,C_{r\,m\,m''}^{2\,\frac{3}{2}\,\frac{5}{2}}\,W^{2\,+}_{m''}
\Bigg)
\,,
\nonu \\
&& [W^{2\, -}_{r}\, ,\, W^{2\, -}_{\rho}]
 = 0\,,
\nonu \\
&& [W^{2\, -}_{r}\, ,\, W^{2\, 1}_{m}]
= \frac{1}{12}\sqrt{(\frac{3}{2}-r)!(\frac{3}{2}+r)!(2-m)!(2+m)!}
 \nonu\\
&&
\times
\Bigg(
\sum_{m''=-\frac{1}{2}}^{\frac{1}{2}}
\frac{1}{\sqrt{2(\frac{1}{2}-r-m)!(\frac{1}{2}+r+m)!}}
\Big(
f_{T \bar{\Psi} \bar{\Psi}}^{2\,\frac{3}{2}\,\frac{1}{2}}
+f_{U \bar{\Psi} \bar{\Psi}}^{2\,\frac{3}{2}\,\frac{1}{2}}
\Big)\,C_{r\,m\,m''}^{2\,\frac{3}{2}\,\frac{1}{2}}\,G^{-}_{m''}
\nonu\\
&&
-
\sum_{m''=-\frac{3}{2}}^{\frac{3}{2}}
\frac{\sqrt{3}}{\sqrt{(\frac{3}{2}-r-m)!(\frac{3}{2}+r+m)!}}
\Big(
f_{T \bar{\Psi} \bar{\Psi}}^{2\,\frac{3}{2}\,\frac{3}{2}}
+f_{U \bar{\Psi} \bar{\Psi}}^{2\,\frac{3}{2}\,\frac{3}{2}}
\Big)\,C_{r\,m\,m''}^{2\,\frac{3}{2}\,\frac{3}{2}}\,W^{2\,-}_{m''}
\nonu\\
&&
-
\sum_{m''=-\frac{5}{2}}^{\frac{5}{2}}
\frac{10\,\sqrt{2}\,c_{22,3}}{\sqrt{3(\frac{5}{2}-r-m)!(\frac{5}{2}+r+m)!}}\,
\Big(
f_{T \bar{\Psi} \bar{\Psi}}^{2\,\frac{3}{2}\,\frac{5}{2}}
+f_{U \bar{\Psi} \bar{\Psi}}^{2\,\frac{3}{2}\,\frac{5}{2}}
\Big)\,C_{r\,m\,m''}^{2\,\frac{3}{2}\,\frac{5}{2}}\,W^{2\,-}_{m''}
\Bigg)
\,,
\nonu \\
&& [W^{2\, 1}_{m}\, ,\, W^{2\, 1}_{n}]
 = 
\frac{1}{48}\sqrt{(2-m)!(2+m)!(2-n)!(2+n)!} 
\nonu\\
&&
\times
\Bigg(
\sum_{m''=-1}^{1}
\frac{2}{3\sqrt{(1-m-n)!(1+m+n)!}}\,
\Big(
(\lambda-2)f_{T T T}^{2\,2 \,1}
-
(\lambda+1)f_{U U U}^{2\,2\,1}
\Big)\,C_{r\,m\,m''}^{2\,2\,1}\,L_{m''}
\nonu\\
&&
+
\sum_{m''=-1}^{1}
\frac{2}{3\sqrt{(1-m-n)!(1+m+n)!}}\,
\Big(
f_{T T T}^{2\,2 \,1}
-
f_{U U U}^{2\,2\,1}
\Big)\,C_{r\,m\,m''}^{2\,2\,1}\,W^{2\,0}_{m''}
\nonu\\
&&
-
\sum_{m''=-3}^{3}
\frac{40\,c_{22,3}}{7\sqrt{(3-m-n)!(3+m+n)!}}\,
\Big(
(\lambda-4)f_{T T T}^{2\,2 \,3}
-
(\lambda+3)f_{U U U}^{2\,2\,3}
\Big)\,C_{r\,m\,m''}^{2\,2\,3}\,W^{3\,1}_{m''}
\nonu\\
&&
+
\sum_{m''=-3}^{3}
\frac{4\,c_{22,3}\,c_{23,4}}{\sqrt{3(3-m-n)!(3+m+n)!}}\,
\Big(
f_{T T T}^{2\,2 \,3}
-
f_{U U U}^{2\,2\,3}
\Big)\,C_{r\,m\,m''}^{2\,2\,3}\,W^{4\,0}_{m''}
\Bigg),
\label{finalanticommappendix}
 \eea
 where the $\la$-dependent
 structure constants in \cite{FL} are given by
\bea
f_{TTT}^{1\,1\,1} & = & -2\sqrt{2},
\qquad f_{UUU}^{1\,1\,1}
=
-2\sqrt{2},
\nonu \\
f_{T\Psi\Psi}^{1\,\frac{3}{2}\,\frac{1}{2}} & = &
\frac{1}{\sqrt{3}}(\lambda+1)(\lambda-2), \qquad
f_{T\Psi\Psi}^{1\,\frac{3}{2}\,\frac{3}{2}}=\frac{1}{\sqrt{15}}(\lambda-8), \nonu \\
f_{T\Psi\Psi}^{1\,\frac{3}{2}\,\frac{5}{2}} & = & -\frac{3}{\sqrt{5}},
\qquad
f_{U\Psi\Psi}^{1\,\frac{3}{2}\,\frac{1}{2}}=-\frac{1}{\sqrt{3}}(\lambda+1)(\lambda-2), \nonu \\
f_{U\Psi\Psi}^{1\,\frac{3}{2}\,\frac{3}{2}} & = &
-\frac{1}{\sqrt{15}}(\lambda+7), \qquad
f_{U\Psi\Psi}^{1\,\frac{3}{2}\,\frac{5}{2}}=\frac{3}{\sqrt{5}},
\nonu \\
f_{T\bar{\Psi}\bar{\Psi}}^{1\,\frac{3}{2}\,\frac{1}{2}} & = &
-\frac{1}{\sqrt{3}}(\lambda+1)(\lambda-2), \qquad
f_{T\bar{\Psi}\bar{\Psi}}^{1\,\frac{3}{2}\,\frac{3}{2}}=\frac{1}{\sqrt{15}}(\lambda-8), \nonu \\
f_{T\bar{\Psi}\bar{\Psi}}^{1\,\frac{3}{2}\,\frac{5}{2}} & = &
\frac{3}{\sqrt{5}},
\qquad
f_{U\bar{\Psi}\bar{\Psi}}^{1\,\frac{3}{2}\,\frac{1}{2}}=\frac{1}{\sqrt{3}}(\lambda+1)(\lambda-2), \nonu \\
f_{U\bar{\Psi}\bar{\Psi}}^{1\,\frac{3}{2}\,\frac{3}{2}} & = &
-\frac{1}{\sqrt{15}}(\lambda+7), \qquad
f_{U\bar{\Psi}\bar{\Psi}}^{1\,\frac{3}{2}\,\frac{5}{2}}=-\frac{3}{\sqrt{5}},
\nonu \\
f_{TTT}^{1\,2\,2} & = & -2\sqrt{6}, \qquad
f_{UUU}^{1\,2\,2}=-2\sqrt{6},
\nonu \\
f_{\Psi\bar{\Psi} T}^{\frac{3}{2}\,\frac{3}{2}\,0} & = &
\frac{1}{2}\lambda(\lambda+1)(\lambda-2), \qquad
f_{ \Psi \Psi T}^{\frac{3}{2}\,\frac{3}{2}\,1}=\frac{1}{2\sqrt{5}}(\lambda+1)(\lambda-8), 
\nonu\\
f_{\Psi \Psi T}^{\frac{3}{2}\,\frac{3}{2}\,2} & = &
-\frac{1}{2}(\lambda+4), \qquad
f_{ \Psi \Psi T}^{\frac{3}{2}\,\frac{3}{2}\,3}=-\frac{3}{2\sqrt{5}}, 
\nonu\\
f_{\Psi\bar{\Psi} U}^{\frac{3}{2}\,\frac{3}{2}\,0} & = &
\frac{1}{2}(\lambda+1)(\lambda-1)(\lambda-2), \qquad
f_{\Psi \Psi U}^{\frac{3}{2}\,\frac{3}{2}\,1}=\frac{1}{2\sqrt{5}}(\lambda-2)(\lambda+7), 
\nonu\\
f_{\Psi \Psi U}^{\frac{3}{2}\,\frac{3}{2}\,2} & = &
-\frac{1}{2}(\lambda-5), \qquad
f_{ \Psi \Psi U}^{\frac{3}{2}\,\frac{3}{2}\,3}=-\frac{3}{2\sqrt{5}}, 
\nonu \\
f_{T \Psi \Psi}^{2\,\frac{3}{2}\,\frac{1}{2}}&=&
-\frac{1}{\sqrt{5}}(\lambda+1)(\lambda-2)(\lambda-3), \qquad
f_{T \Psi \Psi}^{2\,\frac{3}{2}\,\frac{3}{2}}=\frac{1}{\sqrt{5}}(\lambda+4)(\lambda-3), \nonu \\
f_{T \Psi \Psi}^{2\,\frac{3}{2}\,\frac{5}{2}} & = &
\sqrt{\frac{3}{35}}(\lambda-11), 
\qquad
f_{U \Psi \Psi}^{2\,\frac{3}{2}\,\frac{1}{2}}=\frac{1}{\sqrt{5}}(\lambda+1)(\lambda+2)(\lambda-2), \nonu \\
f_{U \Psi \Psi}^{2\,\frac{3}{2}\,\frac{3}{2}} & = &
-\frac{1}{\sqrt{5}}(\lambda+2)(\lambda-5), \qquad
f_{U \Psi \Psi}^{2\,\frac{3}{2}\,\frac{5}{2}}=-\sqrt{\frac{3}{35}}(\lambda+10), 
\nonu \\
f_{T \bar{\Psi} \bar{\Psi}}^{2\,\frac{3}{2}\,\frac{1}{2}}&=&
-\frac{1}{\sqrt{5}}(\lambda+1)(\lambda-2)(\lambda-3), \qquad
f_{T \bar{\Psi} \bar{\Psi}}^{2\,\frac{3}{2}\,\frac{3}{2}}=-\frac{1}{\sqrt{5}}(\lambda+4)(\lambda-3), \nonu \\
f_{T \bar{\Psi} \bar{\Psi}}^{2\,\frac{3}{2}\,\frac{5}{2}} & = &
\sqrt{\frac{3}{35}}(\lambda-11), 
\qquad
f_{U \bar{\Psi} \bar{\Psi}}^{2\,\frac{3}{2}\,\frac{1}{2}}=\frac{1}{\sqrt{5}}(\lambda+1)(\lambda+2)(\lambda-2), \nonu \\
f_{U \bar{\Psi} \bar{\Psi}}^{2\,\frac{3}{2}\,\frac{3}{2}} & = &
\frac{1}{\sqrt{5}}(\lambda+2)(\lambda-5), \qquad
f_{U \bar{\Psi} \bar{\Psi}}^{2\,\frac{3}{2}\,\frac{5}{2}}=-\sqrt{\frac{3}{35}}(\lambda+10), 
\nonu \\
f_{T T T}^{2\,2\,1} & = &
6\sqrt{\frac{2}{5}}(\lambda+1)(\lambda-3), \qquad
f_{T T T}^{2\,2\,3}=-12 \sqrt{\frac{2}{5}},
\nonu\\
f_{U U U}^{2\,2\,1} & = &
6\sqrt{\frac{2}{5}}(\lambda+2)(\lambda-2), \qquad
f_{U U U}^{2\,2\,3}=-12 \sqrt{\frac{2}{5}}.
\label{struc}
\eea
The $c_{22,3}$ and $c_{23,4}$ are the structure constants.
Note that 
the mode index $m''$ in the right hand side of
(\ref{finalanticommappendix})
is equal to the sum of
two indices of the left hand side
according to the definition in the footnote
\ref{Cdef}.
For the term of $W^{40}$ in the anticommutator of
higher spin-$\frac{5}{2}$ generator
and the higher spin-$\frac{5}{2}$ generator, the coefficient is
equal to zero because the structure constants $f$ are equal
to each other from (\ref{struc}). Then
there is no contribution for this generator in this anticommutator. This feature also occurs at the last relation of
(\ref{finalanticommappendix})
\footnote{
We can obtain the other (anti)commutators between the
next higher spin generators by using
the structure constants in \cite{FL} like as (\ref{22commutator})
and (\ref{finalanticommappendix}) in order to see the matching
with the corresponding ${\cal N}=2$ higher spin algebra.}.



\end{document}